\documentclass{article} % For LaTeX2e
\usepackage{iclr2025_conference,times}

\usepackage{amsmath,amsfonts,bm}

\def\eqref#1{equation~\ref{#1}}
\def\1{\bm{1}}

\DeclareMathAlphabet{\mathsfit}{\encodingdefault}{\sfdefault}{m}{sl}
\SetMathAlphabet{\mathsfit}{bold}{\encodingdefault}{\sfdefault}{bx}{n}

\newcommand{\E}{\mathbb{E}}

\newcommand{\R}{\mathbb{R}}

\usepackage{hyperref}
\usepackage{url}
\usepackage{graphicx, wrapfig}

\renewcommand{\R}{\mathbb{R}}
\newcommand{\Npix}{N_{\textrm{pix}}}
\newcommand{\N}{\mathbb{N}}
\renewcommand{\E}{\mathbb{E}}
\newcommand{\Ndomains}{N_{\textrm{segm}}}
\newcommand{\eqdef}{:=}
\newcommand{\Nres}{R_{\mathrm{res}}}
\newcommand{\DKL}{\mathrm{D}_{\mathrm{KL}}}

\title{cryoSPHERE: Single-Particle HEterogeneous REconstruction from cryo EM}

\author{Gabriel Ducrocq\\
Division of Statistics and Machine Learning\\
Linköping University, Linköping, Sweden \\
\texttt{gabriel.ducrocq@liu.se}
\And
Lukas Grunewald \\
Department of Chemistry \\
Uppsala University, Uppsala, Sweden \\
\texttt{lukas.grunewald@kemi.uu.se}
\And
Sebastian Westenhoff \\
Department of Chemistry \\
Uppsala University, Uppsala, Sweden \\
\texttt{sebastian.westenhoff@kemi.uu.se}
\And
Fredrik Lindsten\\
Division of Statistics and Machine Learning\\
Linköping University, Linköping, Sweden \\
\texttt{fredrik.lindsten@liu.se}
}

\iclrfinalcopy % Uncomment for camera-ready version, but NOT for submission.
\begin{document}

\maketitle

\begin{abstract}
%The three-dimensional structure of a protein plays a key role in determining its function. Methods like AlphaFold have revolutionized protein structure prediction based only on the amino-acid sequence. However, proteins often appear in multiple different conformations, and it is highly relevant to resolve the full conformational distribution.
%Single-particle cryo-electron microscopy (cryo EM) is a powerful tool for capturing a large number of images of a given protein, frequently in different conformations (referred to as \textit{particles}). The images are, however, very noisy projections of the protein, and traditional methods for cryo EM reconstruction are limited to recovering a single, or a few, conformations.
%In this paper, we introduce cryoSPHERE, a deep learning method that takes as input a nominal protein structure, e.g. from AlphaFold, learns how to divide it into segments, and how to move these as approximately rigid bodies to fit the different conformations present in the cryo EM dataset. This formulation is shown to provide enough constraints to recover meaningful reconstructions of single protein structures. This is illustrated in three examples, two synthetic datasets with varying level of noise and on real dataset, where we show consistent improvements over the current state-of-the-art for heterogeneous reconstruction. We conclude that cryoSPHERE is more resilient to high levels of noise than the state of the art methods.
The three-dimensional structure of proteins plays a crucial role in determining their function. Protein structure prediction methods, like AlphaFold, offer rapid access to a protein’s structure. However, large protein complexes cannot be reliably predicted, and proteins are dynamic, making it important to resolve their full conformational distribution. Single-particle cryo-electron microscopy (cryo-EM) is a powerful tool for determining the structures of large protein complexes. Importantly, the numerous images of a given protein contain underutilized information about conformational heterogeneity. These images are very noisy projections of the protein, and traditional methods for cryo-EM reconstruction are limited to recovering only one or a few consensus conformations.
In this paper, we introduce cryoSPHERE, which is a deep learning method that uses a nominal protein structure (e.g., from AlphaFold) as input, learns how to divide it into segments, and moves these segments as approximately rigid bodies to fit the different conformations present in the cryo-EM dataset. This approach provides enough constraints to enable meaningful reconstructions of single protein structural ensembles. We demonstrate this with two synthetic datasets featuring varying levels of noise, as well as two real dataset. We show that cryoSPHERE is very resilient to the high levels of noise typically encountered in experiments, where we see consistent improvements over the current state-of-the-art for heterogeneous reconstruction.
\end{abstract}

\section{Introduction}\label{sec:intro}
Single-particle cryo-electron microscopy (cryo-EM) is a powerful technique for determining the three-dimensional structure of biological macromolecules, including proteins. In a cryo-EM experiment, millions of copies of the same protein are first frozen in a thin layer of vitreous ice and then imaged using an electron microscope. This yields a micrograph: a noisy image containing 2D projections of individual proteins. The  protein projections are then located on this micrograph and cut out so that  an experiment typically yields $10^4$ to $10^7$  images of size $\Npix\times \Npix$ of individual proteins, referred to as \textit{particles}. Our goal is to reconstruct the possible structures of the proteins given these images.
Frequently, proteins are conformationally heterogeneous and each copy represents a different structure. Conventionally, this information has been discarded, and all of the sampled structures were assumed to be in only one or a few conformations (\emph{homogeneous} reconstruction). Here, we would like to recover all of the structures in a   \emph{heterogeneous} reconstruction. 

Structure reconstruction from cryo-EM presents a number of  challenges. First, each image shows a particle in a different, unknown orientation. Second, because of the way the electrons interact with the protein, the spectrum of the images is flipped and reduced. Mathematically, this corresponds to a convolution of each individual image with the Point Spread Function (PSF). Third, the images typically have a very low signal-to-noise ratio (SNR).
For these reasons, it is very challenging to perform \textit{de novo} cryo-EM reconstruction. Standard methods, produce electron densities averaged over many, if not all conformations \citep{scheres_relion_2012,punjani_cryosparc_2017}, performing discrete heterogeneous reconstruction. More recent methods attempt to extract continuous conformational heterogeneity, e.g.,\ by imposing constraints on the problem through an underlying structure deformed to fit the different conformations present in the dataset, see e.g. \citet{rosenbaum_inferring_2021, zhong_exploring_2021, li_cryostar_2023}. AlphaFold \citep{jumper_highly_2021} and RosettaFold \citep{baek_accurate_2021} can provide such a structure based on the primary sequence of the protein only. In spite of this strong prior, it is still difficult to recover meaningful conformations. The amount of noise and the fact that we observe only 2D projections creates local minima that are difficult to escape \citep{zhong_exploring_2021, rosenbaum_inferring_2021}, leading to unrealistic conformations.

%Because of the high level of noise, it is difficult to recover an atomic structure from the images. Even when imposing a strong prior constraint in the form of an underlying structure that we deform on a residue basis to fit the conformations, the images are so noisy that we can place any residue anywhere in the image, leading to poor conformations.

\begin{figure*}[t]
    \vspace{-1.0cm}
    \centering
    \includegraphics[width=0.8\linewidth]{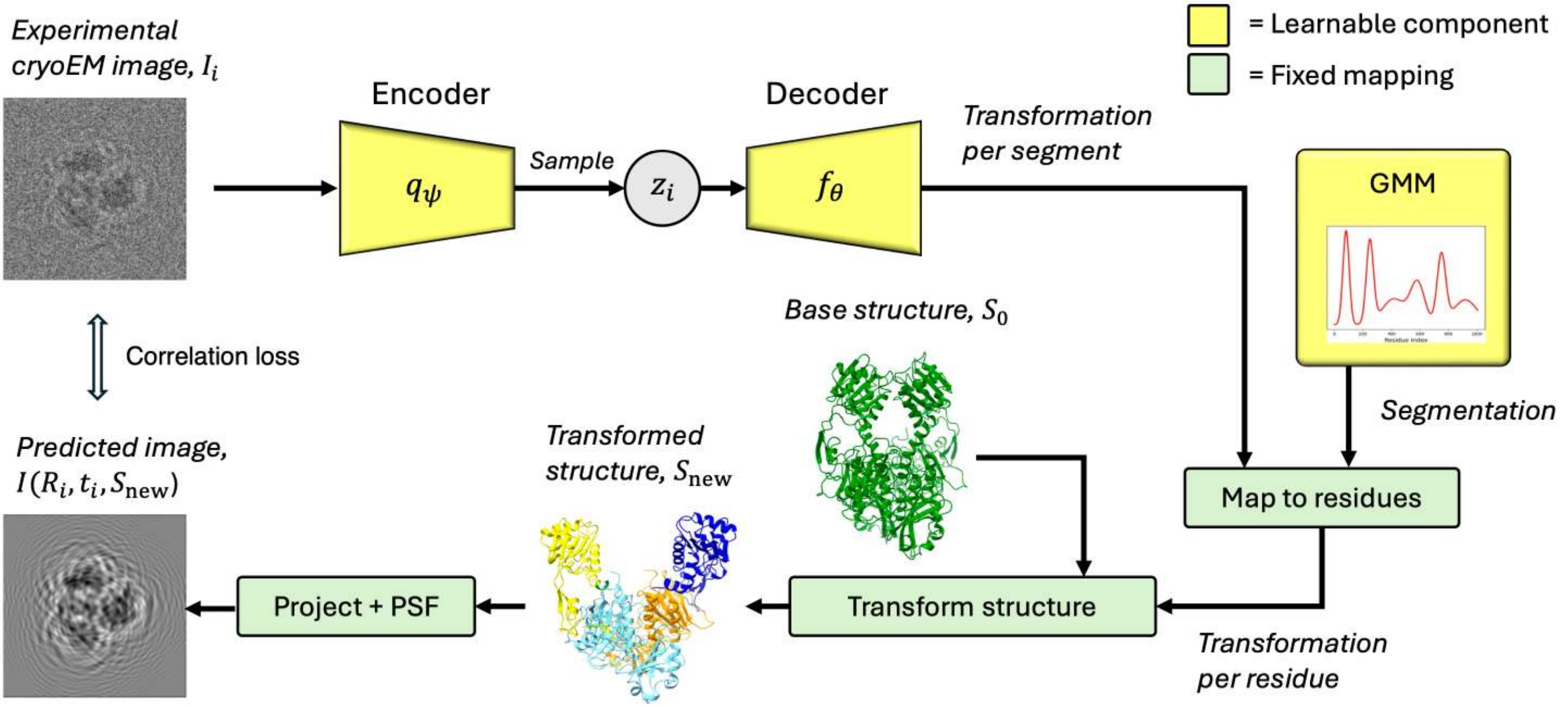}
    \caption{Flow chart of our network. The learnable parts of the model are the encoder, the decoder and the Gaussian mixture. Note that even though the transformations predicted by the decoder are on a per image basis, that is not the case of the Gaussian mixture, which is shared across all particles.}
    \label{fig:network}
    \vspace{-0.5cm}
\end{figure*}

To remedy this, we root our method in the observation that different conformations can often be explained by large scale movements of domains of the protein \citep{mardt_deep_2022}. Specifically, we develop a variational auto-encoder (VAE) \citep{kingma_auto-encoding_2014} that, from a nominal structure and a set of cryo-EM images:
\begin{itemize}
    \color{black}{\item Learns how to divide the amino-acid chain into segments, given a user defined maximum number of segments; see Figure~\ref{fig:150k_mask}.
    The nominal structure can for instance be obtained by AlphaFold \citep{jumper_highly_2021}.}
%    \item Starting from a nominal structure and a set of cryo-EM images, we use a variational auto-encoder (VAE) \citep{kingma_auto-encoding_2014} that learns how to cut an underlying structure into segments, given a user defined maximum number of segments. This structure can be obtained by e.g. AlphaFold \citep{jumper_highly_2021}.
    %\item Our VAE also learns how to approximately rigidly move these domains to fit the different conformations.
    \color{black}{\item For each image, learns approximately rigid transformations of the identified segments of the nominal structure, which effectively allows us to recover different conformations on an image-by-image (single particle) basis.}
\end{itemize}
These two steps happen concurrently, and the model is end-to-end differentiable.
The model is illustrated in Figure~\ref{fig:network}.
The implementation of the model is available on github \footnote{https://github.com/Gabriel-Ducrocq/cryoSPHERE}.

\textcolor{black}{Note that what we call a segment is conceptually different from a domain in the structural biology sense. The domains of a protein play a pivotal role in diverse functions, engaging in interactions with other proteins, DNA/RNA, or ligand, while also serving as catalytic sites that contribute significantly to the overall functionality of the protein, see e.g. \citet{schulz_principles_1979, nelson_lehninger_2017}.  By comparison, the segments we learn do not necessarily have a biological function. However, while not strictly necessary for the function of the method, experiments in Section~\ref{sec:experiments} show that our VAE often recovered the actual domains corresponding to different conformations.}

%The paper is divided into 4 sections. In Section \ref{sec:Background} we introduce the necessary notations, introduce the problem and review existing work. In Section \ref{sec:method} we provide an in-depth description of the method. Section \ref{sec:experiments} subsequently showcases three different experiments. Finally, we conclude our paper in Section \ref{sec:conclusion}.

%\section{Background and method -- cryoSPHERE}\label{sec:method}
\section{Notations and problem formulation}
In what follows, we consider only the $C_{\alpha}$ atoms of the protein. A protein made of a number $\Nres\in \N^\star$ of residues $r_i$ is denoted $S = \{r_{i}\}_{i=1}^{\Nres}$, where the coordinates of residue $i$ are the coordinates of its $C_{\alpha}$ atom.
%
%Each residue is itself a set of atoms: $r_i =\{a_k\}_{k=1,\dots, A_i}$ where $A_i\in \N^\star$ is the number of atoms present in residue $i$. We can equivalently define the structure $S = \{a_{k}\}_{k=1,\dots, A_{\mathrm{total}}}$ in terms of its atoms, where $A_{\mathrm{total}}\in\N^\star$ is the total number of atoms in the structure. We denote by $a_{k}=(x_k, y_k, z_k)^T\in\R^3$ the coordinates of residue $k\in\{1, \dots, A_{\mathrm{total}}\}$.
%This means that in this paper, we limit ourselves to the backbone of the protein: we consider only the $C_{\alpha}$ atoms. 
%
The electron density map of a structure $S$, also called a volume, is a function $V_{S}:\R^{3}\rightarrow \R$, where $V_{S}(x)$ is proportional to the probability density function of an electron of $S$ being present in an infinitesimal region around $x\in\R^{3}$. That is, the expected number of electrons in $B\subseteq\R^3$ is proportional to $\int_B V_{S}(x) \mathrm{d}x$.

Assume we have a set of 2D images $\{I_i\}_{i=1}^{N}$ of size $\Npix \times\Npix$, representing 2D projections of different copies of the same protein in different conformations. Traditionally, the goal of cryo-EM heterogeneous reconstruction has been to recover, for each image $i$, the electron density map $V_{i}$ corresponding to the underlying conformation present in image $i$; see Section~\ref{sec:related_works} for a review of these methods.
%
%
%
%In recent years, some works aimed at recovering, for each image $i$, the underlying structure $S_{i}$ explaining the image. That is, these works try to recover the precise position in $\R^3$ of each residue.
%
However, following recent works, e.g., \citet{rosenbaum_inferring_2021, zhong_exploring_2021}, we aim at recovering, for each image $i$, the underlying structure $S_{i}$ explaining the image. That is, we try to recover the precise position in $\R^3$ of each residue.

\section{Method -- cryoSPHERE}\label{sec:method}
%\subsection{cryoSPHERE}
In this section, we present our method for single-particle heterogeneous reconstruction, denoted cryoSPHERE. The method focuses on structure instead of volume reconstruction. It differs from the previous \citep{rosenbaum_inferring_2021} and concurrent \citep{li_cryostar_2023} works  along this line in the way the movements of the residues are constrained: instead of deforming the base structure on a residue level and then imposing a loss on the reconstructed structure, 
%it moves parts of the protein approximately rigidly by construction.
%
our method learns to decompose the amino-acid chain of the protein into segments and, for each image $I_{i}$, to rigidly move the learnt segments of a base structure $S_0$ to match the conformation present in that image. %, as explained in Section \ref{sec:intro}.
%We can often explain the different conformations of a large protein by large scale movements of its domains \citep{mardt_deep_2022}. Based on this, we develop a method with two goals:
%\begin{enumerate}
%    \item From a given base structure $S_0$, where $S_0$ denotes the set of residue positions as described above, learn the decomposition of the amino-acid chain of the protein into segments.
%    \item For each image $I_{i}$, learn how to rigidly move the learnt segments of $S_0$ to match the different conformations present in the cryo-EM dataset. 
%\end{enumerate}
%
This is motivated by the fact that different conformations of large proteins often can be explained by large scale movements of its domains \citep{mardt_deep_2022}.

The base structure $S_0$ can be obtained using methods like
AlphaFold \citep{jumper_highly_2021} and RosettaFold \citep{baek_accurate_2021},
 %can provide such a structure $S_0$
based on the amino-acid sequence of the protein. In Section \ref{sec:experiments}, we further fit the AlphaFold predicted structure into a volume recovered by a custom backprojection algorithm provided by \citet{zhong_reconstructing_2020}.

We use a type of VAE architecture, see Figure \ref{fig:network}. %for a depiction of how the information flows in our network. 
We map each image to a latent variable by a stochastic encoder, which is then decoded %by the decoder which outputs 
to
a rigid body transformation per segment. Based on these transformations and the segment decomposition, the underlying structure $S_0$ is deformed, posed and turned into a volume that is used to create a projected image. This image is then compared to the input image. After that, the backward pass updates the parameters of the encoder, decoder and Gaussian mixture. We now describe the details of our model.

\subsection{Image formation model}
To compute the 2D projection of the protein structure $S$, we first estimate its 3D electron density map $V$:
\begin{equation}\label{eq:structToMap}
V_S(r) \eqdef \sum_{a\in S} A_a\exp\left({-\dfrac{||r - a||^2}{2\sigma^{2}}}\right)
\end{equation}
where $A_a$ is the average number of electrons per atom in residue $a$, $r\in\R^{3}$ and $\sigma=2$ by default. Hence, the protein's electron density is approximated as the sum of Gaussian kernels centered on its $C_\alpha$ atoms.
From these density maps, we then compute an image projection $I\in\R^{\Npix\times \Npix}$ as:
\begin{equation}\label{eq:imageFormationStructure}
I (R, t, S)(r_x,r_y) = g * \int_{\R} V_{R S + t}(r)dr_z,
%+ \eta
\end{equation}
where %$\eta\sim\mathcal{N}(0, \sigma_\mathrm{noise}^2)$ is a Gaussian noise term, 
$(r_x, r_y)\in\R^{2}$ are the coordinates of a pixel, $r_z\in\R$ is the coordinate along the $z$ axis, $R\in SO(3)$ is a rotation matrix and $t\in\R^{3}$ is a translation vector. The abuse of notation $R S + t$ means that every atom of $S$ is rotated according to $R$ and then translated according to $t$. The image is finally convolved with the point spread function (PSF) $g$, which in Fourier space is the contrast transfer function (CTF), see \citet{vulovic_image_2013}.
Note that the integral can be computed exactly for our choice of approximating the density map as a sum of Gaussian kernels, which significantly reduces the computing time.
%In practice, we avoid approximating the integral in Equation~\eqref{eq:imageFormationStructure}. The density map, being a sum of Gaussian p.d.f, allows for an exact calculation by integrating over the last marginal. This significantly reduces the computing time.

\subsection{Maximum likelihood with variational inference}\label{subsec:vae}
To learn a distribution of the different conformations, we hypothesize that the conformation seen in image $I_i$ depends on a latent variable $z_i\in\R^{L}$, with prior $p(z_i)$.  
Let $f_{\theta}(S_0,z)$ be a function which, for a given base structure $S_0$ and latent variable $z$, outputs a new transformed structure $S$. This function depends on a set of learnable parameters $\theta$. 
Then, the conditional likelihood of an image $I^{\star}\in\R^{\Npix\times\Npix}$ with a pose given by a rotation matrix $R$ and a translation vector $t$ is modeled as $p_{\theta}(I^{\star} | R, t, S_0, z) = \mathcal{N}( I^{\star}|I(R, t, f_{\theta}(S_{0}, z)), \sigma_\mathrm{noise}^{2})$, where 
$\sigma_\mathrm{noise}^{2}$ is the variance of the observation noise.
The marginal likelihood is thus given by
\begin{equation}
\label{eq:marginalLikelihood}
p_{\theta}(I^{\star} | R, t, S_0) = \int p_{\theta}(I^{\star} | R, t, S_0, z) p(z)dz.
\end{equation}
\textcolor{black}{In practice, the pose $(R,t)$ of a given image is unknown. However, following similar works \citep{zhong_exploring_2021, li_cryostar_2023}, we suppose that we can estimate $R$ and $t$ to sufficient accuracy using off-the-shelf methods \citep{scheres_relion_2012, punjani_cryosparc_2017}.
}

Directly maximizing the likelihood (\ref{eq:marginalLikelihood}) is infeasible because one needs to marginalize over the latent variable. For this reason, we adopt the VAE framework, conducting variational inference on $p_{\theta}(z | I^\star)\propto p_{\theta}(I^\star | z) p(z)$, and simultaneously performing maximum likelihood estimation on the parameters $\theta$.
%We aim to recover $p_{\theta}(z | I^\star)\propto p_{\theta}(I^\star | z) p(z)$, which is not straightforward due to the potentially complex likelihood. In addition, we seek to perform maximum likelihood estimation on the parameters $\theta$. 

Let $q_{\psi}(z|I^{\star})$ denote an approximate posterior distribution over the latent variables. We can then maximize the evidence lower-bound (ELBO):
\begin{align}\label{eq:lowerbound}
\mathcal{L}(\theta, \psi) = %;I^{\star}) = 
%\E_{q_{\psi}(z|I^{\star})}[\log p_{\theta}(I^{\star} | z)] \\ - \DKL(q_{\psi}(z|I^{\star}) || p(z))
\E_{q_{\psi}}[\log p_{\theta}(I^{\star} | z)]  - \DKL(q_{\psi}(z|I^{\star}) || p(z))
\end{align}
which lower bounds the log-likelihood $\log p_{\theta}(I^{\star})$. Here $\DKL$ denotes the Kullback-Leibler (KL) divergence. In this framework $f_{\theta}$ is called the decoder and $q_{\psi}(z|I^\star)$ the encoder.

%\subsection{Architectural choices.}
%We now describe the choice of decoder, encoder and prior.
\subsection{Segment decomposition}

To handle the often very low SNR encountered in cryo-EM data, we regularize the transformation of the structure produced by the decoder by restricting it to transforming whole segments of the protein.
We fix a maximum number of segments $\Ndomains\in\{1, \dots, \Nres\}$ and we represent the decomposition of the protein by a stochastic matrix $G\in\R^{\Nres\times \Ndomains}$. The rows of $G$ represent "how much of each residue belongs to each segment", and our objective is to ensure that each residue \emph{primarily} belongs to one segment, that is:
\begin{equation}\label{eq:conditionMask}
\begin{split}
\forall i \in\{1,\dots, \Nres\}, & \;\exists m^\star \in\{1,\dots, \Ndomains\} \\
\mathrm{s.t} & \sum_{m\neq m^\star} G_{im} \ll 1
\end{split}
\end{equation}
We also aim for the segments to respect the sequential structure of the amino acid chain, and the model to be end-to-end differentiable. \textcolor{black}{Without end-to-end differentiability, we could not apply the reparameterization trick and we would have to resort to Monte Carlo estimation of the gradient of the segments, which has a higher variance, see e.g.\ \citet{mohamed_monte_2019}}.

\begin{wrapfigure}{l}{5.5cm}
    \centering
    \vspace{-0.5cm}
    \includegraphics[width=0.15\columnwidth]{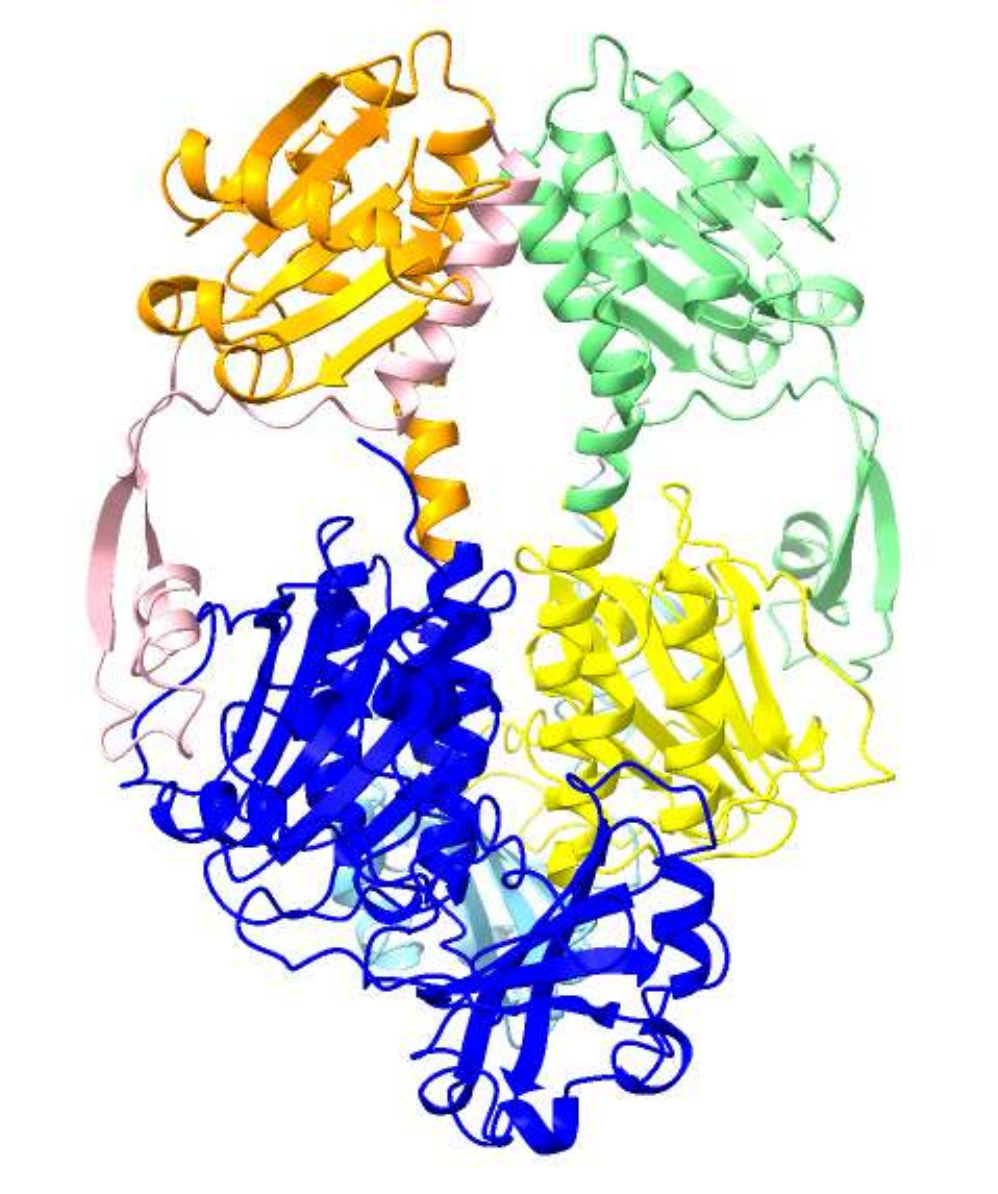}\hspace{0cm}
    \includegraphics[width=0.23\columnwidth]{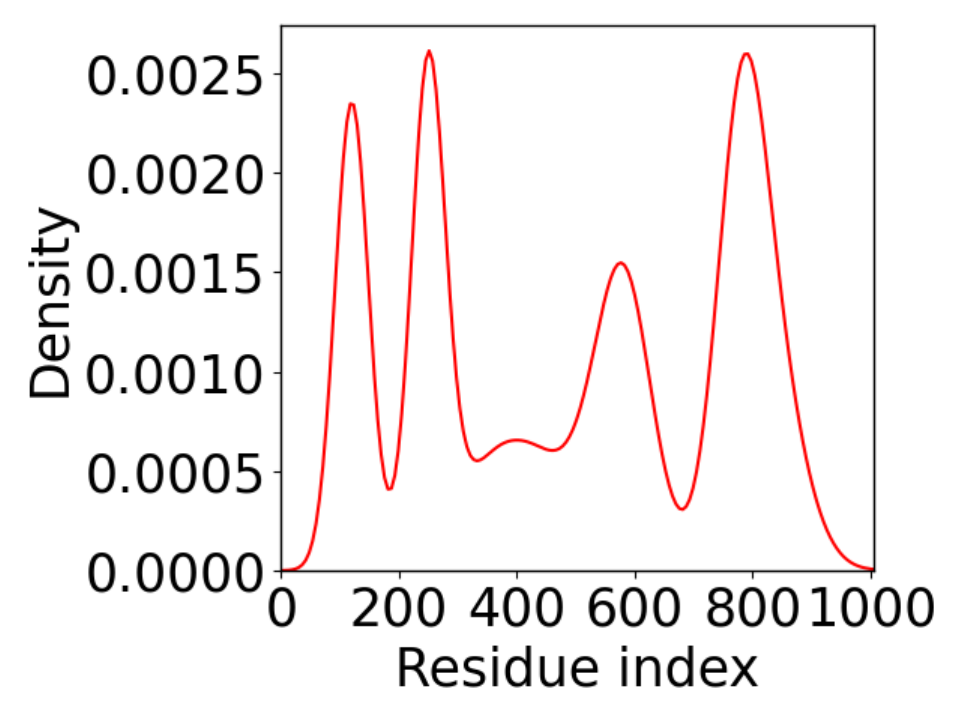}
    \caption{Example of segments recovered with a Gaussian mixture of 6 components.}
    \label{fig:150k_mask}
    \vspace{-1cm}
\end{wrapfigure}

%A decomposition of the structure into segments can be represented by a stochastic matrix $G\in\R^{\Nres\times \Ndomains}$. The rows of $G$ represent "how much of each residue that belongs to each segment". We would also like that each residue mostly belongs to one segment, that is:
%\begin{multline}\label{eq:conditionMask}
%\forall i \in\{1,\dots, \Nres\}, \exists m^\star \in\{1,\dots, \Ndomains\} \\
%\mathrm{s.t} \sum_{m\neq m^\star} G_{im} \ll 1
%\end{multline}
%Furthermore, we want the segments to respect the sequential structure of the amino-acids chain and the model to be end-to-end differentiable.
To meet these criteria, we fit a Gaussian mixture model (GMM) %\citep{hastie_elements_2009} 
with $\Ndomains$ components on the real line supporting the residue indices. %, \textcolor{black}{that is, on} $\R$.
Each component $m$ has a mean $\mu_m$, standard deviation $\sigma_m$ and a logit weight $\alpha_m$. The $\{\alpha_m\}$ are passed into a softmax to obtain the weights $\{\pi_m \}$ of the GMM, ensuring they are positive and summing to one.  
%We define the probability that a residue $i$ belongs to segment $m$ as the probability that the index $i$ belongs to mode $m$, annealed by a temperature $\tau\in\R^{+\star}$:
We further anneal the Gaussian components by a temperature $\tau > 0$, and define the probability that a residue $i$ belongs to segment $m$ as:
\begin{equation}
G_{im} \eqdef \dfrac{\{\phi(i|\mu_m, \sigma^{2}_m)\pi_m)\}^{\tau}}{\sum_{k=1}^{\Ndomains}\{\phi(i|\mu_k, \sigma^{2}_k)\pi_k\}^{\tau}}
\end{equation}
where $\phi(x|\mu, \sigma^{2})$ is the \textcolor{black}{unidimensional} Gaussian probability density function with mean $\mu$ and variance $\sigma^{2}$ and $\tau$ is a fixed hyperparameter. If $\tau$ is sufficiently large, we can expect condition (\ref{eq:conditionMask}) to be verified. See Figure~\ref{fig:150k_mask} for an example of a segment decomposition using a Gaussian mixture.

In this "soft" decomposition of the protein, each residue can belong to more than one segment, allowing for smooth deformations. In addition, the differentiable  architecture is amenable to gradient descent methods, and a well chosen $\tau$ can approximate a "hard" decomposition of the protein. We set $\tau=20$ in the experiment section. \textcolor{black}{In our experience, this segmentation procedure is very robust to different initialization and converges in only a few epochs.}

\newpage
\subsection{Decoder architecture}
The decoder describes the distribution of the images given the latent variables, which include:
\begin{enumerate}
\item One latent variable $z_i\in\R^L$ per image, parameterizing the conformation.
\item The global parameters $\{\mu_m , \sigma_m , \alpha_m\}_{m=1}^{\Ndomains}$ of the GMM describing the segment decomposition.
\end{enumerate}

Given these latent variables and a base structure $S_0$, we parameterize the decoder $f_{\theta}$ in three steps. 
First, a neural network with parameters $\theta$ maps $z_{i}\in\R^L$ to 
a set of rigid body transformations, one for each segment $m=1, \dots, \Ndomains$. The transformation of segment $m$ is represented by a translation vector $\vec{t}_m$ and a unit quaternion $\vec{q}_m$ \citep{vicci_quaternions_2001}, which can further be decomposed into an axis of rotation $\vec{\phi}_m$ and rotation angle $\delta_m$.
%one 3D translation per segment $\{\vec{t}_m\}_{m=1, \dots, \Ndomains}$ and one rotation per segment. The rotation is represented by a unit quaternion $\{\vec{q}_m\}_{m=1, \dots, \Ndomains}$ \citet{vicci_quaternions_2001}, which can be equivalently described as a set of axes of rotation $\{\vec{\phi}_m\}_{m=1, \dots, \Ndomains}$ and corresponding angles of rotations $\{\delta_m \}_{m=1, \dots, \Ndomains}$.
%
Second, given the parameters of the GMM, we compute the matrix $G$. 
Finally, for each residue $i$ of $S_0$, we update the coordinates of all its atoms $\{a_{ik}\}_{k=1}^{A_i}$:
\begin{enumerate}
\item First, $a_{ik}$ is successively rotated around the axis $\vec{\phi}_m$ with an angle $G_{im}\delta_m$ for $m\in\{1,\dots, \Ndomains\}$ to obtain updated coordinates $a_{ik}^{\prime}$.
\item Second, it is translated according to:
\(  
    a_{ik}^{\prime\prime} = a_{ik}^{\prime} + \sum_{j=m}^{N} G_{im} \vec{t}_{m}.
\)
\end{enumerate}
This way, the transformation for a residue incorporate contributions from all segments, proportionally on how much they belong to the segments. 
%in the proportion of "how much of the residue belongs to these segments". 
If condition (\ref{eq:conditionMask}) is met, a roughly rigid motion for each segment can be expected.

\subsection{Encoder and priors}

\begin{wrapfigure}{r}{8cm}
    \centering
    \vspace{-0.5cm}
    \includegraphics[scale=0.25]{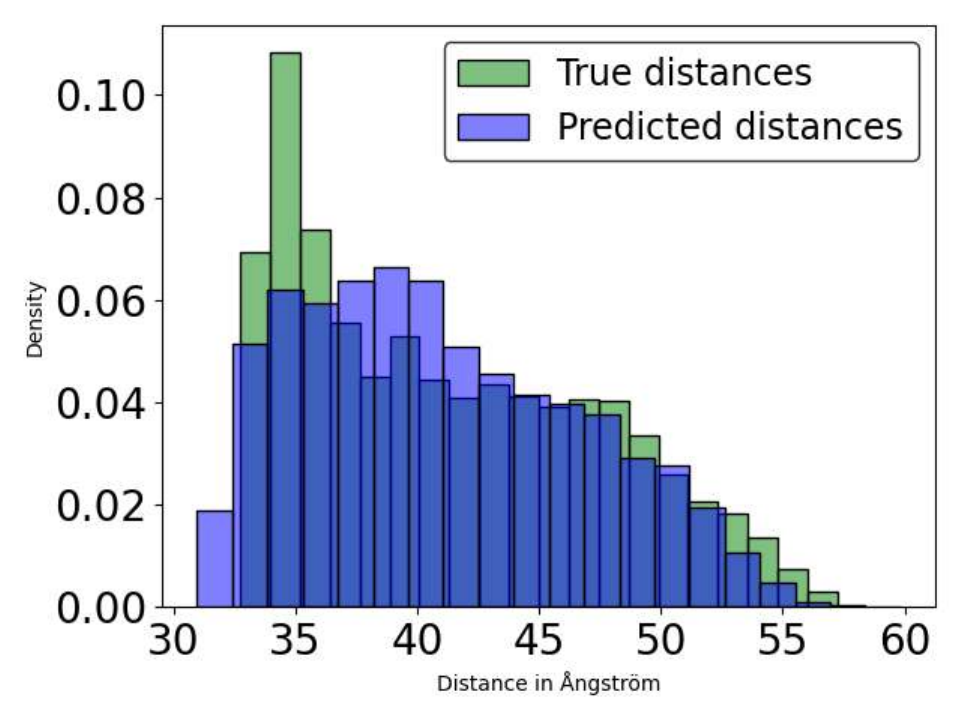}%
    \includegraphics[scale=0.25]{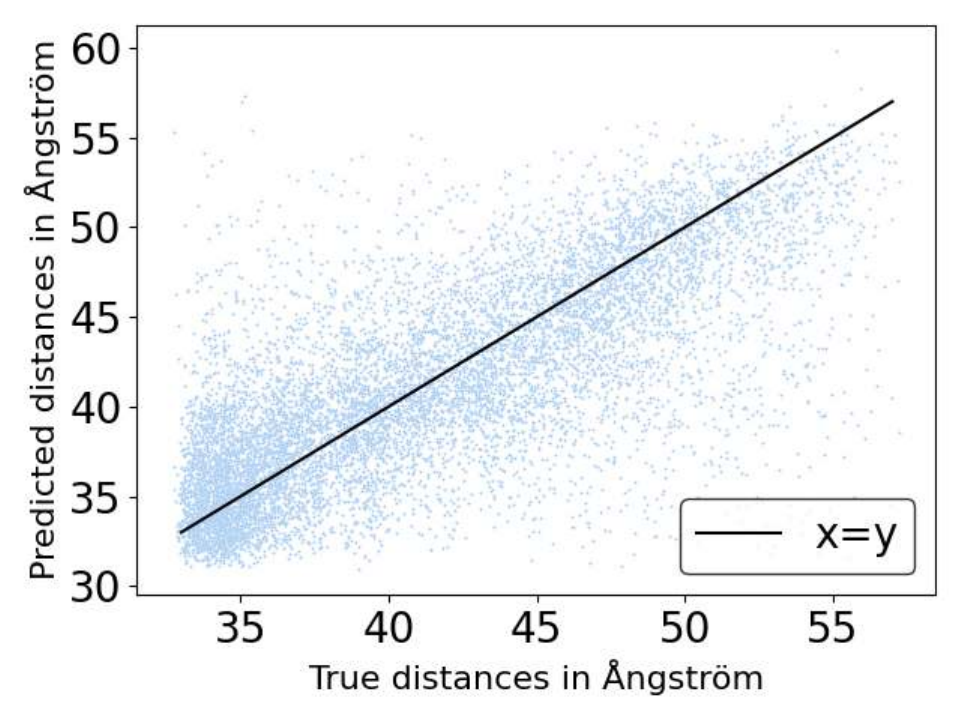}
%    \hspace{0.5cm}
%    \includegraphics[scale=0.15]{figures/experiments/phy_150k/closest.pdf}
%    \hspace{0.5cm}
%    \includegraphics[scale=0.15]{figures/experiments/phy_150k/openest.pdf}
    \caption{MD dataset SNR 0.001. Left: Histograms of the distances of the two upper domains. The true distances are in green. The recovered distances are in blue. Right: Predicted against true distances in Ångström. The black line represent $x=y$.The correlation between the predicted and true distances is 0.73. For the same plot for cryoStar, see Appendix \ref{append:MD} of the supplementary file.}
    \label{fig:150k_histo}
    \vspace{-1.7cm}
\end{wrapfigure}

We follow the classical VAE framework. The distribution $q_{\psi}(y|I^\star)$ is given by a normal distribution $\mathcal{N}(\mu(I^\star), \mathrm{diag}(\sigma^{2}(I^\star)))$ where $\mu\in\R^L$ and $\sigma\in\R_{+}^L$ are generated by a neural network with parameters $\psi$, taking an image $I^\star$ as input.
%We denote by $\psi$ the parameters of this neural network.
Additionally, the approximate posterior distribution on the parameters of the GMM is chosen to be Gaussian and independent of the input image:
\begin{align*}
\mu_m &\sim \mathcal{N}(\nu_{\mu_m}, \beta^{2}_{\mu_m}) \\
\sigma_m &\sim \mathcal{N}(\nu_{\sigma_m}, \beta^{2}_{\sigma_m}) \\
\alpha_m &\sim \mathcal{N}(\nu_{\alpha_m}, \beta^{2}_{\alpha_m})
\end{align*}
where $\{\nu_{\mu_m}, \beta_{\mu_m}, \nu_{\sigma_m}, \beta_{\sigma_m}, \nu_{\alpha_m}, \beta_{\alpha_m}\}_{m=1}^{\Ndomains}$ are parameters that are directly optimized. In practice we use ELU+1 layers for $\sigma_m$ to avoid negative or null standard deviation.

Finally, we assign standard Gaussian priors to both the local latent variable
\(
z_i \sim \mathcal{N}(0, I_{L}),
\)
and the global GMM parameters $\{\mu_m, \sigma_m, \alpha_m, \}_{m=1}^{\Ndomains}$. This reparameterization  \citep{kingma_auto-encoding_2014} is straightforward for a Gaussian distribution. Calculating the KL-divergence between two Gaussian distributions as in \eqref{eq:lowerbound}, is also straightforward.

\subsection{Loss}
Since the images may be preprocessed in unknown ways before running cryoSPHERE, we use a correlation loss between predicted and ground truth image instead of a mean squared error loss, similar to \citep{li_cryostar_2023}:
\begin{equation}
    \mathcal{L}_{\mathrm{corr}} = \dfrac{-I_{i}^\star \cdot I(R_i, t_i, f_{\theta}(S_0, z))}{||I_{i}^\star||\times||I(R_i, t_i, f_{\theta}(S_0, z))||}
\end{equation}

where $\cdot$ denotes the dot product.
The total loss to minimize writes:
\begin{equation}
\mathcal{L}(I, I^{\star}) = \mathcal{L}_{\mathrm{corr}} + \DKL(q_{\psi}(z|I^{\star}) || p(z))
\end{equation}

In our experience, it is unnecessary to add any regularization term to the correlation and KL divergence losses, except for datasets featuring a very high degree of heterogeneity. In that case, we offer the option of adding a continuity loss to avoid breaking the protein and a clashing loss to avoid clashing residues, as it is done in \citep{rosenbaum_inferring_2021, li_cryostar_2023, jumper_highly_2021}. We describe these losses in Appendix \ref{append:loss} of the supplementary file.

\section{Related works}\label{sec:related_works}
Two of the most popular methods for cryo-EM reconstruction, which are \textit{not} based on deep learning, are
RELION \citep{scheres_relion_2012} and cryoSPARC \citep{punjani_cryosparc_2017}. Both methods perform volume reconstruction, hypothesize that $k$ conformations are present in the dataset and perform maximum a posteriori estimation over the $k$ density maps, thus performing discrete heterogeneous reconstruction. Both of these algorithms operate in Fourier space using an expectation-maximization algorithm \citet{dempster_maximum_1977} and are non-amortized: the poses are refined for each image. Other approaches perform continuous heterogeneous reconstruction. For example, 3DVA \citep{punjani_3d_2021-1} uses a probabilistic principal component analysis model to learn a latent space.%The authors take a Bayesian stance and treat the poses and the assignment of the images to one of the $k$ conformations as latent variables. They perform maximum a posteriori estimation over the $k$ density maps of the $k$ conformations as well as the noise variance and prior variance using an EM algorithm.

Another class of methods involve deep learning and typically performs continuous heterogeneous reconstruction using a VAE architecture. Of those that attempt to reconstruct a density map, cryoDRGN \citep{zhong_reconstructing_2020, zhong_cryodrgn2_2021} and CryoAI \citep{levy_cryoai_2022} use a VAE acting on Fourier space to learn a latent space and a mapping that associates a 3D density map with each latent variable. They perform non-amortized and amortized inference over the poses, respectively. Other methods are defined in the image space, e.g. 3DFlex \citep{punjani_3d_2021} and cryoPoseNet \citep{nashed_cryoposenet_2021}. They both perform non-amortized inference over the poses. These methods either learn, for a given image $I_i$, $\{ V_{i}(x_k)\}$ the values at a set of $\Npix^3$ fixed 3D coordinates $\{x_k\}$, representing the volume on a grid (\emph{explicit} parameterization), or they learn 
an actual function $\hat{V}_i:\R^3\rightarrow\R$ in the form of a neural network that can be queried at chosen coordinates (\emph{implicit} parameterization). 
These volume-based methods cannot use external structural restraints or force fields as additional information. This limits their applicability to low SNR data sets, which are frequent in protein cryo EM. 
%The implicit parameterization requires smaller networks but needs an expensive $\Npix^2$ and $\Npix^3$ evaluation for image and volume generation, respectively.

Other deep learning methods attempt to directly reconstruct structures instead of volumes and share a common process: starting from a plausible base structure, obtained with e.g. AlphaFold \citep{jumper_highly_2021}, for each image, they move each residue of the base structure to fit the conformation present in that specific image. These methods differ on how they parameterize the structure and in the prior they impose on the deformed structure or the motion of the residues. For example AtomVAE \citep{rosenbaum_inferring_2021} considers only residues and penalizes the distances between two subsequent residues that deviate too much from an expected value. CryoFold \citep{zhong_exploring_2021} considers the residue centers and their side-chain and also imposes a loss on the distances between subsequent residues and the distances between the residue centers and their side-chain. Unfortunately, due to the high level of noise and the fact that we observe only projections of the structures, these "per-residue transformation" methods tend to be stuck in local minima, \textcolor{black}{ yielding unrealistic conformations unless the base structure is taken from the distribution of conformations present in the images \citep{zhong_exploring_2021}, limiting their applicability on real datasets. Even though AtomVAE \citep{rosenbaum_inferring_2021} could roughly approximate the distribution of states of the protein, it was not able to recover the conformation given a specific image.}

To reduce the bias that the base structure brings, DynaMight \citep{schwab_dynamight_2023} fits pseudo-atoms in a consensus map with a neural network directly. Similar to our work, several other methods constrain the atomic model to rigid body motions.
For example e2gmm \citep{chen_deep_2021, chen_integrating_2023}
deform a nominal structure $S_0$ based on how much its residues are close to a learnt representation $S_{\mathrm{small}}$ of $S_0$. This is similar to our GMM, except that their takes place in $\R^3$ and is not used to perform rigid body motion. Instead, they ask the user to define the segmentation in a later step. This is in contrast to cryoSPHERE, which learns the motion and the segmentation concurrently.
Using  DynaMight \citep{schwab_dynamight_2024}, \citet{chen_improving_2024} developed a focused refinement on patches of the GMM representation of the protein. These patches are learnt using $k$-means on the location of residues and do not depend on the different conformations of the data set. This in contrast to cryoSPHERE where the learning of the segments of the protein is tightly linked to the change of conformation. Concurrently to our work, \citet{li_cryostar_2023} developed cryoStar which learns to translate each residue independently using a variational auto-encoder. They enforce the local rigidity of the motion of the protein by imposing a similarity loss between the base structure and the deformed structure as well as a clash loss. %This similarity loss is made of three terms. First, an elastic network loss that penalizes the differences between the true distances of every pair of residues which are closer than 12Å in the nominal structure and their predicted counterparts. Second, a continuity loss that is similar to the elastic network loss but for subsequent residues in the primary structure. Third, a clashing loss, that penalizes the distances inferior to 4Å between two residues. This is to be contrasted with cryoSPHERE that enforces the local rigidity of the motion by moving learnt segments of the protein instead enforcing it through a loss term.
The interested reader can see \citet{donnat_deep_2022} for an in-depth review of deep learning methods for cryo-EM reconstruction.

\textcolor{black}{The reconstruction methods relying on an atomic model, such as cryoStar, DynaMight or cryoSPHERE offer the possibility to the user to provide prior information via this atomic model. They also offer the possibility of deforming the protein according to chemical force fields. This is not the case of the methods performing volume reconstruction without such an atomic model.}

\begin{wrapfigure}{r}{10cm}
    %\centering
    \vspace{-0.6cm}
    \includegraphics[scale=0.15]{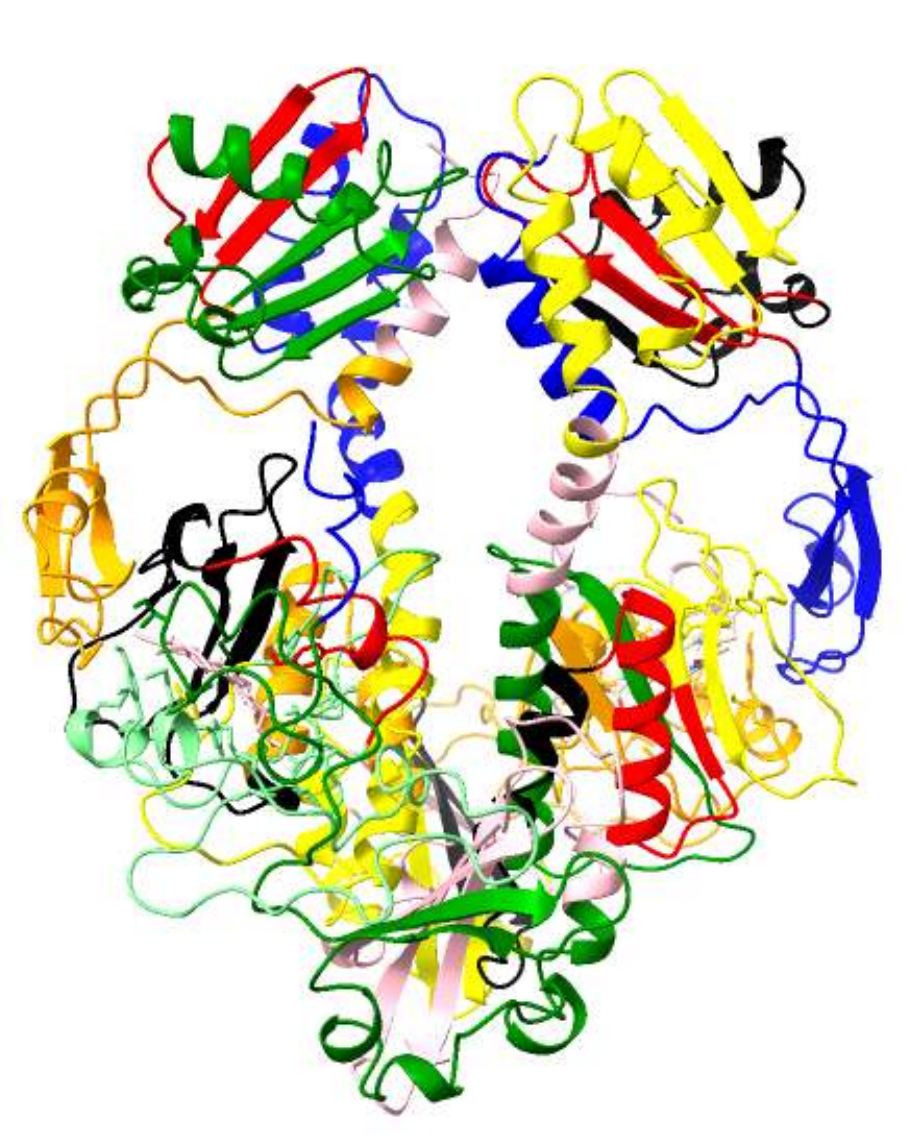}
    \includegraphics[scale=0.23]{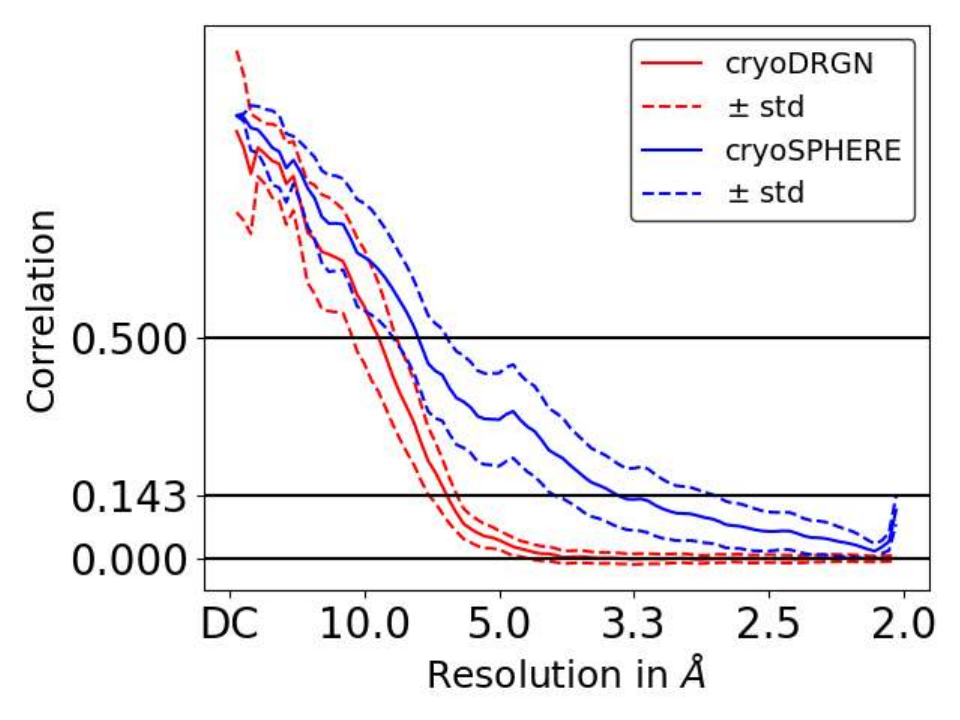}
    \includegraphics[scale=0.23]{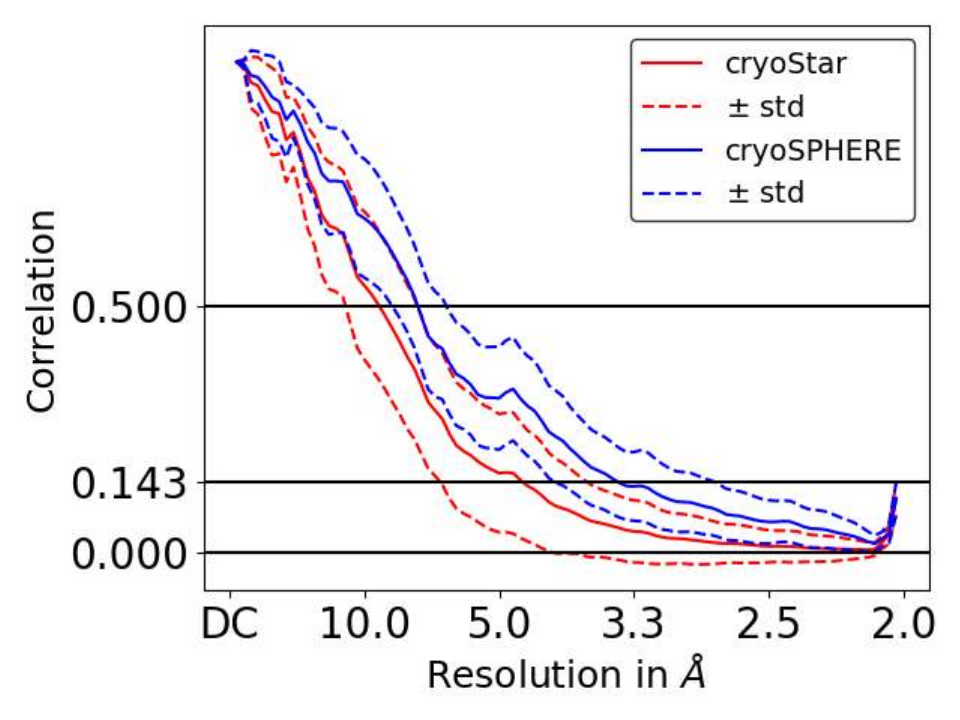}
\caption{MD dataset. Left: cryoSPHERE Recovered segments. The colors denotes different contiguous domains. Middle and right: mean FSC comparison +/- one standard deviation, for cryoSphere and cryoDRGN and cryoStar. For a comparison between cryoStar and cryoDRGN, see Appendix \ref{append:MD} in the supplementary file.}
   \label{fig:phy_20_fitted_mask}
   \vspace{-0.2cm}
\end{wrapfigure}

%\begin{figure}[t]
%centering
%    \includegraphics[scale=0.4]{figures/experiments/toy_data_set/big_phy.pdf}
%    \hspace{0.5cm}
%    \includegraphics[scale=0.23]{figures/experiments/toy_data_set/one_mode.pdf}
%    \hspace{1cm}
%    \includegraphics[scale=0.23]{figures/experiments/toy_data_set/second_mode.pdf}
%\caption{Toy dataset. Base structure with the two domains. Note that the predicted segments exactly match the two domains.}
%\label{fig:toy_base}
%\{figure}

\section{Experiments}\label{sec:experiments}
\textcolor{black}{In this section, we test cryoSPHERE on a set of synthetic\footnote{See Appendix~\ref{append:experiments} of the supplementary file for details on how we created the synthetic datasets.} and real datasets with varying level of noise and compare the results to cryoDRGN \citep{zhong_reconstructing_2020} and cryoStar \citep{li_cryostar_2023}. CryoDRGN is a state-of-the-art method for continuous heterogeneous reconstruction, in which the refinement occurs at the level of electron densities, while cryoStar is a structural method similar to ours.} 
To our knowledge, the code for AtomVAE and CryoFold is not available and non-trivial to reimplement. 
For this reason we focus our comparison on the aforementioned methods, which have furthermore reported state-of-the-art performance. 
In Appendix \ref{append:toy}, we demonstrate that cryoSPHERE is able to recover the exact ground truth when it exists. We also discuss its performances with varying SNR and $\Ndomains$ and show how to debias cryoSPHERE results using DRGN-AI or cryoStar volume method in Appendix \ref{append:MD}. Finally, Appendix \ref{append:cost} compares the computational costs of cryoSPHERE and cryoStar.

\subsection{Molecular dynamics dataset: bacterial phytochrome.}\label{subsec:MD}
As a more difficult test case we simulate a continuous motion of a bacterial phytochrome, with PDB entry 4Q0J \citep{burgie_crystallographic_2014}. The trajectory starts at the closed conformation of Figure~\ref{fig:md_trajectory} and ends at the most open conformation on the same figure. It corresponds to a dissociation of the two top parts of the protein. This dataset has a very low SNR of $0.001$. Our base structure is obtained by AlphaFold and is subsequently fitted into a homogeneous reconstruction given by the backprojection algorithm. We train cryoSPHERE with $\Ndomains = 25$, cryoStar, and cryoDRGN for 24 hours each, using the same single GPU. We get one predicted structure per image for cryoSPHERE and cryoStar, that we turn into volumes using (\ref{eq:structToMap}), and one predicted volume per image for cryoDRGN. See Appendix \ref{append:MD} in the supplementary file for details and comparison with different values of $\Ndomains$. Note that since both cryoSPHERE and cryoStar use a nominal structure, we fit the structure we obtained through AlphaFold in the consensus reconstruction obtained by backprojection and use that exact same structure as the nominal one for both methods.
%\begin{figure}[t]
%    \centering
%    \includegraphics[scale=0.3]{figures/experiments/phy_150k/fitted/mask_fitted_af.pdf}
%    \caption{150k dataset: learnt segments on the AlphaFold structure.}
%    \label{fig:150k_mask}
%\end{figure}

%Once again, we generate a base structure of the phytochrome analogously to Section \ref{append:spike}. Further details can be found in Appendix \ref{append:MD}. Finally, we run our method with the fitted AlphaFold structure as our base structure and $\Ndomains = 20$.

Figure \ref{fig:150k_histo} shows the predicted distance between the two upper parts of the protein being dissociated, against the ground truth distance for each image. In spite of the very low SNR, cryoSPHERE roughly recovers the right distribution of distances. More importantly, the correlation between the predicted distance and ground truth distance is $0.74$, showing that cryoSPHERE is able to recover the correct conformation given an image. This is in stark contrast with \citet{rosenbaum_inferring_2021} who could not recover the conformation conditionally on an image. 
In addition, our model has learnt to separate the two mobile top domains from the fix bottom one, as shown by the segment decomposition in Figure~\ref{fig:phy_20_fitted_mask}. Appendix~\ref{append:MD} in the supplementary file shows the same figures for cryoStar.

We plot the mean of the FSC curves between the predicted volumes and the corresponding ground truth volumes in Figure \ref{fig:phy_20_fitted_mask}, for cryoSPHERE, cryoDRGN and cryoStar. CryoSPHERE performs better than both cryoDRGN and cryoStar at both the $0.5$ and $0.143$ cutoffs. %There are few reasons behind this.
We attribute this to three key properties.
Firstly, we fit our base structure into a consensus reconstruction. This step corrects the position of the medium-scale elements of the base structure that could have been misplaced, boosting the FSC of cryoSPHERE at the $0.5$ cutoff.
%Firstly, we fit our base structure into a volume reconstructed from a homogeneous reconstruction method. This aligns the non-moving parts of the protein with the right places, as provided by the dataset. Without this step, it is likely that some medium-scale part of the protein would have been misplaced, potentially reducing the performance of our algorithm at the $0.5$ cutoff. This step comes at almost no extra cost since it relies on the backprojection algorithm.
%
Secondly, acting directly on the structure level offers a finer resolution than cryoDRGN given the level of noise. Figure~\ref{fig:volumes_cryodrgn_snr_0_001} shows that cryoDRGN underestimated the opening of the protein and sometimes gives very noisy volumes. That explain why we outperform cryoDRGN at the $0.143$ cutoff. Finally, cryoSPHERE is rigidly moving larger segments of the protein. This provide a better resistance to high levels of noise and overfitting compared to moving each residue individually like cryoStar does, providing a possible explanation to the improvement compared to cryoStar at the $0.143$ cutoff.

%Secondly, deforming a good AlphaFold structure offers finer details than cryoDRGN. This accounts for the better performance of cryoSPHERE over cryoDRGN at the $0.143$ cutoff. Figure \ref{fig:volumes_cryodrgn_snr_0_001} shows that cryoDRGN can underestimate the level of dissociation and struggles to provide high resolution volumes. In some instances, this method renders a lot of noise. We do not have these problems with cryoSPHERE.

%Finally, cryoSPHERE is moving entire segments of the protein as rigid bodies, which offers a better resistance to the high level of noise of this dataset. On the contrary, cryoStar moves every residue individually and is therefore prone to overfitting. This is illustrated by the datasets with SNR $0.01$ and $0.001$ in Appendix \ref{append:MD}: we tend to perform similarly to cryoStar on high SNR, but better on lower SNR.

\vspace{-0.2cm}
\subsection{EMPIAR 10180}\label{empiar10180}
We now demonstrate that cryoSPHERE is applicable to real data as well as large proteins. We run cryoSPHERE on EMPIAR-10180 \citet{plaschka_structure_2017}, comprising $327\thinspace490$ images of a pre-catalytic spliceosome with $13\thinspace941$ residues, making it a computationally heavy dataset to tackle. We use the atomic model by \citet{plaschka_structure_2017} (PDB: 5NRL).

\begin{figure}
\vspace{-0.8cm}
    \centering
    \includegraphics[scale=0.18]{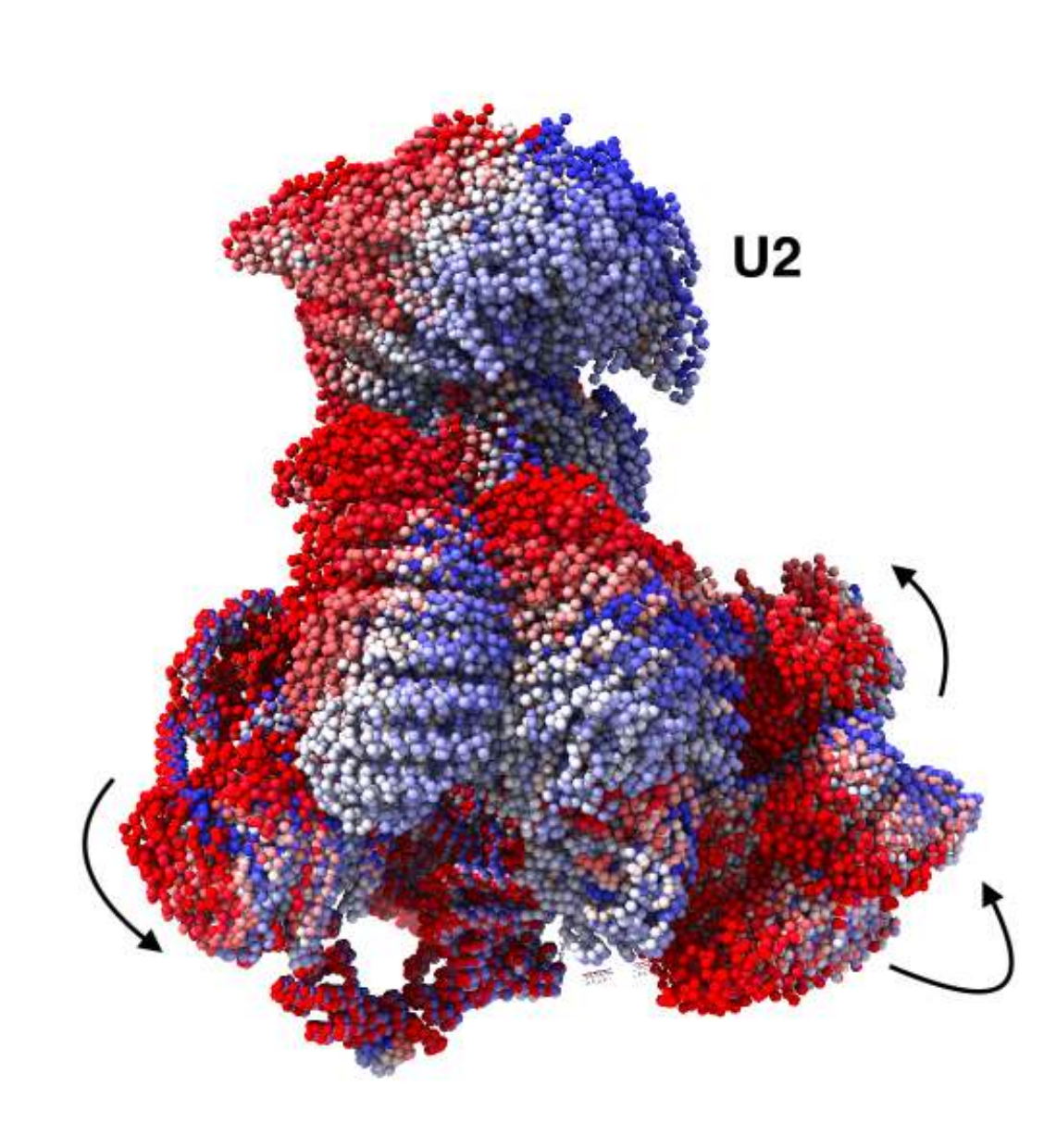}
    \includegraphics[scale=0.18]{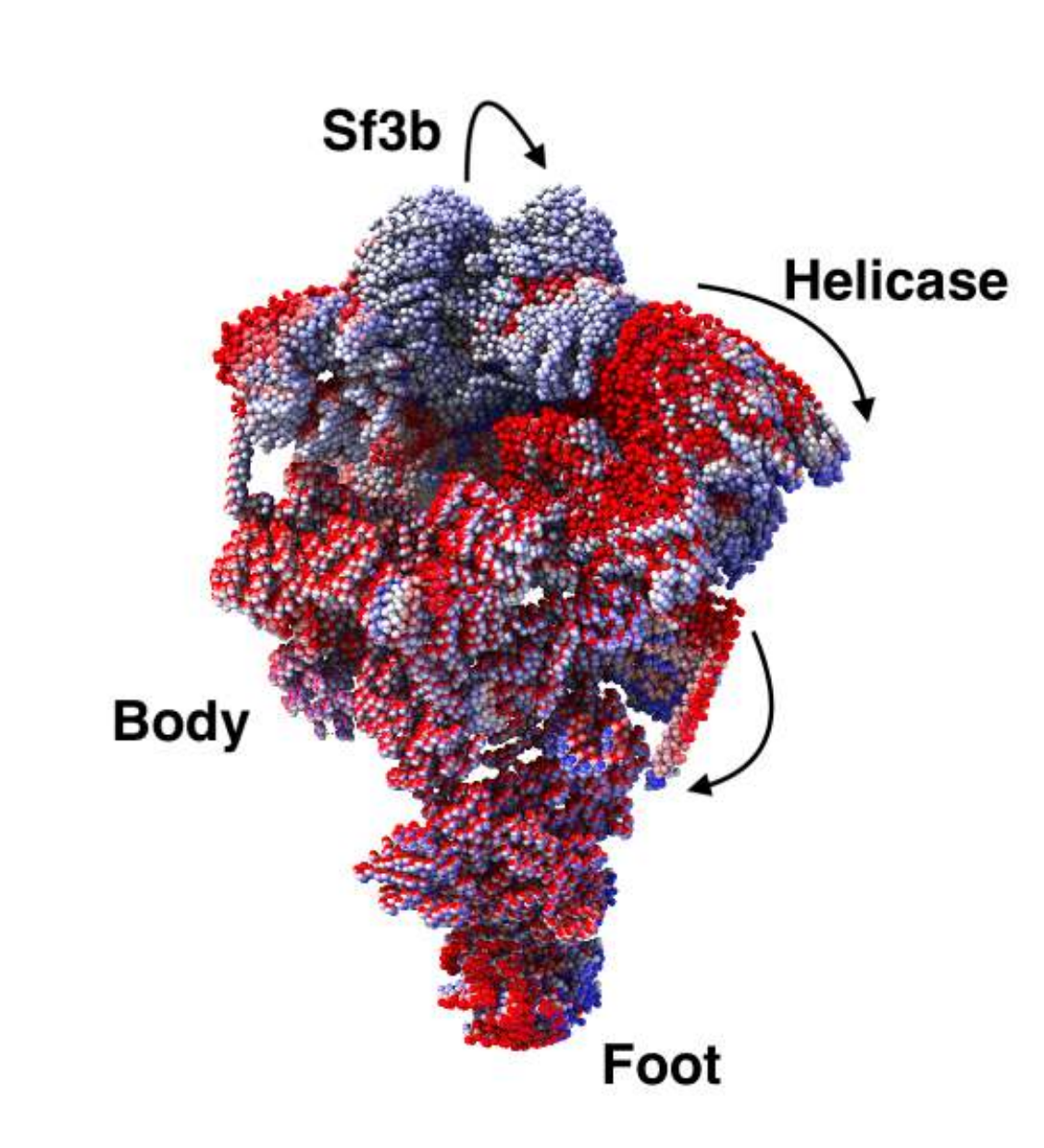}
    \includegraphics[scale=0.21]{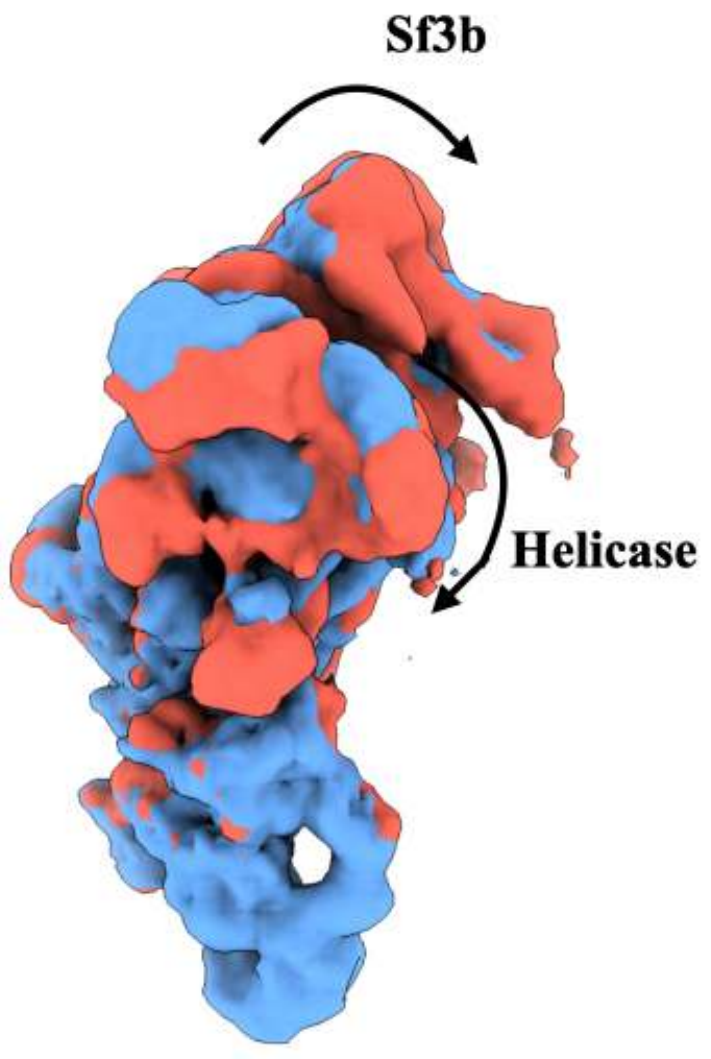}
    \includegraphics[scale=0.21]{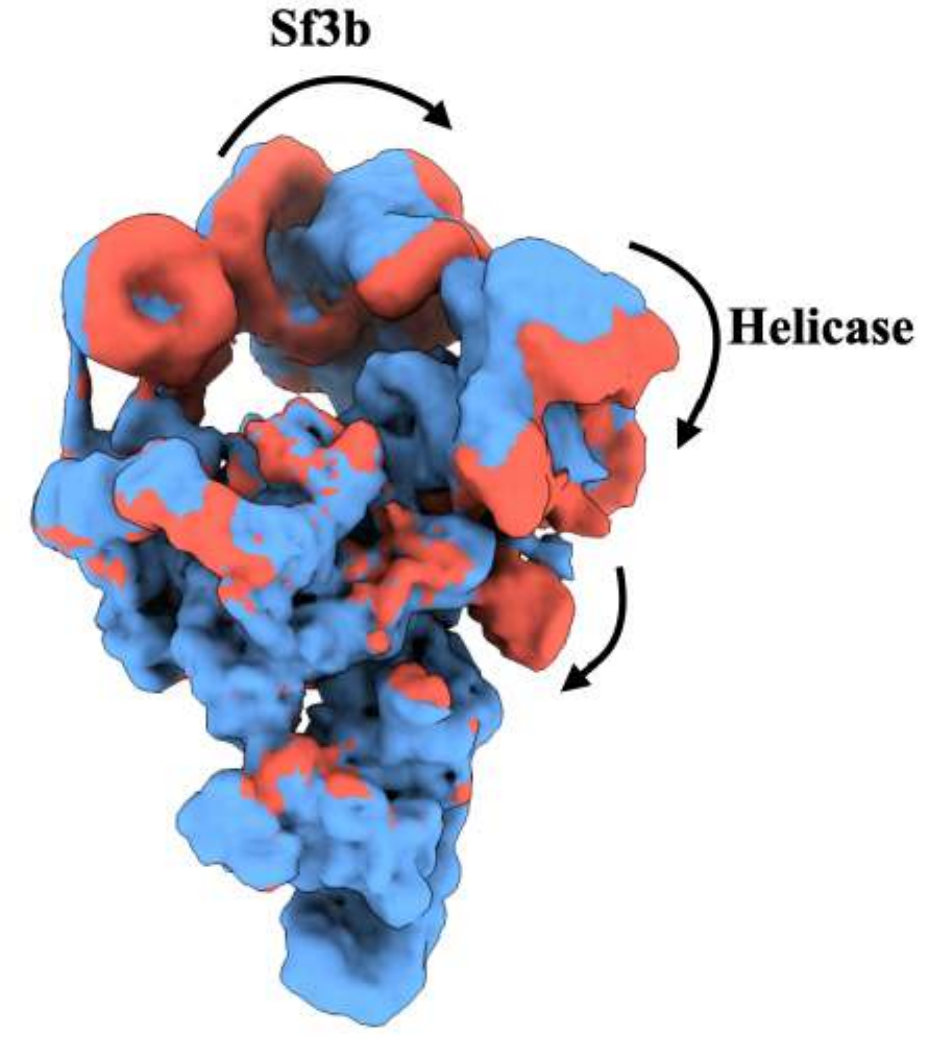}
    \vspace{-0.4cm}
    \caption{EMPIAR10180. Left and middle left: different views of the structures corresponding to the red dots of Figure \ref{fig:empiar10180_latent_space}. The motion goes from red (left in the first principal component) to white to blue (right of the principal component). Only the $C_\alpha$ atoms are shown. Right and middle right: different views of two volumes recovered by training DRGN-AI on the latent space of cryoSPHERE. The U2 domain disappears on the volume because of a compositional heterogeneity.}
    \label{fig:empiar10180}
    \vspace{-0.6cm}
\end{figure}

Figure \ref{fig:empiar10180} shows a set of ten structures taken evenly along the first principal component of the latent space. To interrogate if these structures contain bias from the structural constraints, we perform a volume reconstruction step similar to cryoStar Phase II, see Figure \ref{fig:empiar10180}.

Traversing the first principal component shows that the Sf3b domain gets incurvated down while the helicase move closer to the foot of the protein. This is in line with the literature \citep{li_cryostar_2023, plaschka_structure_2017}. The motion of the protein also brings the alpha helix of the Spp381 domain closer to the foot, as corroborated by \citet{li_cryostar_2023}. Comparison between the recovered structures and volumes (Figure \ref{fig:empiar10180_drgnai}) shows similar movements, indicating a small amount of bias from the structural constraints.  In addition, the absence of density corresponding to the U2 domain in the volume indicates that it there is compositional heterogeneity that cryoSPHERE could not detect, see Figure \ref{fig:empiar10180}. We provide a movie of the motion and more structures and volumes in appendix \ref{append:empiar10180} in the supplementary file.

\subsection{EMPIAR-12093}\label{sec:empiar12093}
We now tackle the recently published EMPIAR-12093 \citep{bodizs_cryo-em_2024}. This dataset comprises two sets of images: one non-activated (Pfr) and one activated (Pr). These dataset are very challenging because of the high level of noise and heterogeneity of the protein, especially in the Pfr dataset. Traditional methods like cryoSparc \citep{punjani_cryosparc_2017} or cryoDRGN \citep{zhong_reconstructing_2020, zhong_cryodrgn2_2021} fail at reconstructing the upper part of the protein, see \citet{bodizs_cryo-em_2024} and Appendix \ref{append:EMPIAR-12093}.

\begin{figure}
\vspace{-1.0cm}
    \centering
    \includegraphics[width=0.9\linewidth]{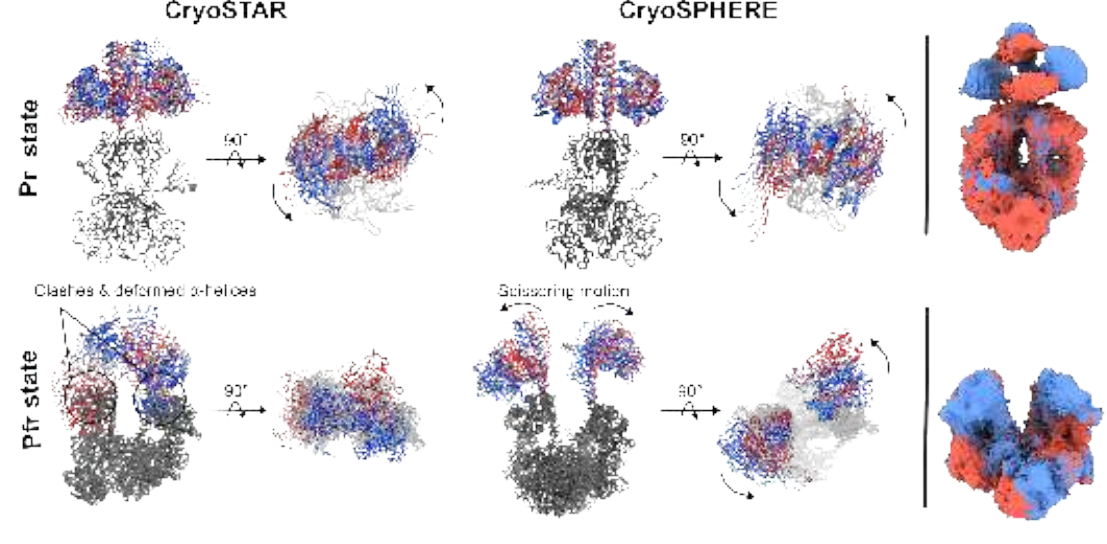}
    \vspace{-0.5cm}
    \caption{EMPIAR12093. Left of the black line: Ten structures sampled along PC1, for cryoSPHERE and cryoStar. Right of the black line: examples of volumes reconstructed by training the cryoStar volume method on the latent space of cryoSPHERE for debiasing. The blue and red volumes correspond to the first and last volumes along PC1. Top row corresponds to Pr; bottom row to Pfr.}
    \label{fig:empiar12093}
    \vspace{-0.5cm}
\end{figure}

Figure \ref{fig:empiar12093} shows principal component 1 traversal for cryoStar and cryoSPHERE. For Pr, both methods are in strong agreement and reveal a rotation of the upper domain around its axis, while the lower part remains stationary. This aligns with previous studies \citep{wahlgren_structural_nodate, malla_photoreception}).

\begin{wrapfigure}{l}{5.5cm}
\vspace{-0.6cm}
    \includegraphics[width= \linewidth]{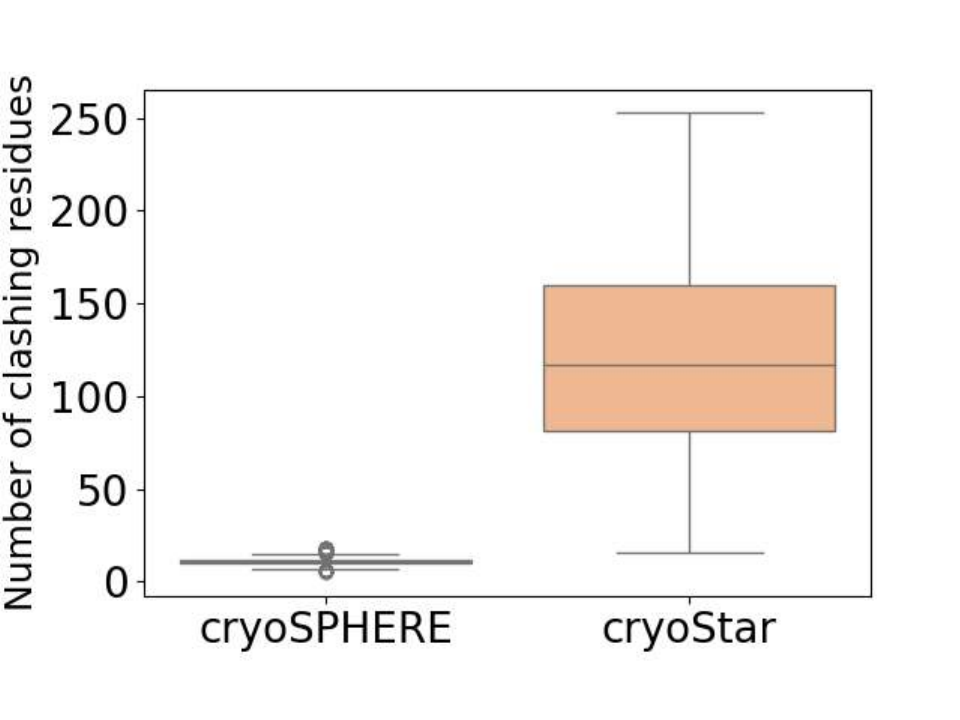}
    \vspace{-0.5cm} %% haha, just had the same idea... will cut some text. I will apply this to all figures
    \caption{EMPIAR12093. Distribution of the number of clashes for 2000 randomly chosen structures Pfr dataset, for cryoSPHERE and cryoStar. Two non contiguous residues are said to be clashing if their distance is less than 4 Å.}
    \label{fig:empiar12093_clashes}
    \vspace{-0.8cm}
\end{wrapfigure}

The Pfr dataset showcases an even lower SNR and more dynamical protein: the protein opens up completely. From consensus reconstructions alone, one could suspect that the upper domains are cut off in the sample preparation procedure. However, the protein is complete in Pr (light-activated) structure and the photocycle is reversible,\citep{takala_signal_2014} suggesting that this is not the case and that strong conformational heterogeneity that is at play. 

For Pfr cryoStar is unable to produce physically plausible results: the top part of the protein appears disordered and shows a random motion. In addition, cryoStar does not recover the "scissoring" motion of protein, which is thought to be active \citep{bodizs_cryo-em_2024}. 
On the contrary, cryoSPHERE gives a high level of motion in a structured manner and recovers the "scissors" opening of the protein. Without any clashes (Fig. \ref{fig:empiar12093_clashes}). \citep{bodizs_cryo-em_2024}.

Analysis of the dataset on phytochromes illustrates the scope and limitations of the different methods. Pure image-based methods (i.e. cryo DRGN) already fail on the Pr state with its intermediate disorder, while cryoSTAR and cryoSPHERE succeed in obtaining reasonable reconstructions (Figure \ref{fig:empiar12093}). For the Pfr state it becomes evident that cryoSTAR struggles with the high noise and large motions encoded in the dataset. Its deformation-based approach results in unphysical motions along the first principal component, often leading to structural clashes. In contrast, cryoSPHERE handles the noise effectively, producing physically plausible large-scale motions in both the upper and lower domains, see Figure \ref{fig:empiar12093_clashes}, the  supplementary movies and  \ref{append:EMPIAR-12093}. We assign this superior performance to the higher degree of structural constraints that are used in cryoSPHERE compared to cryoSTAR. 

We also performed debiasing of cryoSPHERE with a volume method and show examples of reconstructed volumes in Figure \ref{fig:empiar12093}.
For Pr, recovered densities are visible for the entire protein and confirm the dynamics of the upper domains, confirming the absence of compositional heterogeneity and a minimum of bias due to structural constraints. However, for Pfr, meaningful density of the upper (dynamic) part of the protein cannot be recovered, because the signal level in the averaged density is too low. Thus, for this most dynamic protein case, volume-based debiasing is not possible, despite the fact that the structure based cryoSPHERE finds solutions that fit the data set.  

\section{Discussion}
CryoSPHERE presents several advantages compared to other methods for volume and structure reconstruction. 
\vspace{-0.3cm}
\paragraph{Efficiency in Deformation:} Deforming a base structure into a density map avoids the computationally expensive $\Npix^2$ evaluation required by a decoder neural network in methods implicitly parameterising the grid, such as \citet{zhong_cryodrgn2_2021, levy_cryoai_2022}. Furthermore, direct deformation of a structure directly avoids the need for subsequent fitting into the recovered density map.
\vspace{-0.5cm}
\paragraph{Reduced Dimensionality and Noise Resilience:} Learning one rigid transformation per segment, where the number of segments is much smaller than the number of residues, reduces the dimensionality of the problem. This results in a smaller neural network size compared to approaches acting on each residues, such as \citet{rosenbaum_inferring_2021}. Rigidly moving large portions of the protein corresponds to low-frequency movements, less prone to noise pollution than the high-frequency movements associated with moving each residue independently. 
%likely to be polluted by noise. This is in contrast to moving each residue independently, which corresponds to high frequency movements that are more difficult to recover given the low SNR.
In addition, since our goal is to learn one rotation and one translation per segment, a latent variable of dimension $6\times\Ndomains$ is, in principle, a sufficiently flexible choice to model any transformation of the base structure. Choosing the latent dimension is more difficult for volume reconstruction methods such as \citep{zhong_cryodrgn2_2021}.
\vspace{-0.3cm}
\paragraph{Interpretability:} CryoSPHERE outputs segments along with one rotation and one translation per segment, providing valuable and interpretable information. Practitioners can easily interpret how different parts are moving based on the transformations the network outputs. This interpretability is often challenging for deep learning models such as \citet{zhong_cryodrgn2_2021, rosenbaum_inferring_2021}.  

Section \ref{sec:experiments} and Appendix \ref{append:MD} demonstrate cryoSPHERE's capability to recover conformational heterogeneity while performing structure reconstruction. The division into $\Ndomains$ is learned from the data and only marginally impacts the FSC to the ground truth. Moreover,  cryoSPHERE recovers the correct motion for the entire range of $\Ndomains$ values and is able to keep the minimum necessary number of domains when the user sets it too high (Appendix \ref{append:toy}).

\paragraph{Structural restraints allow interpretation of low SNR datasets:} It is evident that structural restraints  as implemented in cryoSPHERE (this work) and cryoSTAR provide additional restraints that pure volume methods (i.e. cryoDRGN) lack, thus giving better reconstructions for high noise data sets.  The additional restraints may introduce bias, which needs to be alleviated  using a backprojection algorithm. This, combined with cryoSPHERE's latent space, achieves better $0.5$ cutoffs than cryoDRGN, indicating its effectiveness in resolving conformational heterogeneity and debiasing the results. If such a volume is unavailable, simply increasing $\Ndomains$ can reduce the bias. As a note of caution we find that for most dynamic protein studies here (the Pfr state of the phytochrome), we find that volume-based debiasing fails because of the very low electron density levels in the reconstructions. Here, other metrics should be developed in the future. 

\paragraph{Summary:} Our study opens up for significant advancements in predicting protein ensembles and dynamics, critically important for unraveling the complexity of biological systems. By predicting all-atom structures from cryo-EM datasets through more realistic deformations, our work lays the foundation for extracting direct insights into thermodynamic and kinetic properties. This work is an important milestone in showing that one can learn a segmentation of the protein that is intimately linked to the change of conformation of the underlying protein, in an end-to-end fashion. In the future, we anticipate the ability to predict rare and high-energy intermediate states, along with their kinetics, a feat beyond the reach of conventional methods such as molecular dynamics simulations. 

\textcolor{black}{It would be interesting to assess how much our segmentation correlates with bottom-up segmentation into domains conducted on the “omics” scale, see e.g.\ \citet{lau_merizo_2023}. To achieve this quantitatively, we would need many examples of moving segments from cryo-EM investigations to match the millions of segments from the “omics” studies. Therefore, we leave this investigation to later work.}

\newpage
%\section{Conclusion}\label{sec:conclusion}
%We presented cryoSPHERE, a model that goes beyond volume reconstruction and targets structure reconstruction from cryo-EM data. It learns how to segment the amino-acid sequence of the protein and how to move the corresponding parts approximately rigidly to fit into the different images.

%We leave to future work to integrate a more realistic image formation model, which will enable our method to run on real data.

 \section*{Acknowledgement}
This work was financially supported by the Wallenberg AI, Autonomous Systems and Software Program (WASP) 
and the Data-Driven Life Science Program (DDLS)
funded by the Knut and Alice Wallenberg Foundation
through the WASP-DDLS collaboration,
the Swedish Research Council (project no: 2020-04122, 2024-05011),
and
the Excellence Center at Linköping--Lund in Information Technology (ELLIIT).
Our computations were enabled by the Berzelius resource at the National Supercomputer Centre, provided by the Knut and Alice Wallenberg Foundation.
 
%G.D is financially supported by the Wallenberg AI, Autonomous Systems and Software Program (WASP) and the Data-Driven Life Science (DDLS) program through the WASP-DDLS collaboration.
%This project would not have been possible without the computing resources provided by the Knut and Alice Wallenberg Foundation at the National Supercomputer Centre (Berzelius resource). 

We thank Claudio Mirabello at the National Bioinformatics Infrastructure Sweden at SciLifeLab for providing us access to his AlphaFold installation on Berzelius
and
Nancy Pomarici for providing input files and explanation of the metadynamics simulation. 

\section{Reproducibility statement}
As part of the current paper, we provide a github link to the source code in Section \ref{sec:intro}. We also describe in detail how we generate the synthetic datasets in Appendix \ref{append:MD} and the hyperparameters chosen to run cryoStar, cryoSPHERE and cryoDRGN in Appendix \ref{append:experiments} for each of the experiments.

\bibliography{iclr2025_conference}
\bibliographystyle{iclr2025_conference}

\newpage
\appendix
\section{Method}\label{append:method}
In this appendix we provide more details on cryoSPHERE.
\subsection{Loss}\label{append:loss}
In our experience it is not necessary to add any regularization term to the loss Equation \ref{eq:lowerbound}, except for the datasets featuring a very high level of conformational heterogeneity. In that case, we offer the option to add a continuity loss, which prevents the network from breaking the protein, as well as a clashing loss, which prevents clashing between different residues. The idea of a continuity loss has been introduced in \citep{jumper_highly_2021} and exploited in cryoEM in \citep{li_cryostar_2023, rosenbaum_inferring_2021}. The clashing loss has also been introduced in \citep{jumper_highly_2021} and has been exploited in \citep{li_cryostar_2023}.

For two subsequent residues belonging to a same chain, we define the continuity loss as:
\begin{equation}
    \mathcal{L}_{\mathbf{cont}} = \dfrac{1}{N_\mathbf{cont}} \sum_{i}^{N_{\mathbf{cont}}} ||d_{i} - \hat{d}_{i} ||^2
\end{equation}
where $N_\mathbf{cont}$ is the number of pairs of residues that are subsequent in the entire protein, $d_{i}$ is the distance between the two residues in pair $i$ in the base structure $\mathcal{S}_{0}$ and $\hat{d}_{i}$ is the predicted distance for the corresponding pair. The form of our continuity loss is similar to \citep{li_cryostar_2023}.

We define the clashing loss as:
\begin{equation}
    \mathcal{L}_{\mathbf{clash}} = \dfrac{1}{N_\mathbf{clash}} \sum_{i}^{N_{\mathbf{clash}}} ||\hat{d}_{i} - k_{\mathrm{clash}} ||^2
\end{equation}
where $N_\mathbf{clash}$ denotes the number of clashing residues in the protein, where two residues are said to clash if $\hat{d}_{i} < k_{\mathrm{clash}}$, with $k_{\mathrm{clash}}=4$ by default. For very large proteins with a high number of residues, computing this clashing loss is impractical. In that case, we compute it for pairs of residues that are distant from $4$ to $10$Å in the base structure $S_0$. Note that our clashing loss takes into account all of the residues. This is not the case of \citet{li_cryostar_2023}, who implements a similar form as our clashing loss for big proteins while they describe the same loss as we do their paper.

\section{Experiments}\label{append:experiments}
In this appendix, we provide more details on the experiments of Section \ref{sec:experiments}.
\textcolor{black}{We followed the same approach to create all the images of the synthetic datasets. We first pose the ground truth structure,  which we then convert into a volume, which we then project into a 2D image according to our image formation model in (\ref{eq:imageFormationStructure}) with $\sigma = 2$. After that, we corrupt all the images according to the same CTF parameters described in Table \ref{tab:ctf}. Finally Gaussian noise is added to achieve different SNRs. Here, SNR is defined as the ratio of the variance of the images to the variance of the noise.
In this context, the poses are assumed to be exactly known. However, since we use a structure $S_0$, which is different from the structures used to generate the datasets (unless stated otherwise), these poses can only be an approximation. This will not be the case of cryoDRGN, for which these poses will indeed be exact, as this method does not use a base structure. Consequently, in this context, the comparison may introduce a bias in favor of cryoDRGN.}

\begin{table}
\begin{center}
\begin{tabular}{ |c | c|}
 \hline
 Parameter & Value\\
 \hline
    dfU & 15301.1 \hspace{0.1cm} Å\\
  dfV & 14916.4 \hspace{0.1cm} Å \\
  dfang & 5.28 \hspace{0.1cm} degrees\\
  spherical aberration & 2.7\hspace{0.1cm} mm\\
  accelerating voltage & 300 \hspace{0.1cm}keV\\
  amplitude contrast ratio & 0.07 \\
  \hline
\end{tabular}\caption{Table of the parameters used to CTF corrupt the generated images. The same values were used for all images of all datasets.}\label{tab:ctf}
\end{center}
\end{table}

\subsection{Toy dataset}\label{append:toy}
For this experiment, we predict the phytochrome structure using AlphaFold multimer \citet{evans_protein_2021} on its
amino acid sequence with the UniProt \citet{the_uniprot_consortium_uniprot_2021} entry Q9RZA4. This protein forms a dimer with 755 residues on each chain. We define two domains for simulation purposes. The first domain comprises the first chain and the first 598 residues of the second chain. The second domain consists of the remaining 157 residues of the second chain. We rotate the second domain around the $(0,1,0)$ axis, sampling $10^{4}$ rotation angles from the Gaussian mixture:
\begin{equation}
0.5\times\mathcal{N}(-\pi /3, 0.04) + 0.5\times\mathcal{N}(-2\pi /3, 0.04)
\end{equation}

\begin{figure*}[t]
\centering
    \includegraphics[scale=0.22]{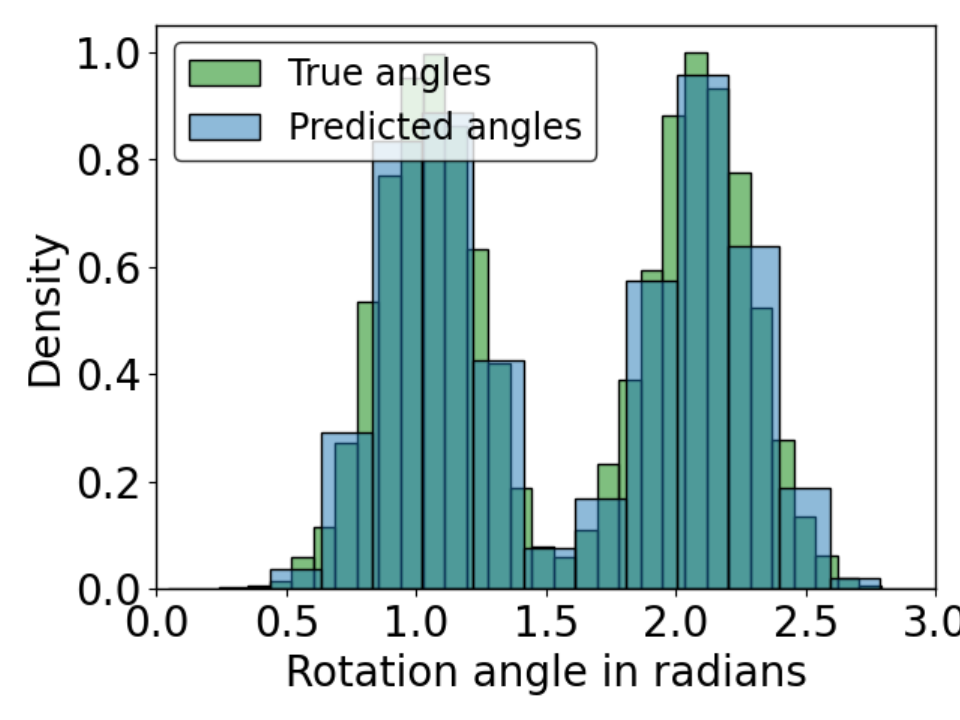}%
    \includegraphics[scale=0.22]{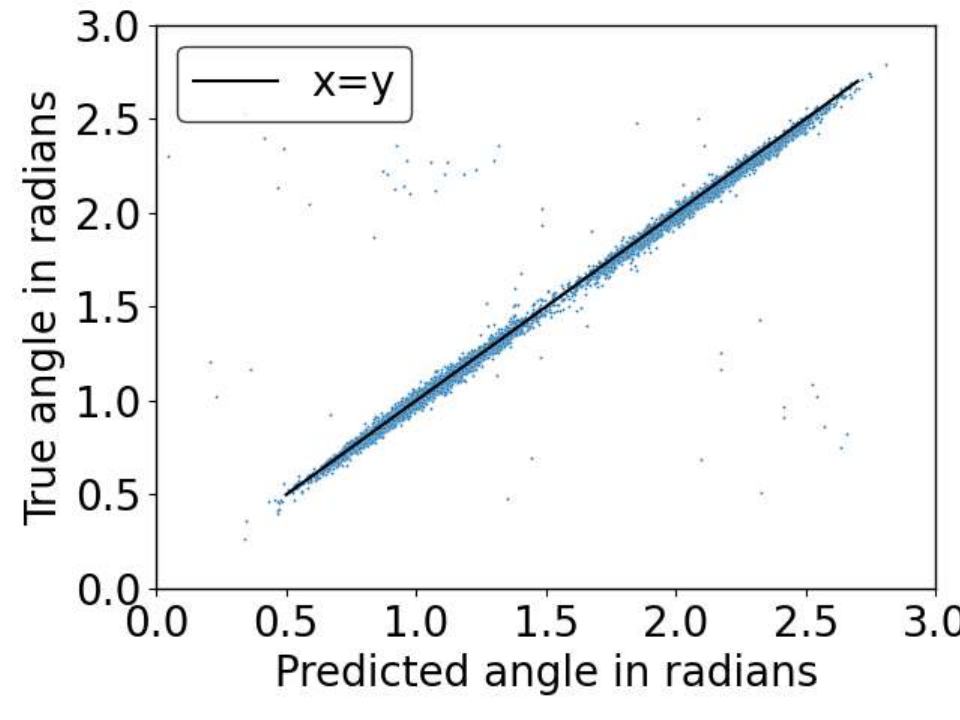}%
    \includegraphics[scale=0.22]{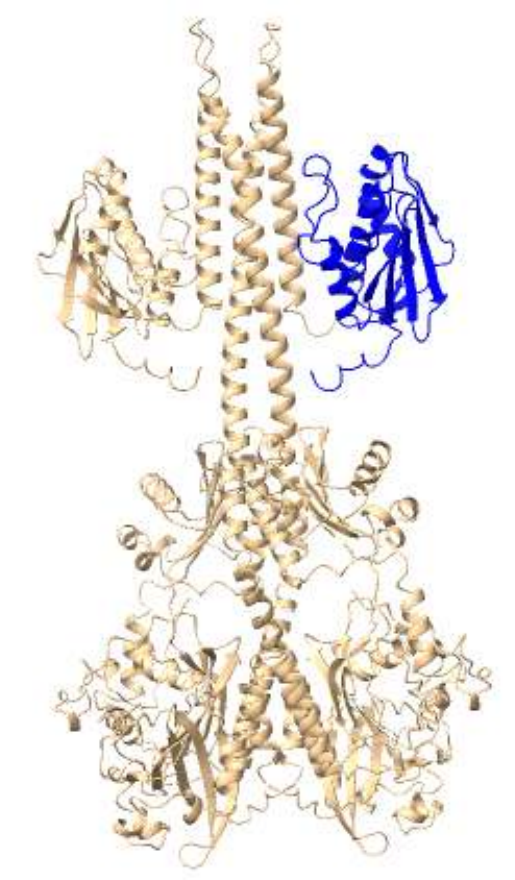}%
    \hspace{0.3cm}
    \includegraphics[scale=0.22]{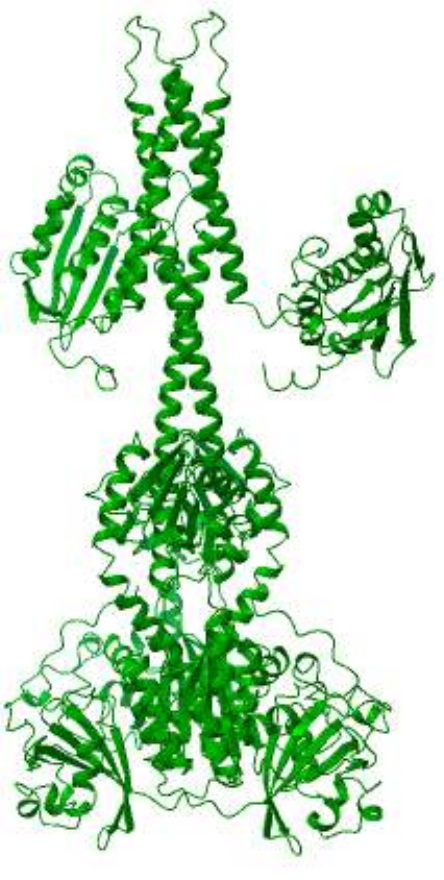}%
    \hspace{0.3cm} \includegraphics[scale=0.22]{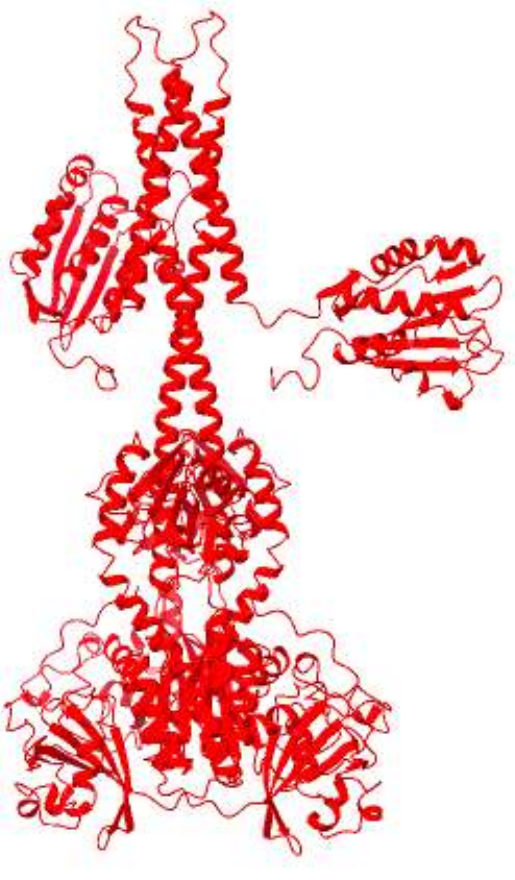}
%    \hspace{0.5cm}
%    \includegraphics[scale=0.23]{figures/experiments/toy_data_set/one_mode.pdf}
%    \hspace{1cm}
%    \includegraphics[scale=0.23]{figures/experiments/toy_data_set/second_mode.pdf}
\caption{Toy dataset. Leftmost: Histograms of the predicted and true angles of rotation in radians. The true angles are in green. The recovered distances are in blue. Left: Predicted against true angles in Ångström. The black line represent $x=y$. Middle: Base structure with the two domains, also used to generate the images. The domain in blue is rotated according to the axis $(0, 1, 0)$ with angles of rotations sampled from the green distribution on the leftmost figure. Note that the segments predicted by cryoSPHERE exactly match the two domains. The fourth segment is in blue and corresponds to the last 157 residues of chain B, matching exactly the ground truth domain. Right: First mode structure. Rightmost: Second mode structure.}
\label{fig:toy_histo}
\end{figure*}

In Figure \ref{fig:toy_histo}, we present the base structure, the decomposition into the two domains as well as the structures corresponding to the deformed base. These deformations represent the mean rotation of each mode.
This gives $10^{4}$ structures. For each structure, we uniformly sample 15 rotation poses and 15 translation poses on $[-10, 10]^{2}$. The structure undergoes rotation, translation, and is then turned into an image according to image formation model \ref{eq:imageFormationStructure}. Subsequently, the images undergo CTF corruption, and noise is added to achieve $\mathrm{SNR}\approx 0.1$. This process generates a total of $150$k images, each of size $\Npix=220$.

We run cryoSPHERE with $\Ndomains=4$ for 48 hours on a single NVIDIA A100 GPU, equivalent to 779 epochs. The encoder has 4 hidden layers of size 2048, 1024, 512, 512 and the decoder has two hidden layers of size 350, 350.We use a learning rate of $0.00003$ for the parameters of the decoder and encoder and a learning rate of $0.0003$ for the segmentation GMM parameters.

Due to computational constraints, the plots in this section are based on only 10000 images, one per conformation.

Testing the segment decomposition, we then run cryoSPHERE by requesting division into $\Ndomains=4$. The program learnt a first and third segment with $0$ residues, a second segment with $1353$ residues and a fourth segment with $157$ residues (Figure \ref{fig:toy_histo}). Thus, cryoSPHERE learnt segments according to the ground truth.

%Moreover, Figure \ref{fig:toy_histo} shows that most of the predicted angles of rotation for the fourth segment are in excellent agreement with the ground truth structural changes.  However,  approximately $12$ \% of the dataset is predicted as belonging to the wrong mode. Figure~\ref{fig:toy_boxplots} in the appendix shows that the norms of the predicted translations for both segments are negligible, also indicating that cryoSPHERE has recovered the ground truth. The same figure shows that we recover the right axis of rotation. In Appendix \ref{append:toy}, we plot the predicted angles against the mean of the latent variable distributions output by the encoder. We observe that cryoSPHERE has recovered the circular motion: the greater the latent mean, the lower the predicted angle, with a quick transition from one mode to the other. This indicates that cryoSPHERE is able to recover the structure of single-particles from cryo-EM data. 

\begin{figure*}[t]
    \centering
    \includegraphics[scale=0.20]{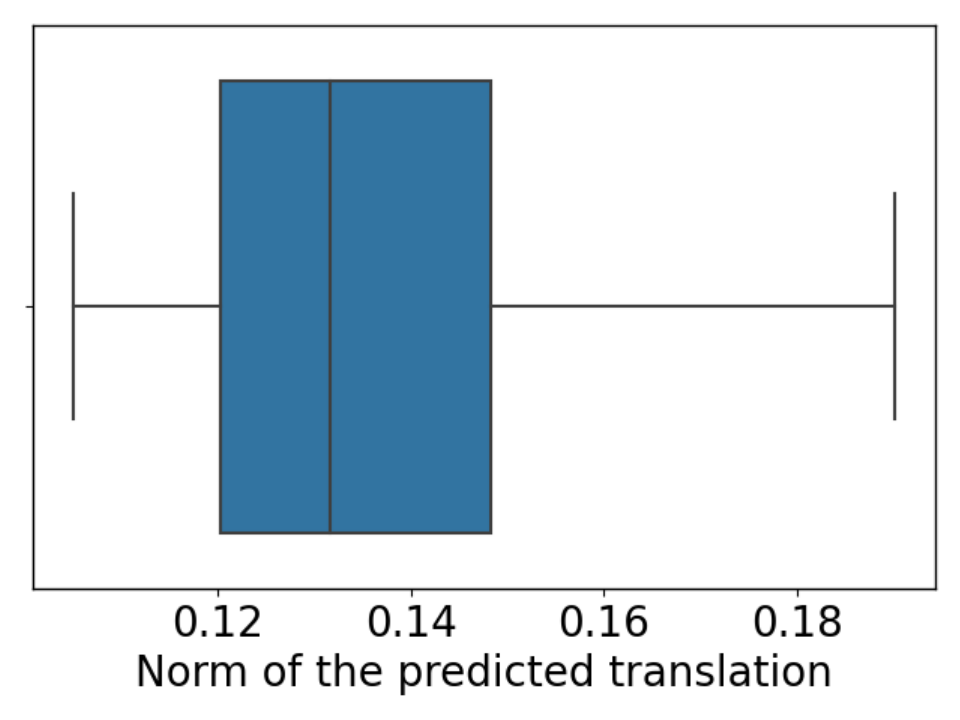}
    \includegraphics[scale=0.20]{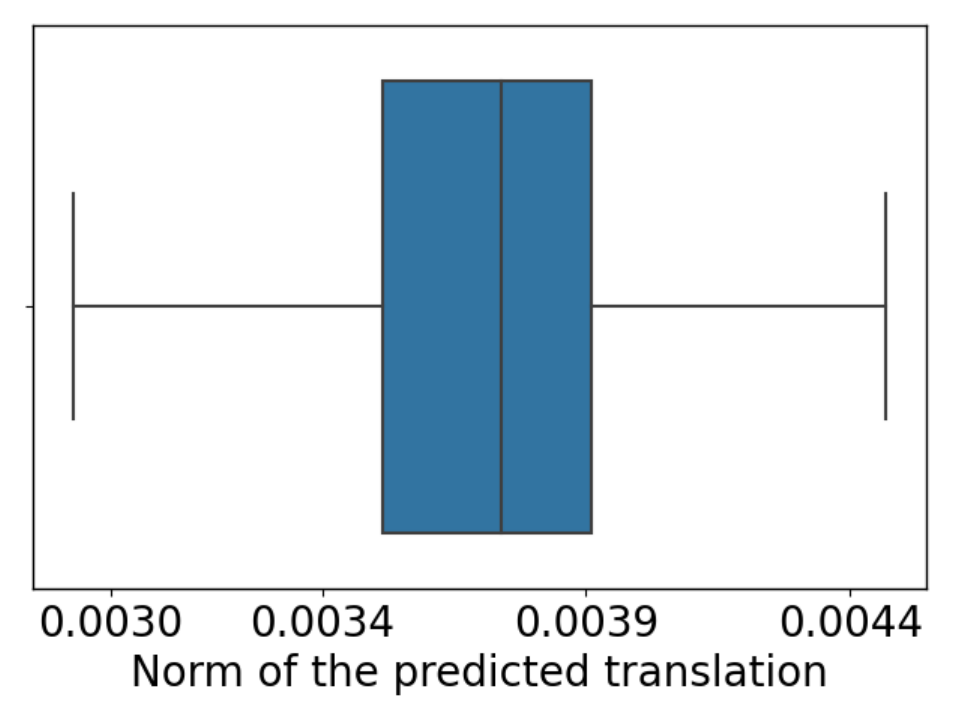}
    \includegraphics[scale=0.20]{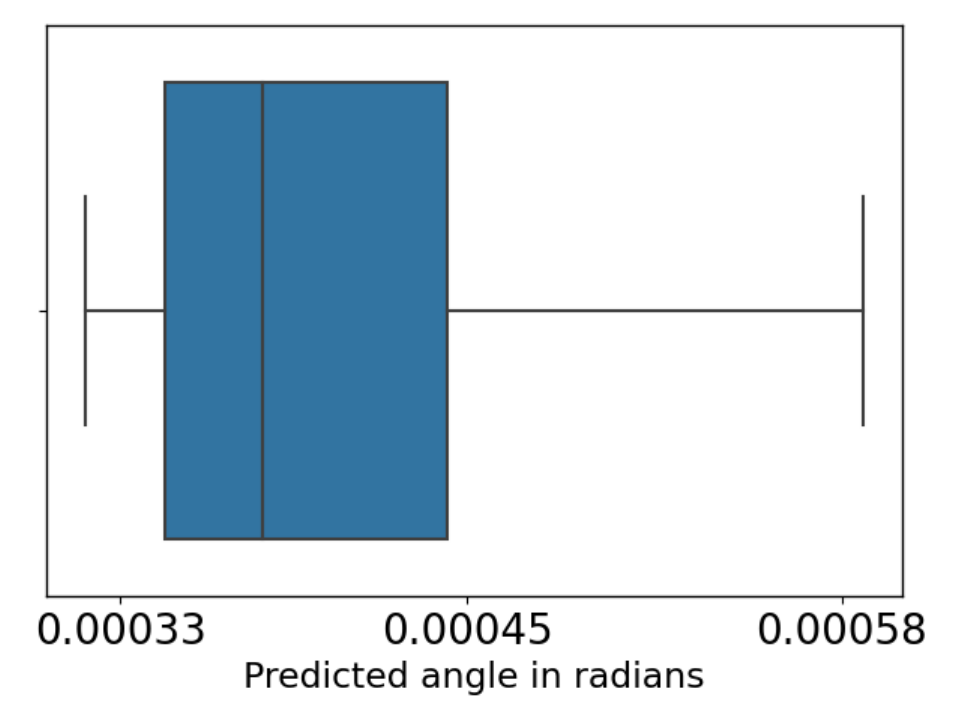}
    \includegraphics[scale=0.20]{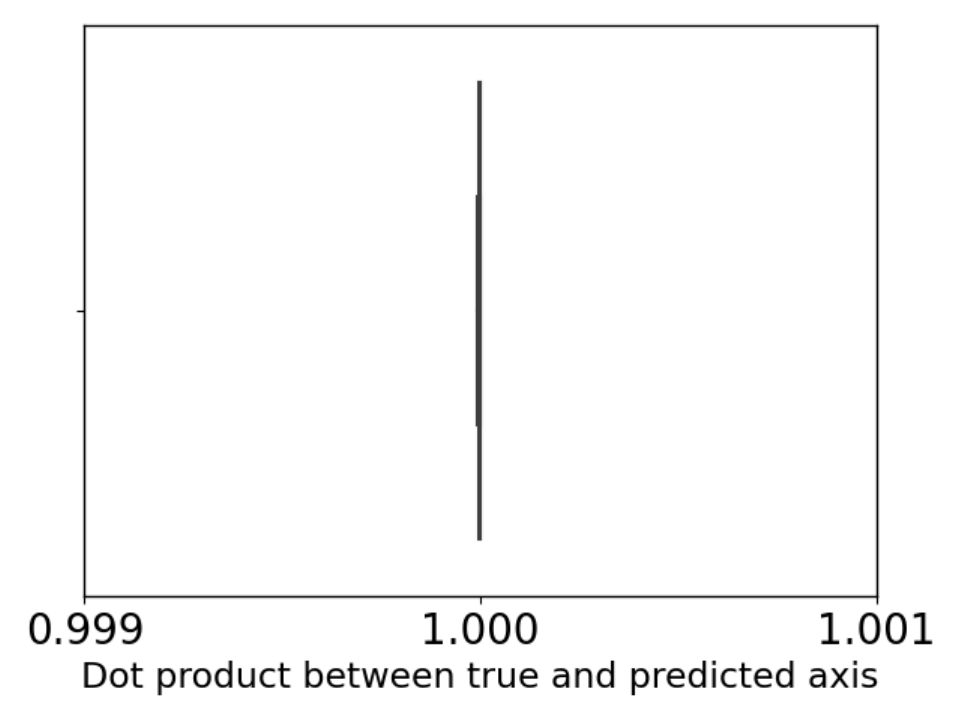}
    \caption{Toy dataset. Leftmost: Boxplot of the norm of the translations predicted for the fourth segment. Middle left: Boxplot of the norm of the translations predicted for the second segment. Middle right: Boxplot of the predicted angle of rotation for the second segment, in radians. Rightmost: Boxplot of the dot product between the predicted axis of rotation for the fourth segment and the true axis of rotation. CryoSPHERE recovers the right axis of rotation almost perfectly.}
    \label{fig:toy_boxplots}
\end{figure*}

Moreover, Figure \ref{fig:toy_histo} shows that most of the predicted angles of rotation for the fourth segment are in excellent agreement with the ground truth structural changes. In addition, the predicted translations for both segments are close to 0, the predicted axis of rotation of the moving segments is close to $(0,1,0)$ and the predicted rotation angles for non moving segment are null, see Figure~\ref{fig:toy_boxplots}. 

\begin{figure}
    \centering
    \includegraphics[scale=0.6]{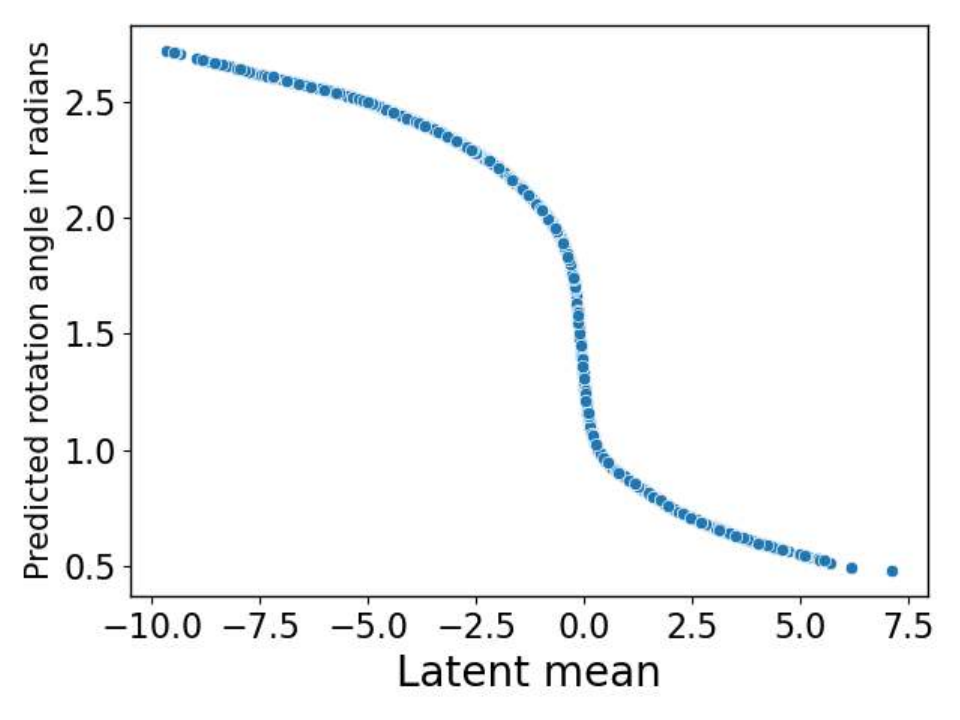}
    \caption{Toy dataset. Predicted angle against latent mean for cryoSPHERE. Note that for clarity $0.3$ percent of the points were removed.}
    \label{fig:toy_ang_latent}
\end{figure}

Finally, Figure \ref{fig:toy_ang_latent} illustrates the predicted angles against the latent means, demonstrating that the model effectively learns rotational motion.

\subsection{Molecular dynamics dataset}\label{append:MD}
\begin{figure}
    \centering
    \includegraphics[scale=0.15]{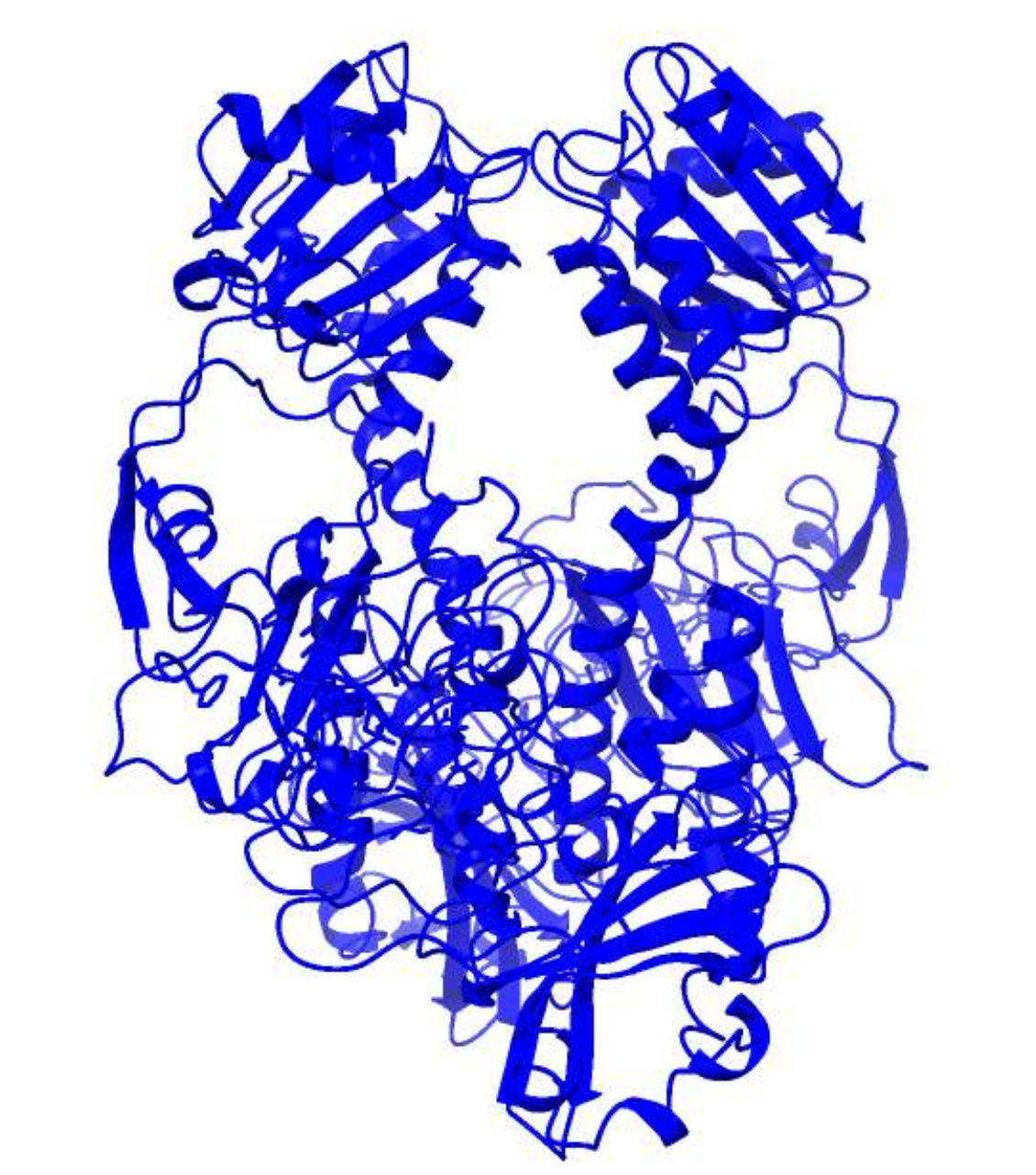}
    \includegraphics[scale=0.15]{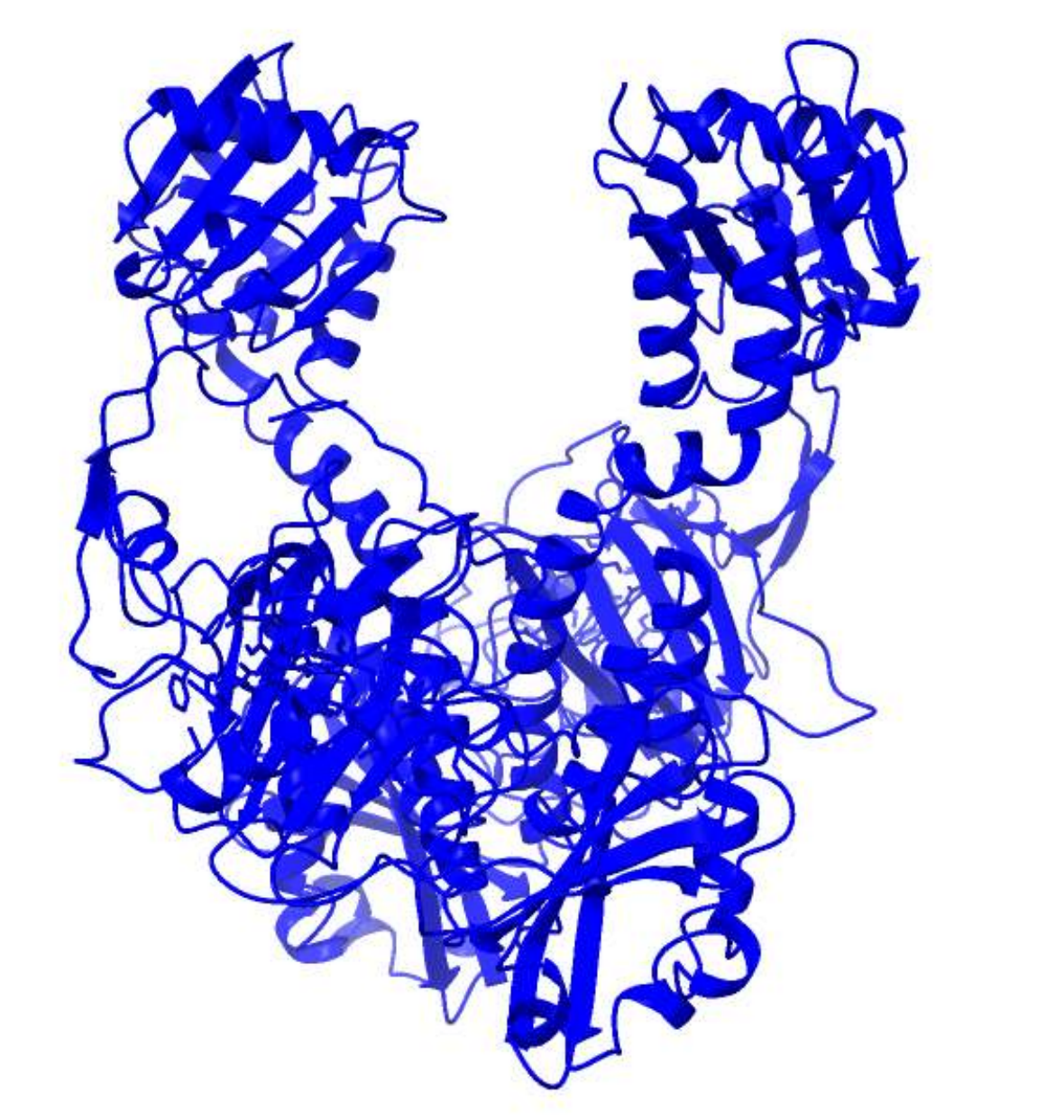}
    \includegraphics[scale=0.15]{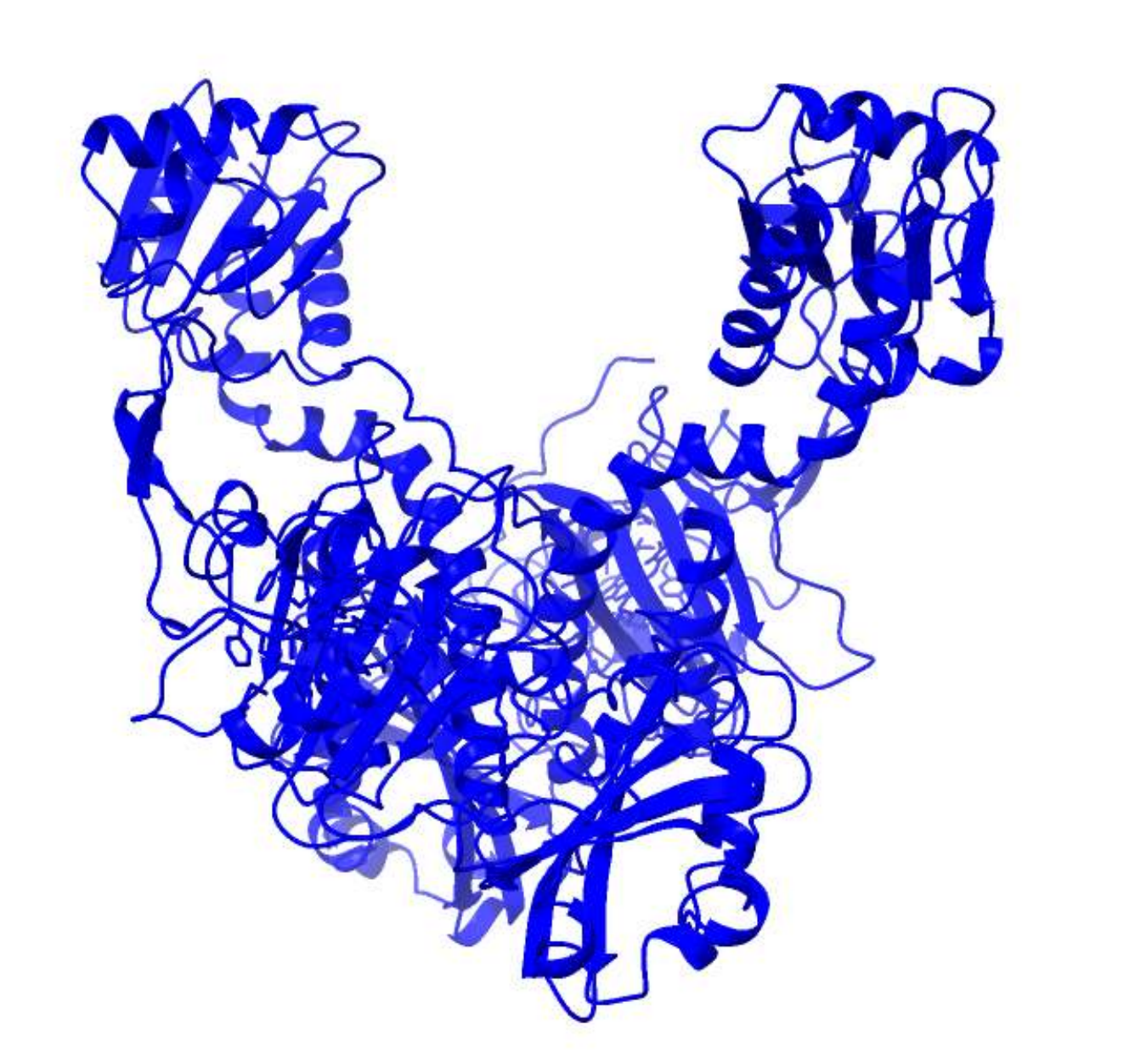}
    \includegraphics[scale=0.15]{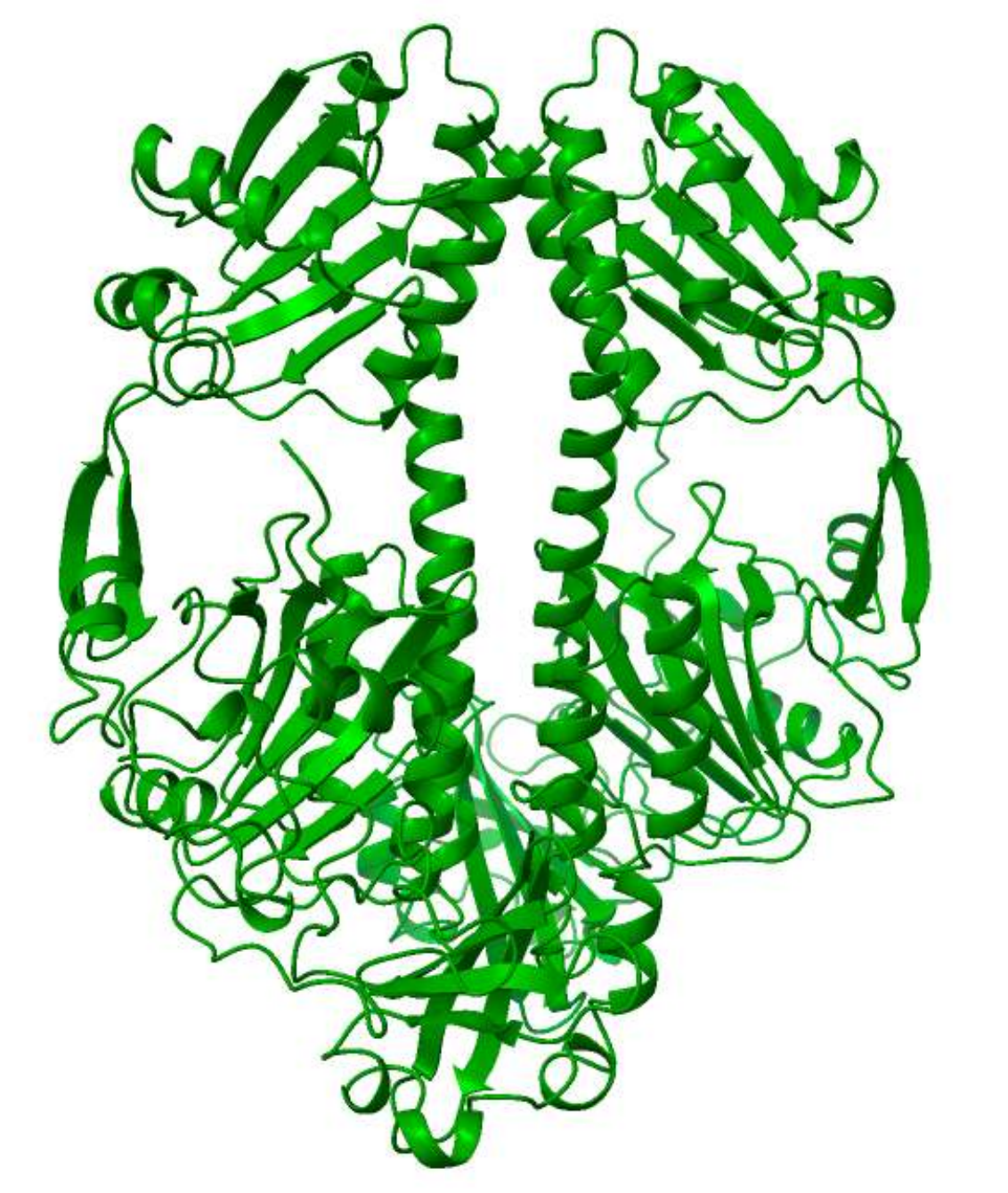}
    \caption{From left to right. 1/ The starting close conformation of the MD simulation. 2/ A medium-open structure arising from the MD simulation. 3/ An open conformation towards the end of the MD simulation. 4/ The AlphaFold structure used as a base structure.}
    \label{fig:md_trajectory}
\end{figure}

We take the structure of a phytochrome with PDB ID 4Q0J \citet{burgie_crystallographic_2014} and define two domains: residues 321 to 502 of the first chain and residues 321 to 502 of the second chain. To simulate the dissociation process of the two upper domains, we perform Metadynamics\citet{barducci_metadynamics_2011} simulations in GROMACS
\citet{abraham_gromacs_2015, pronk_gromacs_2013} with the PLUMED 2 implementation \citet{tribello_plumed_2014}. The collective variable chosen is the distance
between the self-defined centers of mass (COMs) of the upper domains (residues 321-502 of chain A and
B). A 100 ns simulation is conducted using the NpT ensemble, maintaining pressure control through the
Parrinello-Rahman barostat. Gaussian deposition occurs every 5000 steps, featuring a height of 0.1
kJ/mol and a width of 0.05 nm. Afterwards, we extract $10^4$ structures along the dry trajectory. See Figure \ref{fig:md_trajectory} for examples of structures. The closed conformation is the starting conformation of the MD simulation and the most open one corresponds to the end.

%We now test cryoSPHERE with a more difficult dataset. This time, we simulate a continuous motion of a bacterial phytochrome, with PDB entry 4Q0J \citet{burgie_crystallographic_2014}. The trajectory starts at the close conformation of Figure \ref{fig:md_trajectory} and end at the most open conformation on the same figure. This dimer has 503 residues per chain. We define two upper domains corresponding to the residues 321-502 from both chain A and B. Then, we run a Molecular Dynamics (MD) simulation to simulate a dissociation of these two domains, as depicted in Figure \ref{fig:structures} in Appendix \ref{append:MD}. We take $10^4$ structures along the trajectory and sample $15$ poses uniformly for each structure. Based on this, we generate three datasets, with Gaussian noise added to give a SNR of $0.1$, $0.01$, and $0.001$ respectively. We report the results for SNR $0.001$ here and refer to Appendix \ref{append:MD} for more details and results on the two other datasets.

%\begin{figure}
%    \centering
%    \includegraphics[scale=0.15]{figures/experiments/phy_150k/comparison20.pdf}
%    \includegraphics[scale=0.15]{figures/experiments/phy_150k/comparison6118.pdf}
%    \includegraphics[scale=0.15]{figures/experiments/phy_150k/comparison8480.pdf}
%    \includegraphics[scale=0.15]{figures/experiments/phy_150k/comparison10000.pdf}
%    \caption{Examples of predicted structures with the corresponding ground truth structure. Ground truth is in blue. Predicted is in green.}
   %\label{fig:150k_comparison_structures}
%\end{figure}

For each structure, we sample $15$ rotation poses uniformly together with translations uniformly on $[-10, 10]^{2}$. This results in a dataset of $150$k images. We use $\Npix = 190$, and each pixel is of size $1$Å. We finally add Gaussian noise to create three datasets: a $\mathrm{SNR}=0.1$, a $\mathrm{SNR}=0.01$ and $\mathrm{SNR}=0.001$ dataset. The results for $\mathrm{SNR}=0.001$ are described in Section \ref{subsec:MD} and this appendix describes the results for the $\mathrm{SNR}=0.1$ and $\mathrm{SNR}=0.01$ datasets.

 For cryoSPHERE on all three datasets, the encoder is a 4-hidden-layer neural network with fully connected hidden layers of dimension of $512, 256, 64, 64$. The decoder is a 2-hidden-layer neural network with fully connected hidden layers of dimension  $512, 512$. We set the batch size to $128$ and use the ADAM optimizer \citet{kingma_adam_2017} with a learning rate of $0.00003$ for the encoder and decoder parameters, while we set a learning rate of $0.0003$ for the GMM segmentation parameters. The latent dimension is set to $8$ and we run the program with $\Ndomains=10$, $\Ndomains=20$ and $\Ndomains=25$  for all datasets. We train for 24 hours on a single NVIDIA A100 GPU.

We use the default parameters for both cryoStar and cryoDRGN but disable the structural loss of cryoStar, except for the elastic network loss. We train both methods for 24 hours on the same GPU as cryoSPHERE. 
To maintain consistency when comparing volumes generated from cryoDRGN, our methods, cryoStar and the ground truth, we convert the structures (ground truth, predicted with cryoStar and predicted with cryoSPHERE) into volumes using the same image formation model employed to generate the dataset, see \ref{eq:structToMap}. Since cryoStar also proposes a volume method similar to cryoDRGN, after training the structure method of cryoStar for 24 hours, we train their volume methods for the same amount of time. We report the results for this volume method in this appendix.

Since cryoSPHERE and cryoStar are structural methods, we can compute the predicted distance between the two domains defined earlier for each image, both for cryoStar and cryoSPHERE.

For computational efficiency, the FSC plots and distances plots are not based on all structures: only one image per structure is used to compute the distances. Consequently, the distance plots are based on 10k images. For the computation of the FSC curves, we select only 1000 images evenly distributed among these 10000 structures.

\subsubsection{SNR $0.1$}\label{subsubsec:0.1}
%We plot a set of structures that cryoSPHERE predicts together with the corresponding ground truth structure in Figure \ref{fig:150k_comparison_structures}. We see an excellent agreement of the predicted opening with the ground truth.
%As explained in section \ref{subsec:MD}, Figure \ref{fig:fsc_curves} shows that we outperform cryoDRGN in terms of resolution at both the $0.5$ and $0.143$ cutoffs. This of course depends on the base structure we are using.
This subsection describes the results for cryoSPHERE, cryoStar and cryoDRGN on our molecular dynamics simulation dataset with SNR $0.1$. We use $\Ndomains=25$ in this section.

\begin{figure}
    \centering
\includegraphics[width=0.33\textwidth]{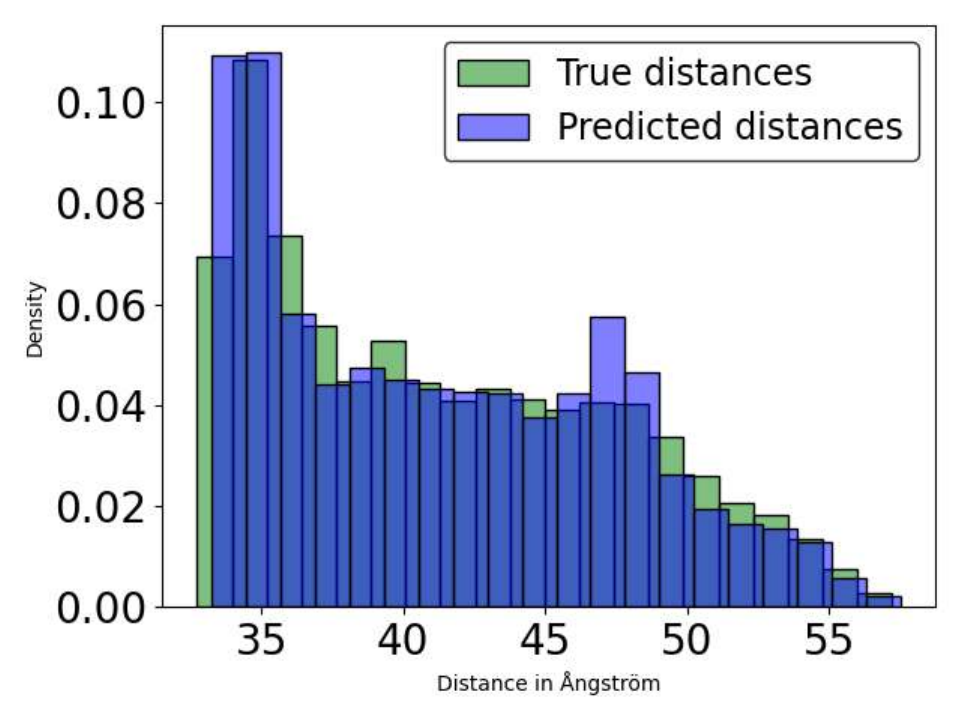}
\includegraphics[width=0.33\textwidth]{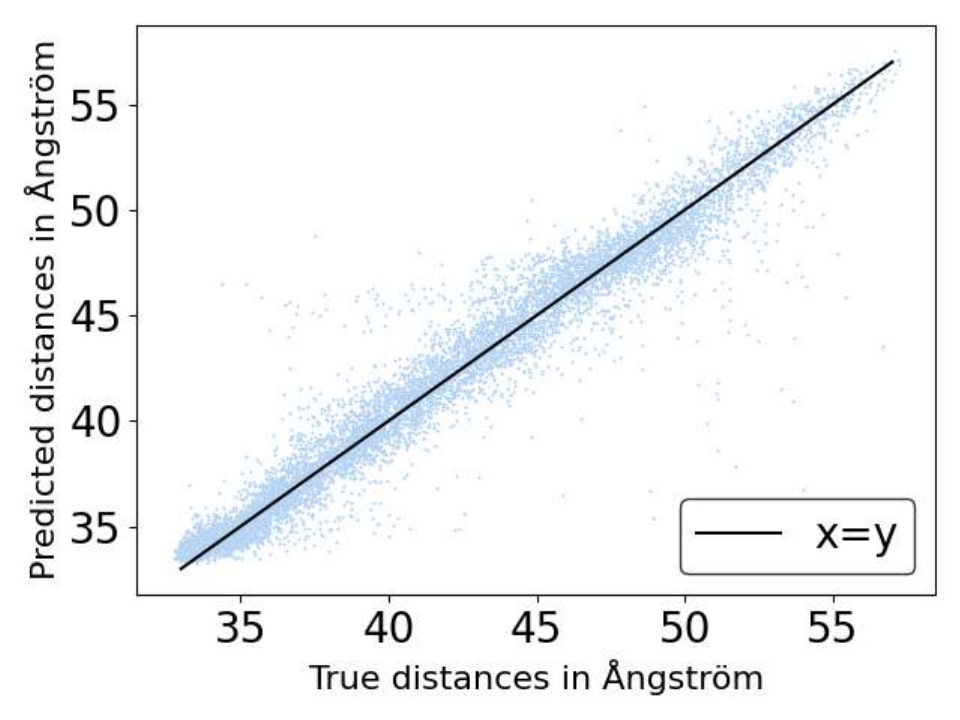}%
\includegraphics[width=0.22\textwidth]{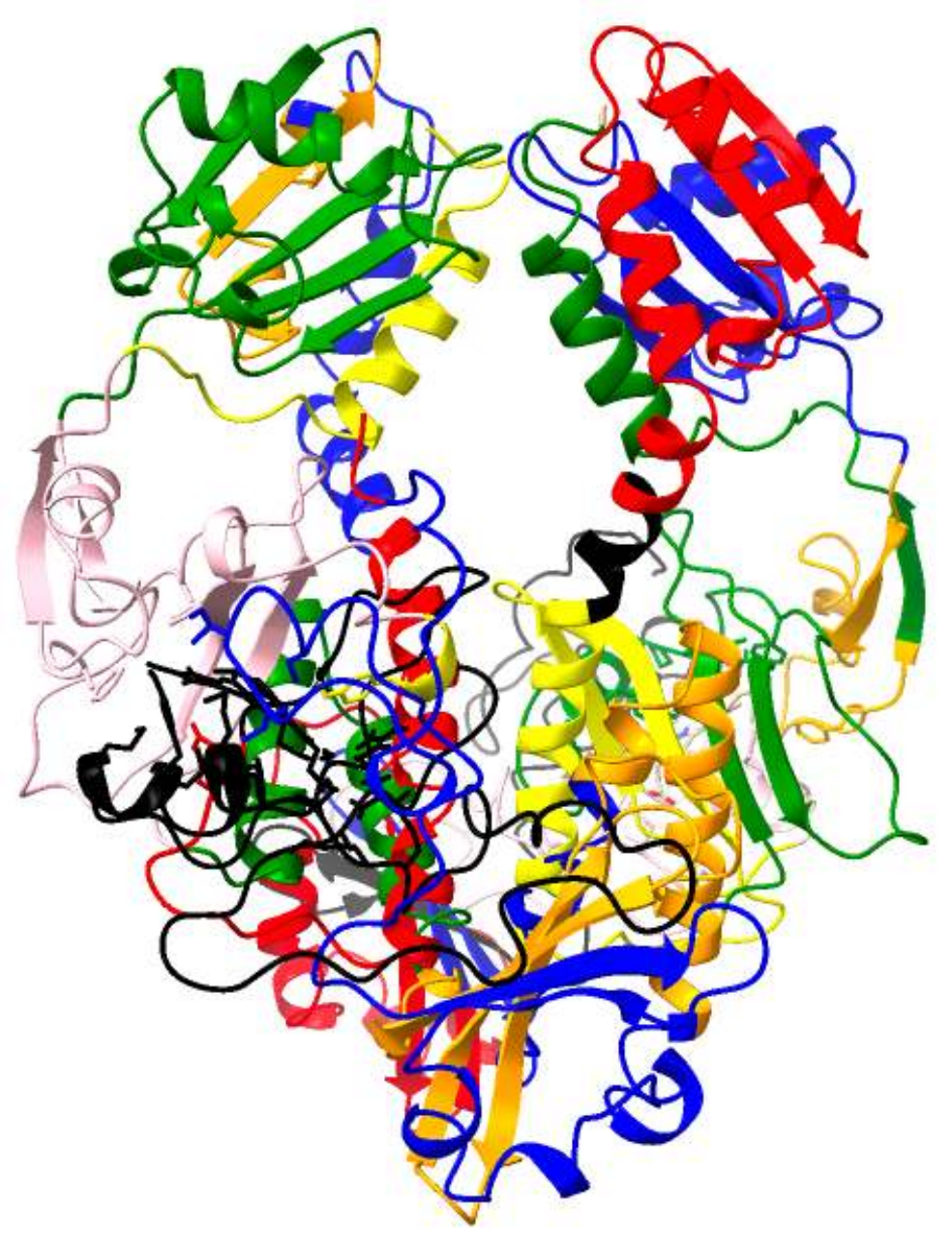}%
\caption{Results for cryoSPHERE on SNR $0.1$ with $\Ndomains=25$. Left: distribution of distances predicted by cryoSPHERE compared to the ground truth distribution. Middle: true versus predicted distances for cryoSPHERE. Right: segments decomposition.}
    \label{fig:snr_0_1_cryosphere}
\end{figure}

\begin{figure}
    \centering
    \includegraphics[width=0.33\textwidth]{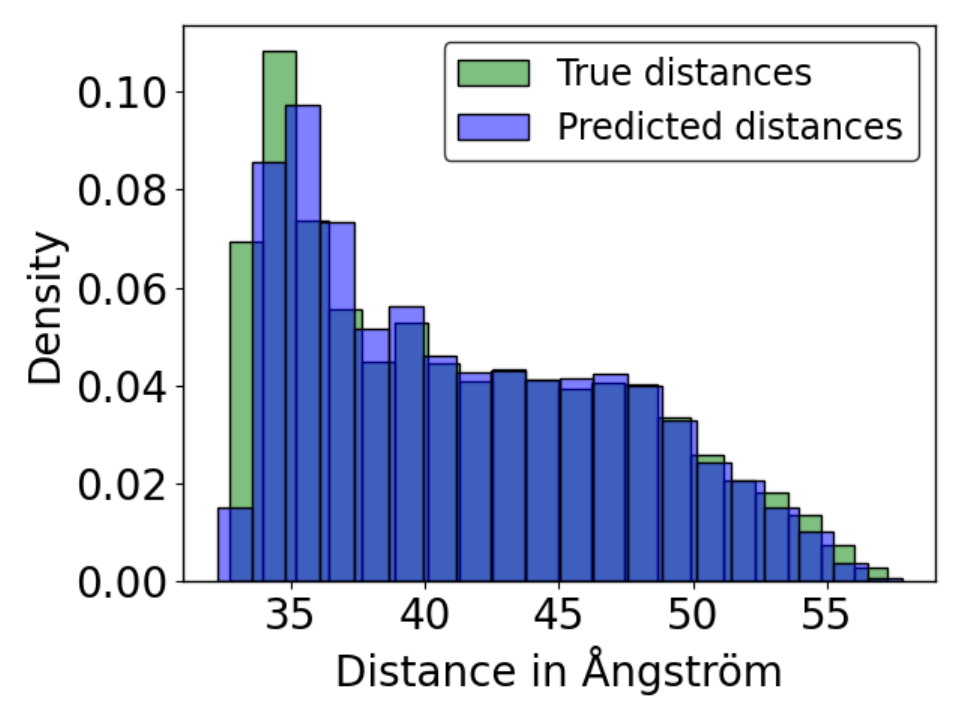}%
    \includegraphics[width=0.33\textwidth]{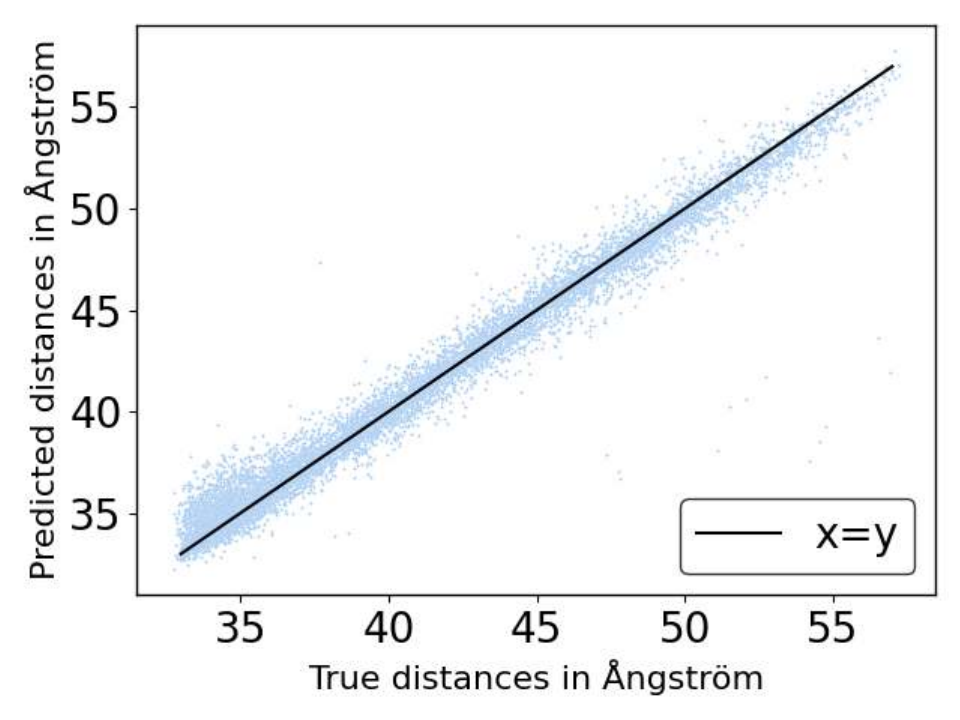}%
    \caption{Results for cryoStar on SNR $0.1$. Left: distribution of distances predicted by cryoStar compared to the ground truth distribution. Right: true versus predicted distances for cryoStar}
    \label{fig:snr_0_1_cryostar}
\end{figure}

Figures \ref{fig:snr_0_1_cryosphere} and \ref{fig:snr_0_1_cryostar} show that both cryoSPHERE and cryoStar recover the ground truth distribution of distances very well. They are also able to identify the correct conformation conditionally on an image, as illustrated by the predicted versus true distances plot.

\begin{figure}
    \centering
\includegraphics[scale=0.24]{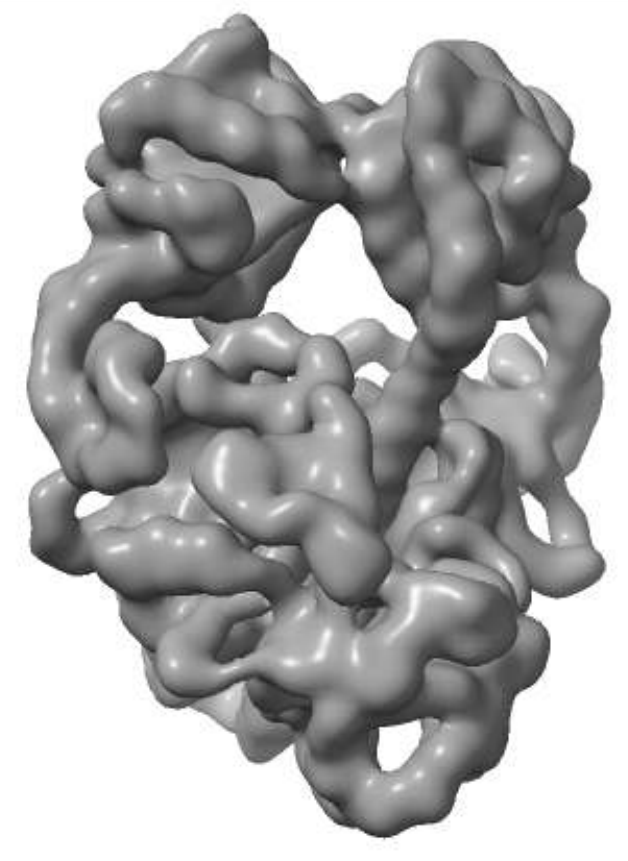}\hspace{0.5cm}    
\includegraphics[scale=0.24]{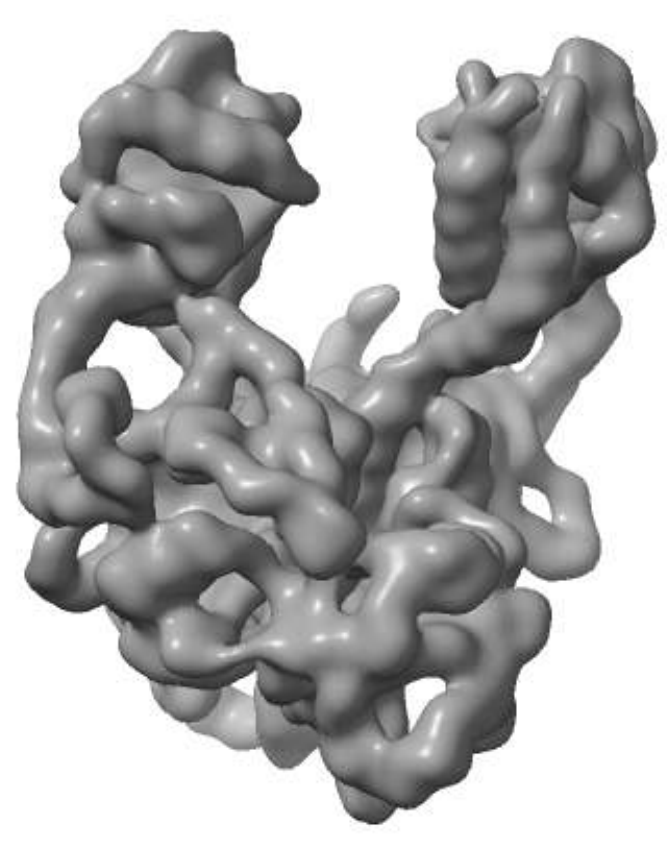}\hspace{0.5cm}  
\includegraphics[scale=0.24]{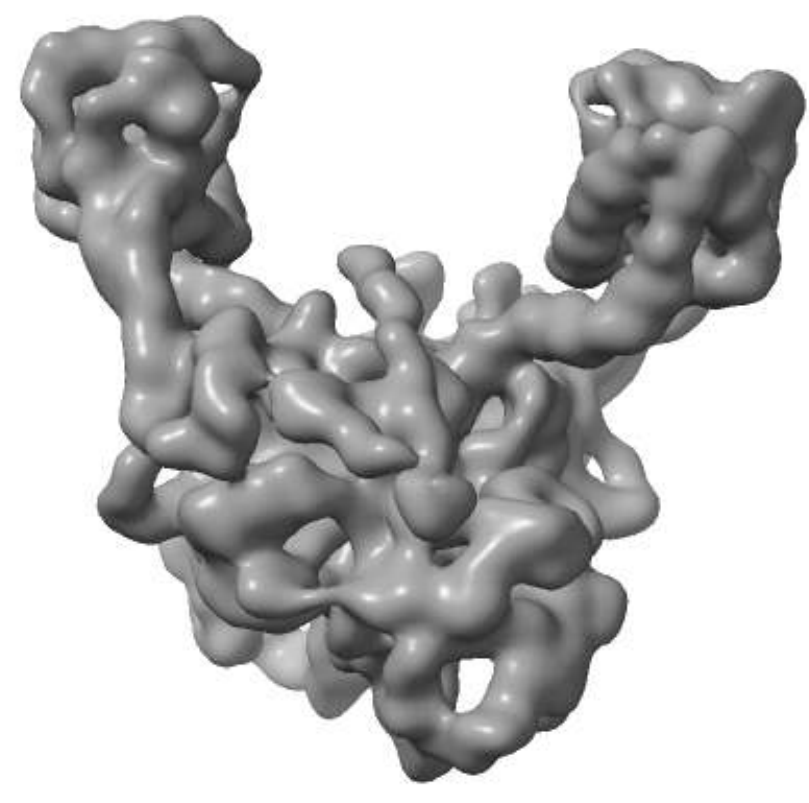} \\    
\includegraphics[scale=0.2]{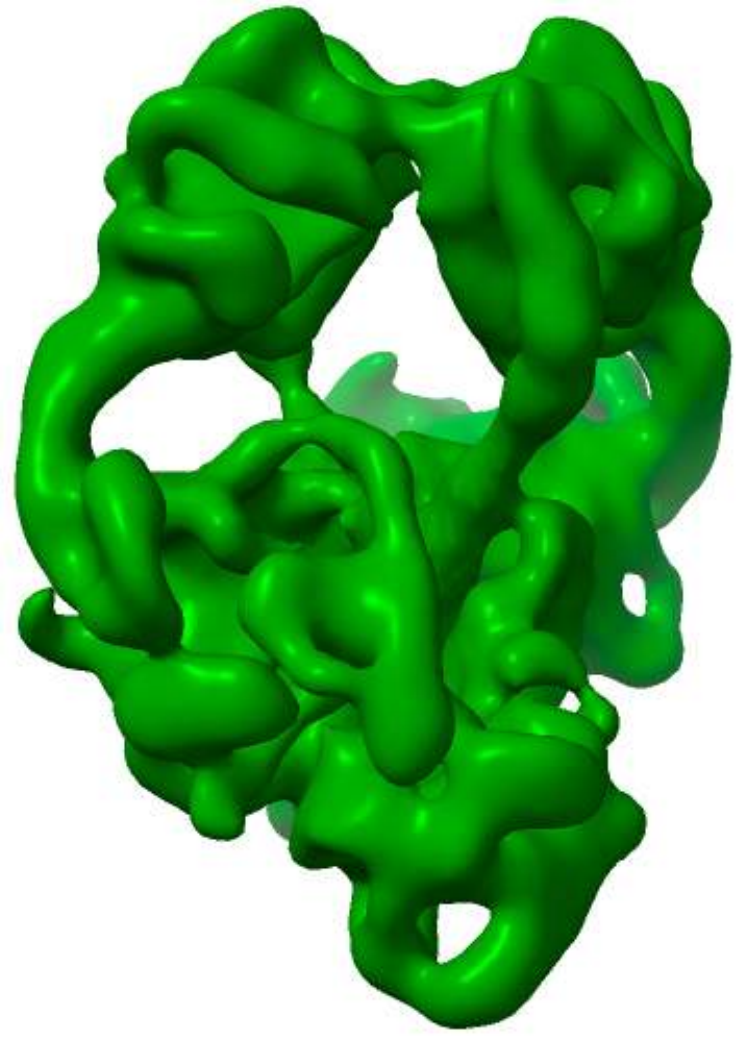}\hspace{0.5cm} 
\includegraphics[scale=0.2]{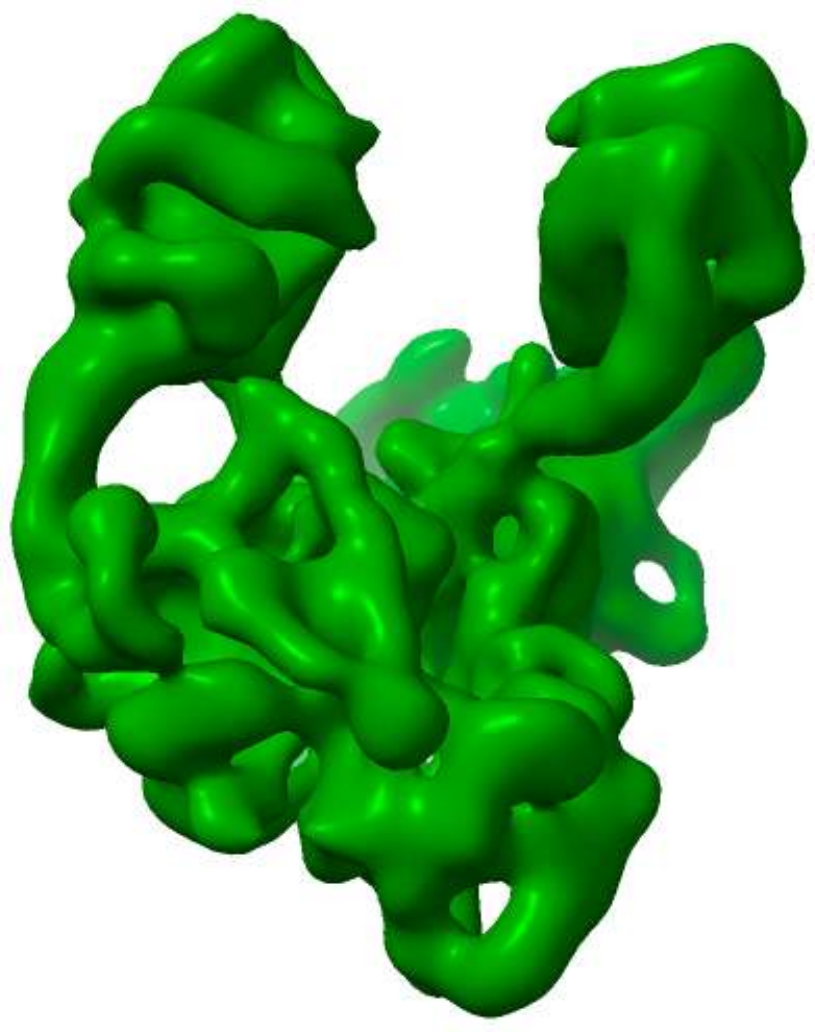}\hspace{0.5cm} 
\includegraphics[scale=0.2]{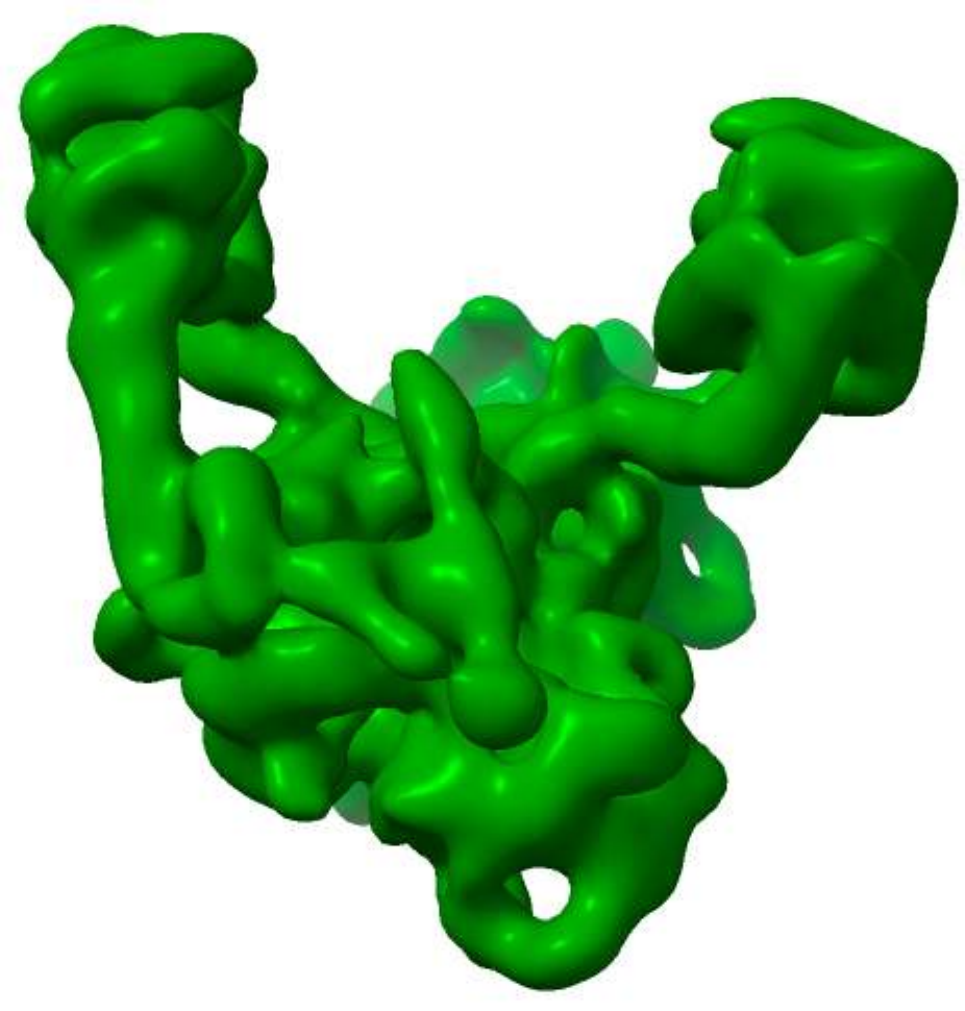}%
\caption{SNR $0.1$. Example volumes produced by the volume method of cryoStar with the corresponding ground truth. Top: ground truth. Bottom: corresponding volumes predicted by the volume method of cryoStar.}
    \label{fig:snr_0_1_cryostar_volume_method}
\end{figure}

\begin{figure}
    \centering
    \includegraphics[width=0.4\textwidth]{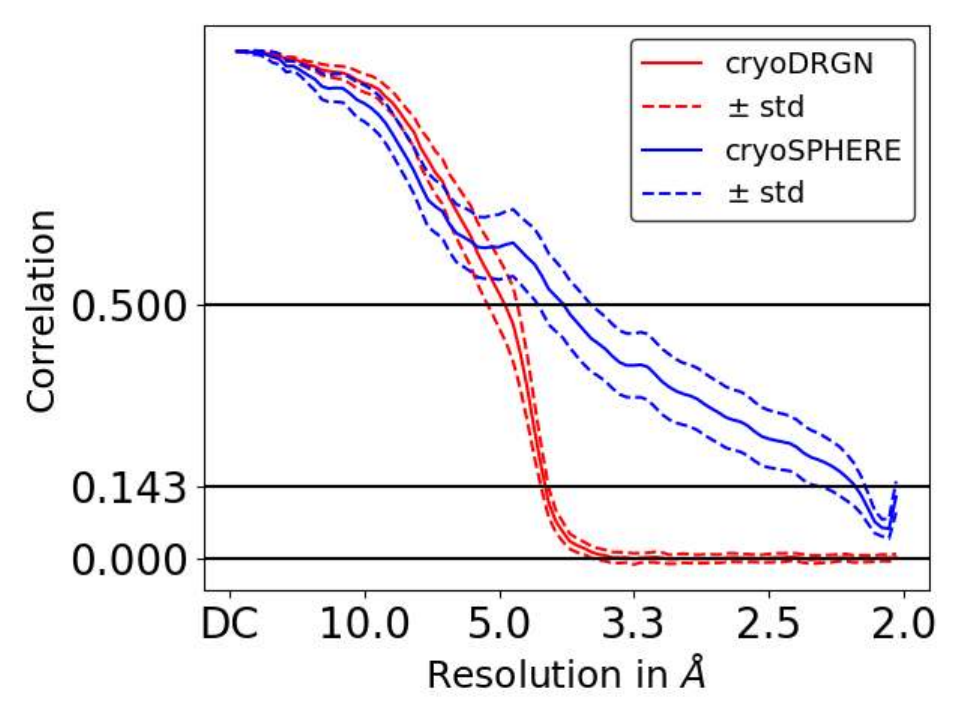}
    \includegraphics[width=0.4\textwidth]{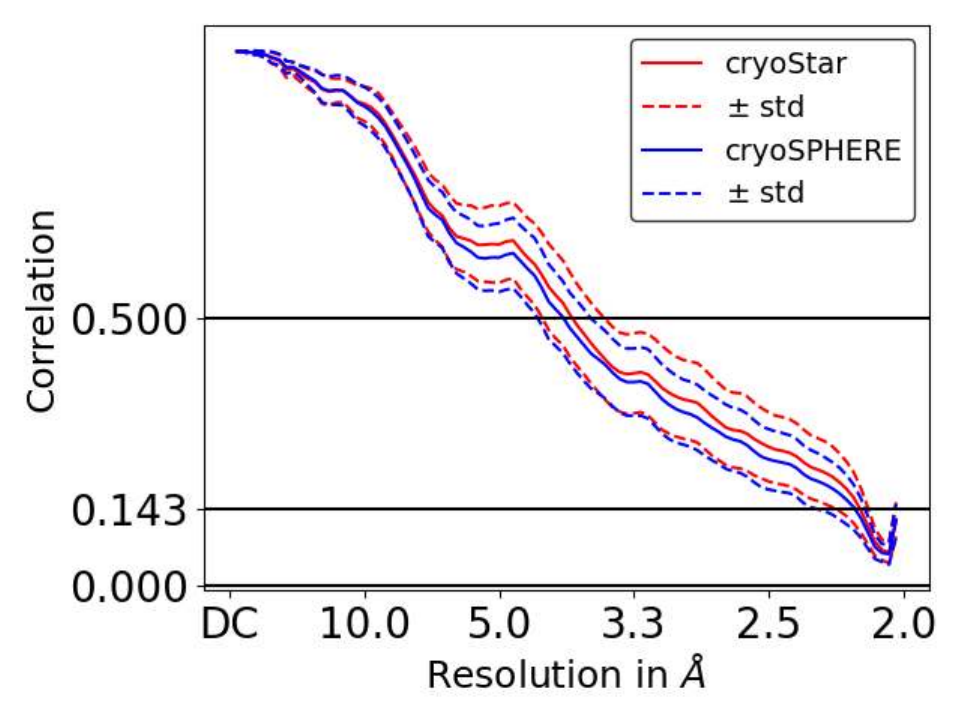}\\
    \includegraphics[width=0.4\textwidth]{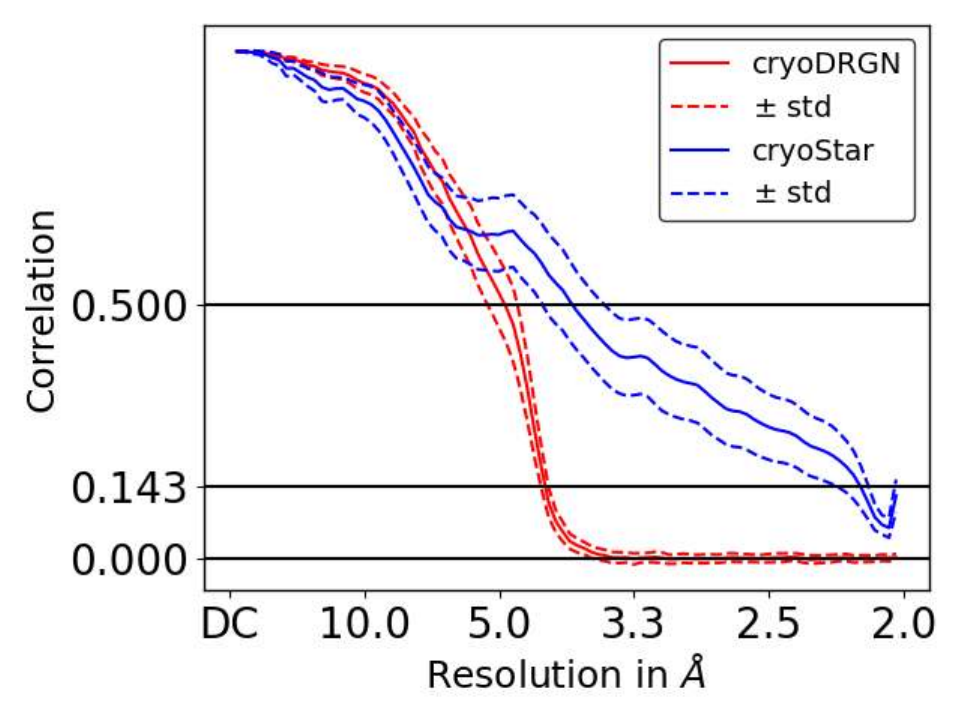}
    \includegraphics[width=0.4\textwidth]{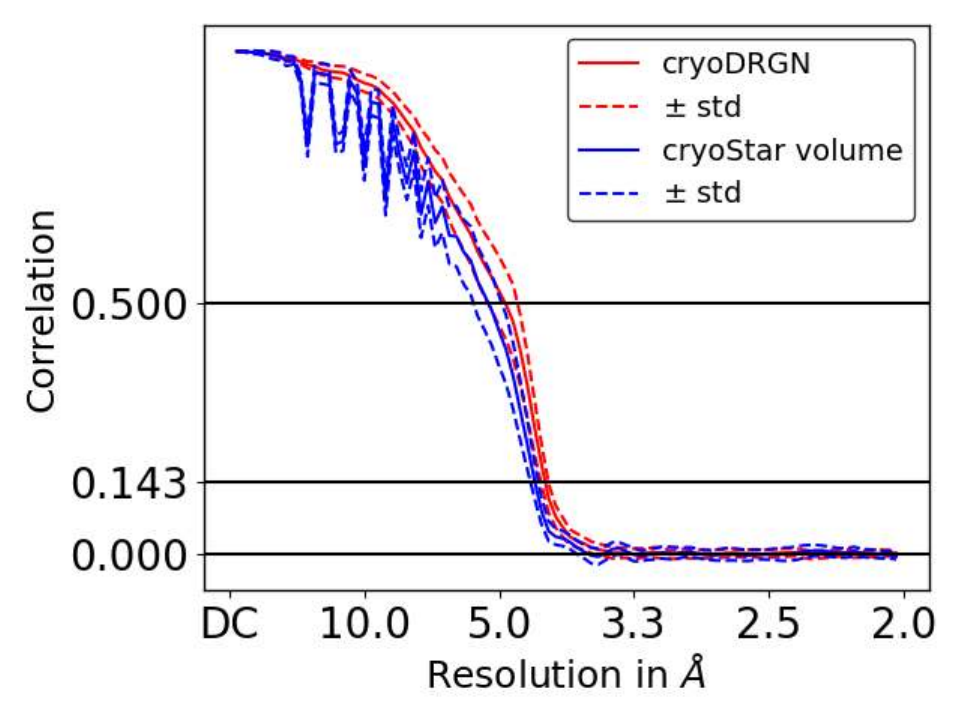}
    \caption{SNR $0.1$. Mean Fourier shell correlation ($\pm$ std) comparison for cryoSPHERE with $\Ndomains = 25$, cryoStar and cryoDRGN. Top left: cryoSPHERE versus cryoDRGN. Top right: cryoSPHERE vs cryoStar. Bottom left: cryoDRGN vs cryoStar. Bottom right: cryoStar volume method vs cryoDRGN}
    \label{fig:snr_0_1_fsc_comparison}
\end{figure}

Figure \ref{fig:snr_0_1_fsc_comparison}, shows that both cryoStar and cryoSPHERE outperform cryoDRGN at both the $0.5$ and $0.143$ cutoffs. It seems cryoStar slightly outperforms cryoSPHERE at both cutoffs. This might be because the SNR is rather high, hence moving each residue individually offers a greater flexibility than moving segments, while the risk of overfitting is low.
In addition, we can see that the volume method of cryoStar perform similarly, if not worse, than cryoDRGN. That seems to indicate that this volume method does not benefit from the information gained by the structural method. Figure \ref{fig:snr_0_1_cryostar_volume_method} shows three examples of volumes predicted by the volume method of cryoStar, together with the corresponding ground truth.

\begin{figure}
    \centering
    \includegraphics[width=0.28\textwidth]{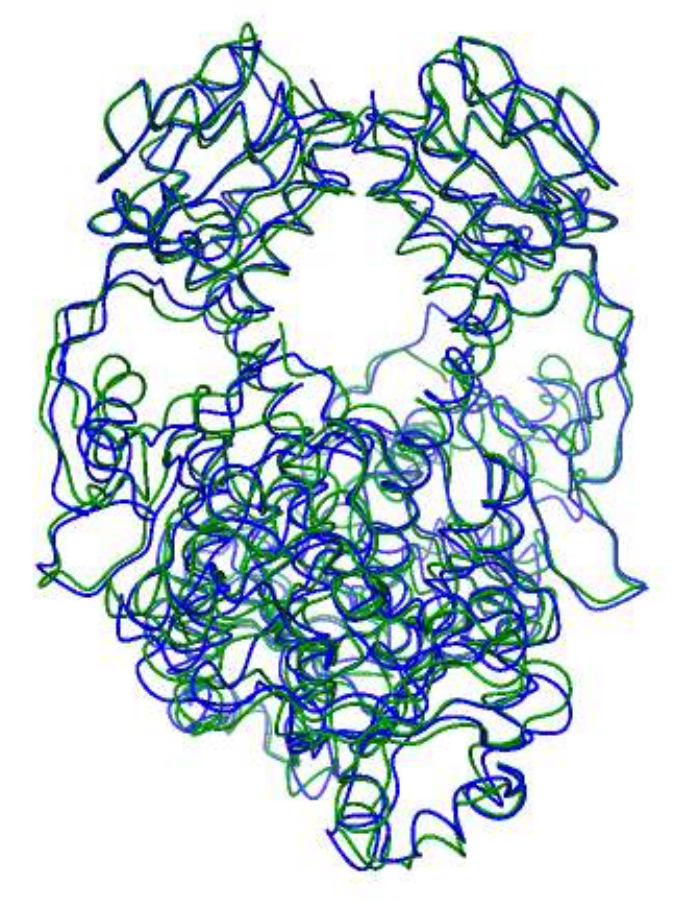}
    \includegraphics[width=0.28\textwidth]{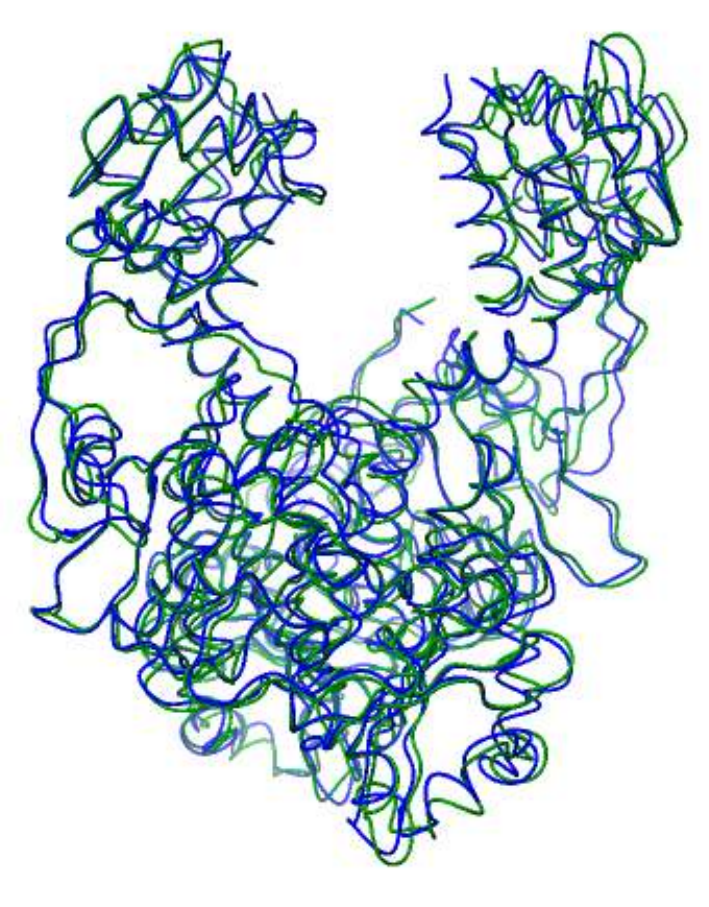}
    \includegraphics[width=0.35\textwidth]{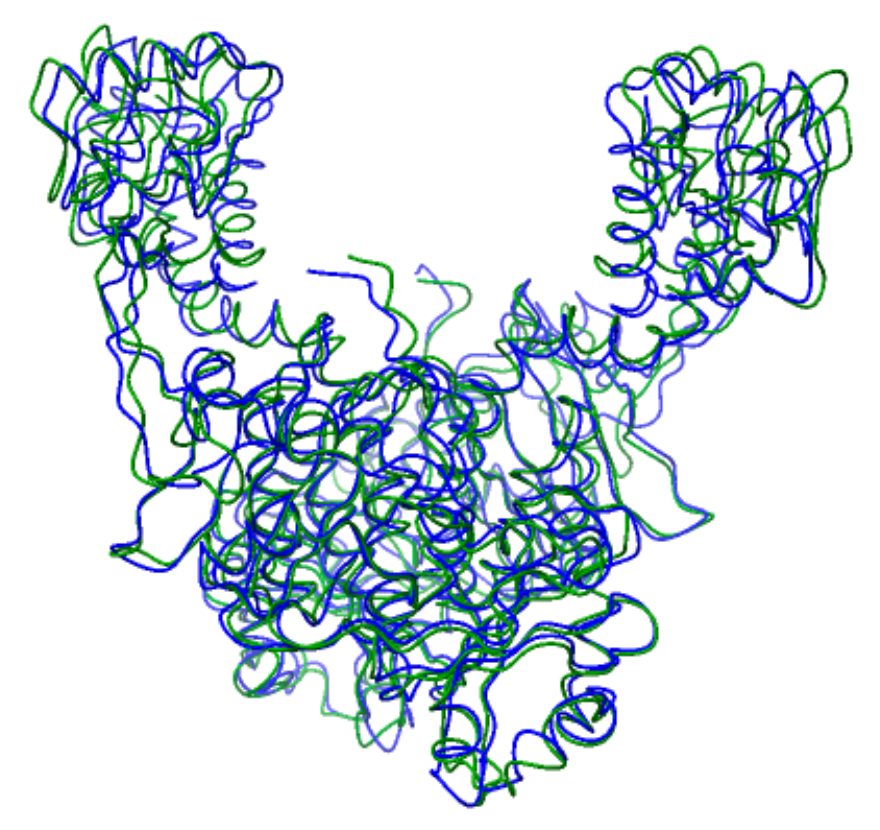}
    \caption{SNR $0.1$. Examples of reconstructed structures by cryoSHERE. Blue is predicted, green is ground truth. CryoSPHERE is able to recover the right conformation. Left to right: image number 11, 5001 and 9999.}
    \label{fig:structures_cryoSPHERE_snr_0_1}
\end{figure}

\begin{figure}
    \centering
    \includegraphics[width=0.28\textwidth]{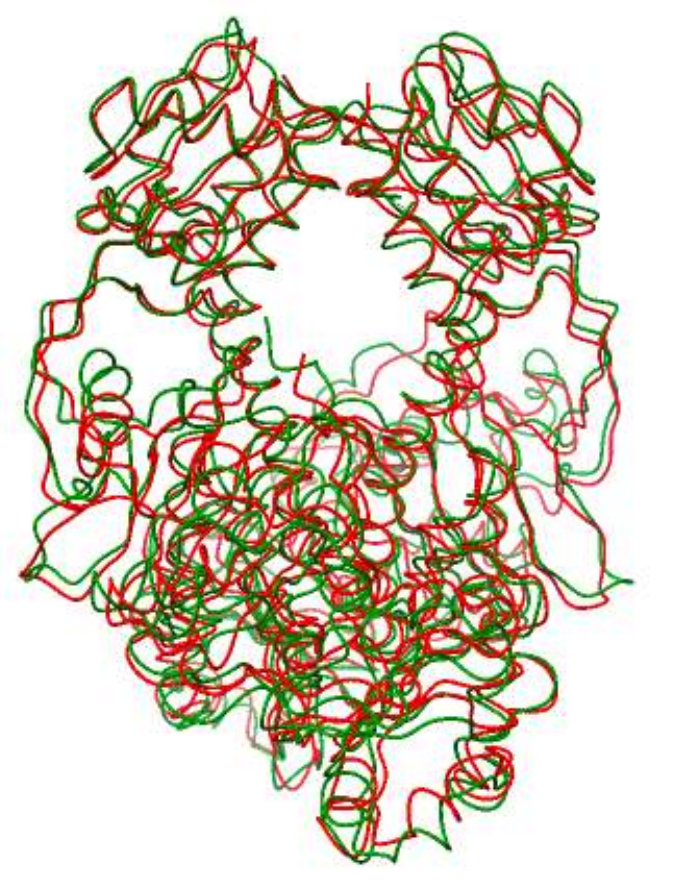}
    \includegraphics[width=0.28\textwidth]{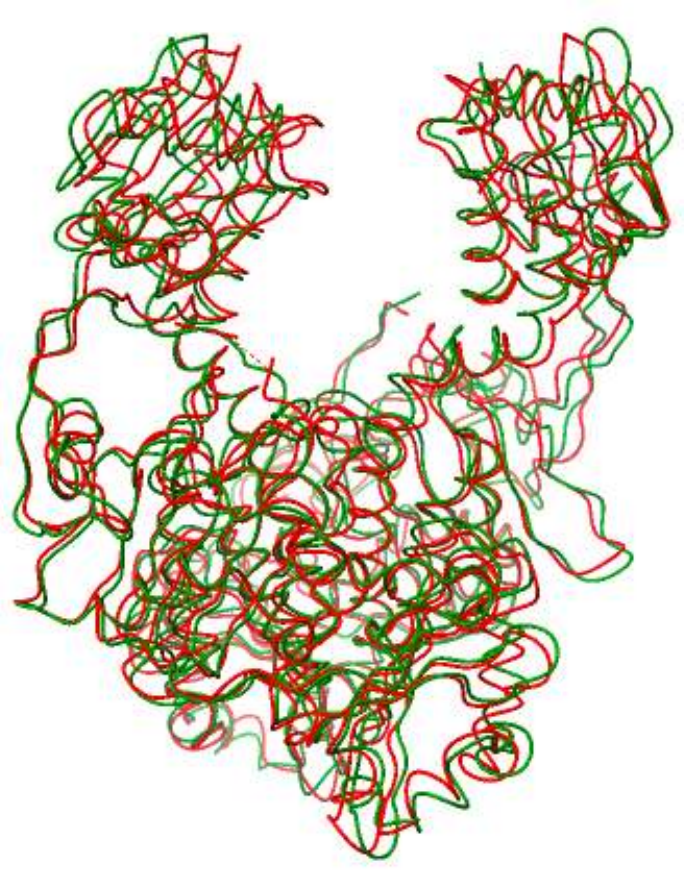}
    \includegraphics[width=0.35\textwidth]{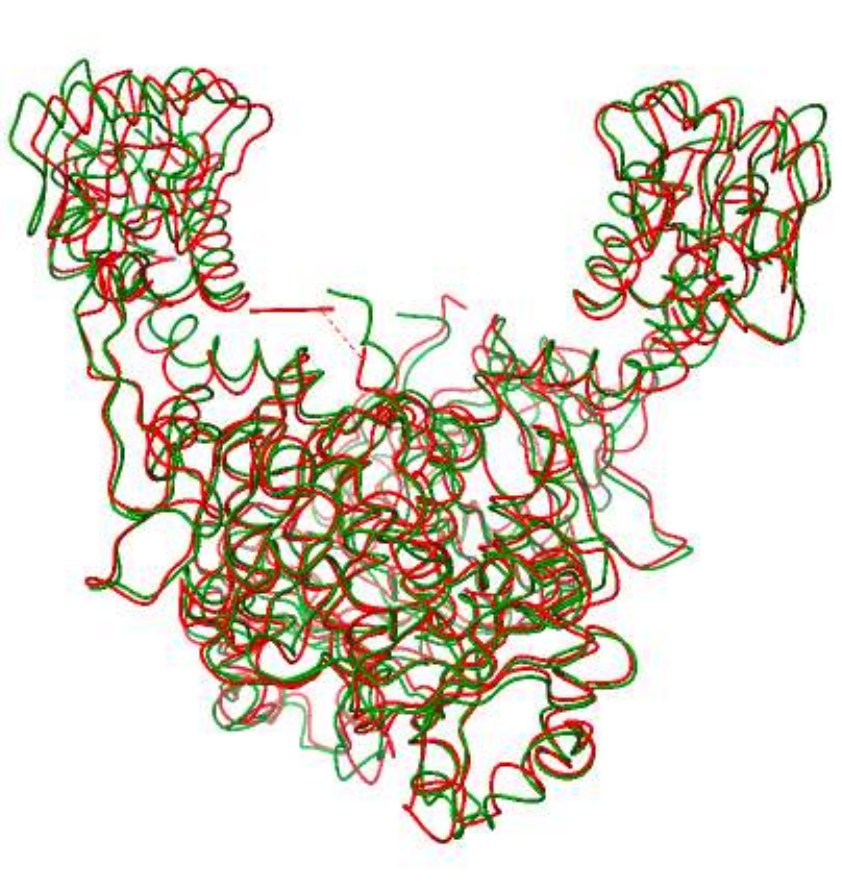}
    \caption{SNR $0.1$. Examples of reconstructed structures by cryoStar. Red is predicted, green is ground truth. cryoStar is able to recover the right conformation. Left to right: image number 11, 5001 and 9999.}
    \label{fig:structures_cryoStar_snr_0_1}
\end{figure}

Figure \ref{fig:structures_cryoSPHERE_snr_0_1} and \ref{fig:structures_cryoStar_snr_0_1} show a set of predicted structures compared to the ground truth for cryoSPHERE and cryoStar. Both methods are able to recover the ground truth almost perfectly.

\begin{figure}
    \centering
    \includegraphics[width=0.2\textwidth]{figures/experiments/phy/ground_truth_1.pdf}
    \hspace{0.5cm}
    \includegraphics[width=0.2\textwidth]{figures/experiments/phy/ground_truth_500.pdf}
    \hspace{0.5cm}
    \includegraphics[width=0.24\textwidth]{figures/experiments/phy/ground_truth_999.pdf}\\
    \includegraphics[width=0.2\textwidth]{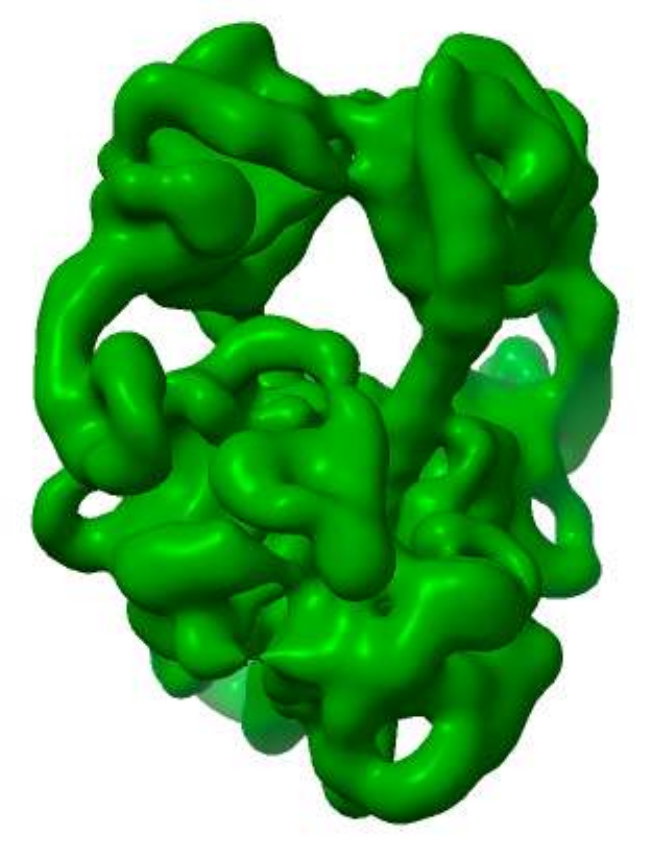}\hspace{0.5cm}
    \includegraphics[width=0.2\textwidth]{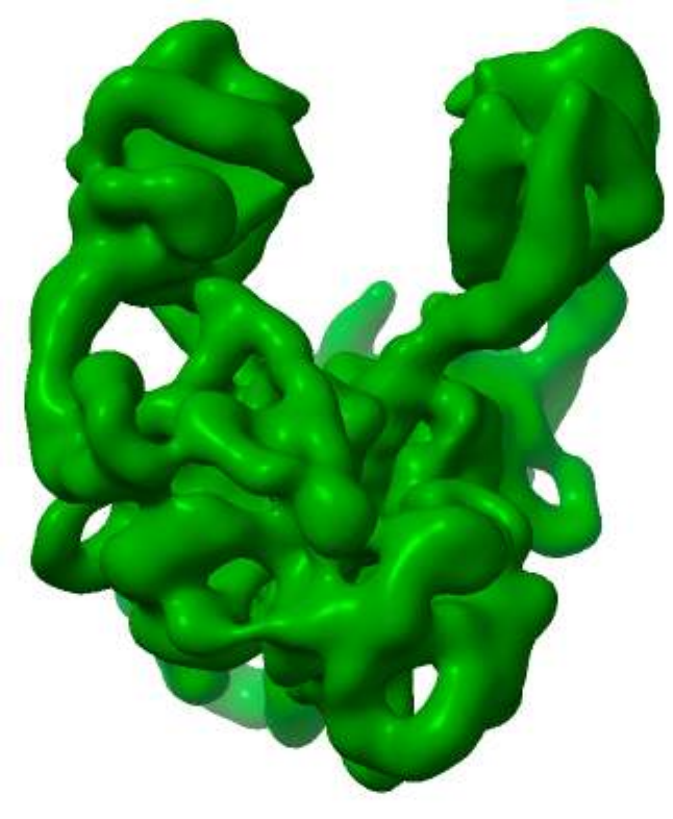}\hspace{0.5cm}
    \includegraphics[width=0.24\textwidth]{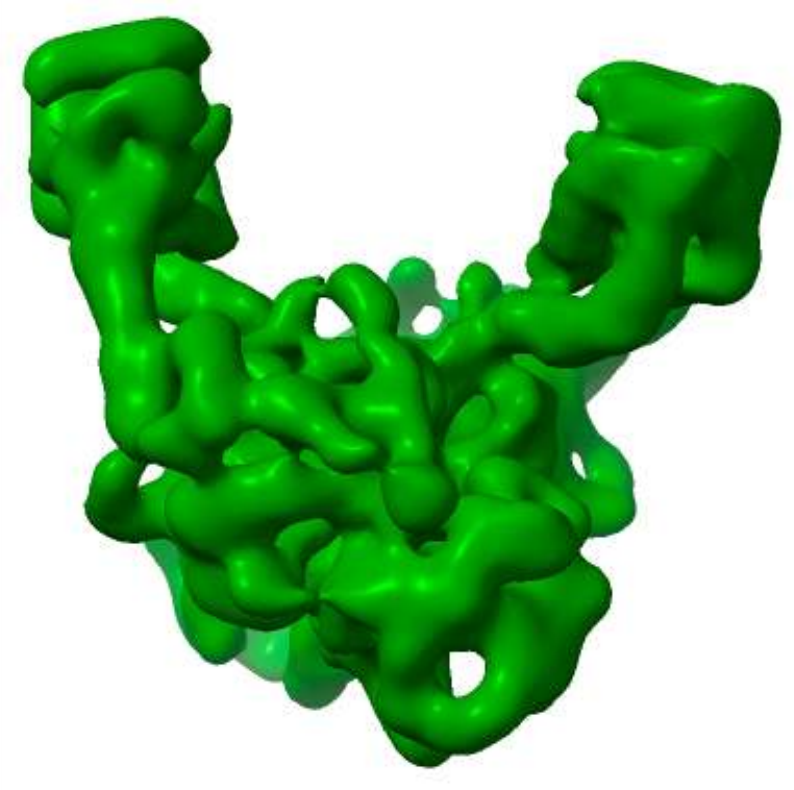}
    \caption{SNR $0.1$. Examples of reconstructed volumes by cryoDRGN. Green is cryoDRGN and gray is the corresponding ground truth volume.}
    \label{fig:volumes_cryodrgn_snr_0_1}
\end{figure}

Figure \ref{fig:volumes_cryodrgn_snr_0_1} shows examples of cryoDRGN predicted volume together with the corresponding ground truth. The method is able to recover the ground truth volumes almost perfectly.

\begin{figure}
    \centering
    \includegraphics[width=0.33\textwidth]{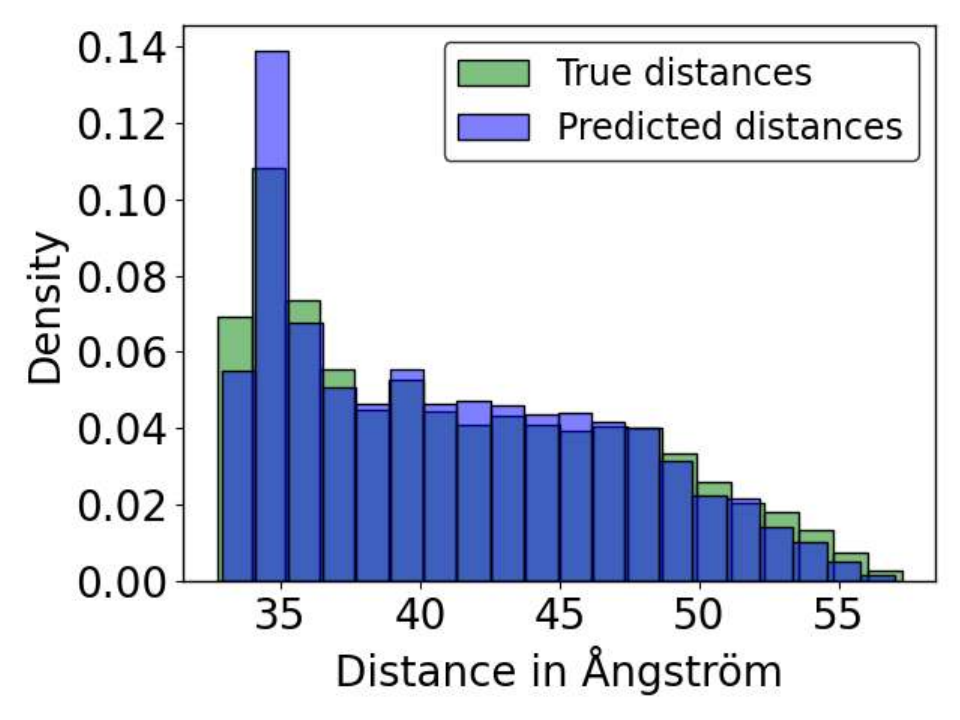}%
    \includegraphics[width=0.33\textwidth]{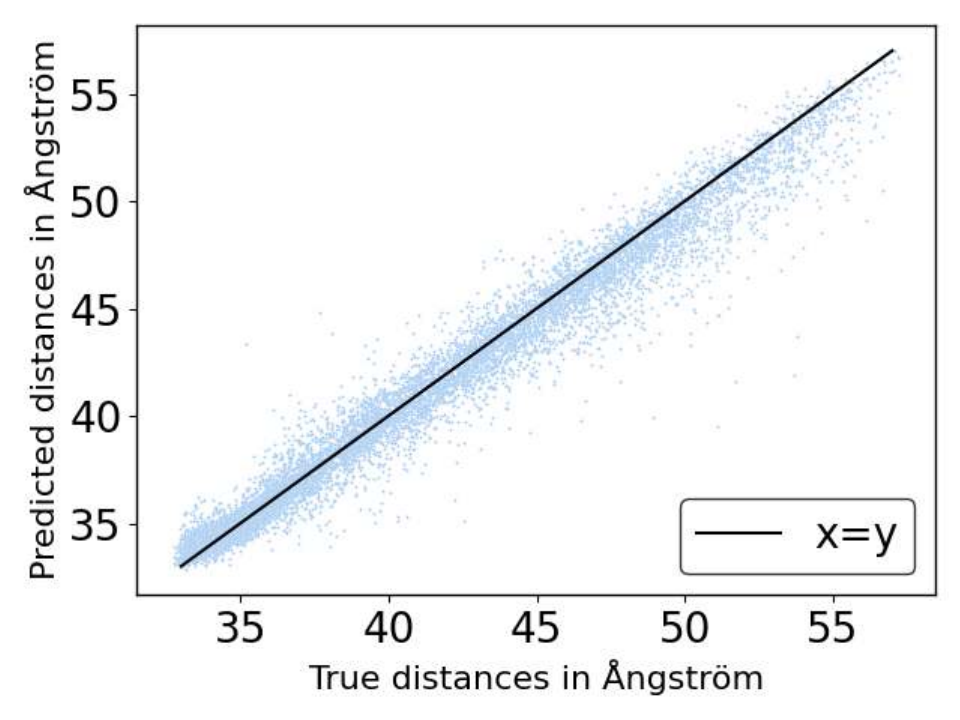}%
  \includegraphics[width=0.33\textwidth]{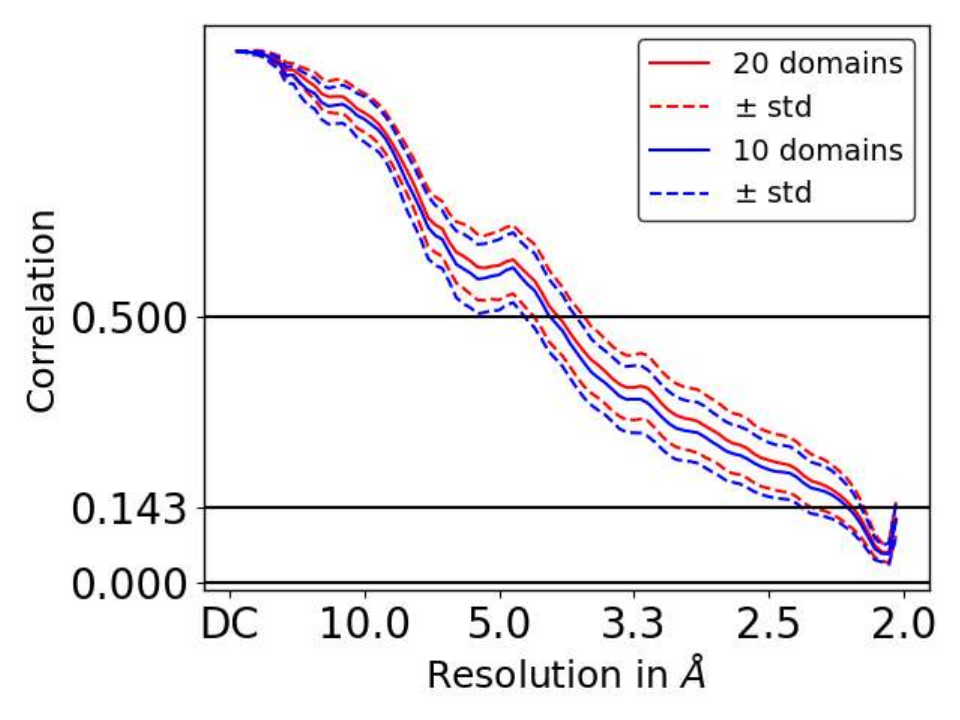}
    \caption{Results for cryoSPHERE on SNR $0.1$ with $\Ndomains=10$. Left: distribution of distances predicted by cryoSPHERE compared to the ground truth distribution. Middle: true versus predicted distances for cryoSPHERE. Right: FSC comparison between cryoSPHERE with $\Ndomains=20$ and $\Ndomains=10$.}
    \label{fig:snr_0_1_cryosphere_10_domains}
\end{figure}

\subsubsection{SNR $0.01$}\label{subsubsec:0.01}
%We plot a set of structures that cryoSPHERE predicts together with the corresponding ground truth structure in Figure \ref{fig:150k_comparison_structures}. We see an excellent agreement of the predicted opening with the ground truth.
%As explained in section \ref{subsec:MD}, Figure \ref{fig:fsc_curves} shows that we outperform cryoDRGN in terms of resolution at both the $0.5$ and $0.143$ cutoffs. This of course depends on the base structure we are using.
This subsection describes the results for cryoSPHERE, cryoStar and cryoDRGN on our molecular dynamics simulation dataset with SNR $0.01$.

\begin{figure}
    \centering
    \includegraphics[width=0.33\textwidth]{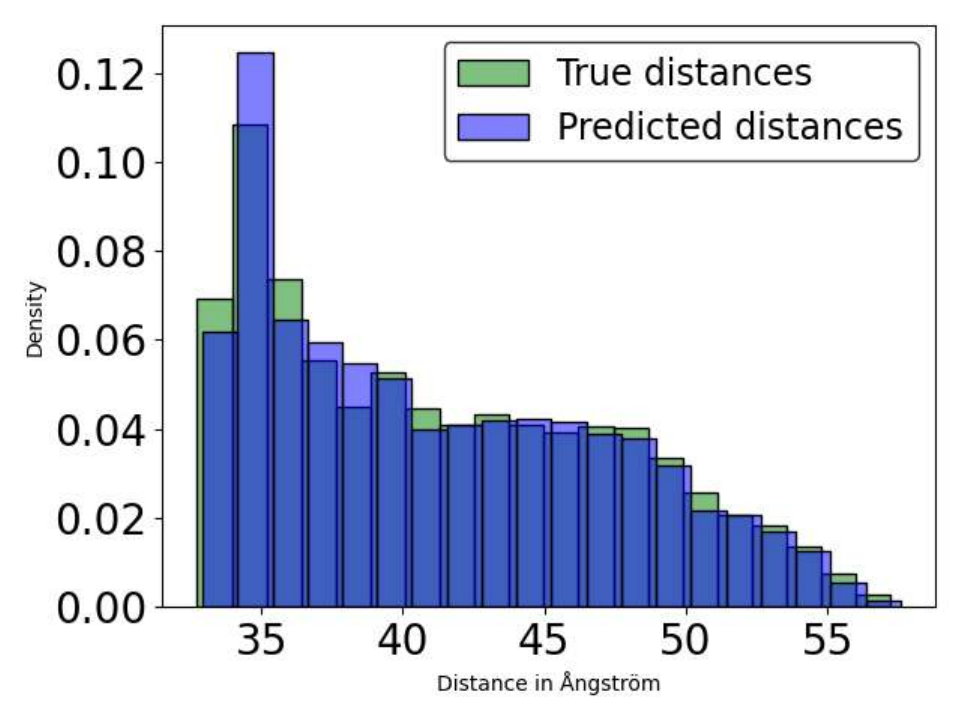}%
    \includegraphics[width=0.33\textwidth]{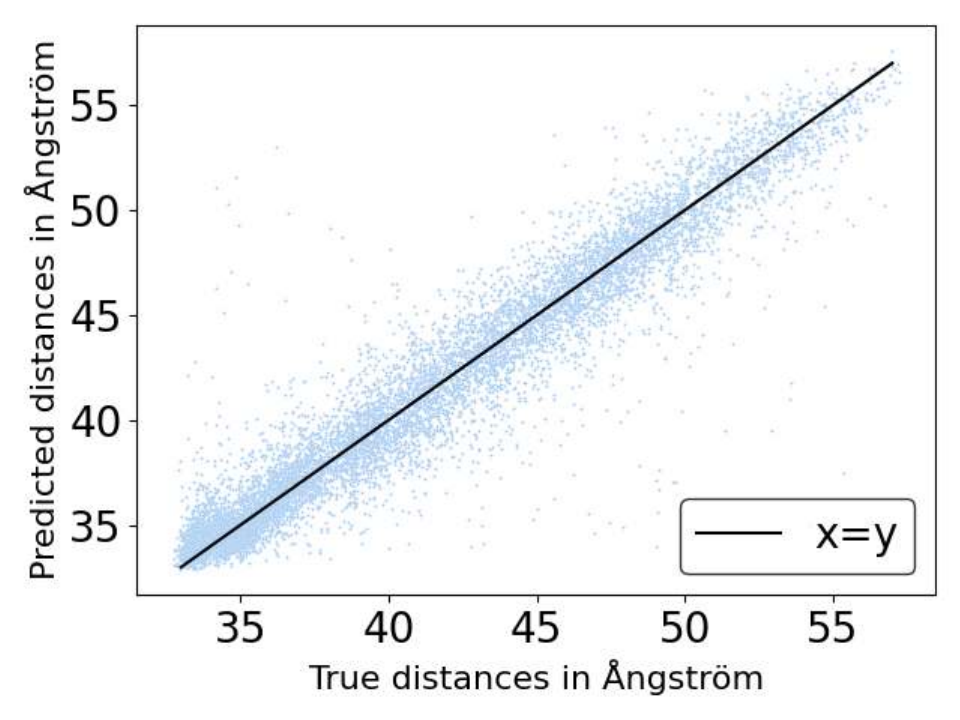}%
    \includegraphics[width=0.2\textwidth]{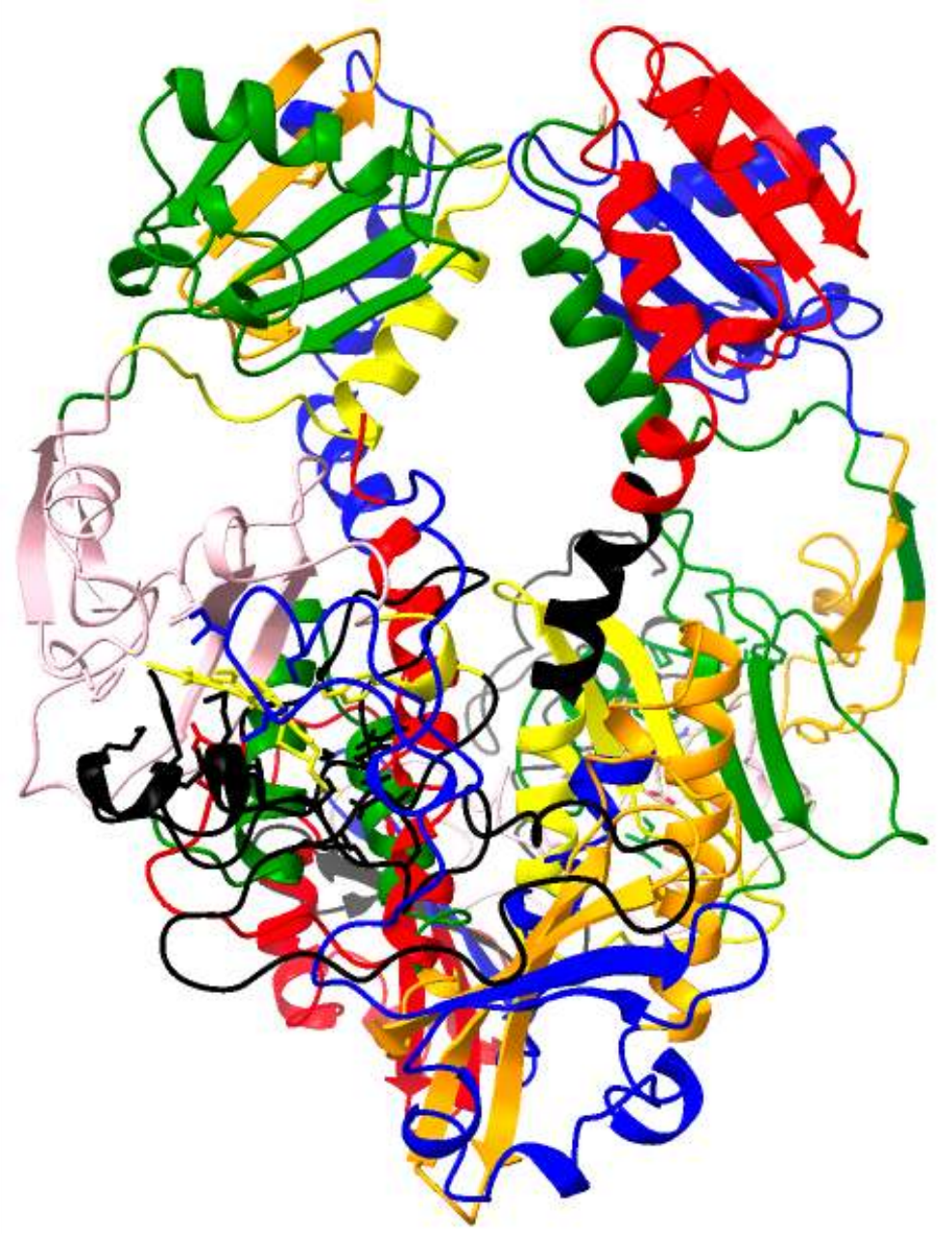}%
    \caption{Results for cryoSPHERE on SNR $0.01$ with $\Ndomains=25$. Left: distribution of distances predicted by cryoSPHERE compared to the ground truth distribution. Middle: true versus predicted distances for cryoSPHERE. Right: segments decomposition.}
    \label{fig:snr_0_01_cryosphere}
\end{figure}

\begin{figure}
    \centering
    \includegraphics[width=0.33\textwidth]{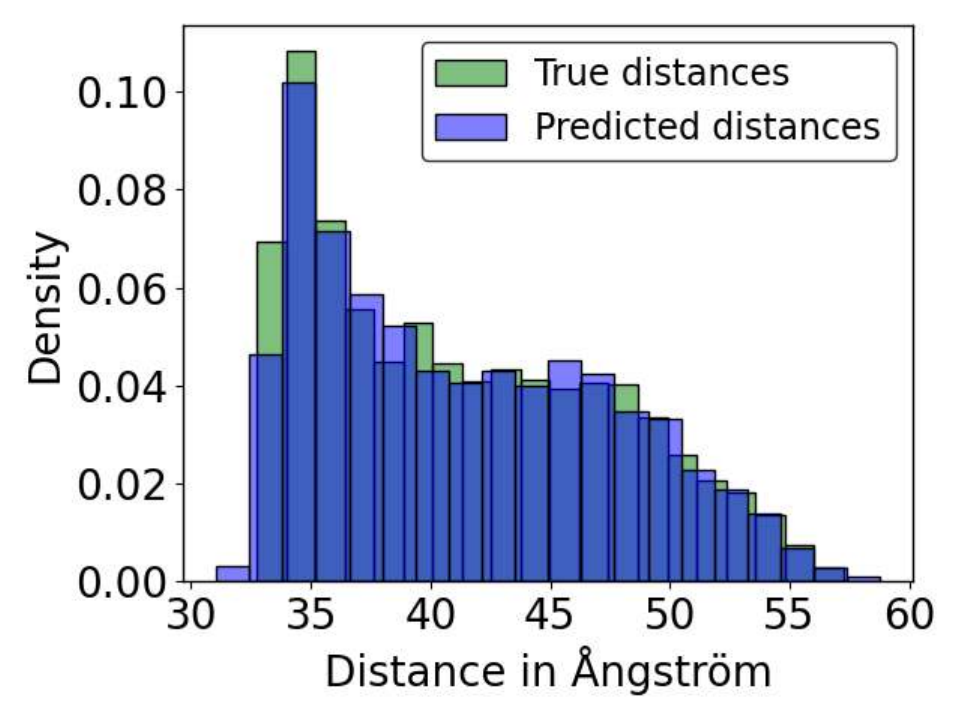}%
    \includegraphics[width=0.33\textwidth]{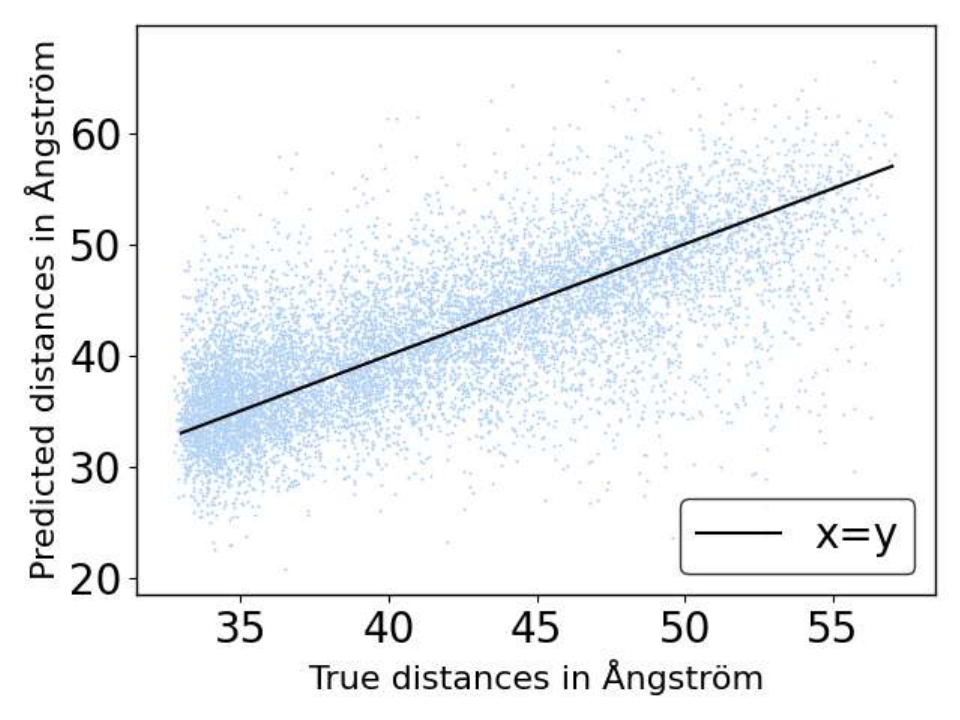}%
    \caption{Results for cryoStar on SNR $0.01$. Left: distribution of distances predicted by cryoStar compared to the ground truth distribution. Right: true versus predicted distances for cryoStar}
    \label{fig:snr_0_01_cryostar}
\end{figure}

Figures \ref{fig:snr_0_01_cryosphere} and \ref{fig:snr_0_01_cryostar} show that both cryoSPHERE and cryoStar are able to recover the ground truth distribution of distance. Also, given an image, both methods are able to recover the correct conformation.

\begin{figure}
    \centering
    \includegraphics[width=0.4\textwidth]{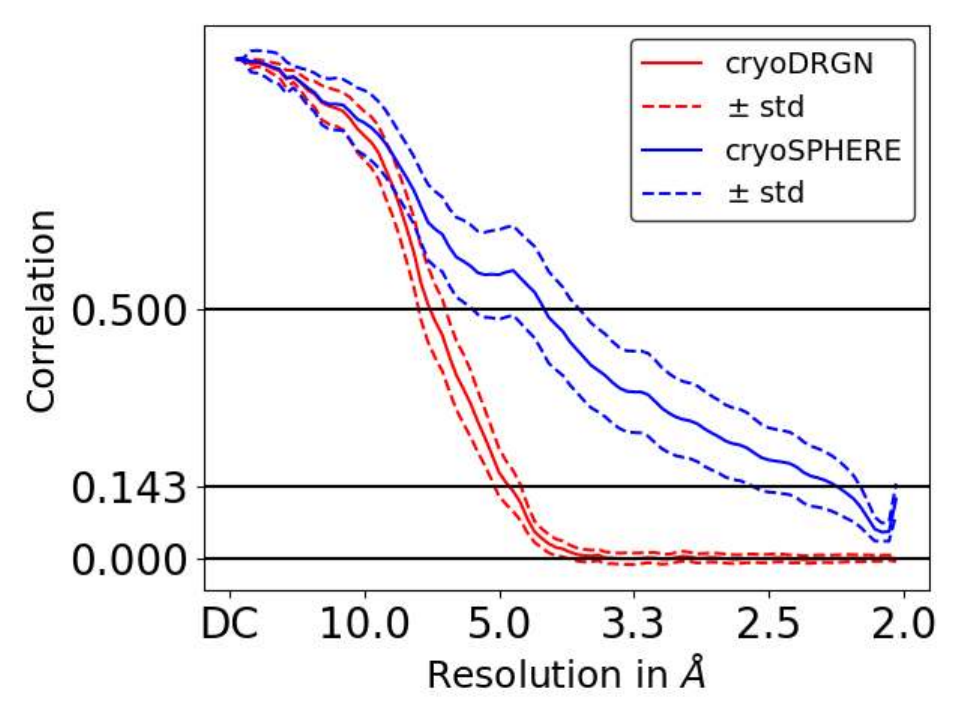}
    \includegraphics[width=0.4\textwidth]{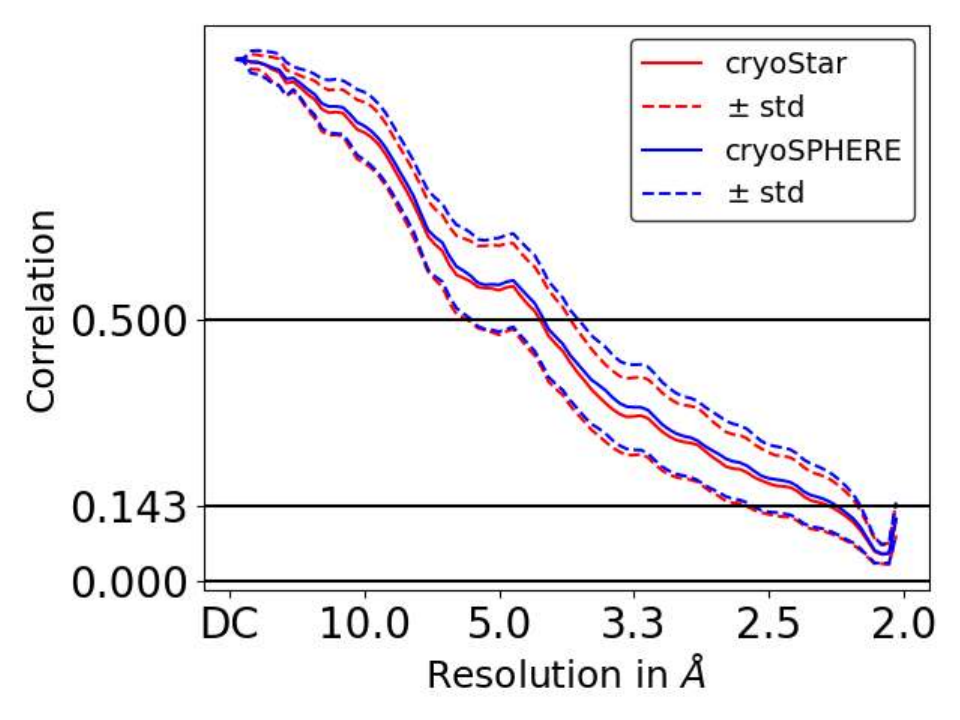}\\
    \includegraphics[width=0.4\textwidth]{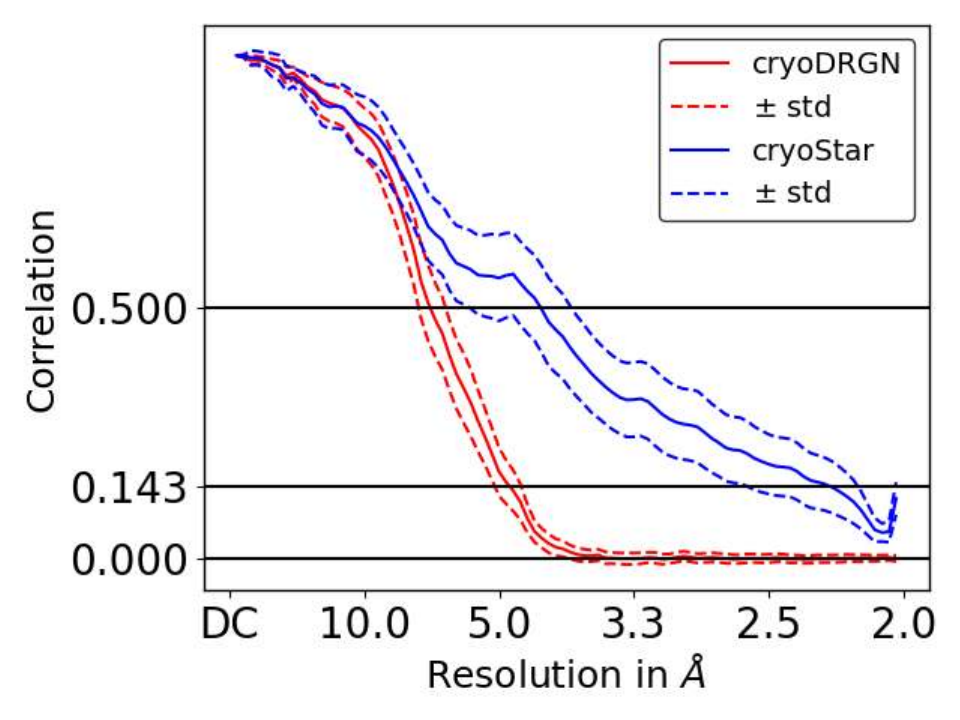}
    \includegraphics[width=0.4\textwidth]{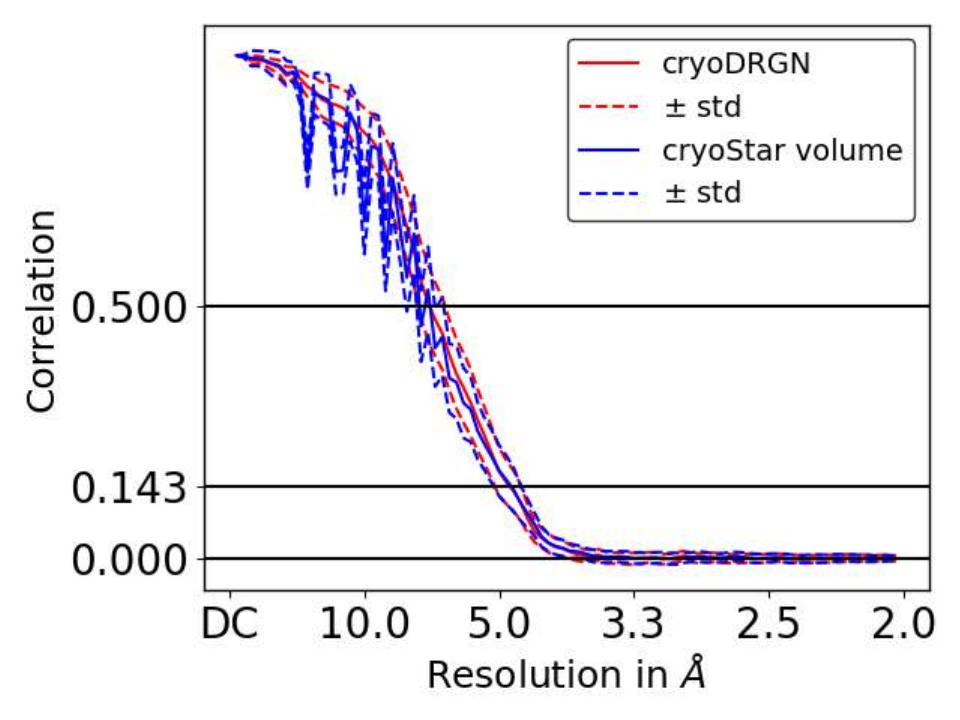}
    \caption{SNR $0.01$. Mean Fourier shell correlation ($\pm$ std) comparison for cryoSPHERE with $\Ndomains=25$, cryoStar and cryoDRGN. Top left: cryoSPHERE versus cryoDRGN. Top right: cryoSPHERE vs cryoStar. Bottom left: cryoDRGN vs cryoStar. Bottom right: cryoStar volume method vs cryoDRGN.}
   \label{fig:snr_0_01_fsc_comparison}
\end{figure}

\begin{figure}
    \centering
\includegraphics[scale=0.3]{figures/experiments/phy/ground_truth_1.pdf}\hspace{0.5cm}    
\includegraphics[scale=0.3]{figures/experiments/phy/ground_truth_500.pdf}\hspace{0.5cm}  
\includegraphics[scale=0.3]{figures/experiments/phy/ground_truth_999.pdf} \\    
\includegraphics[scale=0.2]{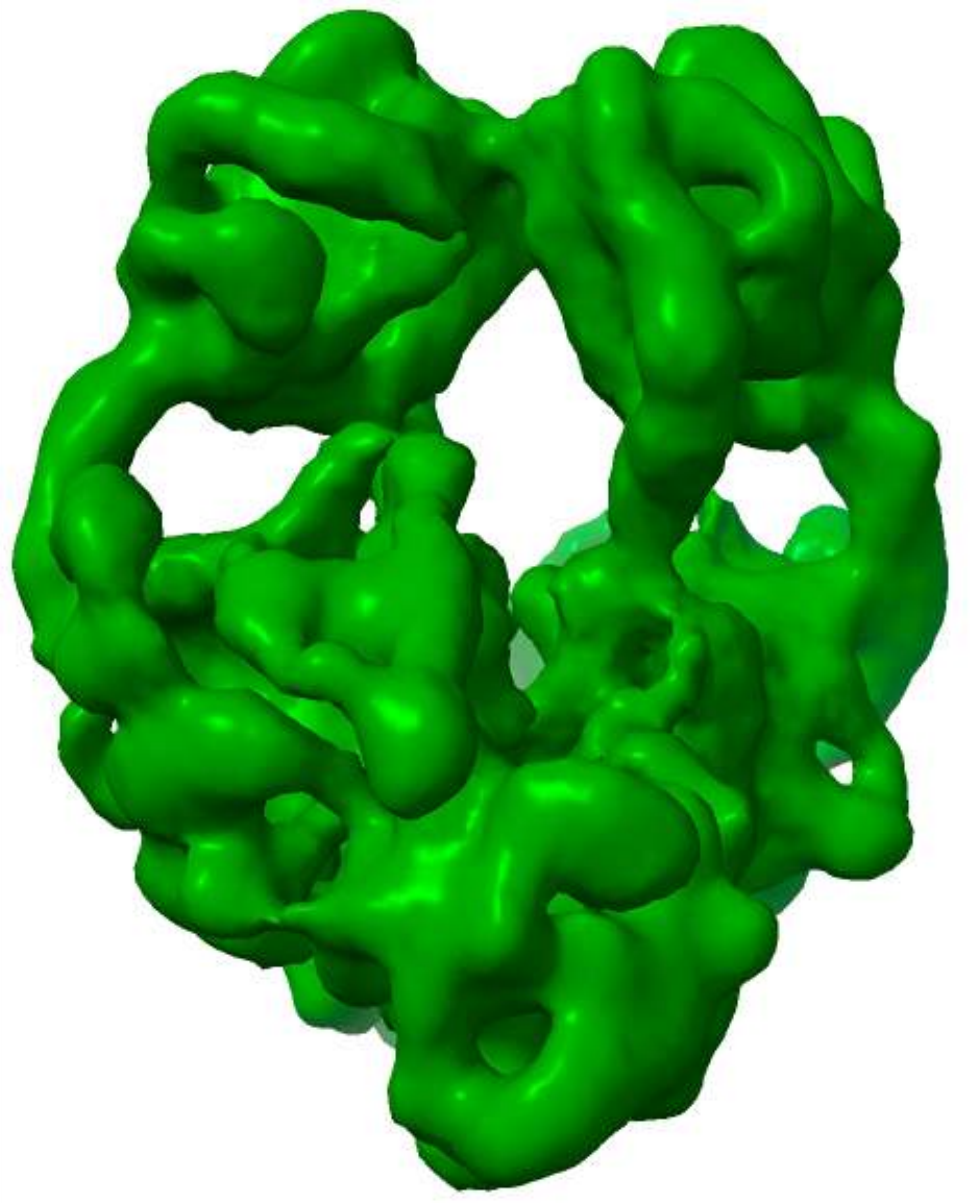}\hspace{0.5cm} 
\includegraphics[scale=0.2]{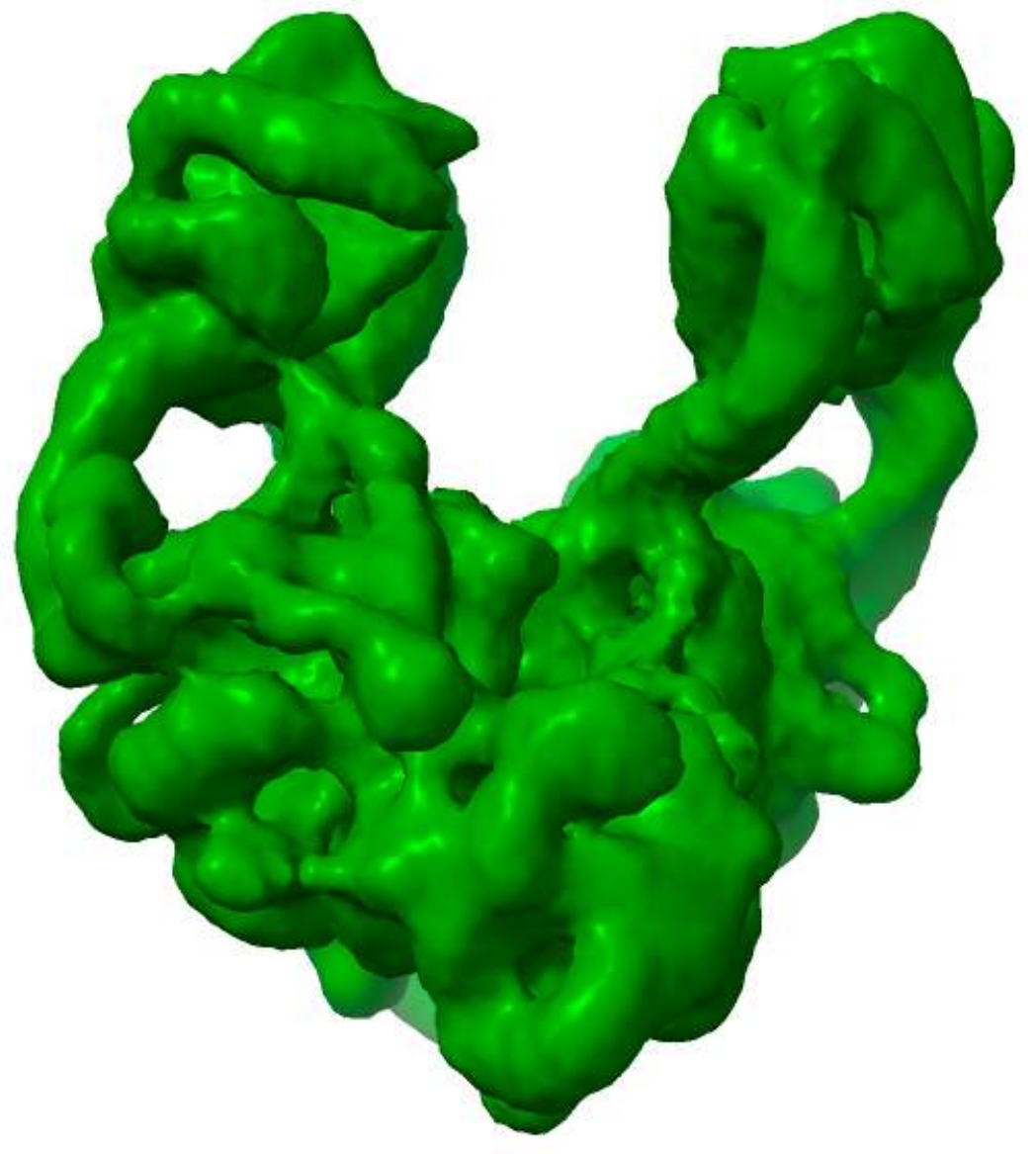}\hspace{0.5cm} 
\includegraphics[scale=0.2]{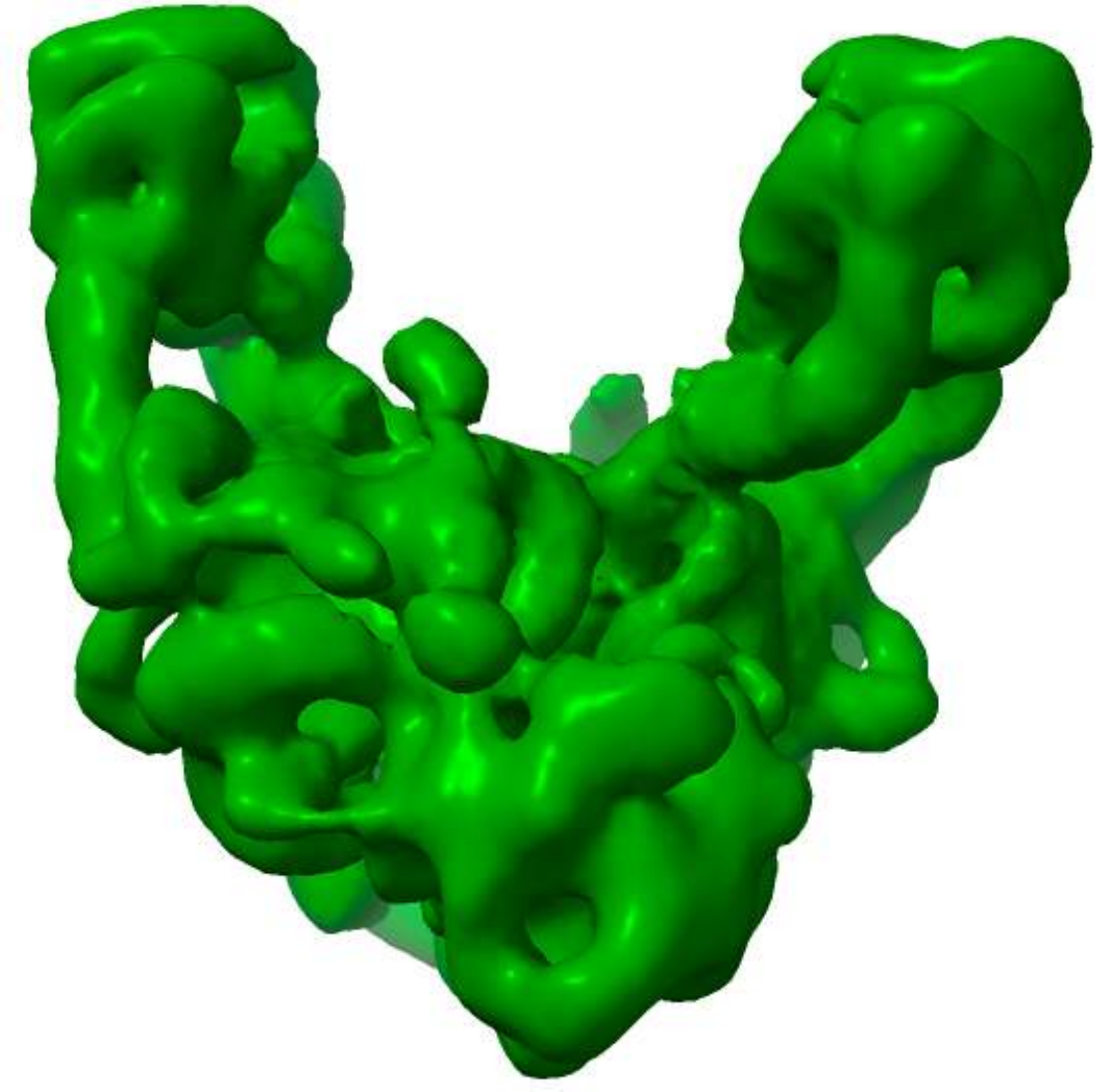}%
\caption{SNR $0.01$. Example volumes produced by the volume method of cryoStar with the corresponding ground truth. Top: ground truth. Bottom: corresponding volumes predicted by the volume method of cryoStar.}
    \label{fig:snr_0_01_cryostar_volume_method}
\end{figure}

Figure \ref{fig:snr_0_01_fsc_comparison} shows that both cryoSPHERE and cryoStar are outperforming cryoDRGN at the $0.5$ and $0.143$ cutoffs while having very similar performances. 
In addition, we can see that the volume method of cryoStar perform similarly, if not worse, than cryoDRGN. That seems to indicate that this volume method does not benefit from the information gained by the structural method. Figure \ref{fig:snr_0_01_cryostar_volume_method} shows three examples of volumes reconstructed with the volume method of cryoStar, together with the corresponding ground truth.

\begin{figure}
    \centering
    \includegraphics[width=0.28\textwidth]{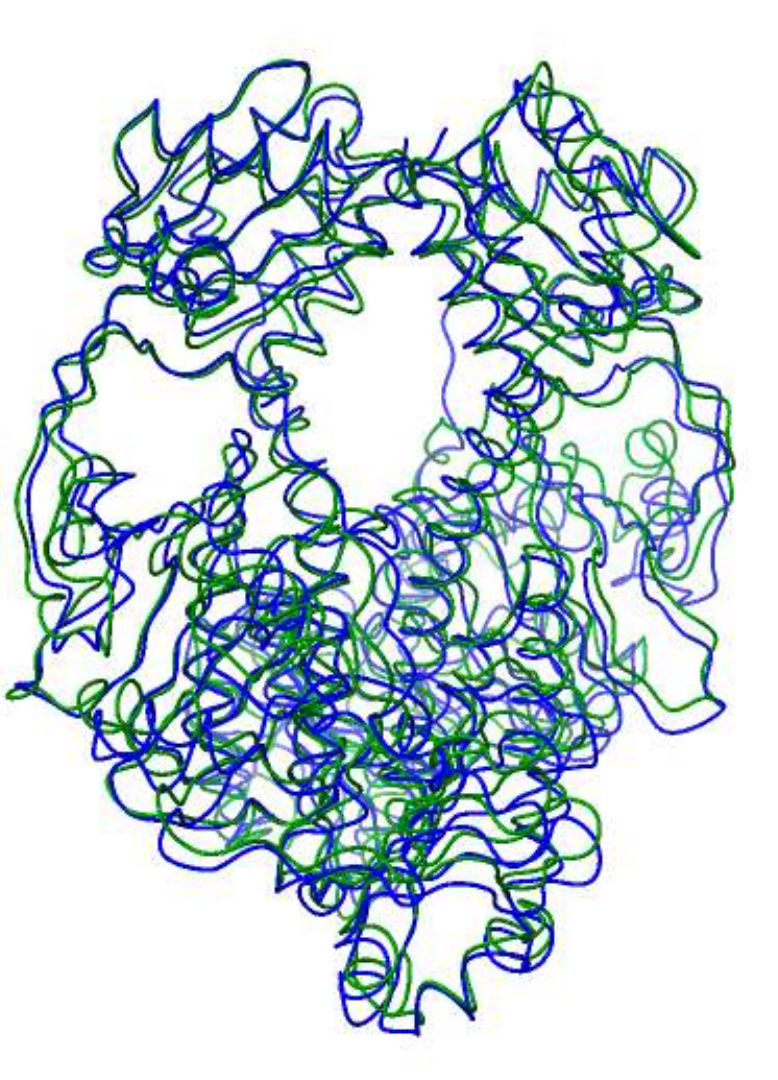}
    \includegraphics[width=0.28\textwidth]{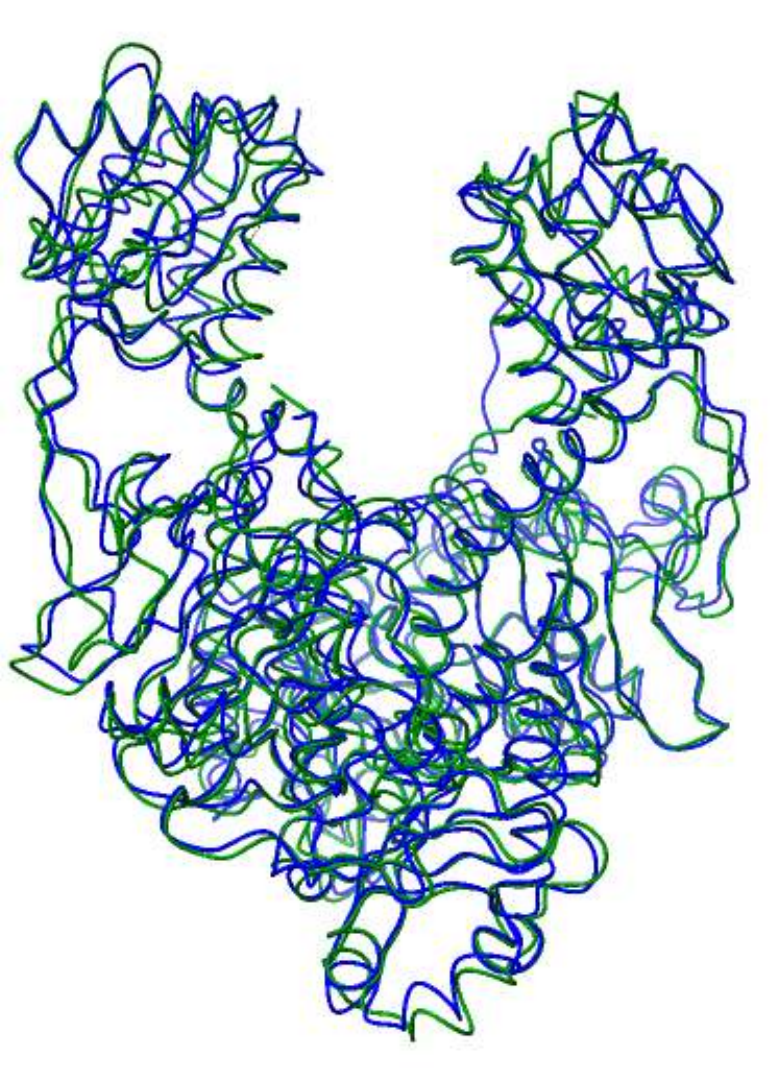}
    \includegraphics[width=0.35\textwidth]{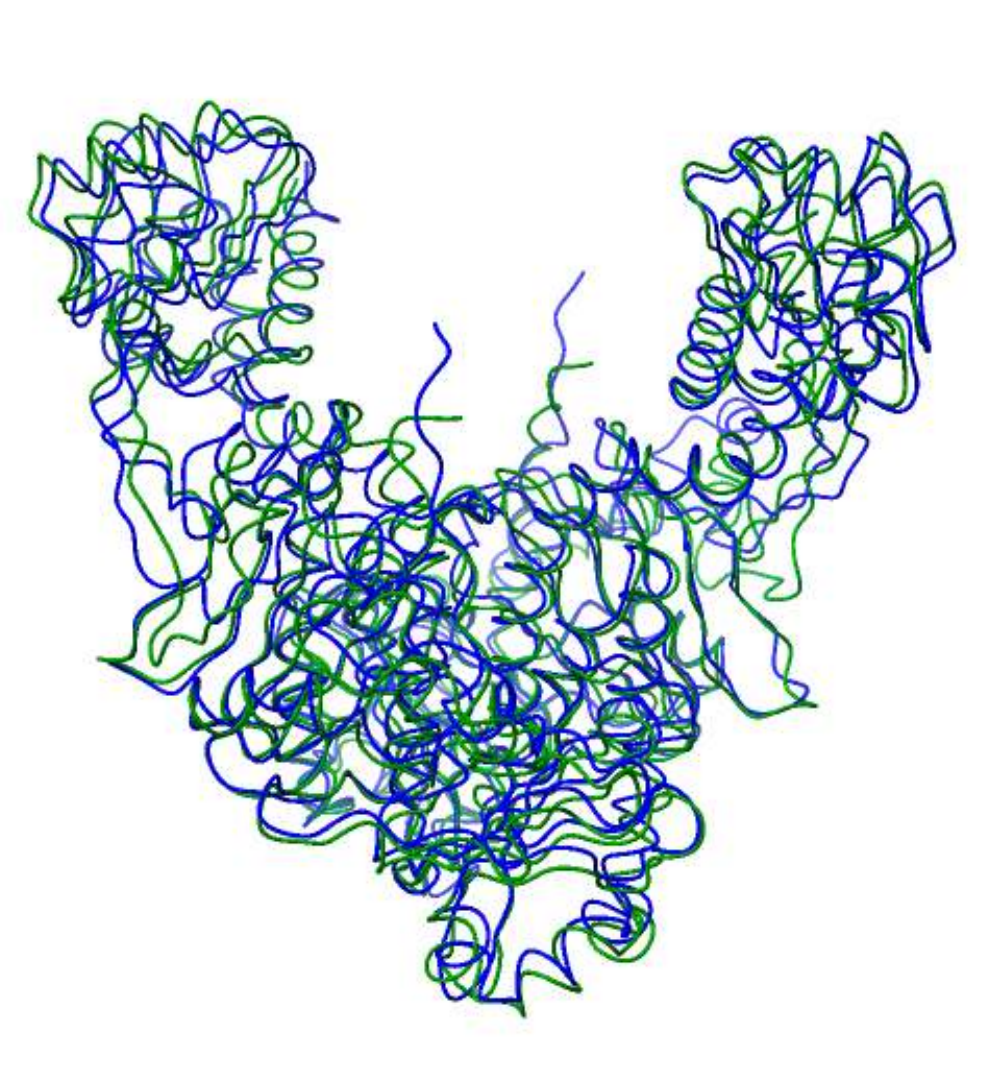}
    \caption{SNR $0.01$. Examples of reconstructed structures by cryoSHERE. Blue is predicted, green is ground truth. cryoSPHERE is able to recover the right conformation. Left to right: image number 11, 5001 and 9999.}
    \label{fig:structures_cryoSPHERE_snr_0_01}
\end{figure}

\begin{figure}
    \centering
    \includegraphics[width=0.28\textwidth]{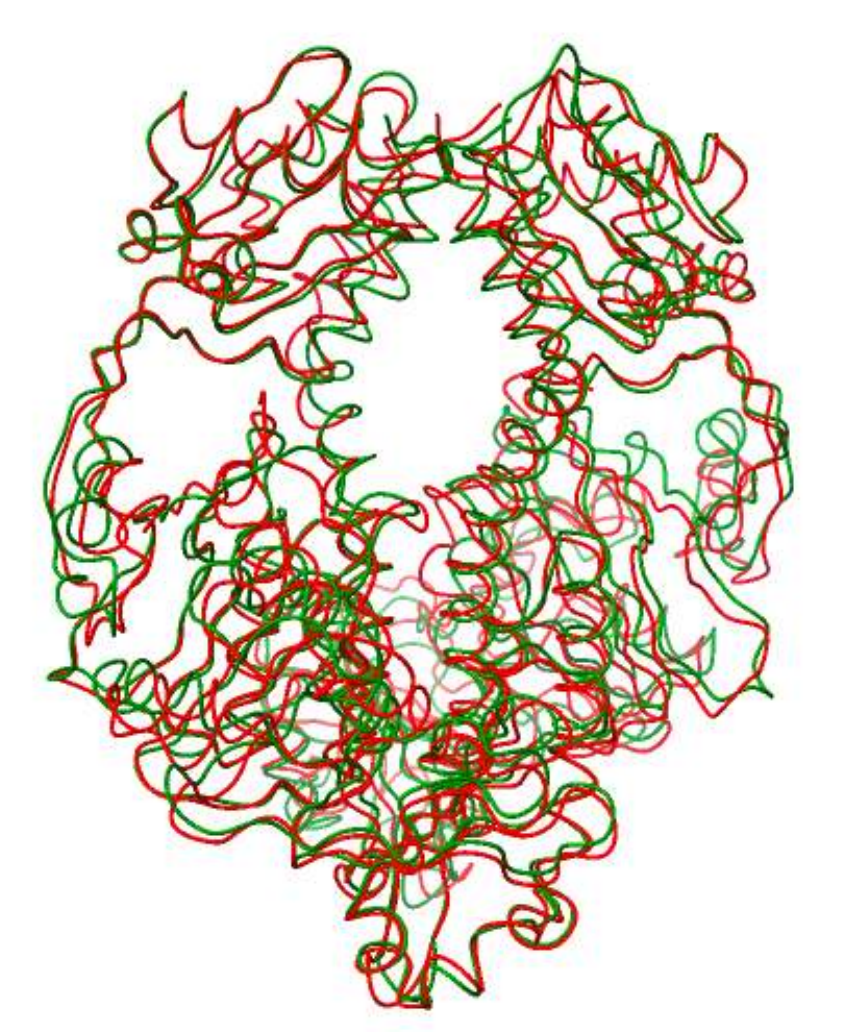}
    \includegraphics[width=0.28\textwidth]{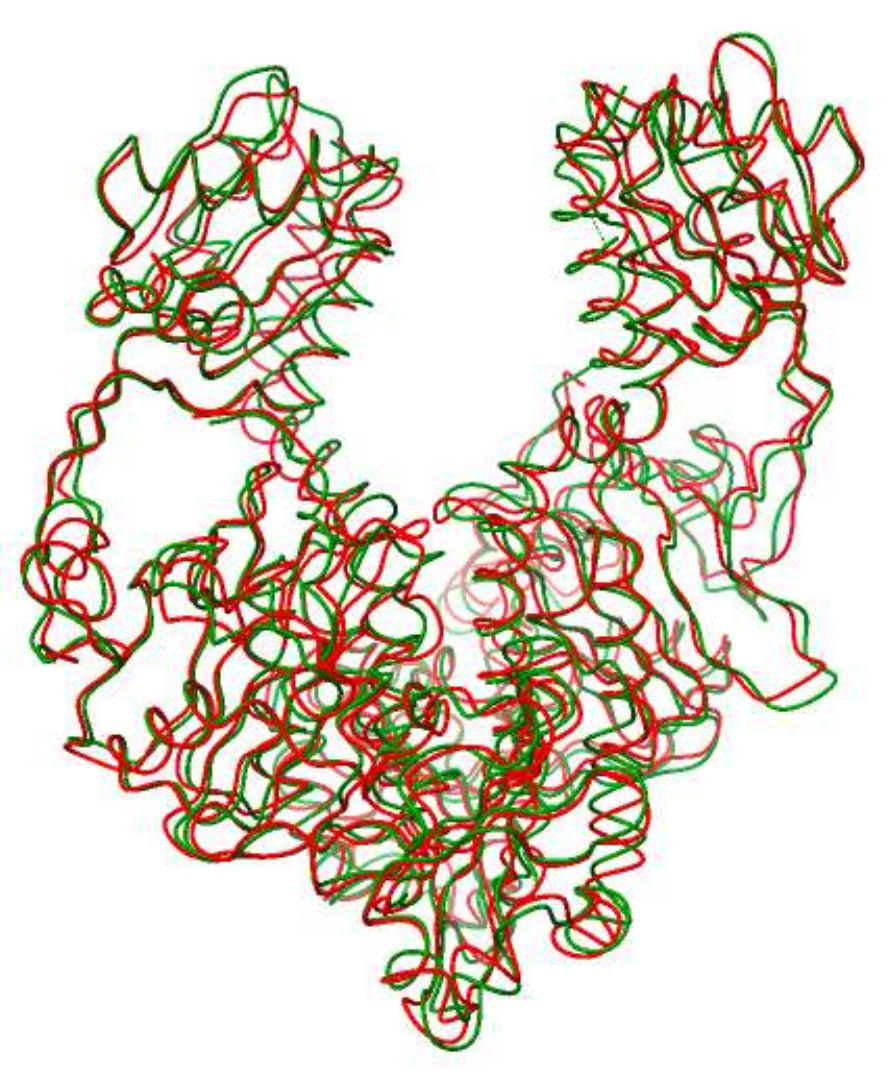}
    \includegraphics[width=0.35\textwidth]{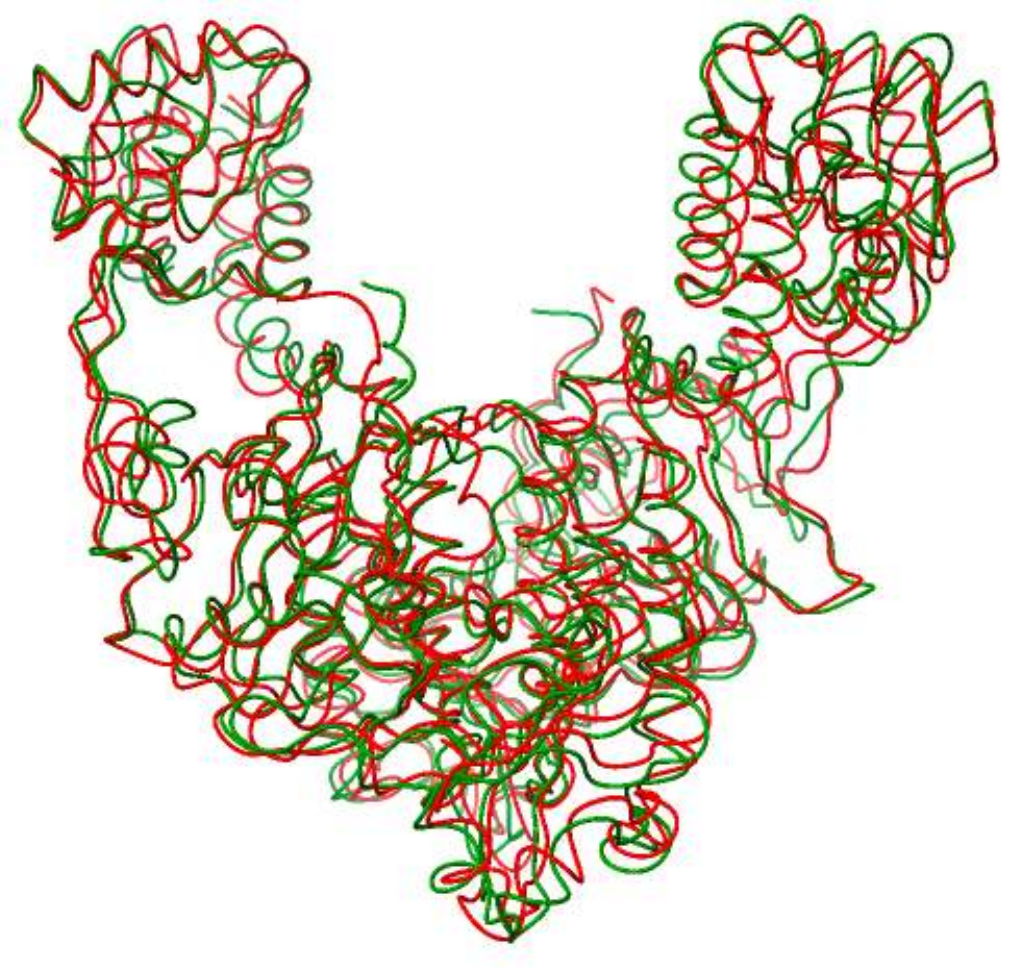}
    \caption{SNR $0.01$. Examples of reconstructed structures by cryoStar. Red is predicted, green is ground truth. cryoStar is able to recover the right conformation. Left to right: image number 11, 5001 and 9999.}
    \label{fig:structures_cryoStar_0_01}
\end{figure}

Figure \ref{fig:structures_cryoSPHERE_snr_0_01} and \ref{fig:structures_cryoStar_0_01} shows a set of predicted structures compared to the ground truth. Both cryoSPHERE and cryoStar are able to recover the ground truth almost perfectly.

\begin{figure}
    \centering
    \includegraphics[width=0.2\textwidth]{figures/experiments/phy/ground_truth_1.pdf}
    \hspace{0.5cm}
    \includegraphics[width=0.2\textwidth]{figures/experiments/phy/ground_truth_500.pdf}
    \hspace{0.5cm}
    \includegraphics[width=0.24\textwidth]{figures/experiments/phy/ground_truth_999.pdf}\\
    \includegraphics[width=0.2\textwidth]{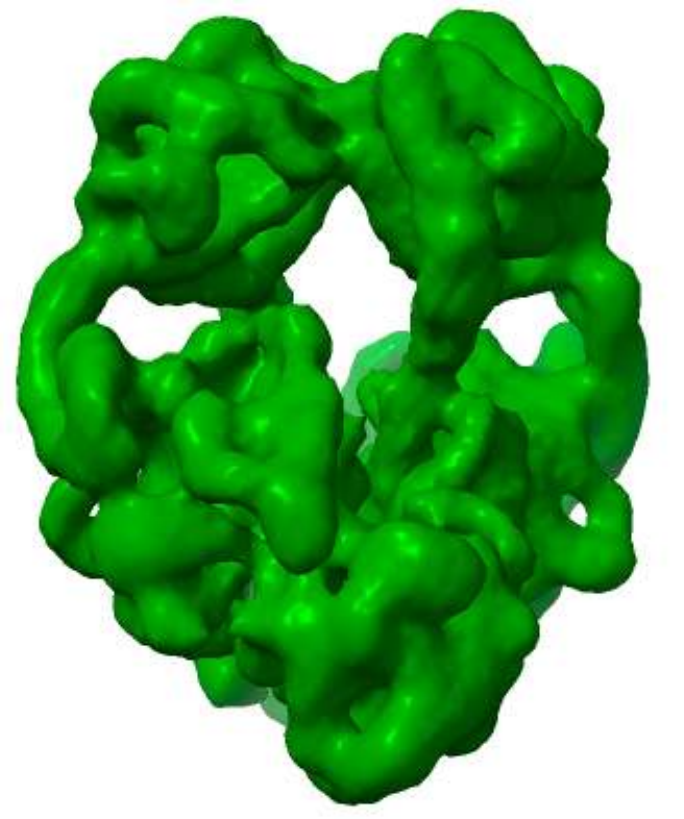}\hspace{0.5cm}
    \includegraphics[width=0.2\textwidth]{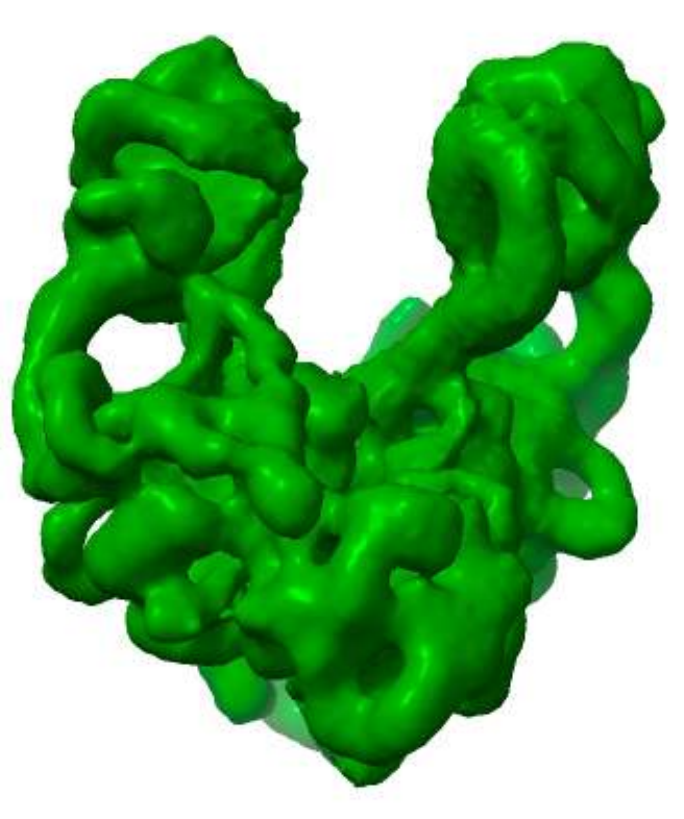}\hspace{0.5cm}
    \includegraphics[width=0.24\textwidth]{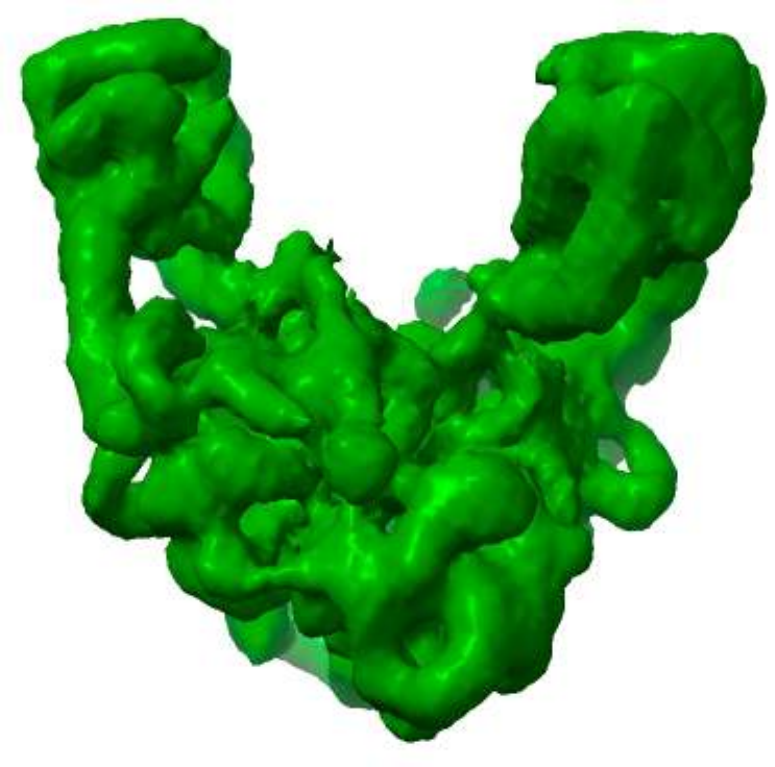}
    \caption{SNR $0.01$. Examples of reconstructed volumes by cryoDRGN. Green is cryoDRGN and gray is the corresponding ground truth volume.}
    \label{fig:volumes_cryodrgn_snr_0_01}
\end{figure}

Finally, Figure \ref{fig:volumes_cryodrgn_snr_0_01} shows examples of cryoDRGN predicted volume together with the corresponding ground truth. The method is able to recover the ground truth volumes almost perfectly, with a somewhat lower resolution compared to Figure \ref{fig:volumes_cryodrgn_snr_0_1}.

\subsubsection{SNR $0.001$}\label{subsubsec:0.001}
This subsection complements the results of Section \ref{subsec:MD}.

\begin{figure}
    \centering
    \includegraphics[width=0.31\textwidth]{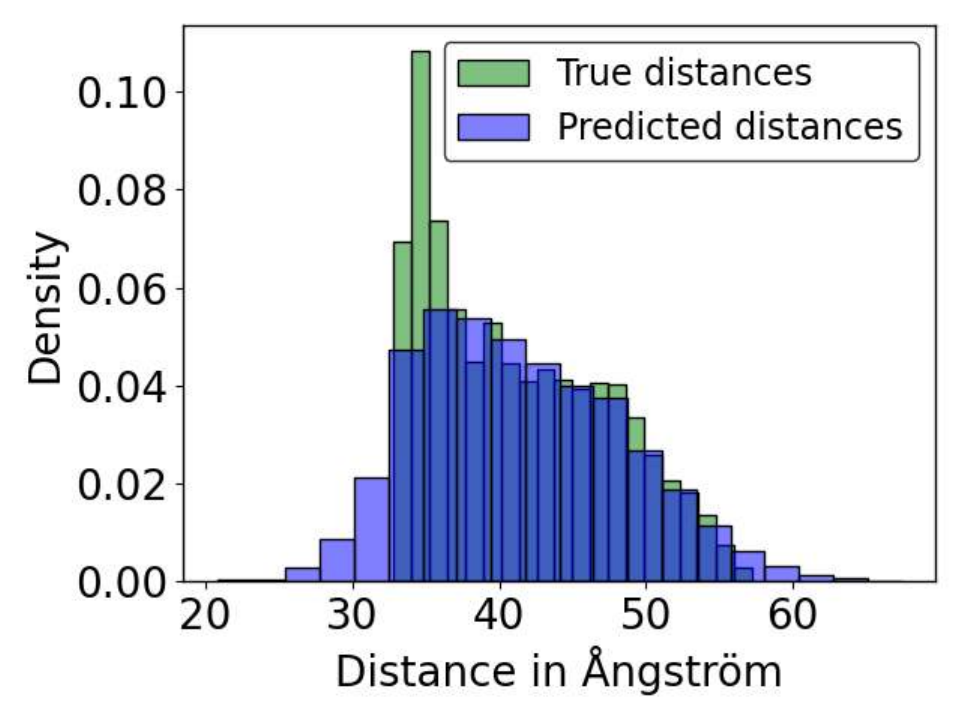}
    \includegraphics[width=0.31\textwidth]{figures/experiments/phy/snr_0_001/cryostar_true_vs_predicted_distance.pdf}
    \includegraphics[width=0.31\textwidth]{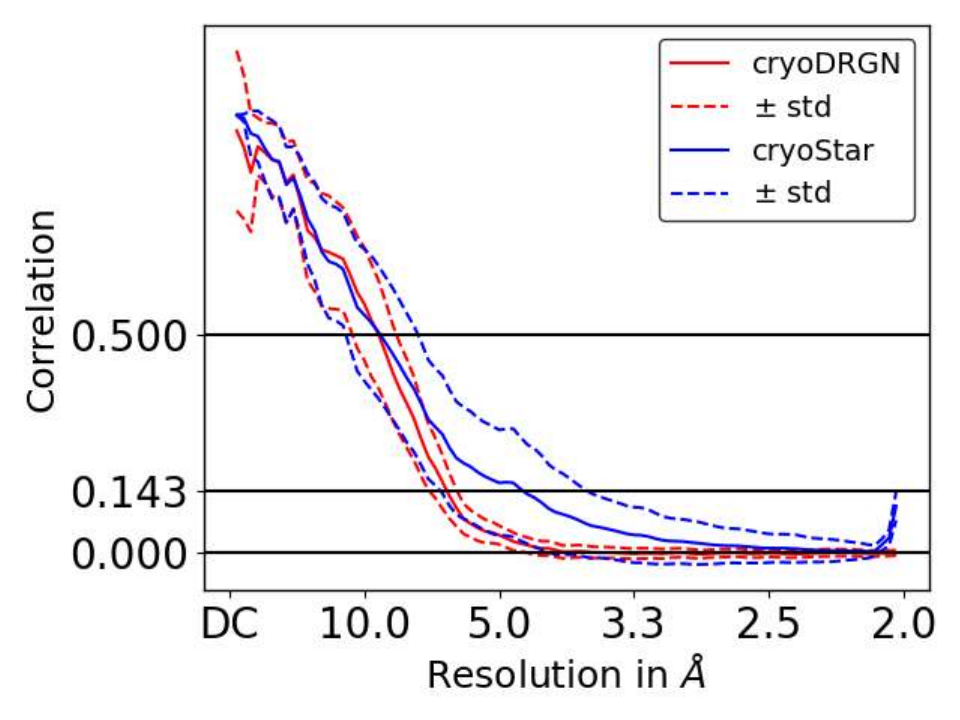}
    \caption{Results for cryoStar on SNR $0.001$. Left: distribution of distances predicted by cryoStar compared to the ground truth distribution. Middle: true vs predicted distances for cryoStar. Right: FSC comparison between cryoStar and cryoDRGN}
    \label{fig:snr_0_001_cryoStar}
\end{figure}

\begin{figure}
    \centering    \includegraphics[width=0.35\textwidth]{figures/experiments/phy/snr_0_001/fsc_cryostar_cryodrgn.pdf}
\includegraphics[width=0.35\textwidth]{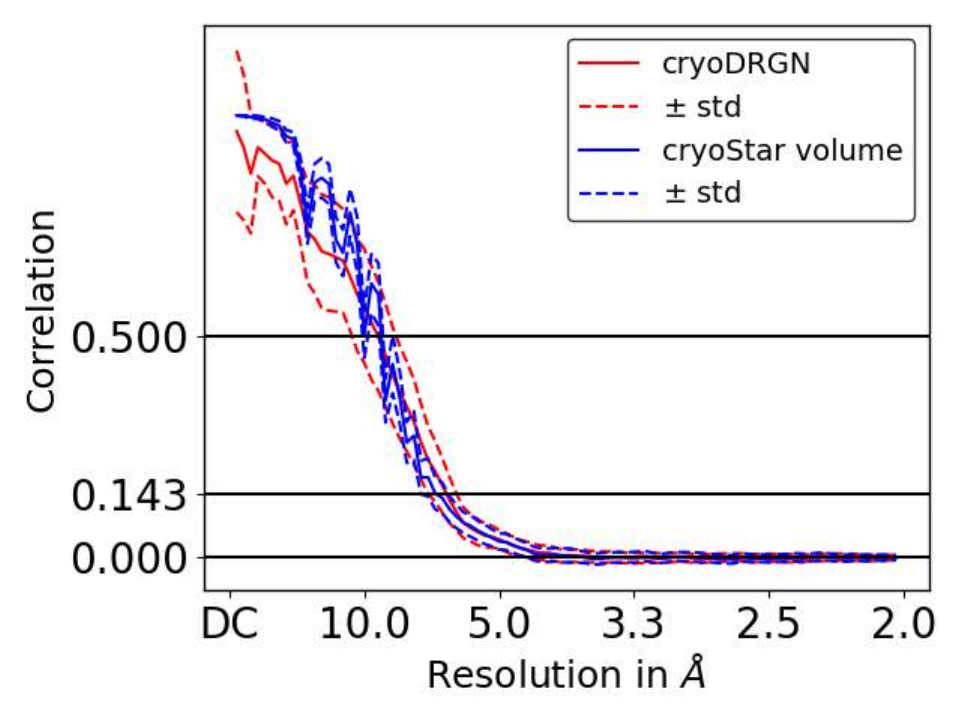}
    \caption{SNR $0.001$. Mean Fourier shell correlation ($\pm$ std) comparison for cryoStar and cryoDRGN. Left: cryoStar versus cryoDRGN. Right: cryoStar volume method vs cryoDRGN.}
   \label{fig:snr_0_001_fsc_comparison}
\end{figure}

\begin{figure}
    \centering
\includegraphics[scale=0.3]{figures/experiments/phy/ground_truth_1.pdf}\hspace{0.5cm}    
\includegraphics[scale=0.3]{figures/experiments/phy/ground_truth_500.pdf}\hspace{0.5cm}  
\includegraphics[scale=0.3]{figures/experiments/phy/ground_truth_999.pdf} \\    
\includegraphics[scale=0.2]{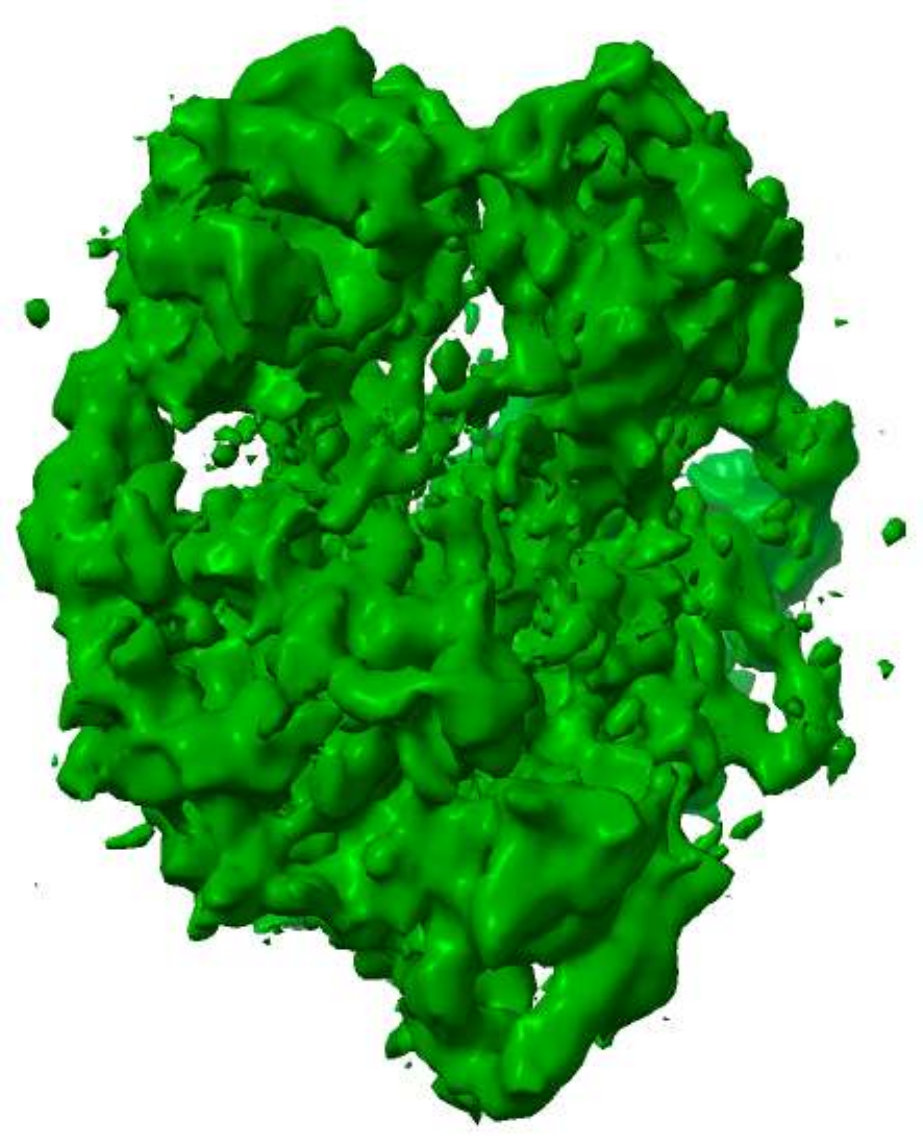}\hspace{0.5cm} 
\includegraphics[scale=0.2]{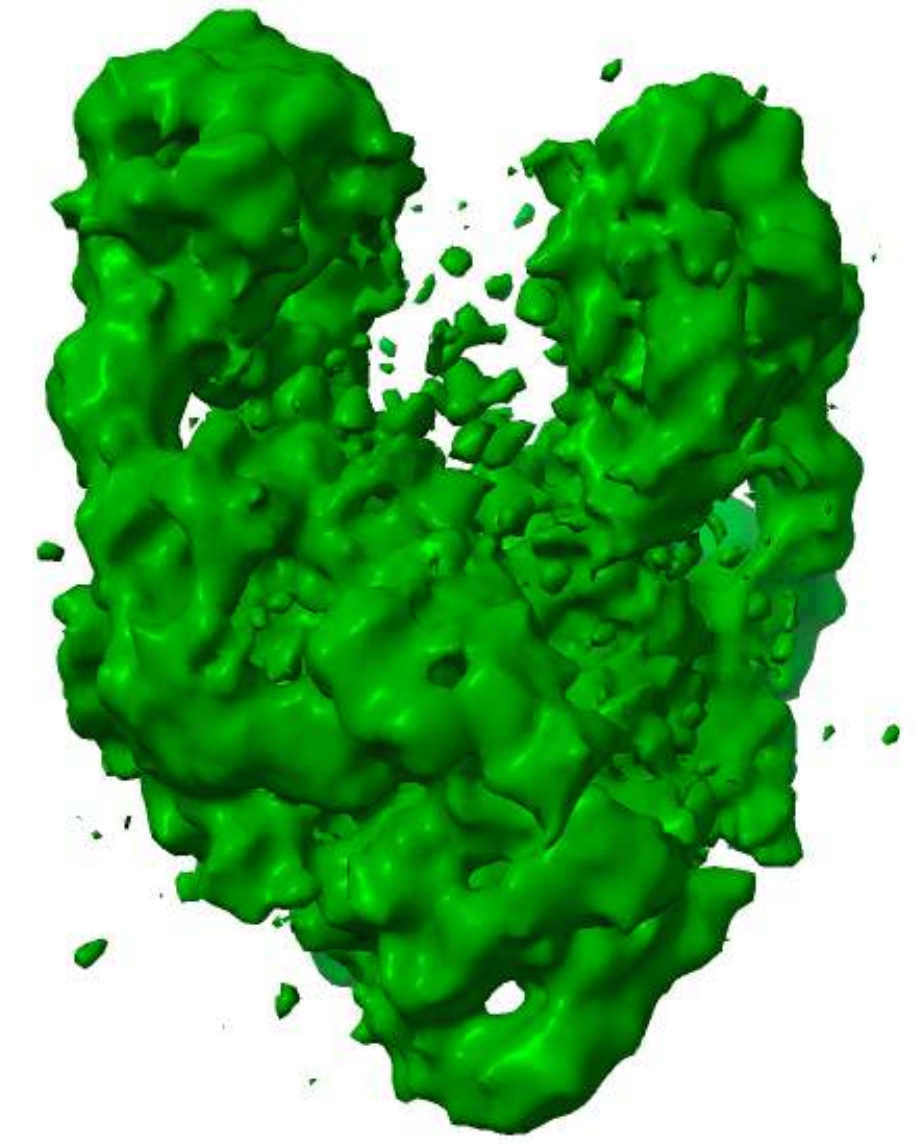}\hspace{0.5cm} 
\includegraphics[scale=0.2]{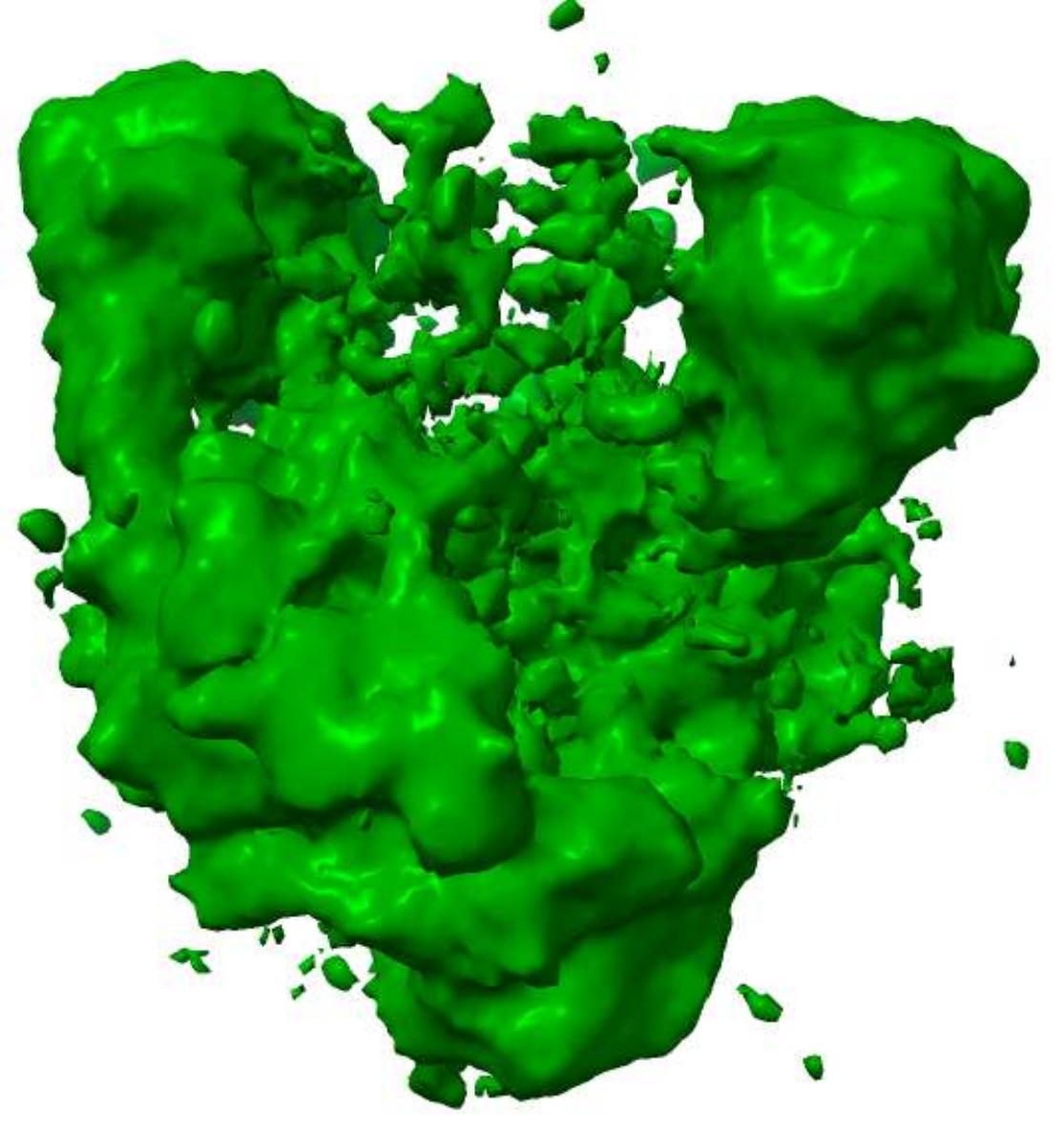}%
\caption{SNR $0.001$. Example volumes produced by the volume method of cryoStar with the corresponding ground truth. Top: ground truth. Bottom: corresponding volumes predicted by the volume method of cryoStar.}
\label{fig:snr_0_001_cryostar_volume_method}
\end{figure}

Figure \ref{fig:snr_0_001_cryoStar} shows that cryoStar is also able to recover the rough distribution of conformations as well as the correct conformation given an image. The comparisons of FSC shows that cryoStar perform similarly to cryoDRGN at the $0.5$ cutoff but better at the $0.143$.
In addition, Figure \ref{fig:snr_0_001_fsc_comparison} shows that cryoStar volume method does not perform significantly better than cryoDRGN, in spite of using the information given by the structural method of cryoStar. Figure \ref{fig:snr_0_001_cryostar_volume_method} shows a set of example volumes reconstructed by the volume method of cryoStar and the corresponding ground truth.

\begin{figure}
    \centering
    \includegraphics[width=0.28\textwidth]{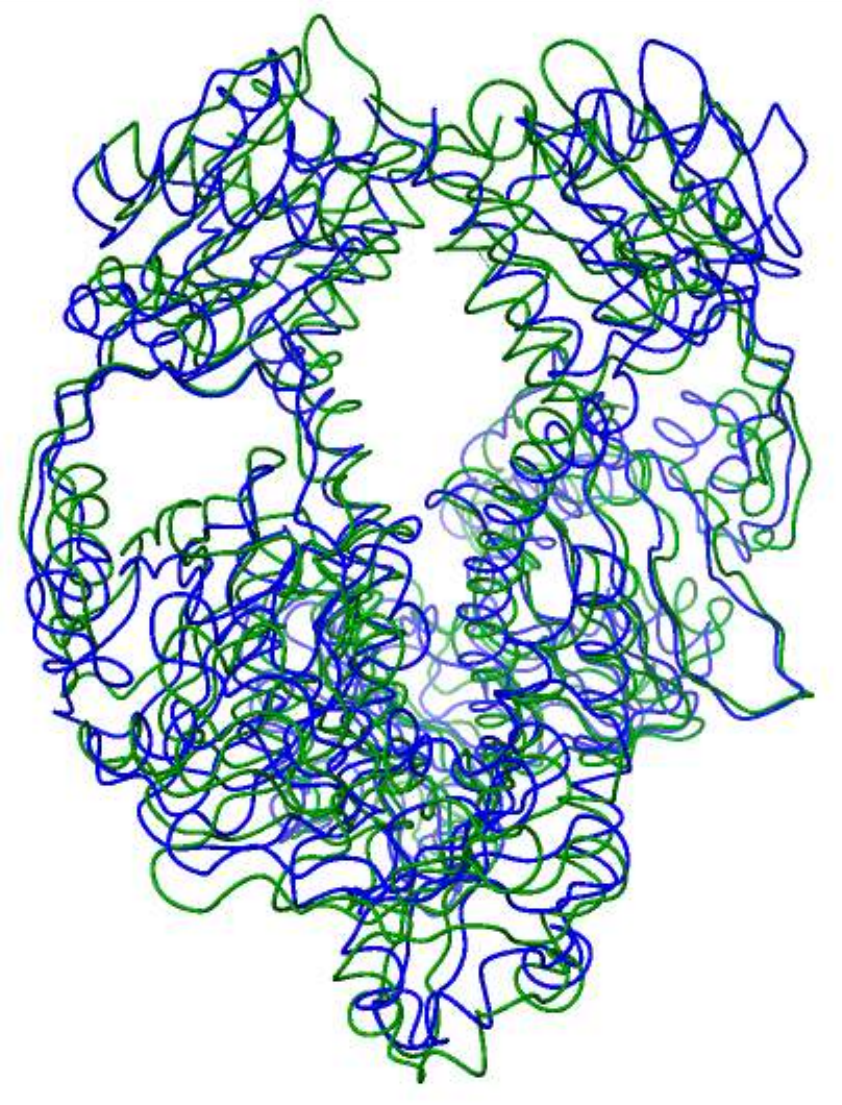}
    \includegraphics[width=0.28\textwidth]{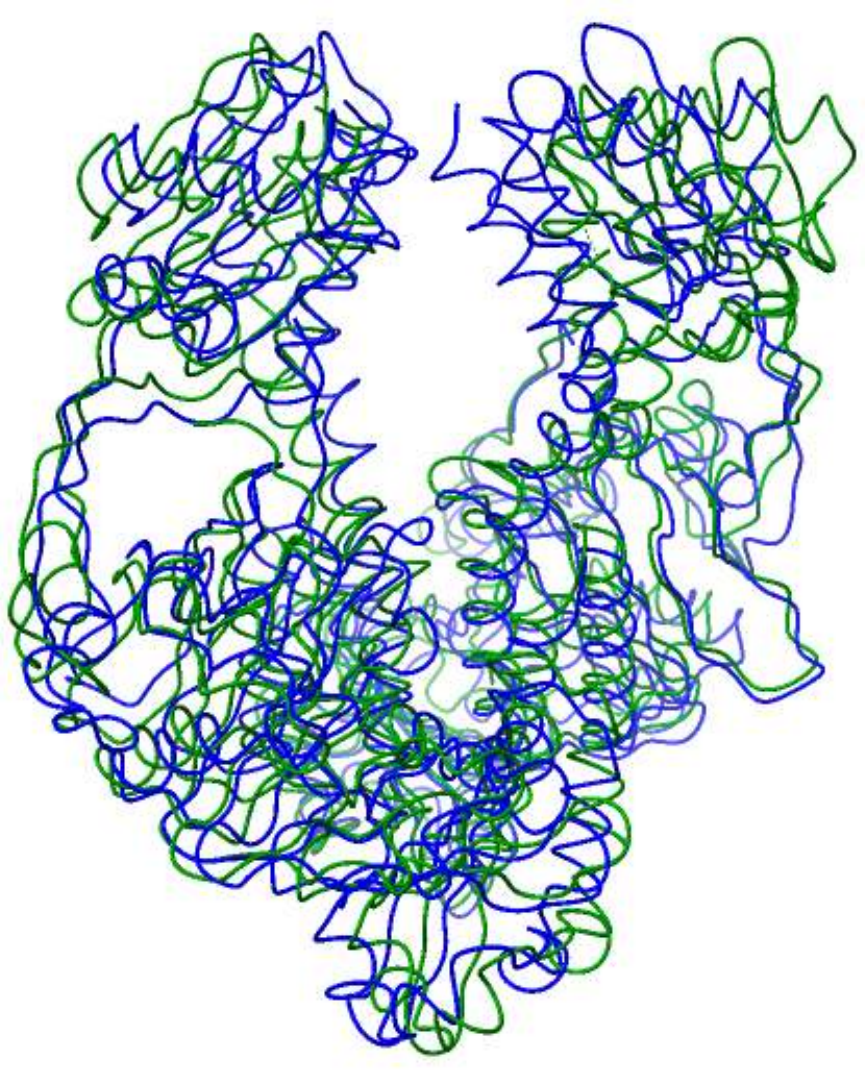}
    \includegraphics[width=0.35\textwidth]{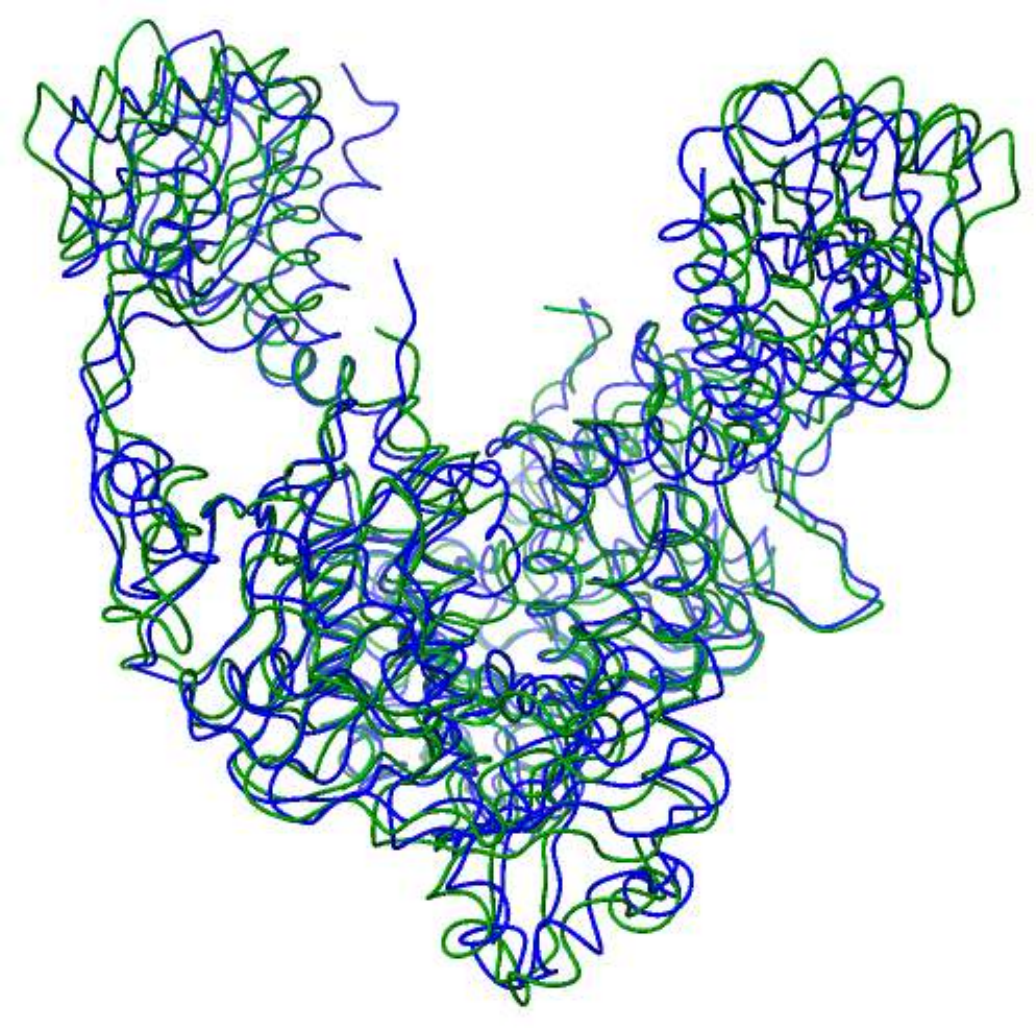}
    \caption{SNR $0.001$. Examples of reconstructed structures by cryoSHERE. Blue is predicted, green is ground truth. CryoSPHERE is able to recover the right conformation. Left to right: image number 11, 5001 and 9999.}
    \label{fig:structures_cryoSPHERE_snr_0_001}
\end{figure}

\begin{figure}
    \centering
    \includegraphics[width=0.28\textwidth]{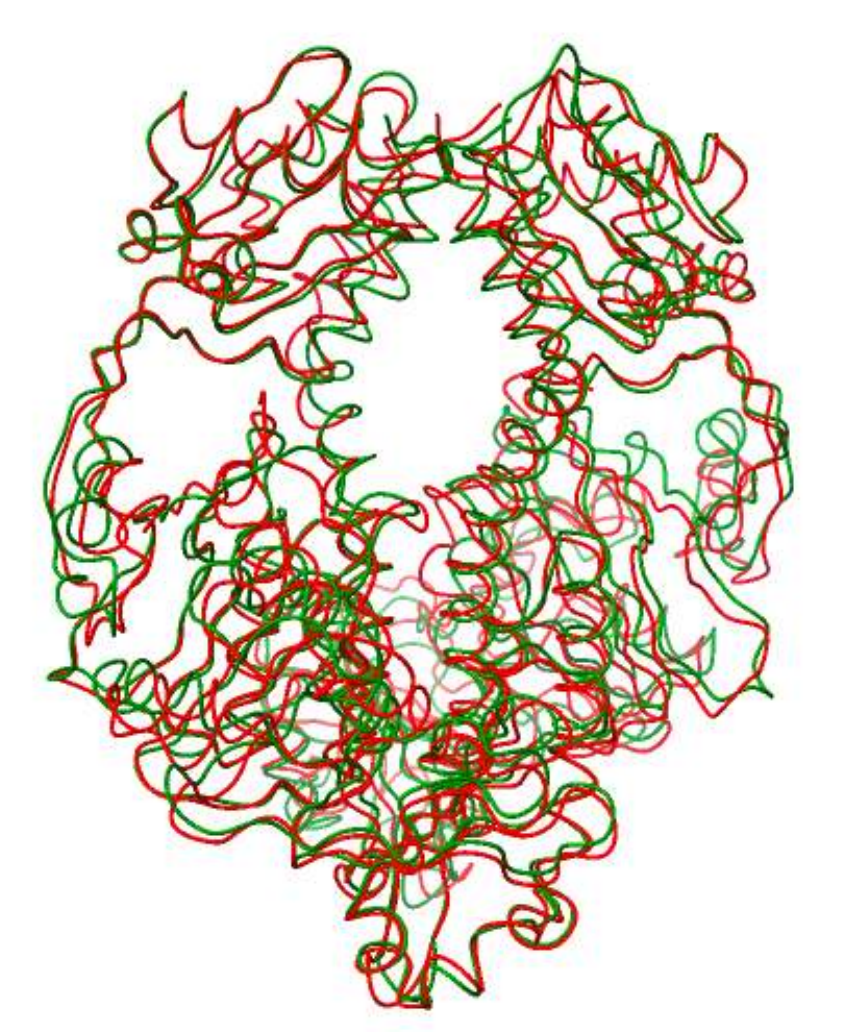}
    \includegraphics[width=0.28\textwidth]{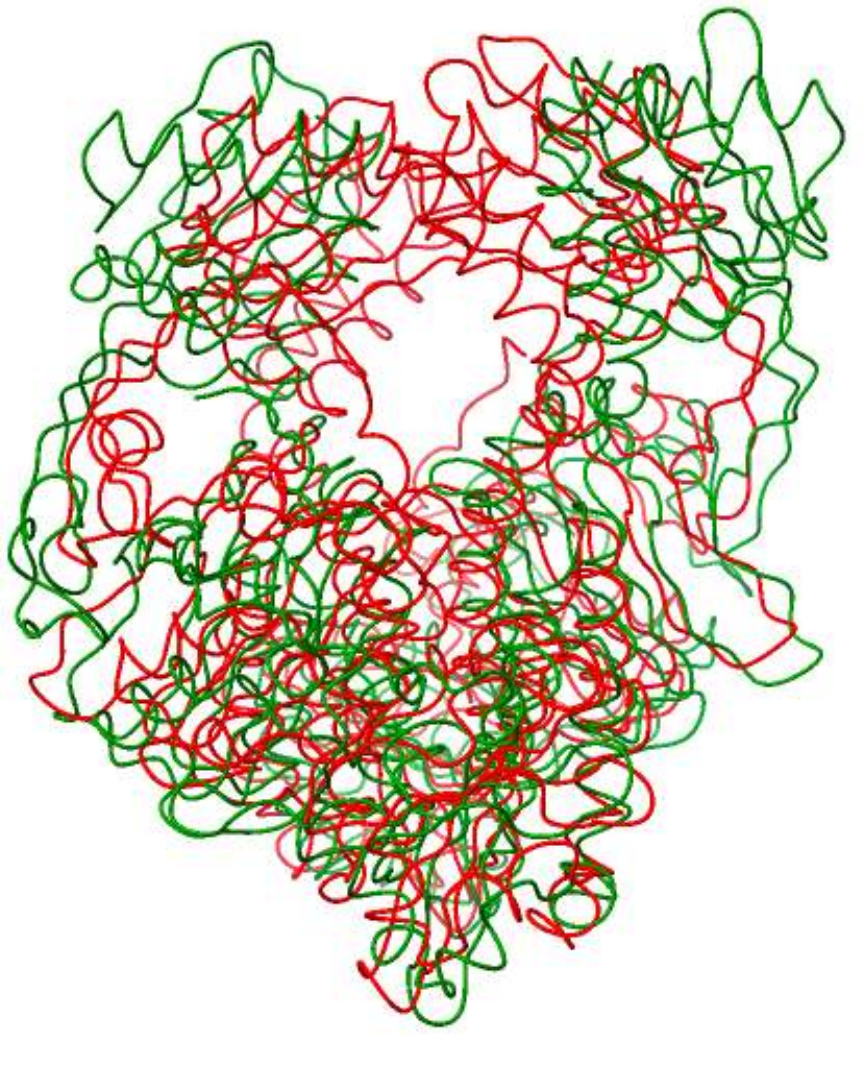}
    \includegraphics[width=0.35\textwidth]{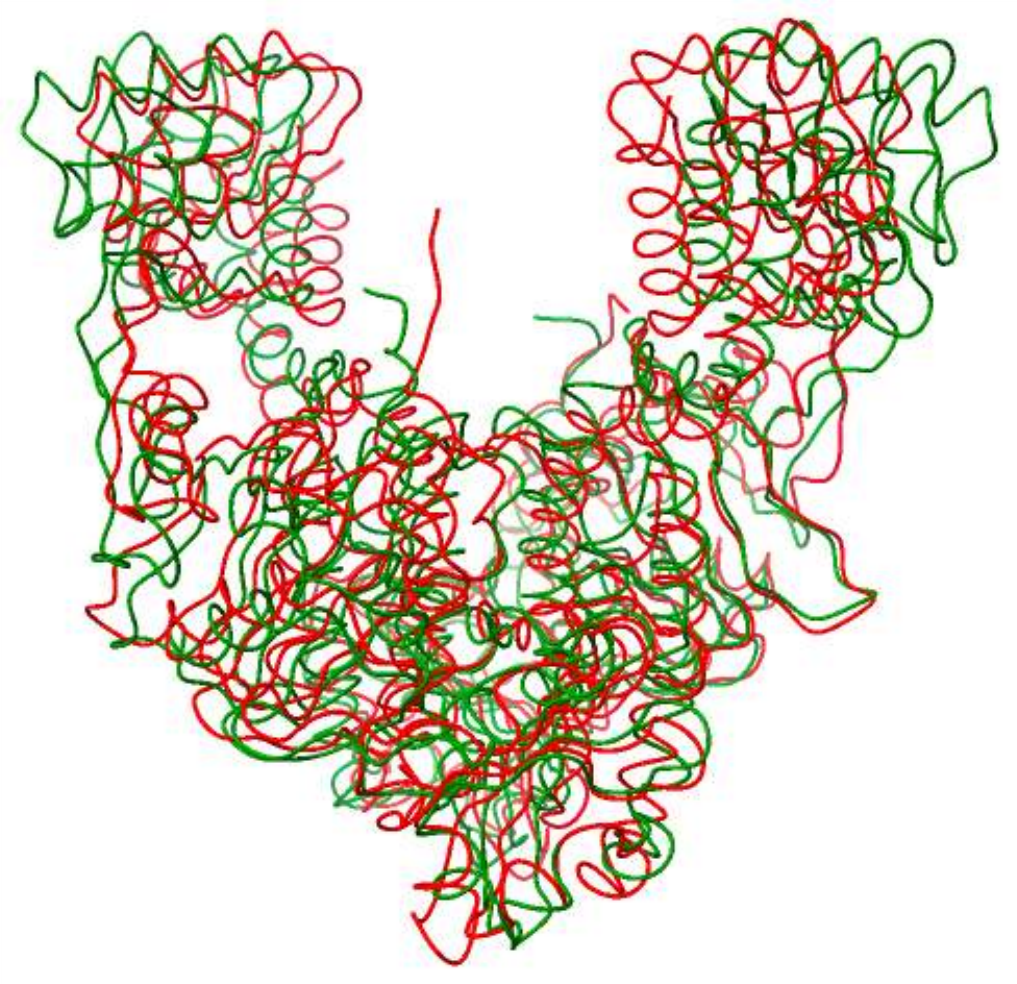}
    \caption{SNR $0.001$. Examples of reconstructed structures by CryoStar. Red is predicted, green is ground truth. CryoStar is able to recover the right conformation. Left to right: image number 11, 5001 and 9999.}
    \label{fig:structures_cryoStar_snr_0_001}
\end{figure}

Figure \ref{fig:structures_cryoSPHERE_snr_0_001} and \ref{fig:structures_cryoStar_snr_0_001} show a set of predicted structures compared to the ground truth. Both cryoSPHERE and cryoStar are able to recover the ground truth, but in a more approximate fashion than for the higher SNR datasets.

\begin{figure}
    \centering
    \includegraphics[scale=0.3]{figures/experiments/phy/ground_truth_1.pdf}
    \hspace{0.1cm}
    \includegraphics[scale=0.3]{figures/experiments/phy/ground_truth_500.pdf}
    \hspace{0.1cm}
    \includegraphics[scale=0.3]{figures/experiments/phy/ground_truth_999.pdf}\\
    \includegraphics[scale=0.3]{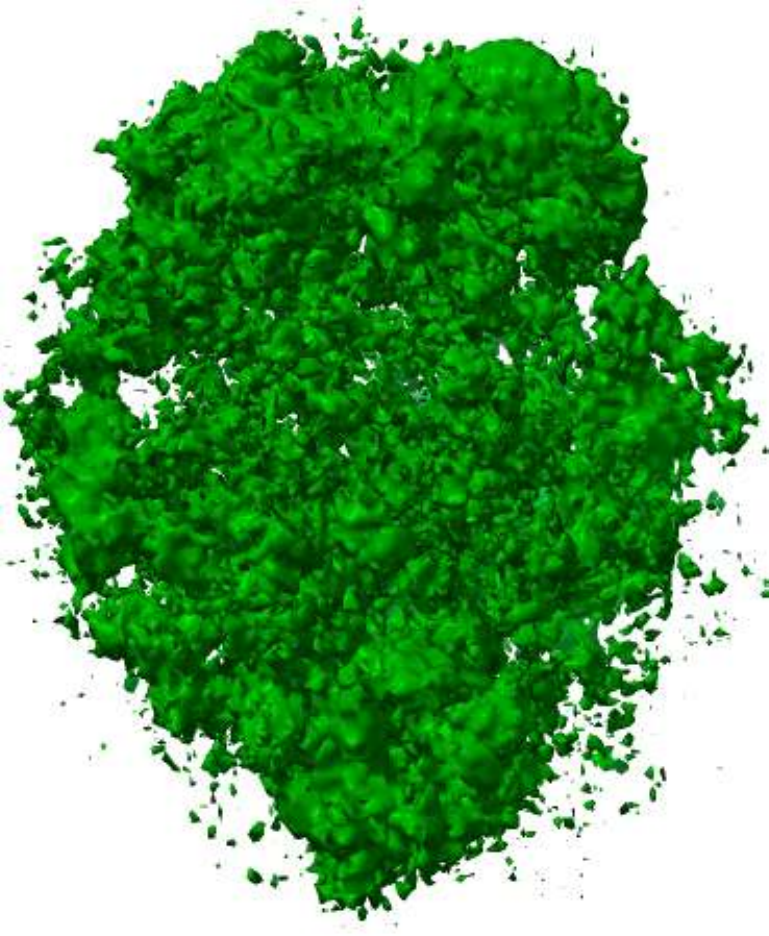}\hspace{0.1cm}
    \includegraphics[scale=0.3]{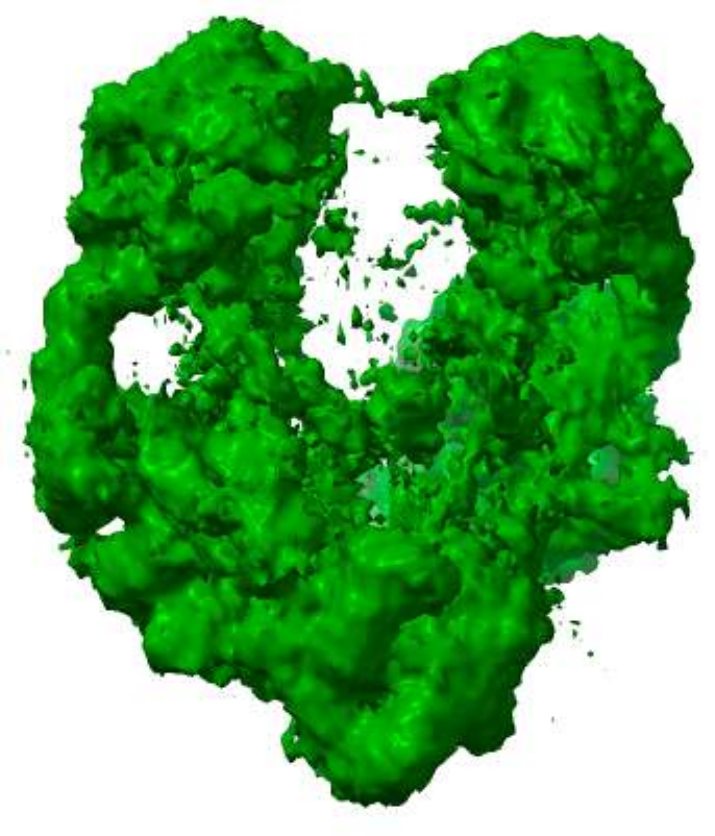}\hspace{0.12cm}
    \includegraphics[scale=0.3]{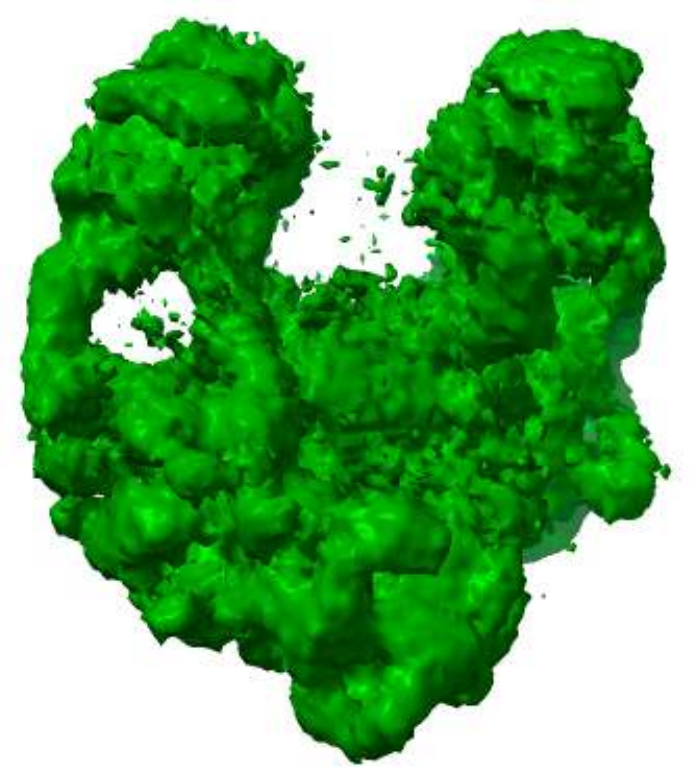}
    \caption{SNR $0.001$. Examples of reconstructed volumes by cryoDRGN. Green is cryoDRGN and gray is the corresponding ground truth volume.}
   \label{fig:volumes_cryodrgn_snr_0_001}
\end{figure}

Figure \ref{fig:volumes_cryodrgn_snr_0_001} shows examples of cryoDRGN predicted volume together with the corresponding ground truth. The method underestimates the opening of the protein and predicts low resolution volumes with a lot of noise. It seems it is overfitting.

\subsubsection{Debiasing cryoSPHERE}
When deforming an atomic model to recover different conformations, one should be careful not to bias the results. CryoStar \citep{li_cryostar_2023} developed a volume method to help assess the bias induced by the atomic model. However, we demonstrate here that this two stages of training are nothing specific to cryoStar and can, in fact, be applied to any structural method just by using DRGN-AI \cite{Levy_drgnai}. 

We remove the last 40 residues of chain B of the 10 000 structures obtained through molecular dynamics simulations, see Appendix \ref{append:MD}. We then follow the exact same process to generate 150k images with SNR 0.01. Again, see Appendix \ref{append:MD}.

\begin{figure}[h]
    \centering
    \includegraphics[width=0.3\textwidth]{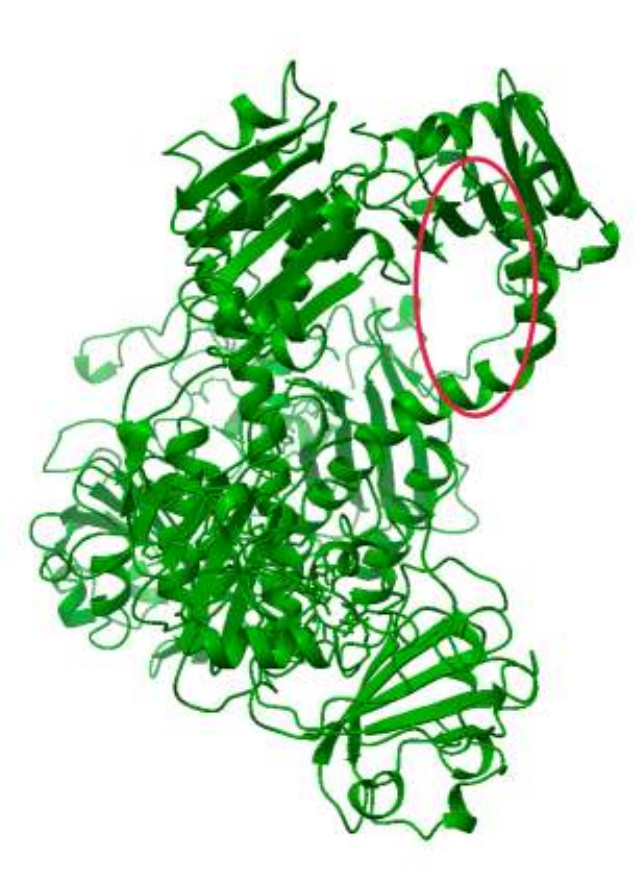}
    \includegraphics[width=0.3\textwidth]{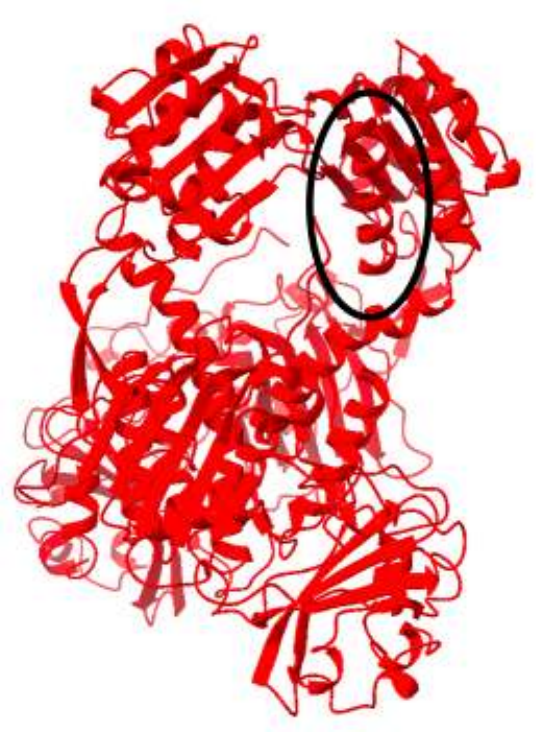}
    \caption{Left: example of a structure used to to generate the images. The missing part is highlighted in a red ellipsoid. Right: the base structure used, with the residues that were removed for the image generation highlighted in a black ellipsoid.}
    \label{fig:debias_base_structure}
\end{figure}

We run cryoSPHERE with $\Ndomains = 20$ for 24 hours using the same base structure as the other experiments of Appendix \ref{append:MD}, that we obtained through AlphaFold. In other words, our base structure has 40 more residues than the structures on the images, see Figure \ref{fig:debias_base_structure}.

\begin{figure}[h]
    \centering
    \includegraphics[width=0.3\textwidth]{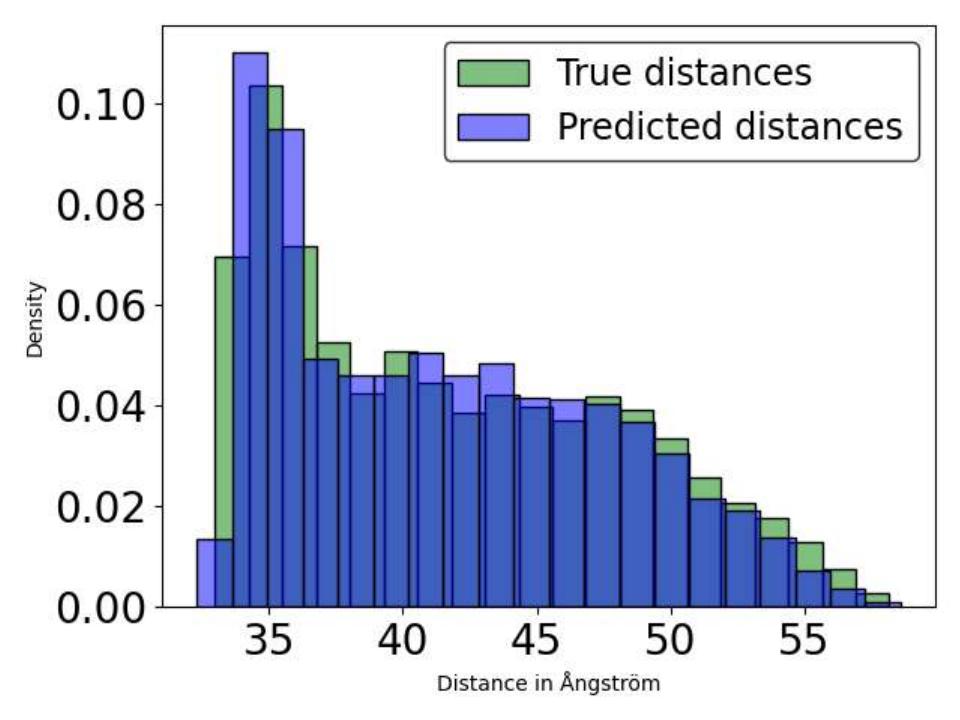}
    \includegraphics[width=0.3\textwidth]{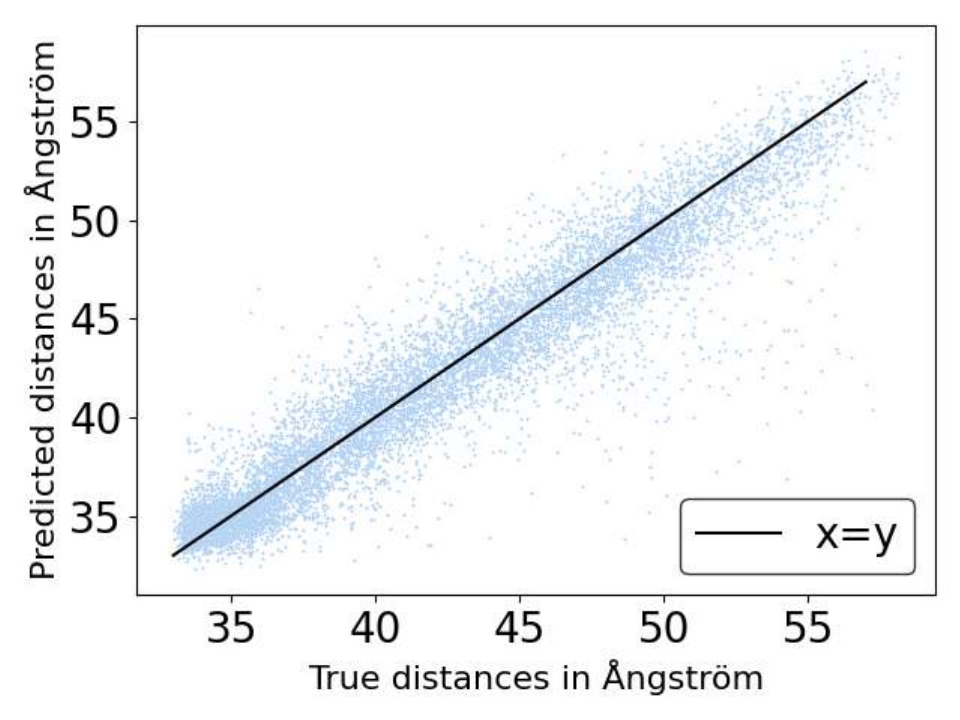}
    \caption{Debiasing the atomic model. Left: Distribution of ground truth and predicted distances. Right: True versus predicted distances.}
    \label{fig:debias_distances}
\end{figure}

\begin{figure}[h]
    \centering
    \includegraphics[width=0.24\textwidth]{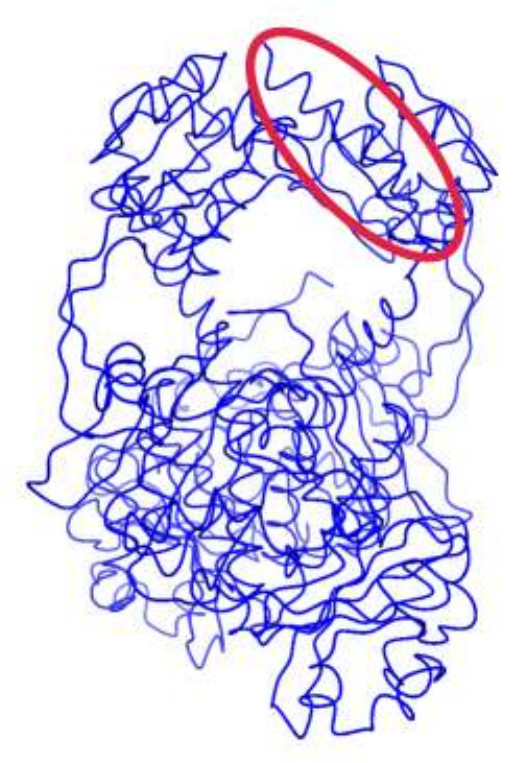}
    \includegraphics[width=0.24\textwidth]{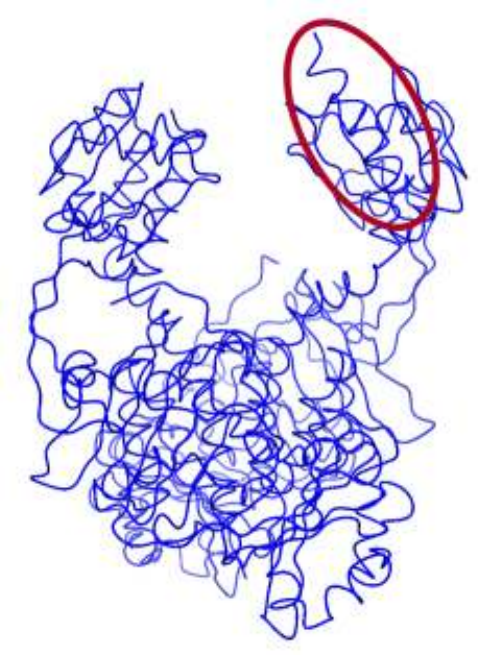}
    \includegraphics[width=0.3\textwidth]{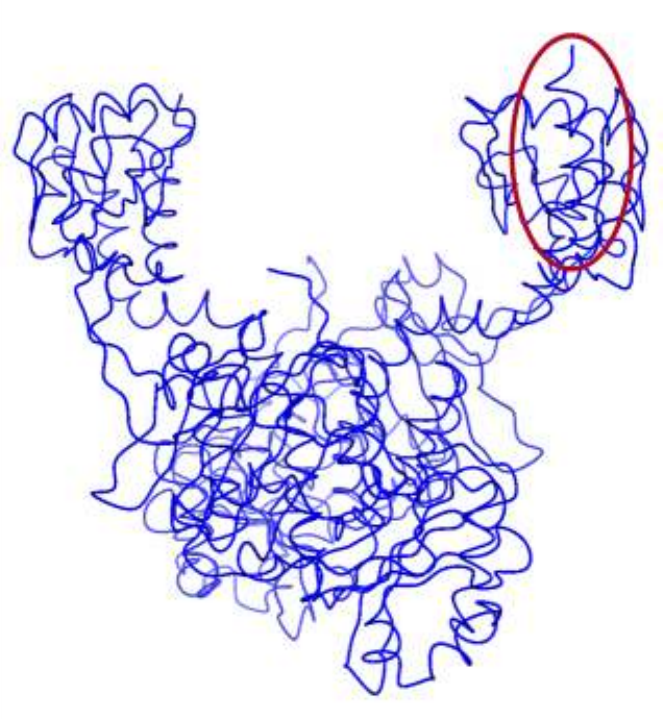}
    \includegraphics[width=0.24\textwidth]{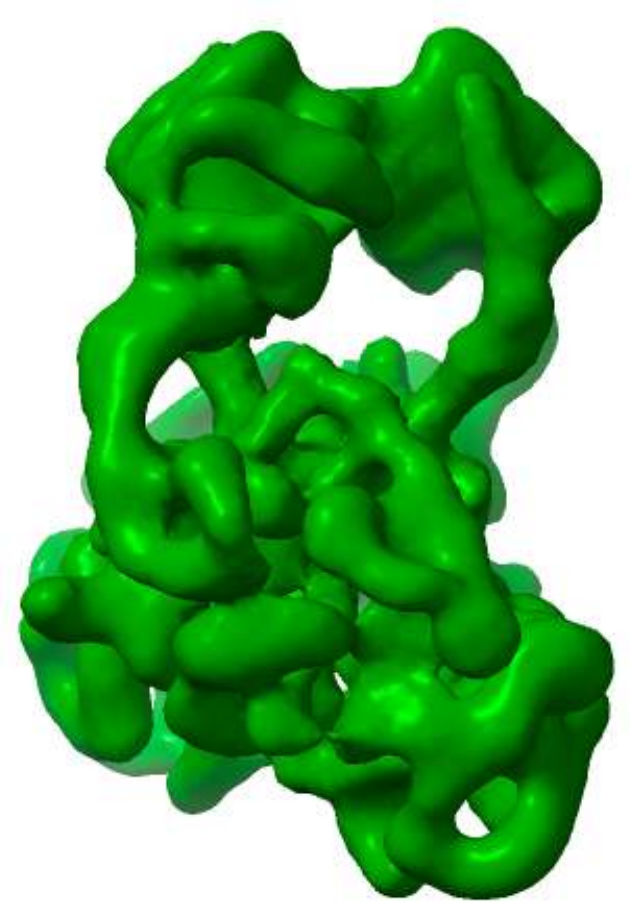}
     \includegraphics[width=0.24\textwidth]{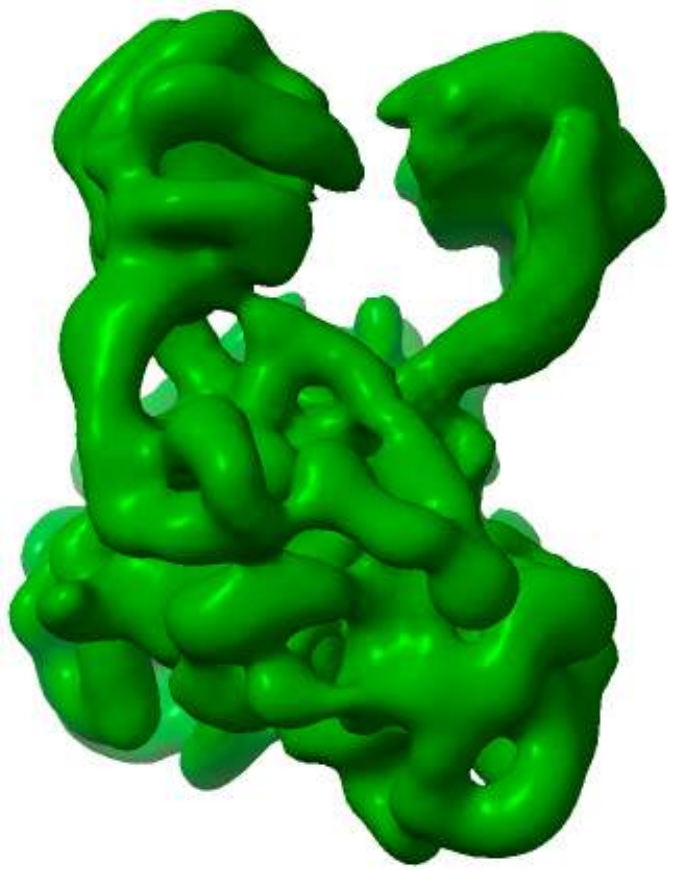}  \includegraphics[width=0.3\textwidth]{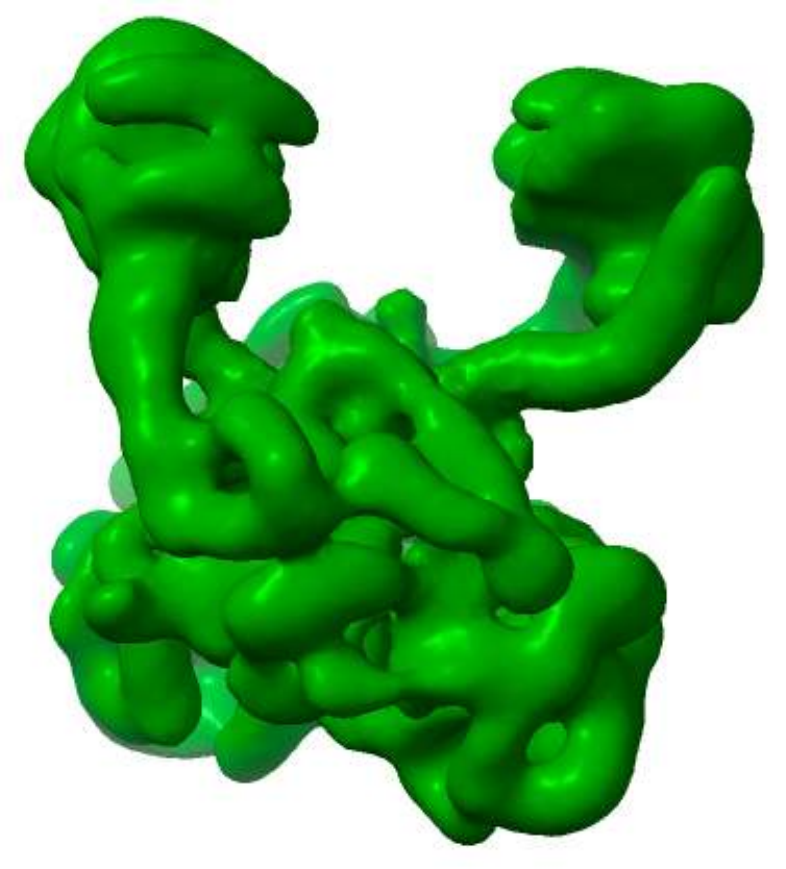}
    \caption{Debiasing the atomic model. Top, from left to right: example of 3 structures predicted by cryoSPHERE. The red ellipses show the alpha helix that is present in the base structure but is not on the images. Bottom, from left to right: corresponding volumes reconstructed by DRGN-AI.}
    \label{fig:debias_predicted_structures}
\end{figure}

This discrepancy does not affect the predicted distances by cryoSPHERE, as shown in Figure \ref{fig:debias_distances}. Figure \ref{fig:debias_predicted_structures} shows 3 structures predicted by cryoSPHERE. Even though the opening motion is correctly recovered, the algorithm tries to remove the missing alpha helix from where there should not be any density. In a sense, cryoSPHERE detects that the atomic model has too many residues, but it cannot temove them, by design. We leave to future work the ability to learn the amplitude of each Gaussian mode of Equation \ref{eq:structToMap}. 
Here, we propose to debias the atomic model by running, in a second step, DRGN-AI with the final latent space of the cryoSPHERE run and fixed poses, similar to what cryoStar proposes. Figure \ref{fig:debias_predicted_structures} shows the reconstructed volumes by DRGN-AI corresponding to the plotted structures predicted by cryoSPHERE. The alpha helix is not present, which permits to detect the bias brought by the base structure for cryoSPHERE.

In terms of computational cost, on a dataset of 150k images of size $190\times 190$, DRGN-AI performs one epoch in 10 minutes while the volume method of cryoStar perform one epoch in 6 minutes.

\subsubsection{Comparison accross SNR for cryoSPHERE, cryoDRGN and cryoStar.}

\begin{figure}[h]
    \centering
    \includegraphics[width=0.3\textwidth]{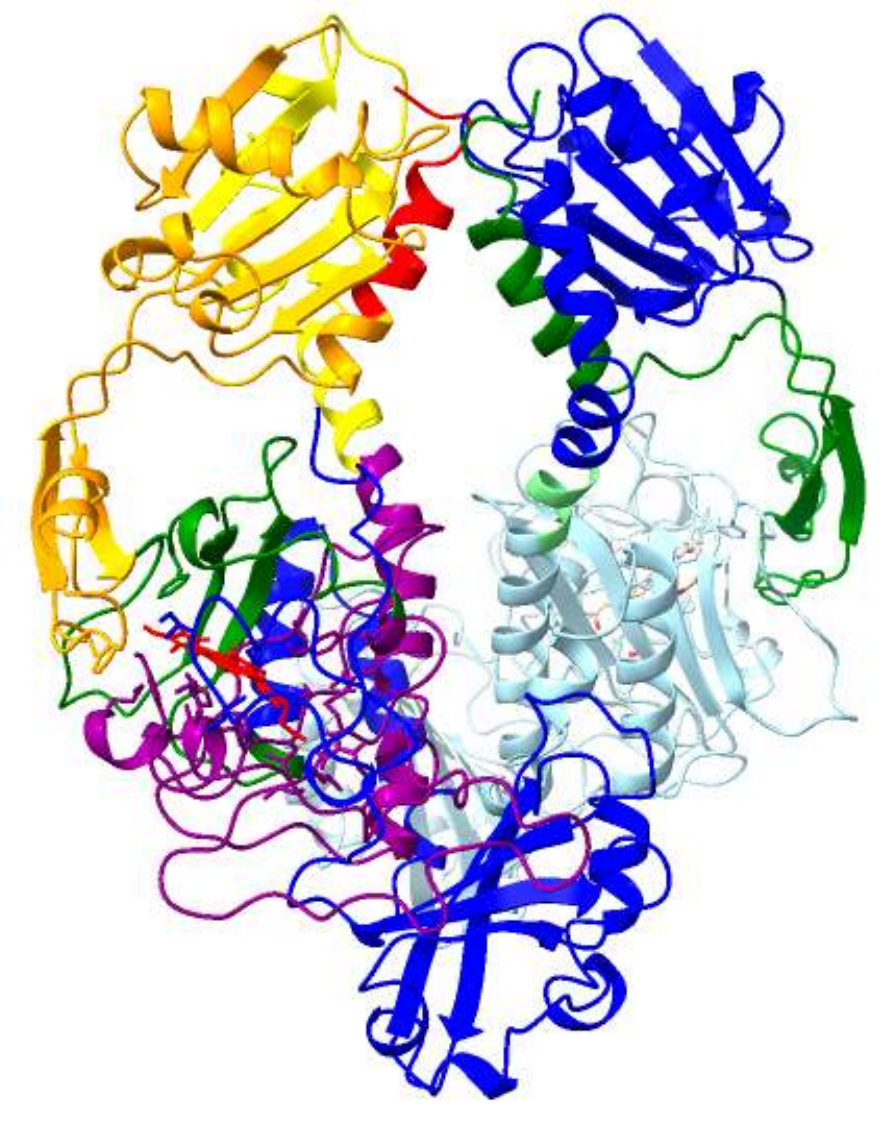}
    \includegraphics[width=0.3\textwidth]{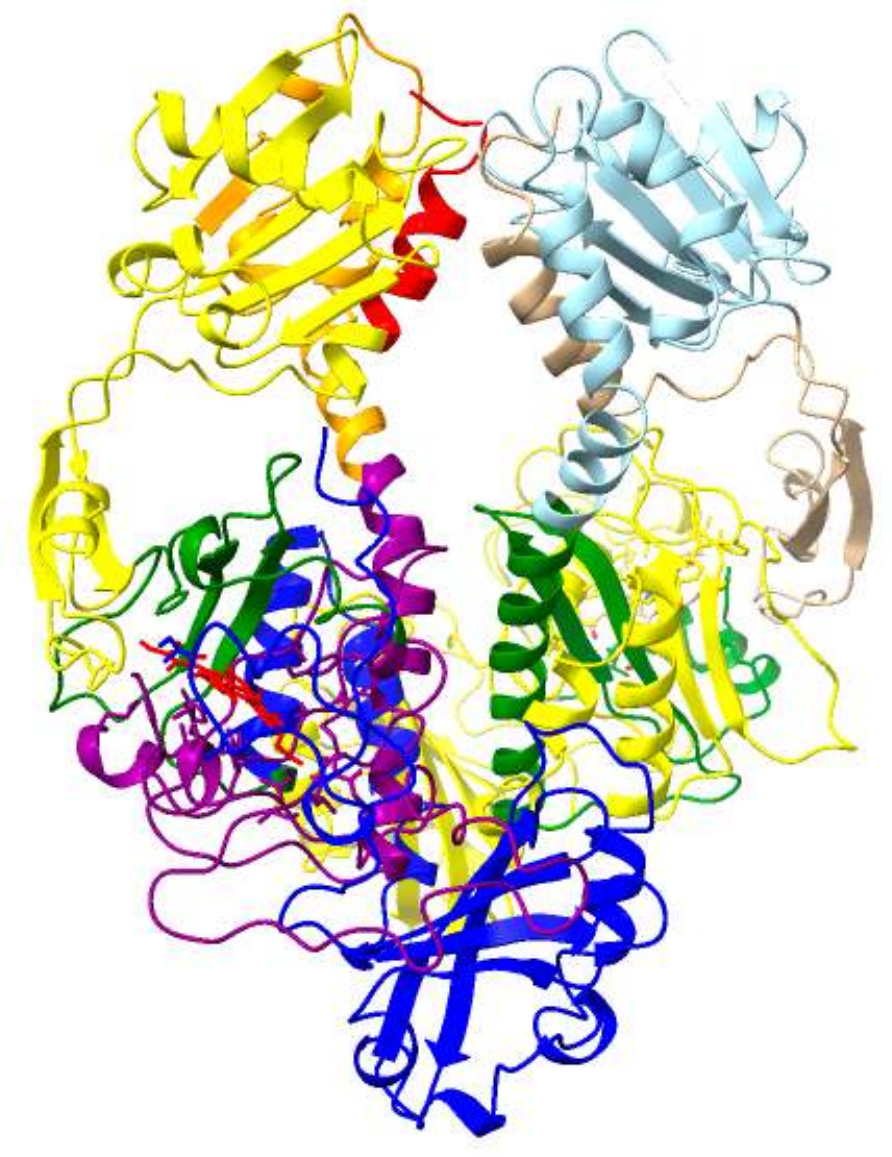}
    \includegraphics[width=0.3\textwidth]{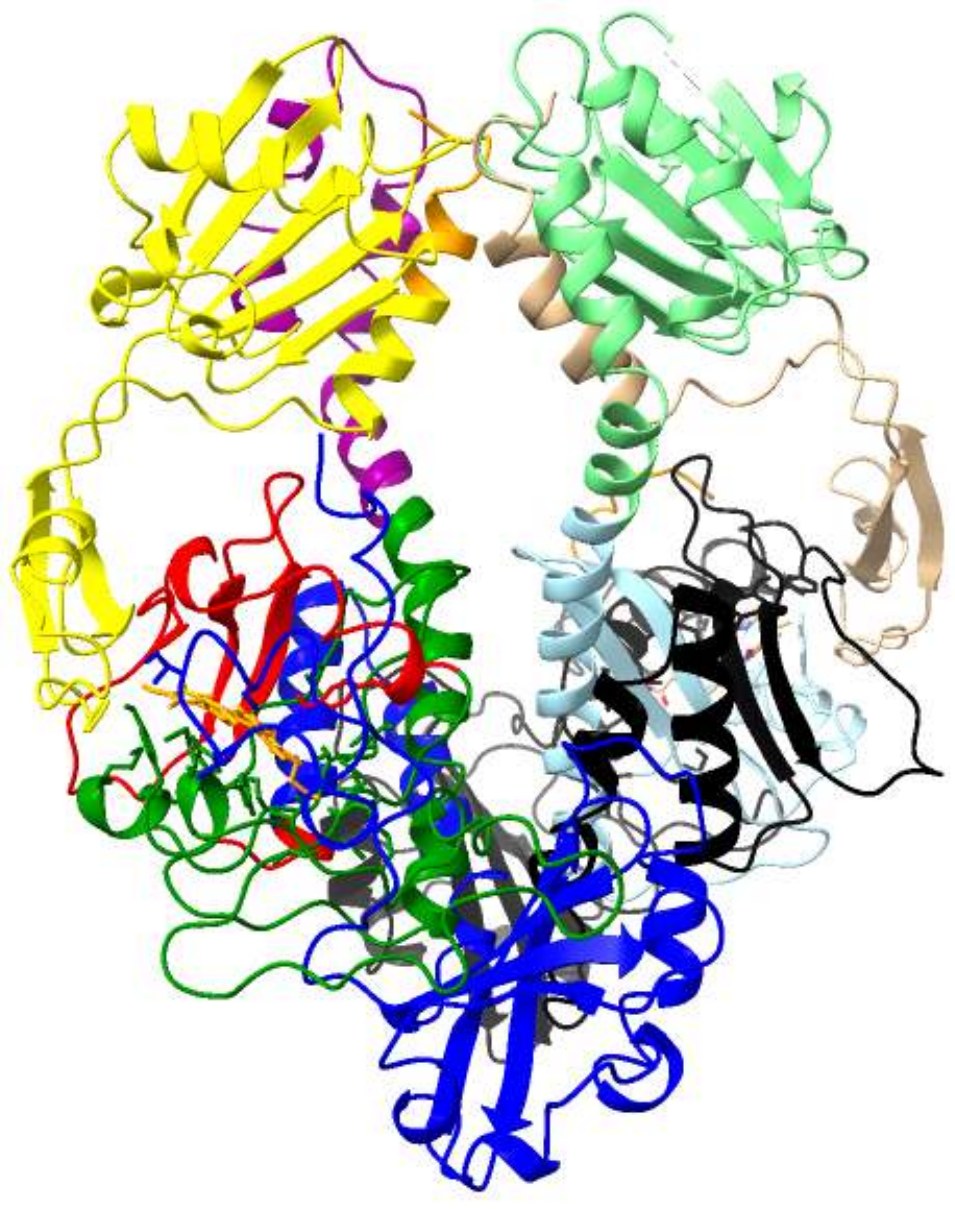}

    \caption{Segmentation of cryoSphere with $\Ndomains=10$. Left: SNR $0.1$. Middle: SNR $0.01$. Right: SNR $0.001$.}
    \label{fig:cryosphere_segmentation_10}
\end{figure}

\begin{figure}[h]
    \centering
    \includegraphics[width=0.3\textwidth]{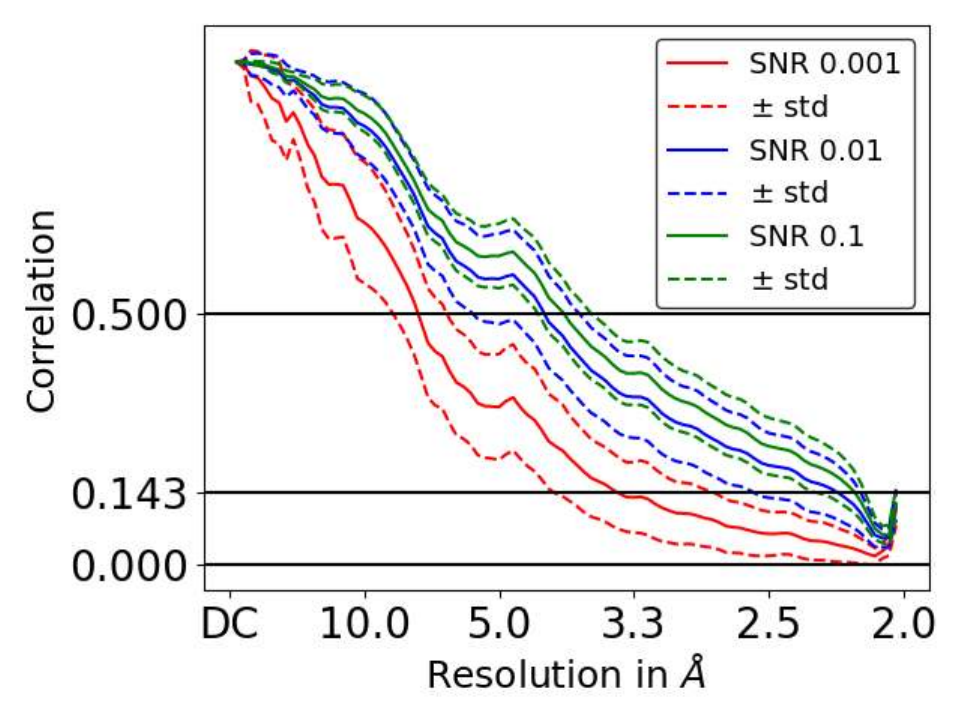}
    \hspace{0.5cm}
    \includegraphics[width=0.3\textwidth]{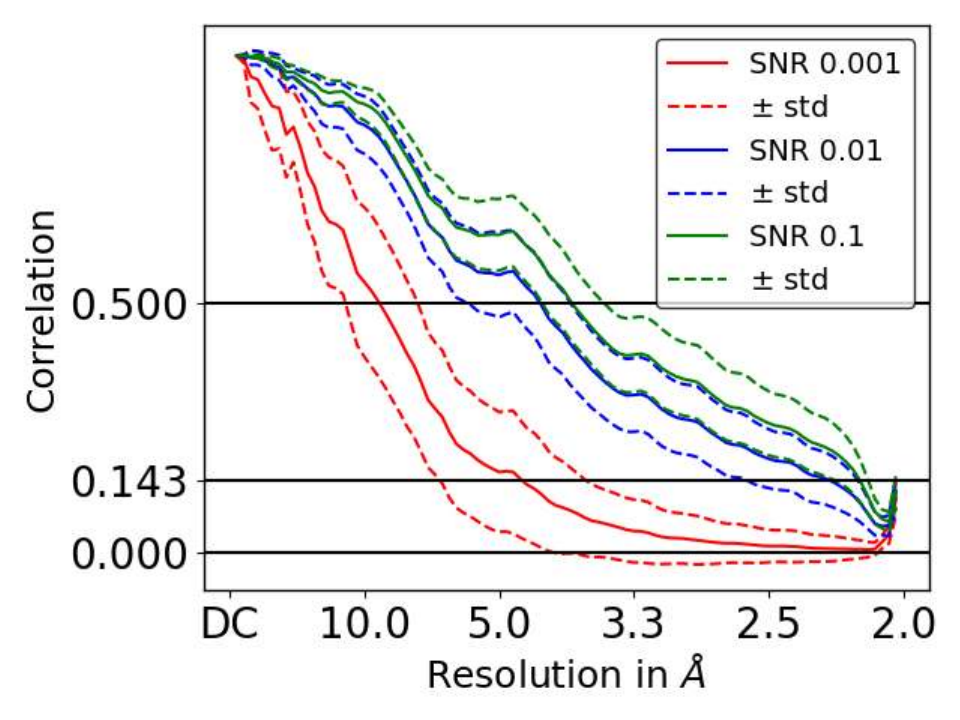}
    \hspace{0.5cm}
    \includegraphics[width=0.3\textwidth]{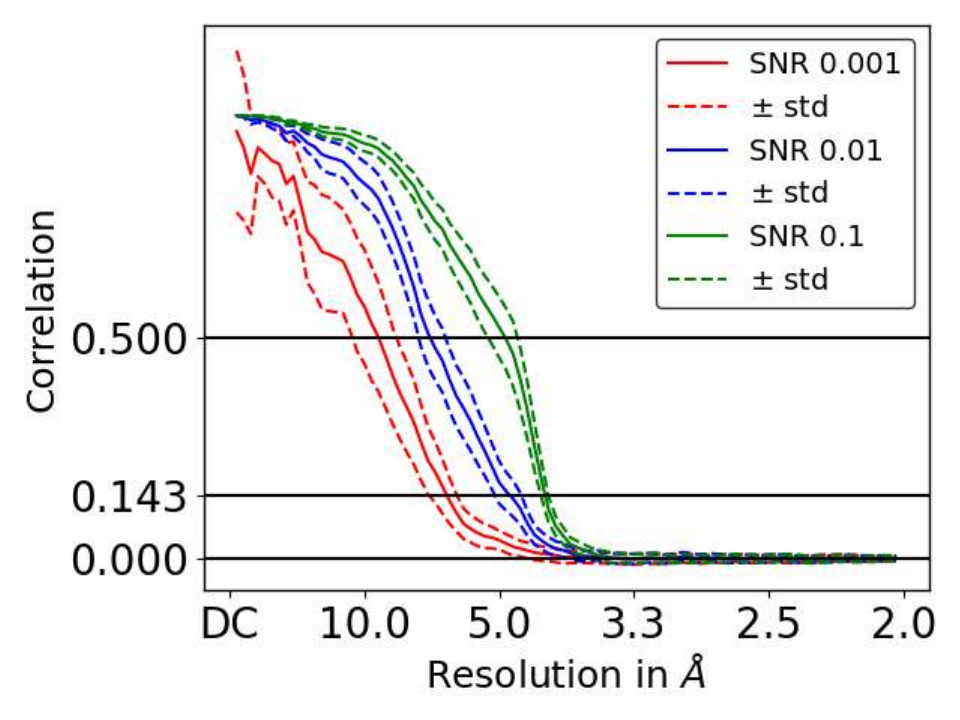}
    \caption{Comparison, for each method, of the Fourier shell correlation for different SNR. Left to right: cryoSPHERE, cryoStar, cryoDRGN.}
    \label{fig:comparison_snr}
\end{figure}

In this section, we look at the change of performance for each method accross the different SNR.

Figure \ref{fig:comparison_snr} shows that all three methods experience a drop in FSC with decreasing SNR. CryoSPHERE and cryoStar show a drop of one standard deviation between SNR $0.1$ and $0.01$ while cryoStar experience a much bigger drop than cryoSPHERE between SNR $0.01$ and $0.001$, which confirms that cryoSPHERE is more resilient to a high level of noise.

CryoDRGN shows a steady decrease in its FSC with decreasing SNR.

The fact that for a SNR of $0.1$ cryoStar is slightly outperforming cryoSPHERE (see Appendix \ref{subsubsec:0.1}), that for a SNR of $0.01$ cryoSPHERE outperforms cryoStar (see Appendix \ref{subsubsec:0.01}), and that for a SNR of $0.001$ cryoSPHERE outperforms cryoStar by one standard deviation (see Section \ref{subsec:MD}) confirms that moving big chunks of the protein as rigid bodies is more resilient to low SNR than moving each residue individually.

\subsubsection{Comparison for different values of $\Ndomains$.}\label{append:different_nsegm}
In this subsection, we compare the results of cryoSPHERE for three different values of $\Ndomains=10, 20, 25$, for the different SNR of the MD dataset in \ref{subsec:MD}.

\begin{figure}[h]
    \centering
    \includegraphics[width=0.3\textwidth]{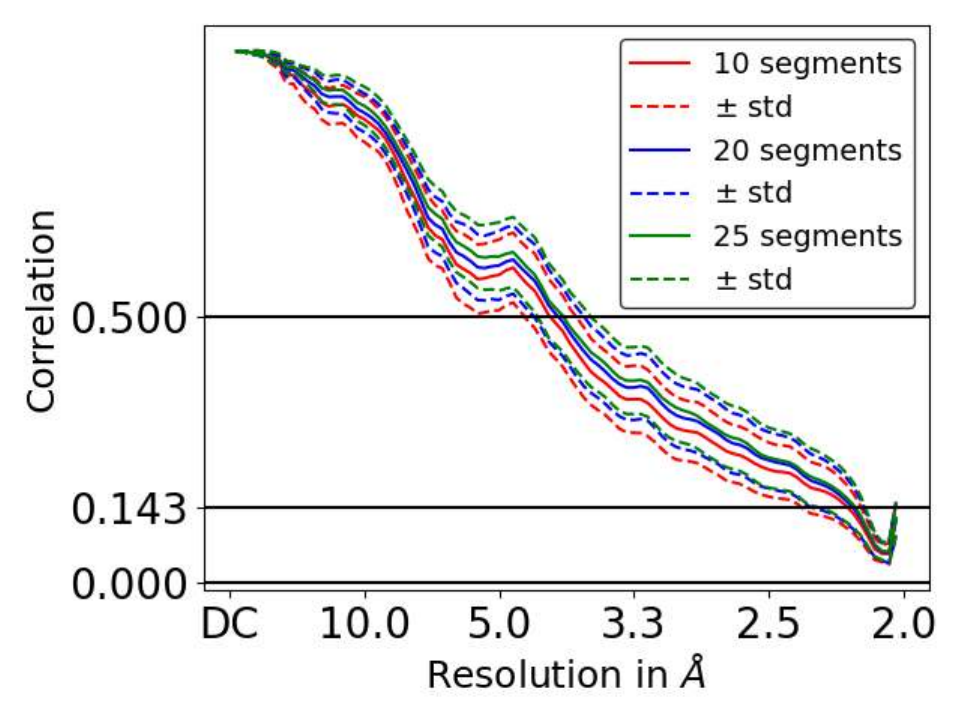}
    \includegraphics[width=0.3\textwidth]{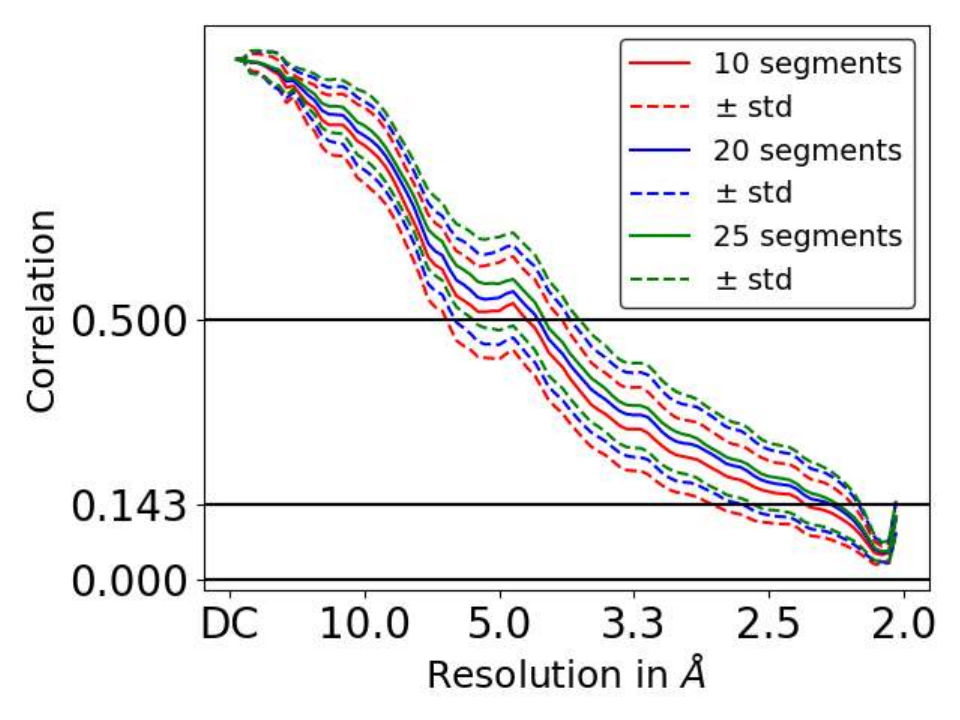}
    \includegraphics[width=0.3\textwidth]{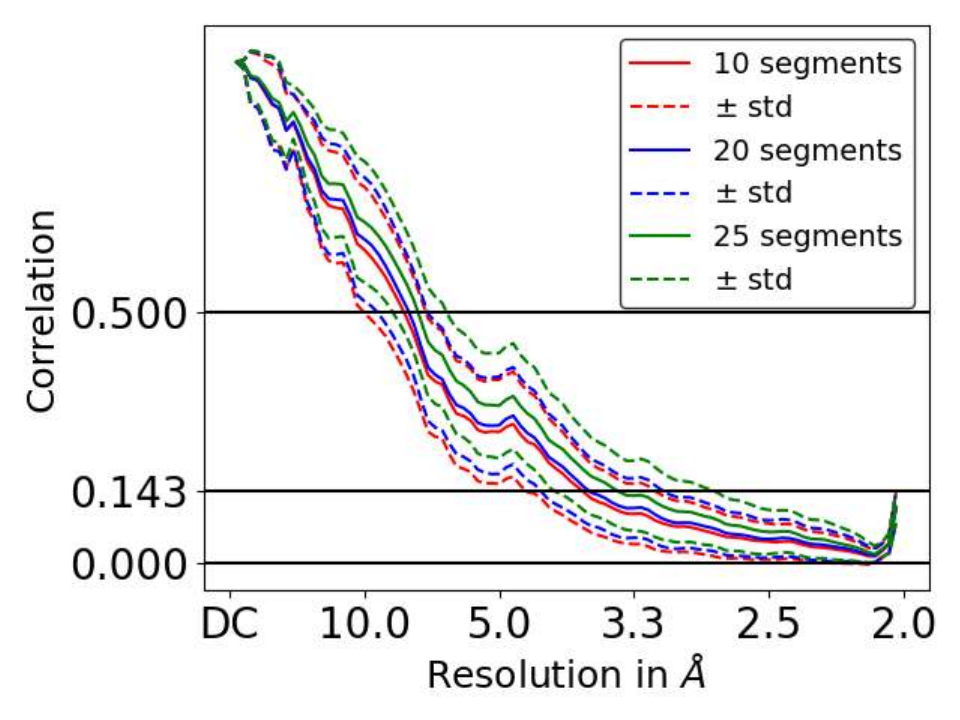}
    
    \caption{FSC curves for different $\Ndomains$ values. From left to right: SNR 0.1, SNR 0.01, SNR 0.001.}\label{fig:n_domains_comparison}
\end{figure}

Figure \ref{fig:n_domains_comparison} shows the evolution of the FSC curves with different values of $\Ndomains$ for different SNR. As we can expect, the high the number of segments, the more flexible cryoSPHERE and the better FSC. The lower the SNR, the greater we gain in FSC by increasing $\Ndomains$. 
This is because with decreasing SNR, the initial fitting of the structure in a consensus reconstruction is less accurate. Hence, the method benefits from a greater flexibility to adjust the protein on a smaller scale.

\begin{figure}[h]
    \centering
    \includegraphics[width=0.3\textwidth]{figures/experiments/phy/snr_0_1/cryosphere_distribution_distances_10_domains.pdf}
    \includegraphics[width=0.3\textwidth]{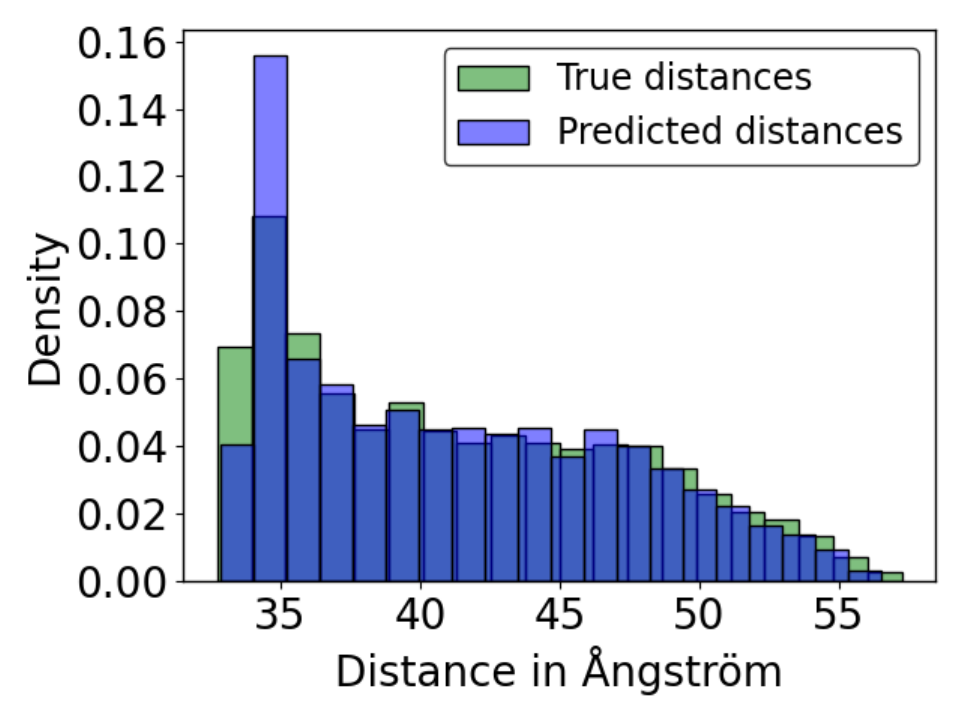}
    \includegraphics[width=0.3\textwidth]{figures/experiments/phy/snr_0_1/cryosphere_distribution_distances_25_domains.pdf}
    \caption{MD dataset, SNR 0.1. Distribution of the predicted distances. From left to right: 10 segments, 20 segments, 25 segments.}\label{fig:distribution_comparison_snr_0_1}
\end{figure}

\begin{figure}[h]
    \centering
    \includegraphics[width=0.3\textwidth]{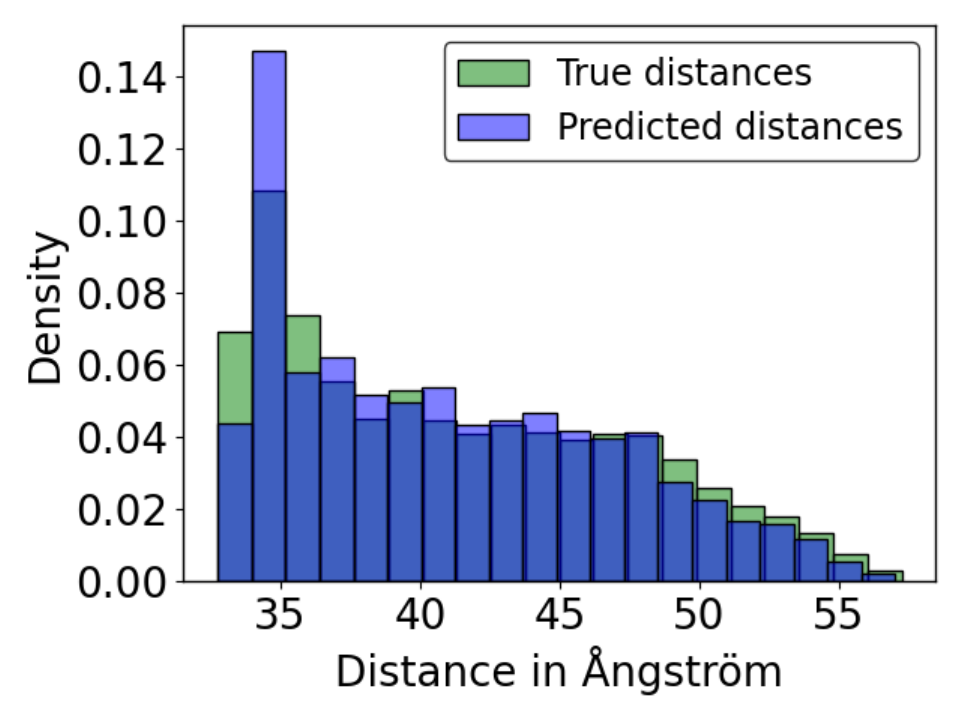}
    \includegraphics[width=0.3\textwidth]{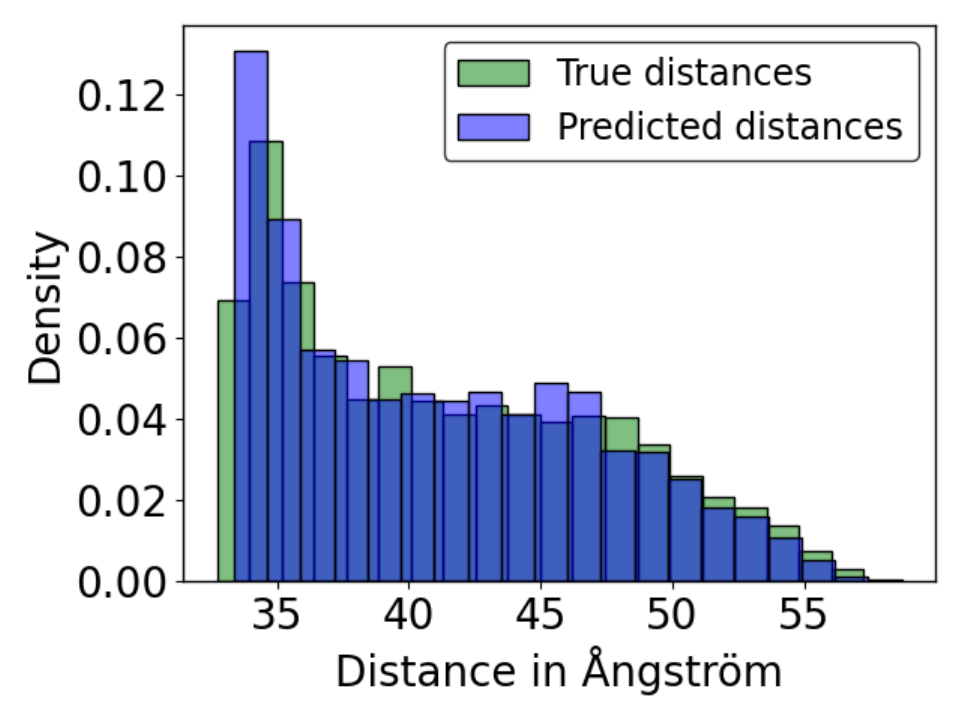}
    \includegraphics[width=0.3\textwidth]{figures/experiments/phy/snr_0_01/cryosphere_distribution_distances_25_domains.pdf}
    \caption{MD dataset, SNR 0.01. Distribution of the predicted distances. From left to right: 10 segments, 20 segments, 25 segments.}\label{fig:distribution_comparison_snr_0_01}
\end{figure}

\begin{figure}[h]
    \centering
    \includegraphics[width=0.3\textwidth]{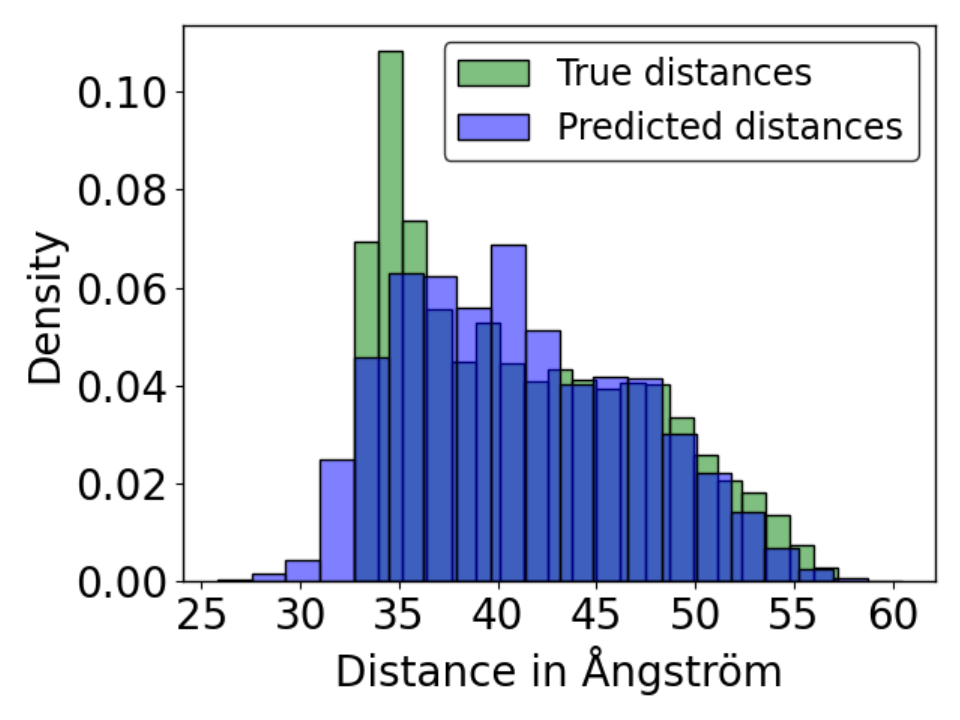}
    \includegraphics[width=0.3\textwidth]{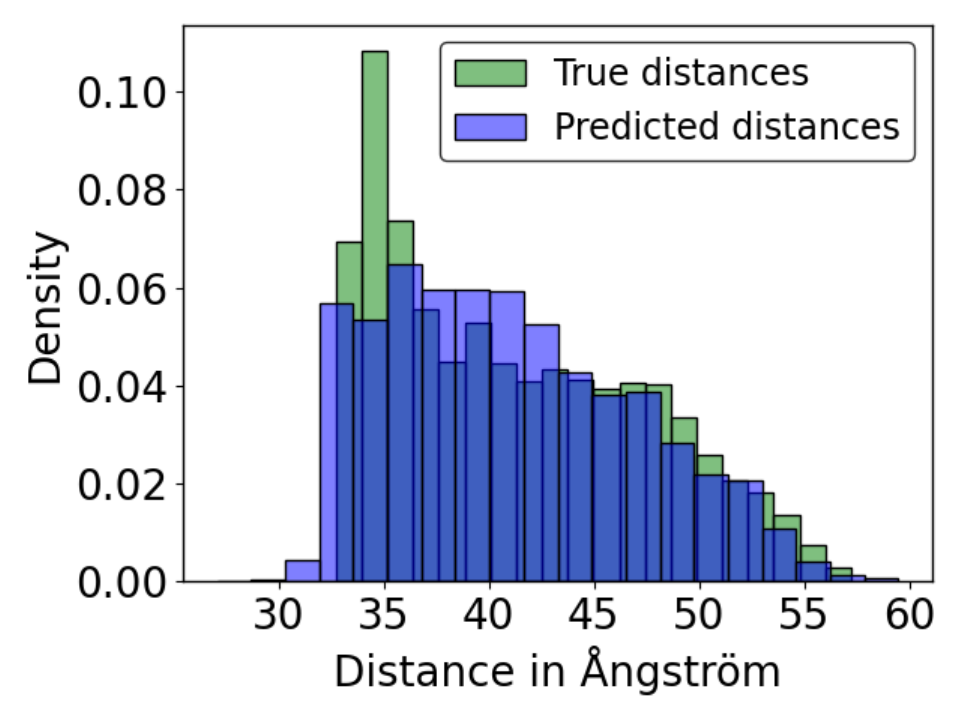}
    \includegraphics[width=0.3\textwidth]{figures/experiments/phy/snr_0_001/cryosphere_distribution_distances_25_domains.pdf}
    \caption{MD dataset, SNR 0.001. Distribution of the predicted distances. From left to right: 10 segments, 20 segments, 25 segments.}\label{fig:distribution_comparison_snr_0_001}
\end{figure}

\begin{figure}[h]
    \centering
    \includegraphics[width=0.3\textwidth]{figures/experiments/phy/snr_0_1/cryosphere_true_vs_predicted_distance_10_domains.pdf}
    \includegraphics[width=0.3\textwidth]{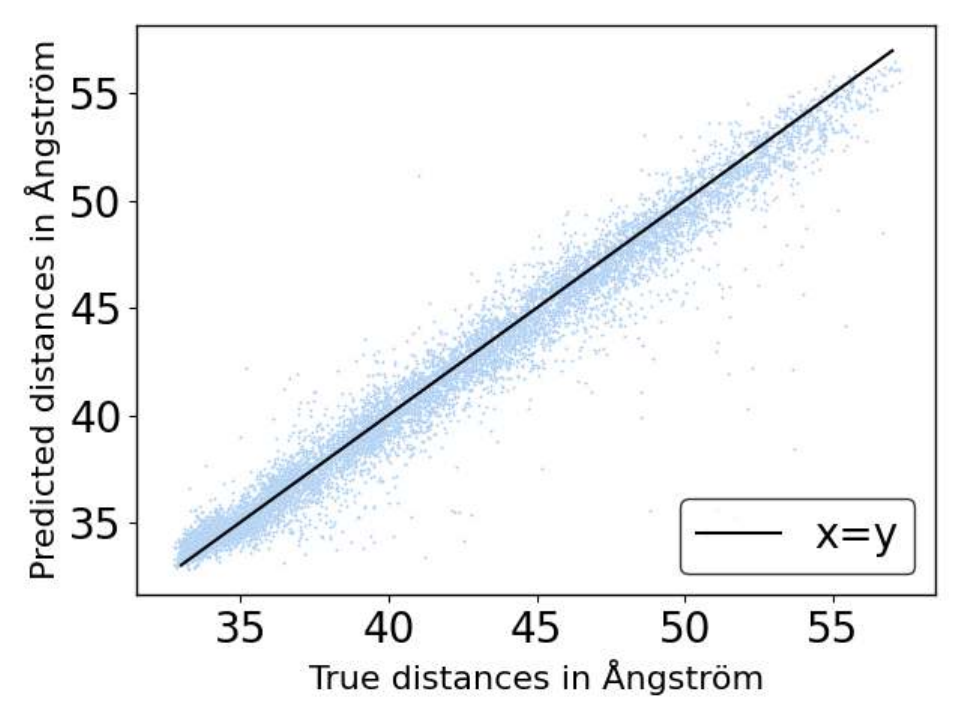}
    \includegraphics[width=0.3\textwidth]{figures/experiments/phy/snr_0_1/cryosphere_true_vs_predicted_distance_25_domains.pdf}
    \caption{MD dataset, SNR 0.1. True versus predicted distances. From left to right: 10 segments, 20 segments, 25 segments.}\label{fig:true_vs_predicted_comparison_snr_0_1}
\end{figure}

\begin{figure}[h]
    \centering
    \includegraphics[width=0.3\textwidth]{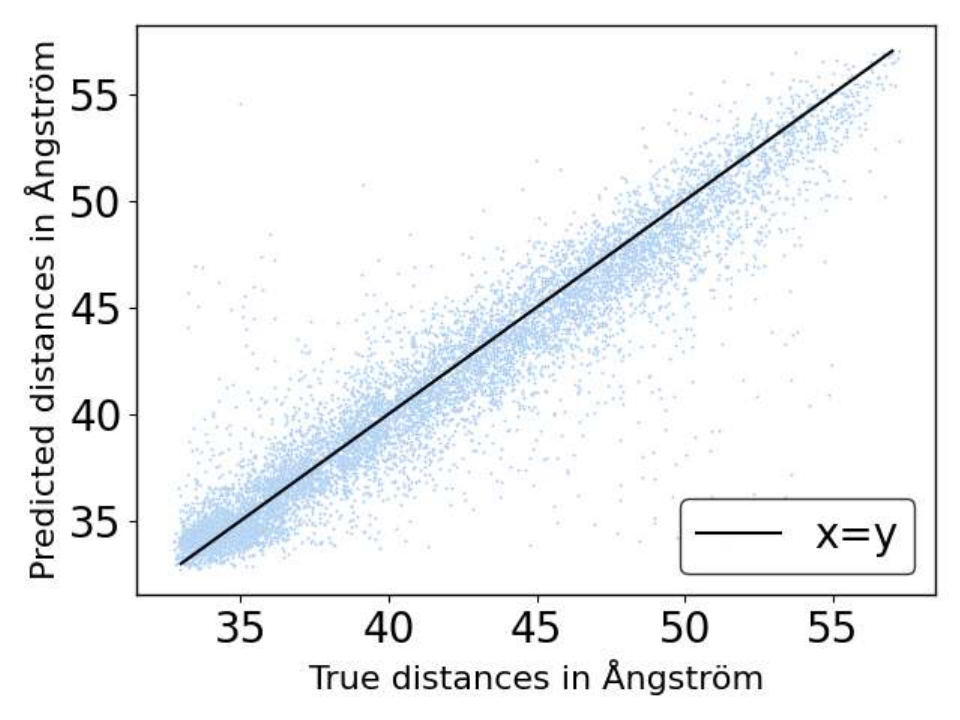}
    \includegraphics[width=0.3\textwidth]{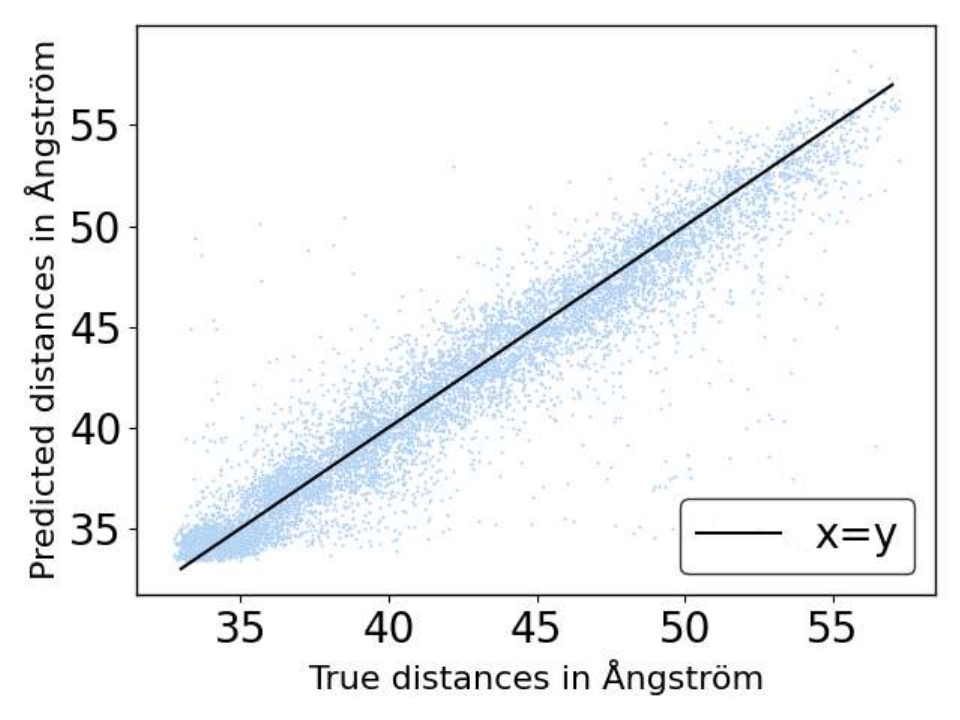}
    \includegraphics[width=0.3\textwidth]{figures/experiments/phy/snr_0_01/cryosphere_true_vs_predicted_distances_25_domains.pdf}
    \caption{MD dataset, SNR 0.01. True versus predicted distances. From left to right: 10 segments, 20 segments, 25 segments.}\label{fig:true_vs_predicted_comparison_snr_0_01}
\end{figure}

\begin{figure}[h]
    \centering
    \includegraphics[width=0.3\textwidth]{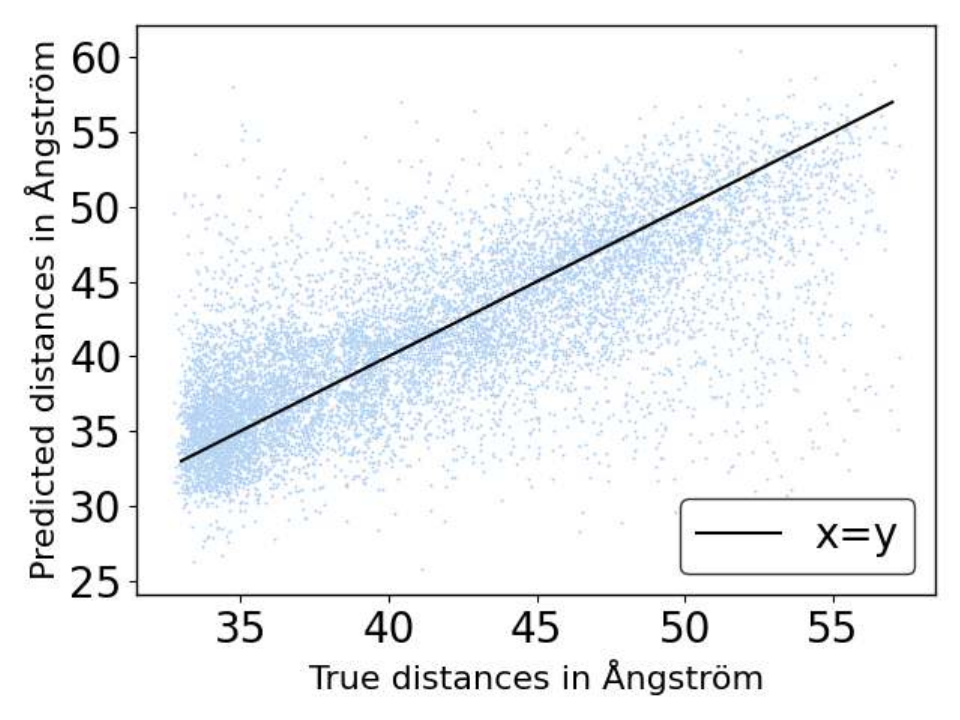}
    \includegraphics[width=0.3\textwidth]{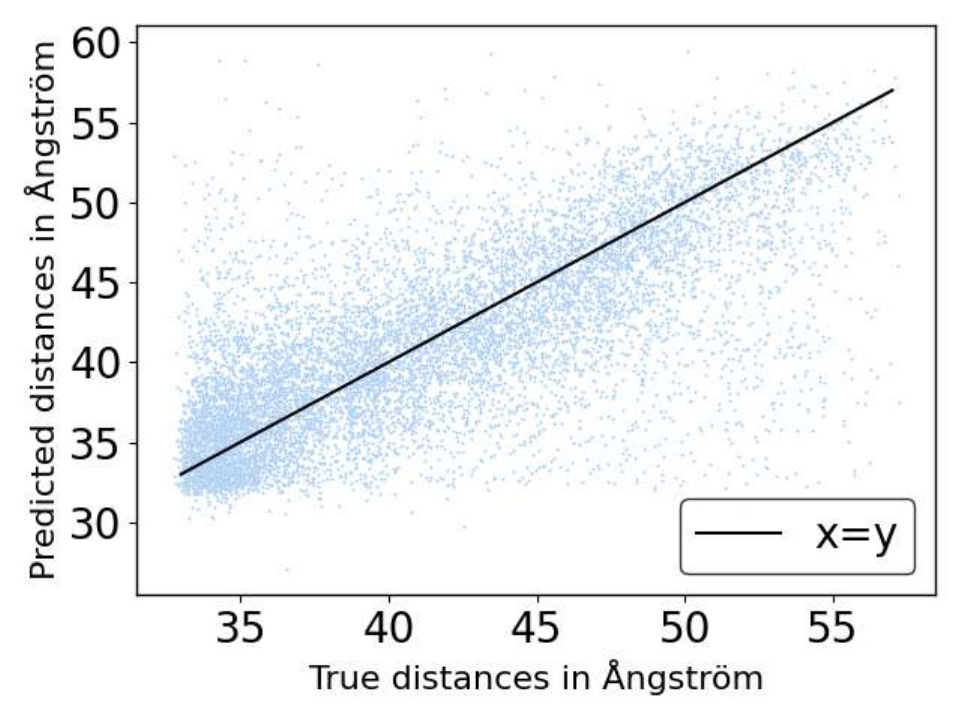}
    \includegraphics[width=0.3\textwidth]{figures/experiments/phy/snr_0_001/cryosphere_true_vs_predicted_distance_25_domains.pdf}
    \caption{MD dataset, SNR 0.001. True versus predicted distances. From left to right: 10 segments, 20 segments, 25 segments.}\label{fig:true_vs_predicted_comparison_snr_0_001}
\end{figure}

Figures \ref{fig:distribution_comparison_snr_0_1}, \ref{fig:distribution_comparison_snr_0_01} \ref{fig:distribution_comparison_snr_0_001},\ref{fig:true_vs_predicted_comparison_snr_0_1},\ref{fig:true_vs_predicted_comparison_snr_0_01},\ref{fig:true_vs_predicted_comparison_snr_0_001} show the distributions of the predicted and true distances and the true versus the predicted distances for each value of SNR and $\Ndomains$. This shows that the choice of $\Ndomains$ is not so critical for cryoSPHERE to work well. The higher $\Ndomains$, the better it is in terms of FSC, but a value of 10 still gives a good performance.

\begin{figure}[h]
    \centering
    \includegraphics[width=0.3\textwidth]{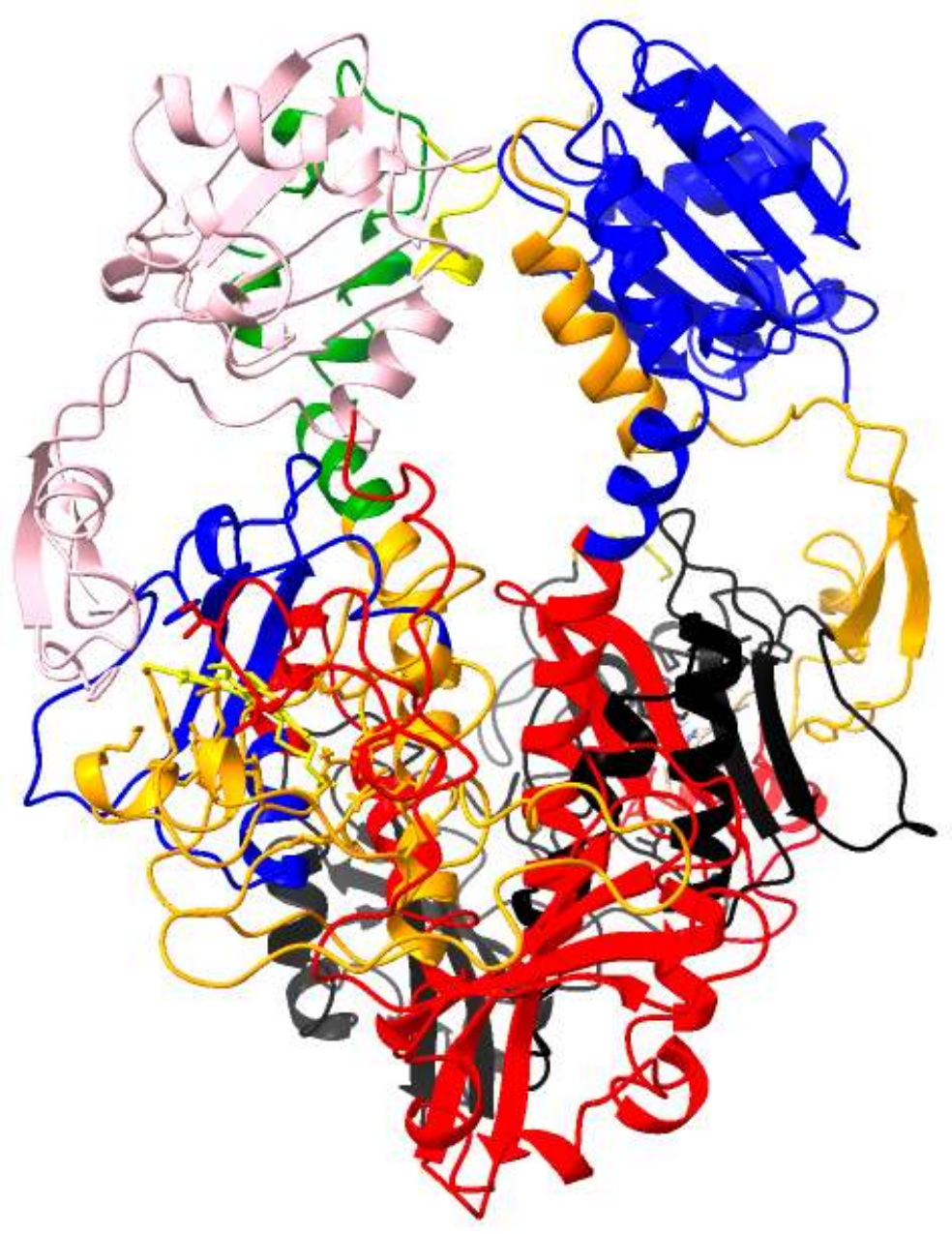}
    \includegraphics[width=0.3\textwidth]{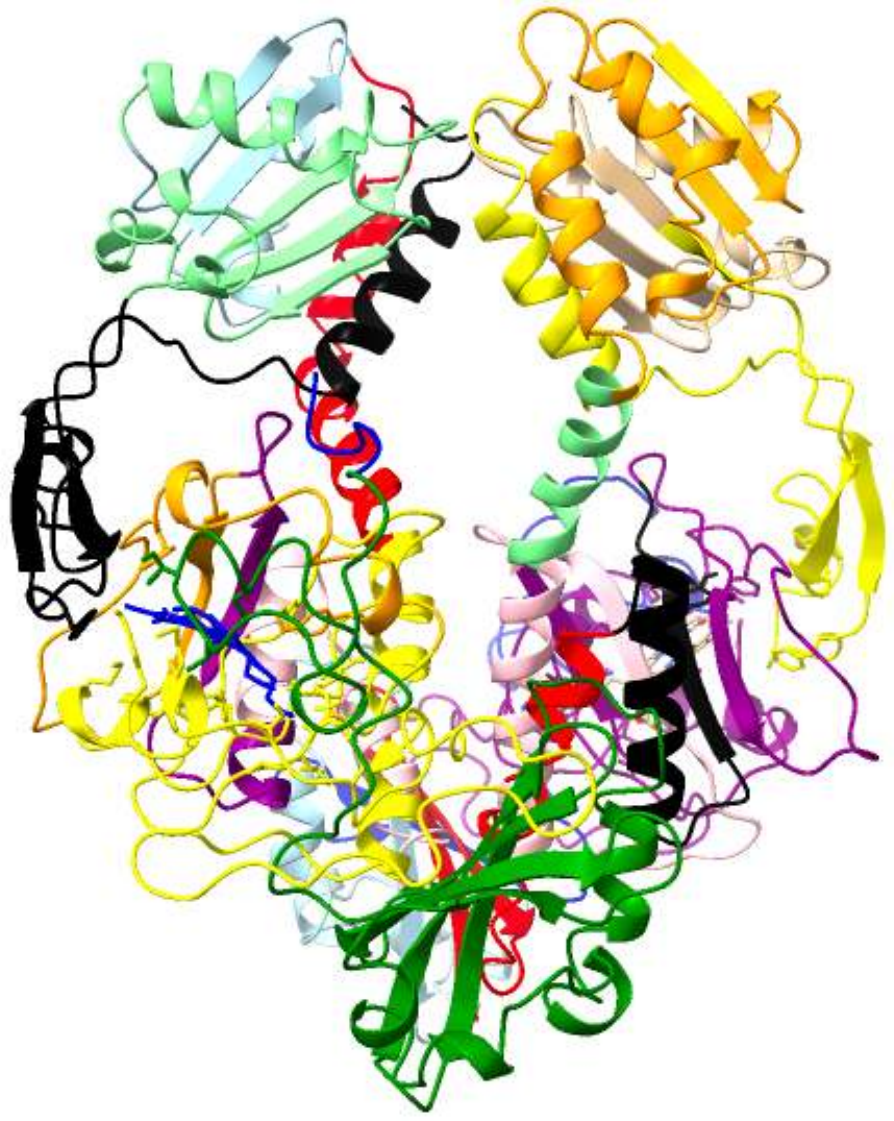}
    \includegraphics[width=0.3\textwidth]{figures/experiments/phy/snr_0_001/segments_25_domains.pdf}
    \caption{MD dataset, SNR 0.001. Segments decomposition. From left to right: 10 segments, 20 segments, 25 segments.}\label{fig:snr_0_001_segments}
\end{figure}

\begin{figure}[h]
    \centering
    \includegraphics[width=0.3\textwidth]{figures/experiments/phy/snr_0_01/segments_snr_0_01_10.pdf}
    \includegraphics[width=0.3\textwidth]{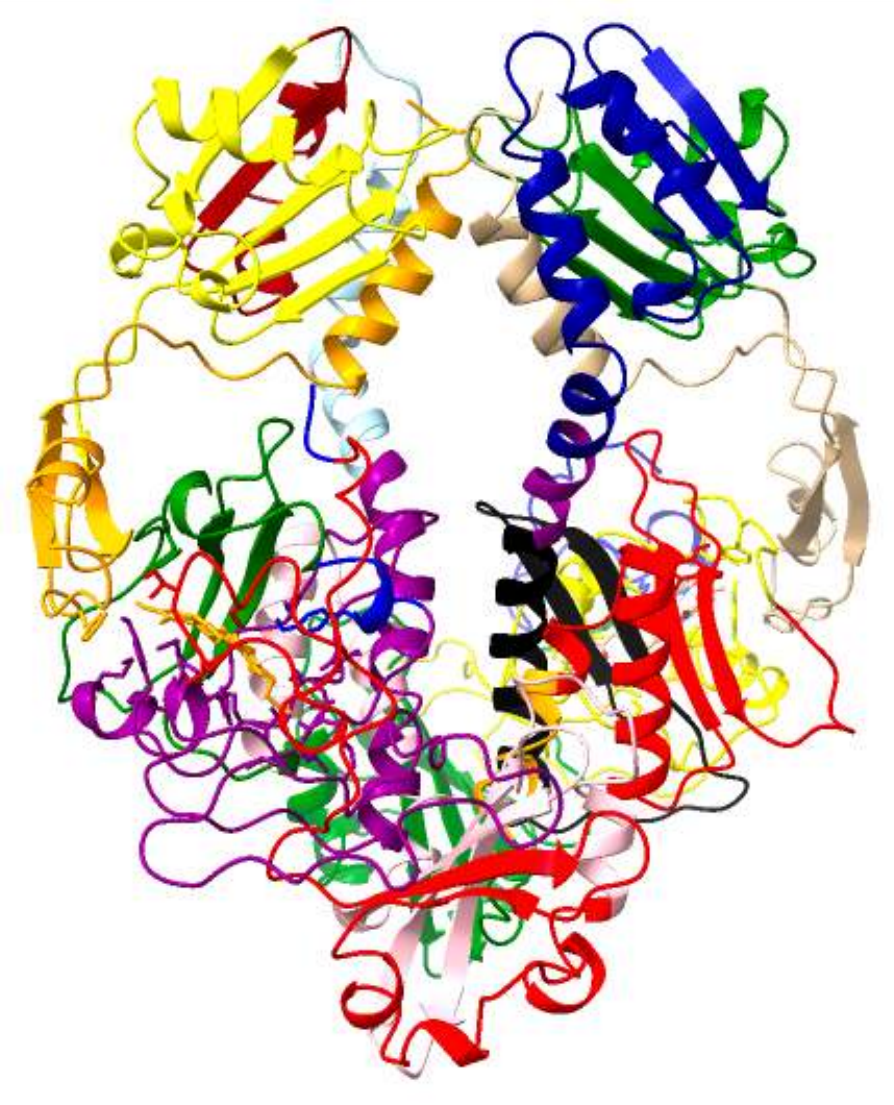}
    \includegraphics[width=0.3\textwidth]{figures/experiments/phy/snr_0_01/25_domains.pdf}
    \caption{MD dataset, SNR 0.01. Segments decomposition. From left to right: 10 segments, 20 segments, 25 segments.}\label{fig:snr_0_01_segments}
\end{figure}

\begin{figure}[h]
    \centering
    \includegraphics[width=0.3\textwidth]{figures/experiments/phy/snr_0_1/segments_snr_0_1_10.pdf}
    \includegraphics[width=0.3\textwidth]{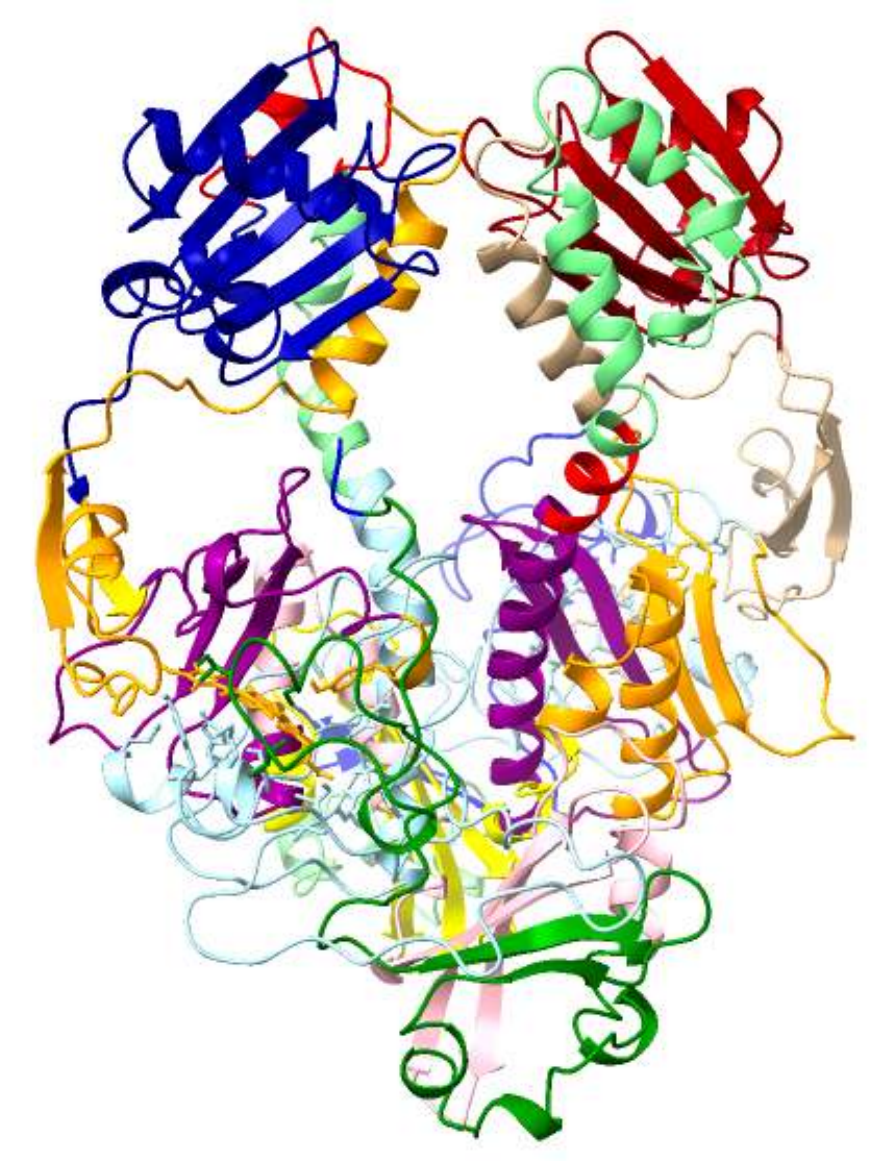}
    \includegraphics[width=0.3\textwidth]{figures/experiments/phy/snr_0_1/25_domains.pdf}
    \caption{MD dataset, SNR 0.1. Segments decomposition. From left to right: 10 segments, 20 segments, 25 segments.}\label{fig:snr_0_1_segments}
\end{figure}

Finally, Figures \ref{fig:snr_0_001_segments},\ref{fig:snr_0_01_segments}and \ref{fig:snr_0_1_segments} show the segment decomposition for different values of $\Ndomains$. CryoSPHERE always learns a segmentation differentiating the two chains and the top and bottom parts of the protein.
\subsection{EMPIAR-10180}\label{append:empiar10180}
This section gives more details on the experiment with the EMPIAR-10180 datasets described in Section \ref{empiar10180}.
The data was processed by Relion, hence the poses and CTF are assumed to be known.
We use an encoder with 4 hidden layers of size 512, 256, 64, 64 and a decoder with 2 hidden layers with size 512, 512.

We train cryoSPHERE with no clashing nor continuity loss, with $\Ndomains=20$. We use the ADAM optimizer with a learning rate of 0.00003 for the parameters of the encoder and decoder, while we set the learning rate to 0.0003 for the parameters of the GMM segmentation.

We low pass filter the images with a bandwith of $23.4$Å, we apply a mask of radius $0.9375$ to the input images and we apply a mask of radius $1$ to the true and predicted images for the computation of the correlation loss.

\begin{figure}
    \centering
    \includegraphics[scale=0.4]{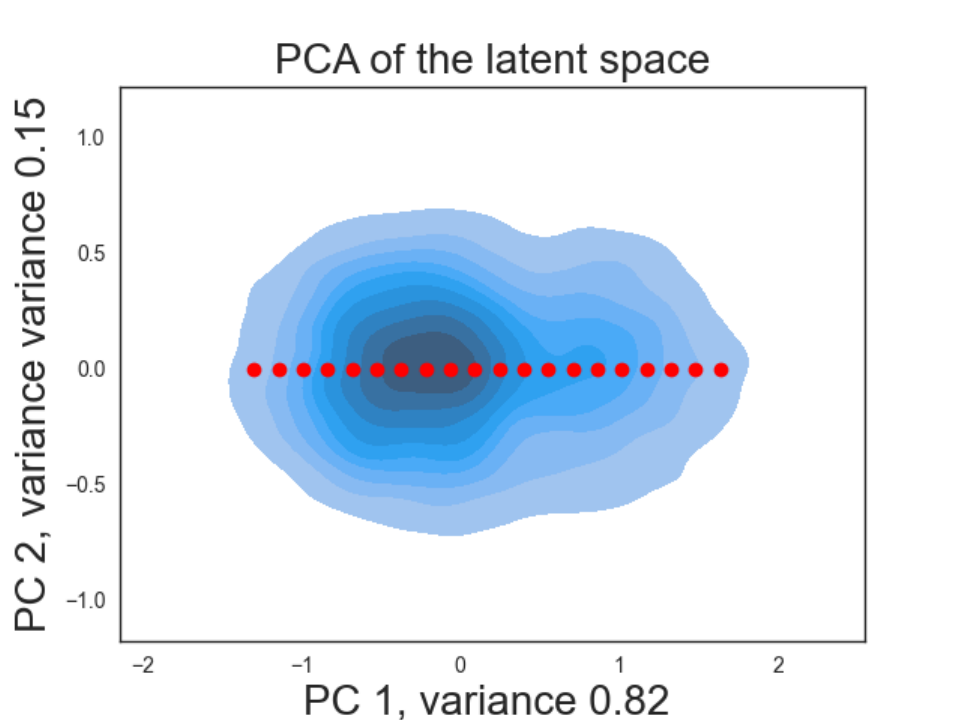}
    \hspace{0.5cm}
    \caption{Empiar 10180: kernel density plot of the first and second principal components of that latent space of cryoSPHERE. The red dots are the point selected for the traversal of the first principal component.}
    \label{fig:empiar10180_latent_space}
\end{figure}

\begin{figure}
    \centering
    \includegraphics[scale=0.15]{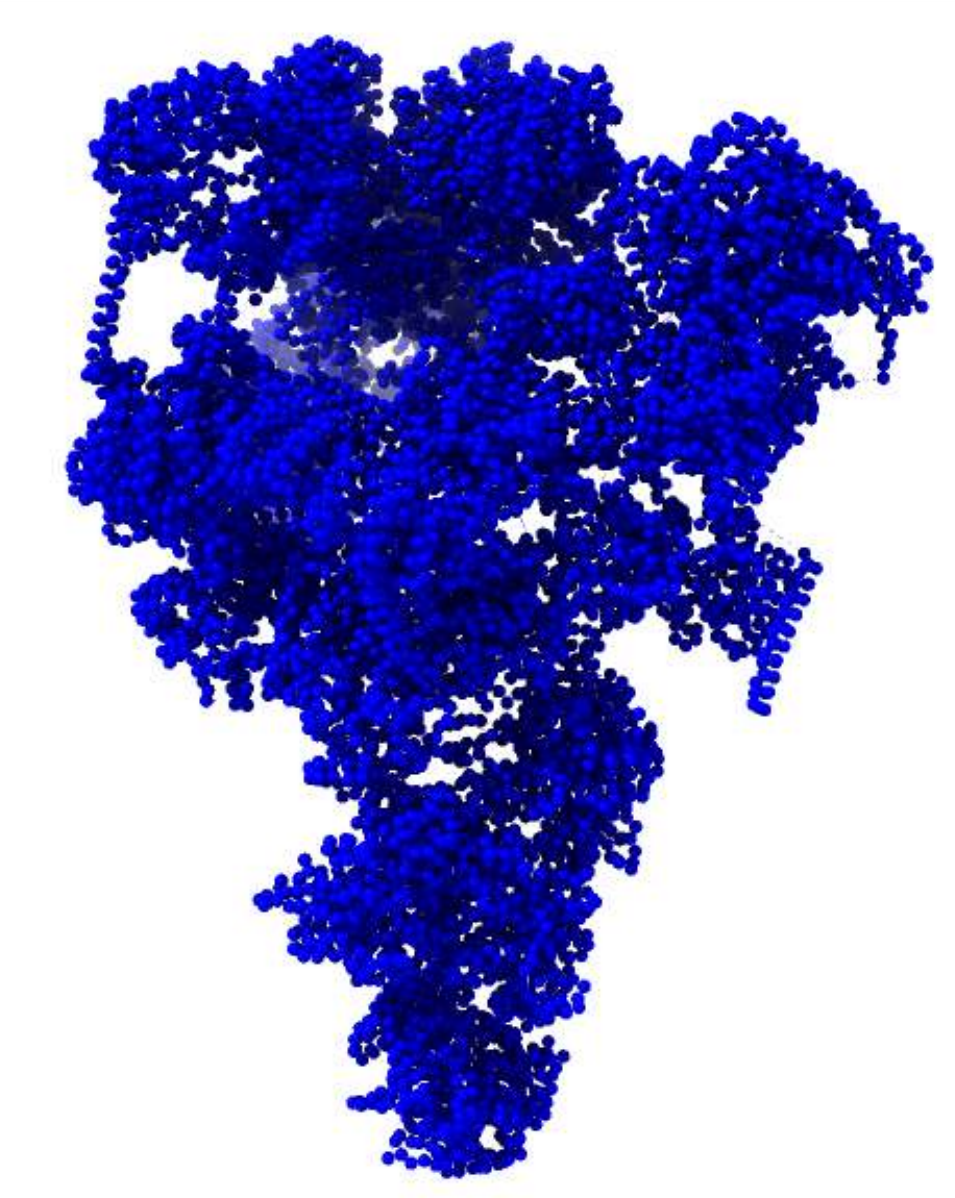}
    \hspace{0.5cm}
    \includegraphics[scale=0.15]{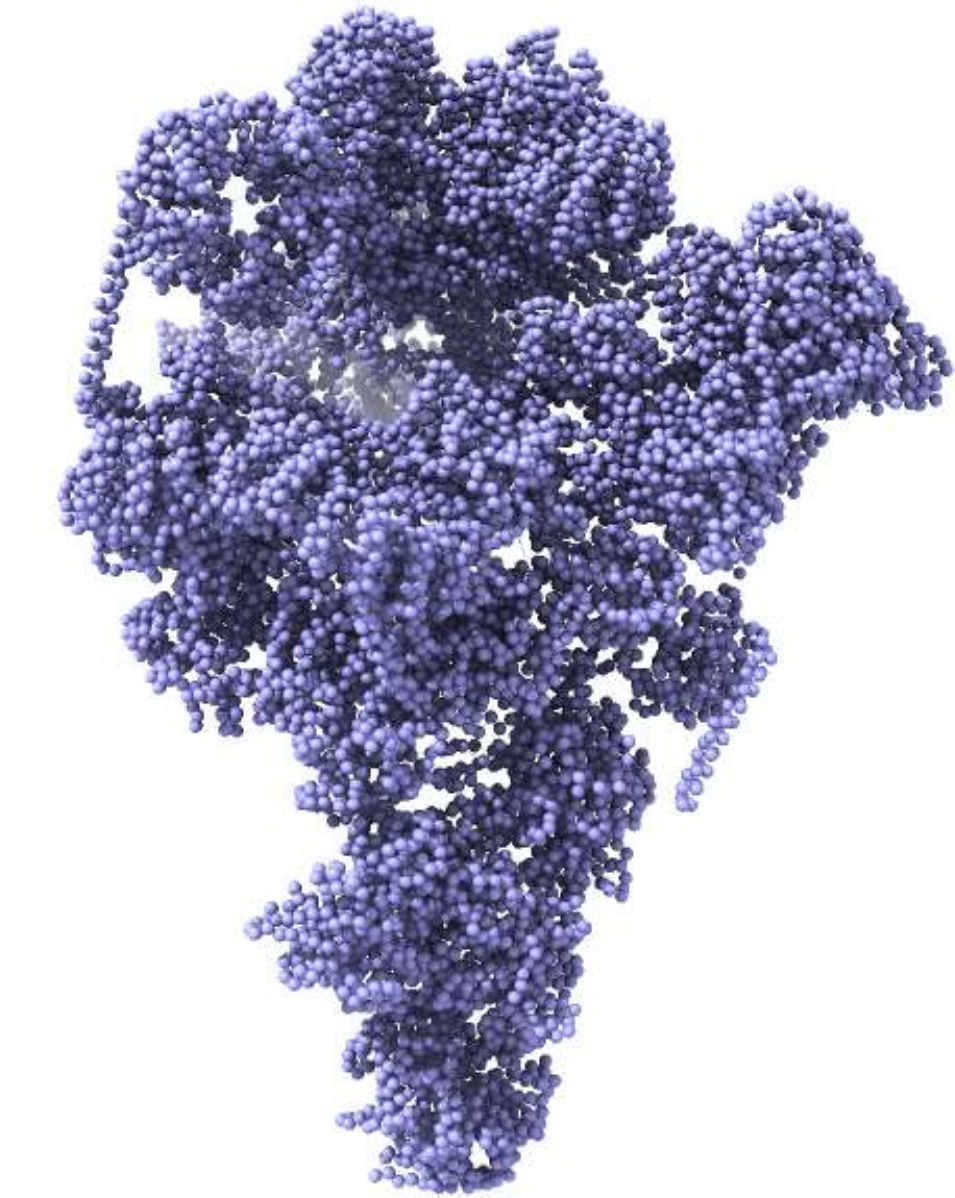}
    \hspace{0.5cm}
    \includegraphics[scale=0.15]{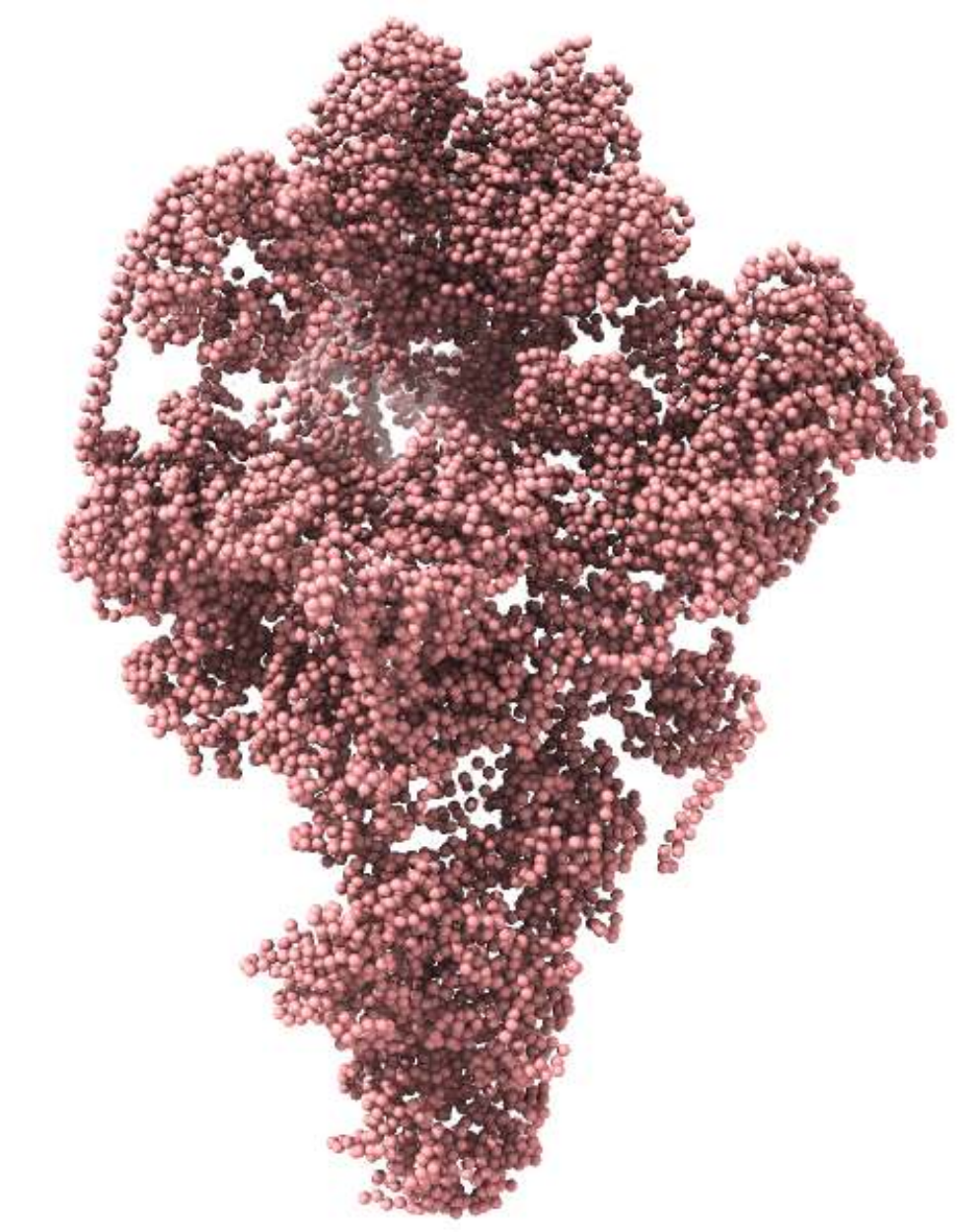}
    \hspace{0.5cm}
    \includegraphics[scale=0.15]{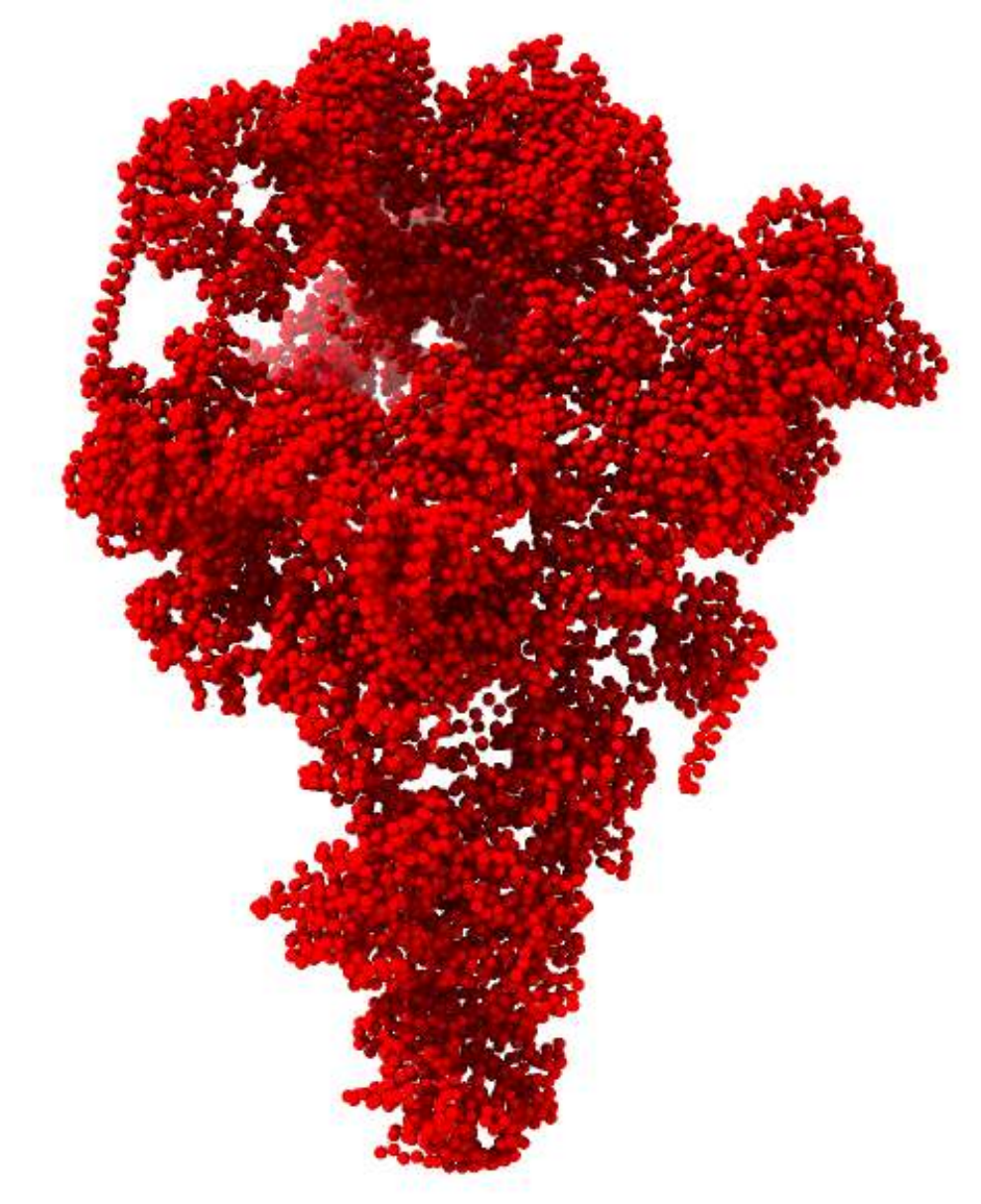}\\
    \includegraphics[scale=0.15]{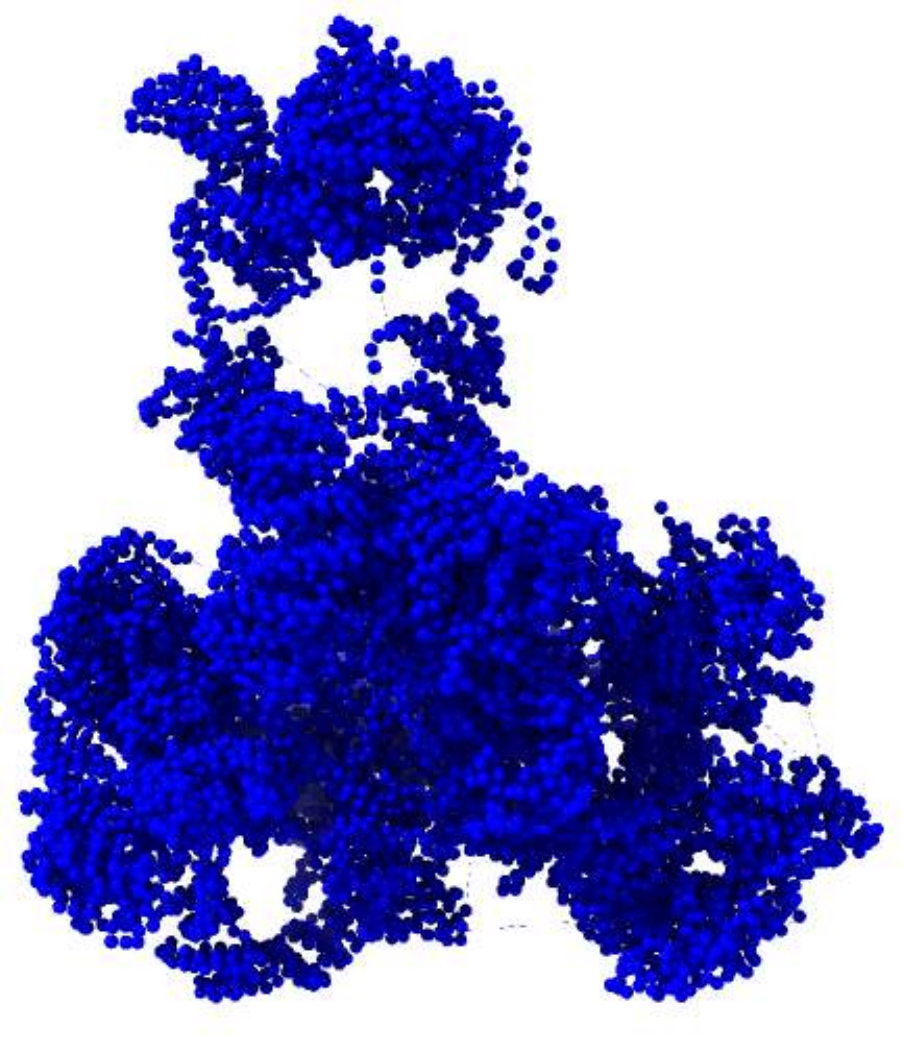}
    \hspace{0.5cm}
    \includegraphics[scale=0.15]{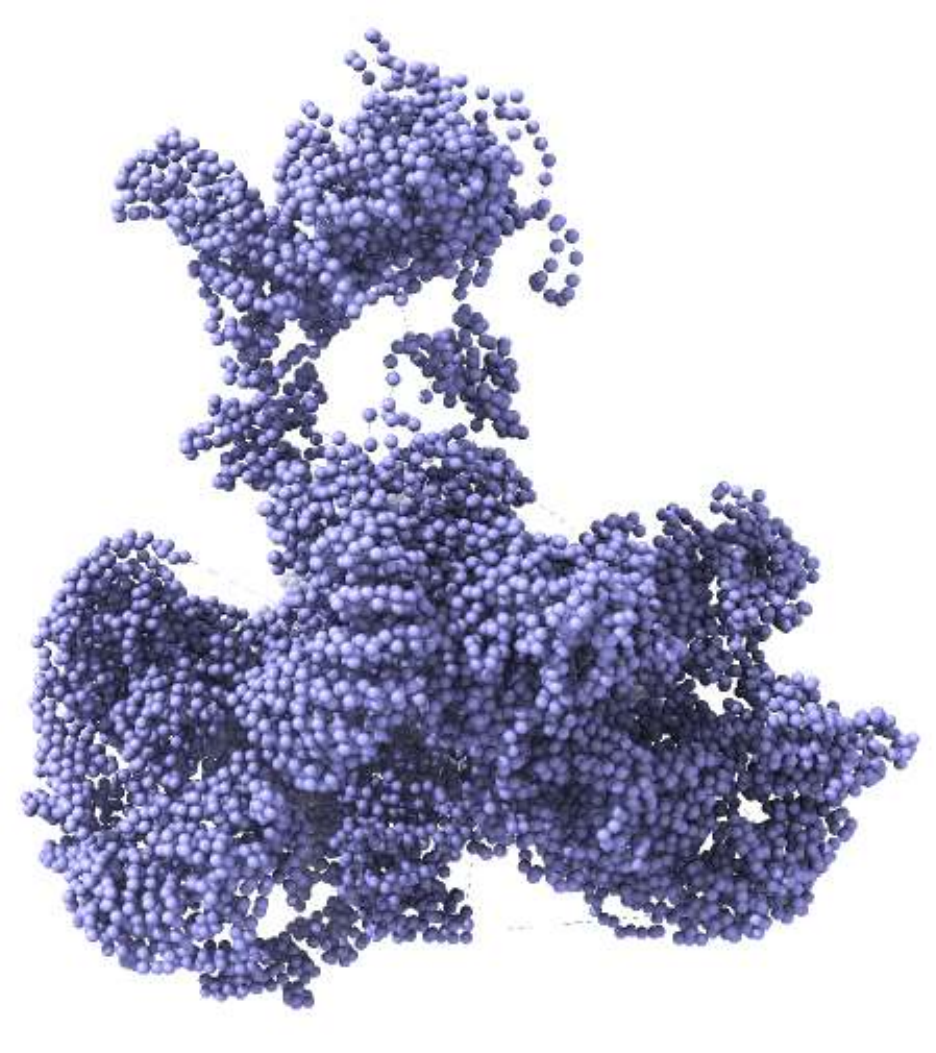}
    \hspace{0.5cm}
    \includegraphics[scale=0.15]{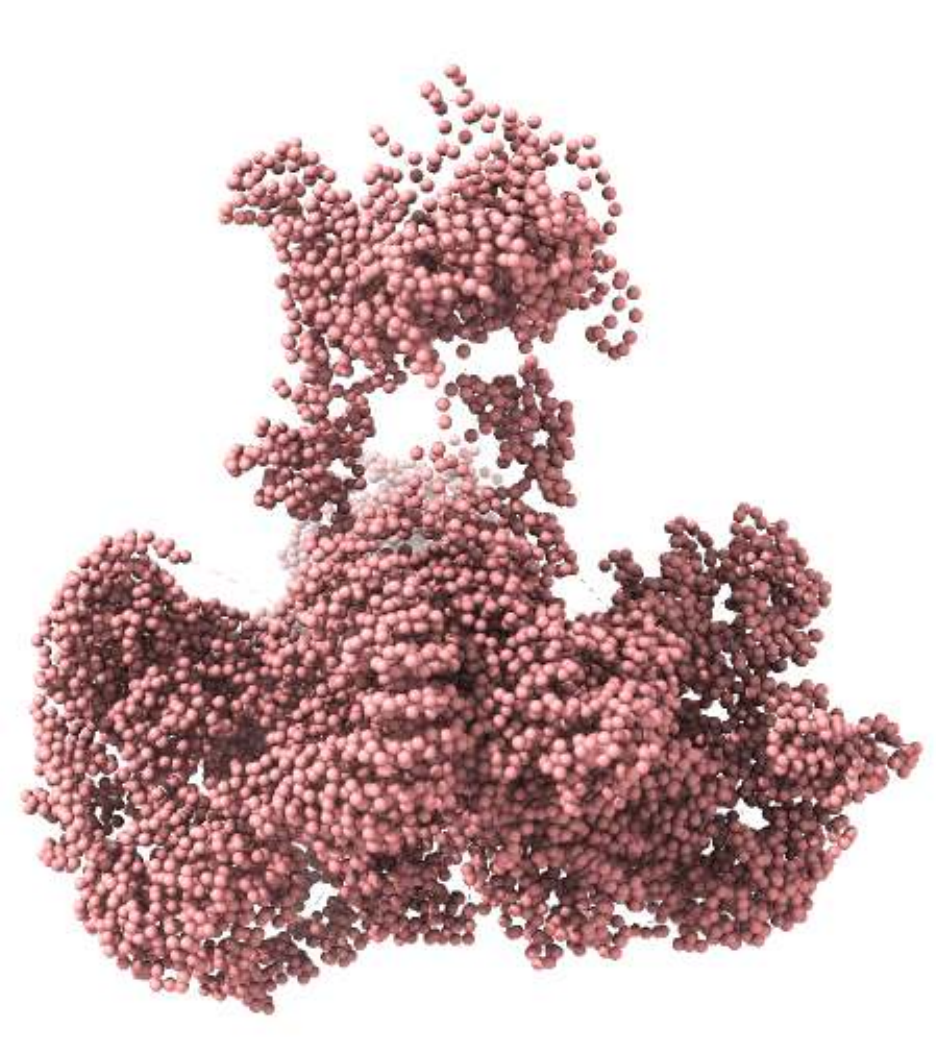}
    \hspace{0.5cm}
    \includegraphics[scale=0.15]{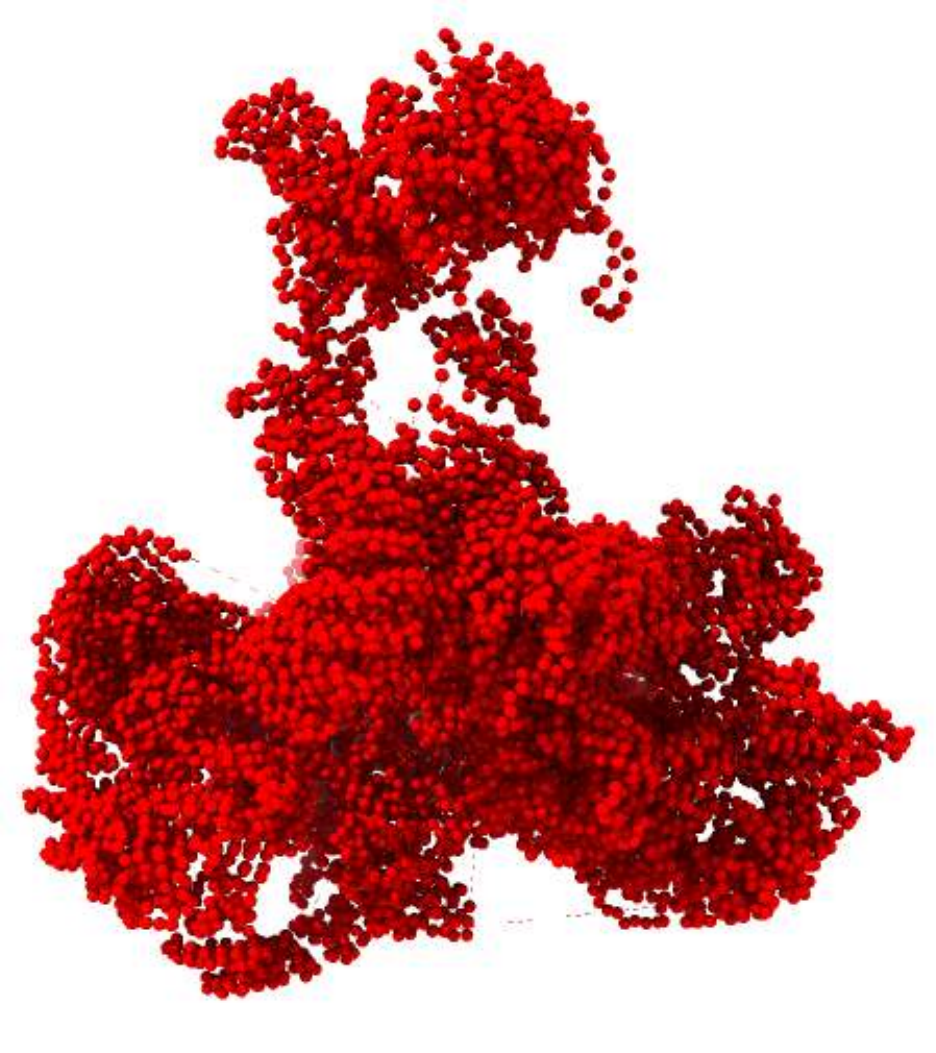}
    \caption{Empiar 10180. Four structures taken along the first principal component, from blue to white to red. Top: view from the "back". Bottom: view from the "top".}
    \label{fig:empiat10180_more_volumes}
\end{figure}

Figure \ref{fig:empiat10180_more_volumes} shows 4 structures taken from the principal component traversal depicted in Figure \ref{fig:empiar10180_latent_space}. The structures contain only the $C_{\alpha}$ atoms. We provide a movie of the traversals of principal component 1 and principal component 2 by clicking \href{https://drive.google.com/file/d/1KSCibHyyM54mP5MCBUb5LPnsvTlTuB2n/view?usp=share_link}{here}.

\begin{figure}
    \centering
    \includegraphics[scale=0.3]{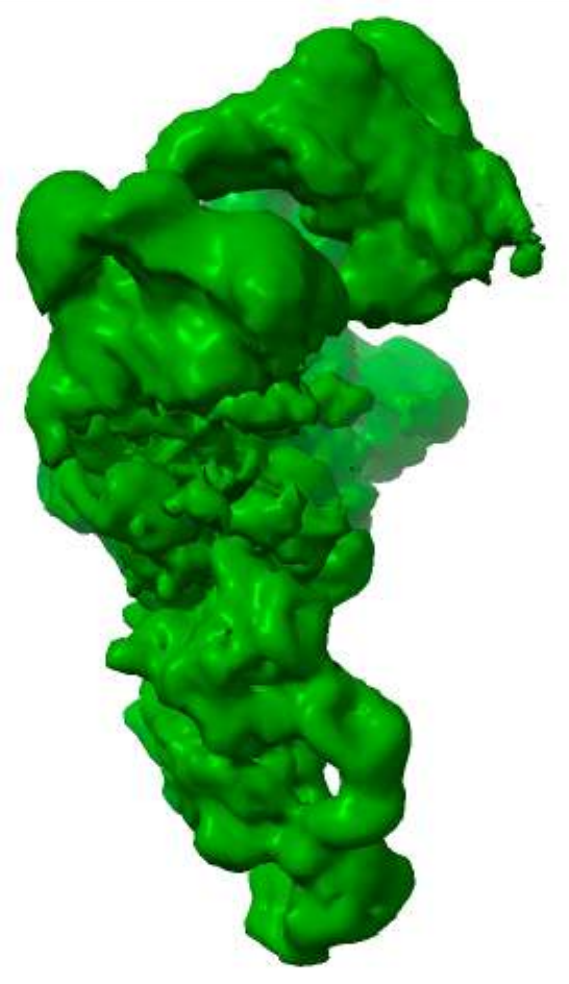}
    \hspace{0.5cm}
    \includegraphics[scale=0.3]{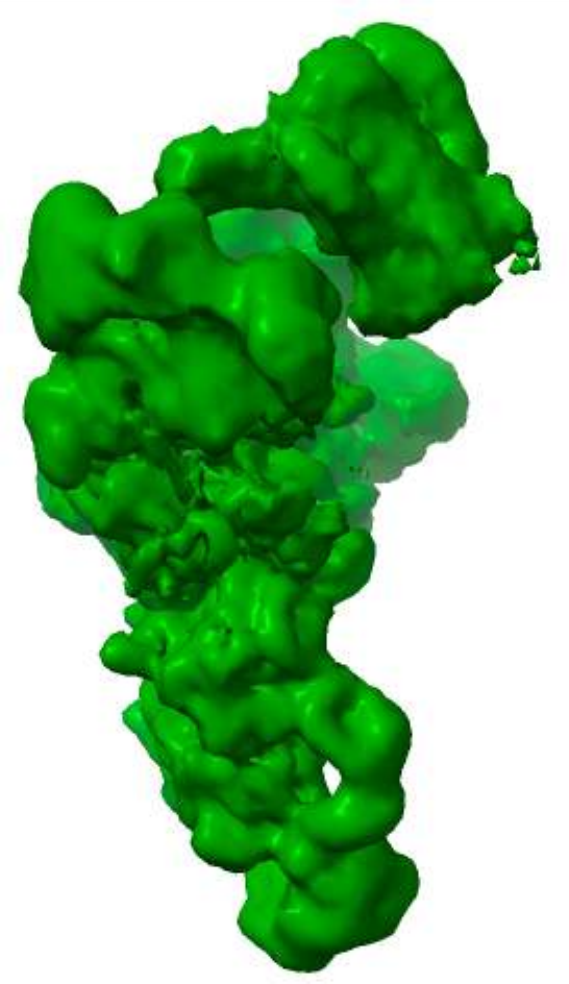}
    \hspace{0.5cm}
    \includegraphics[scale=0.3]{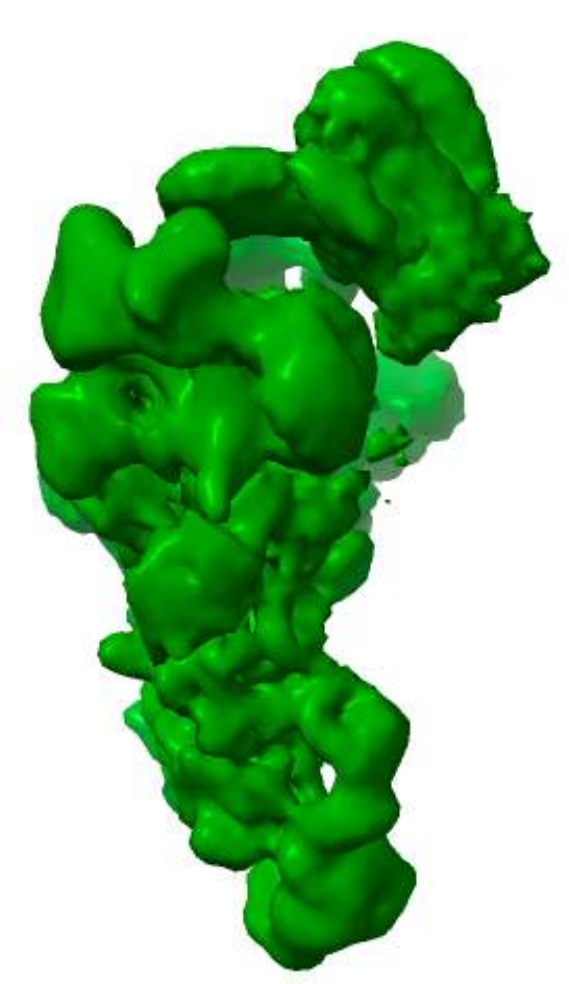}\\
    \includegraphics[scale=0.3]{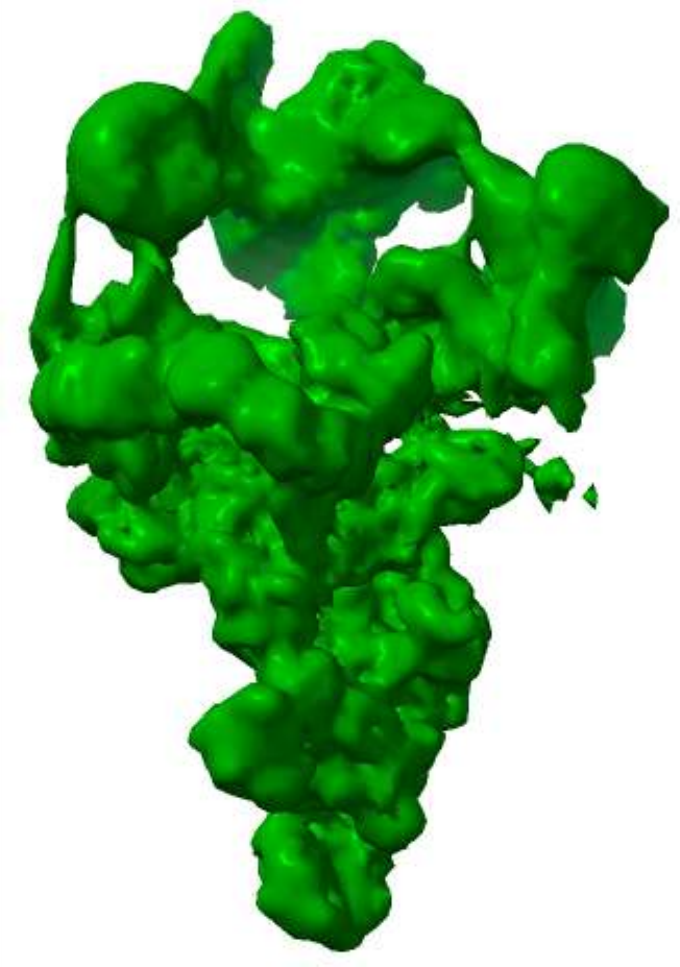}
    \includegraphics[scale=0.3]{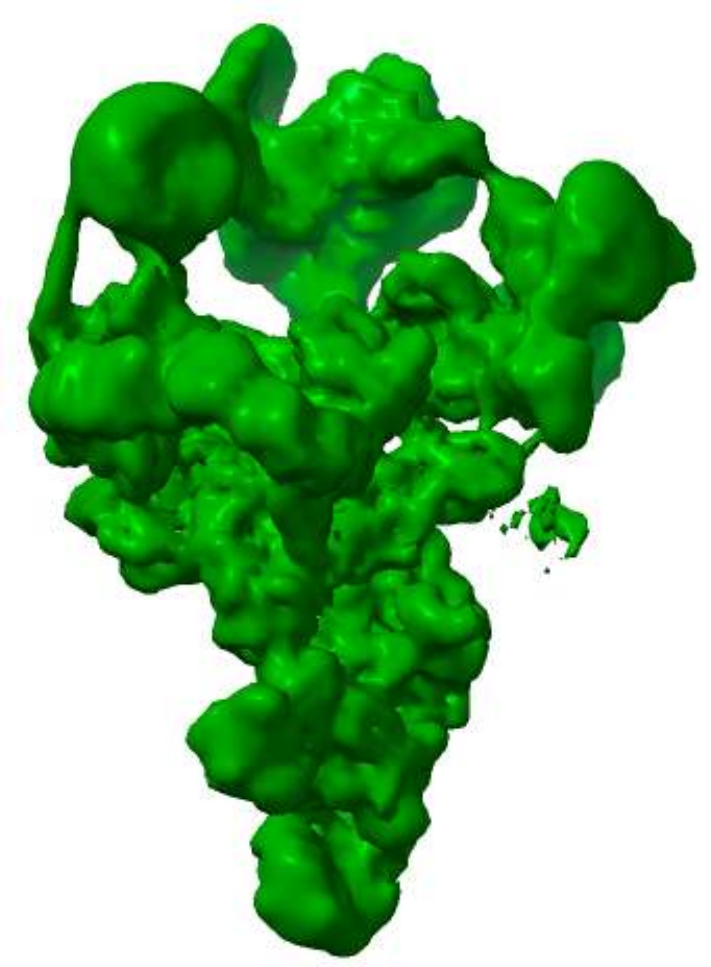}
    \includegraphics[scale=0.3]{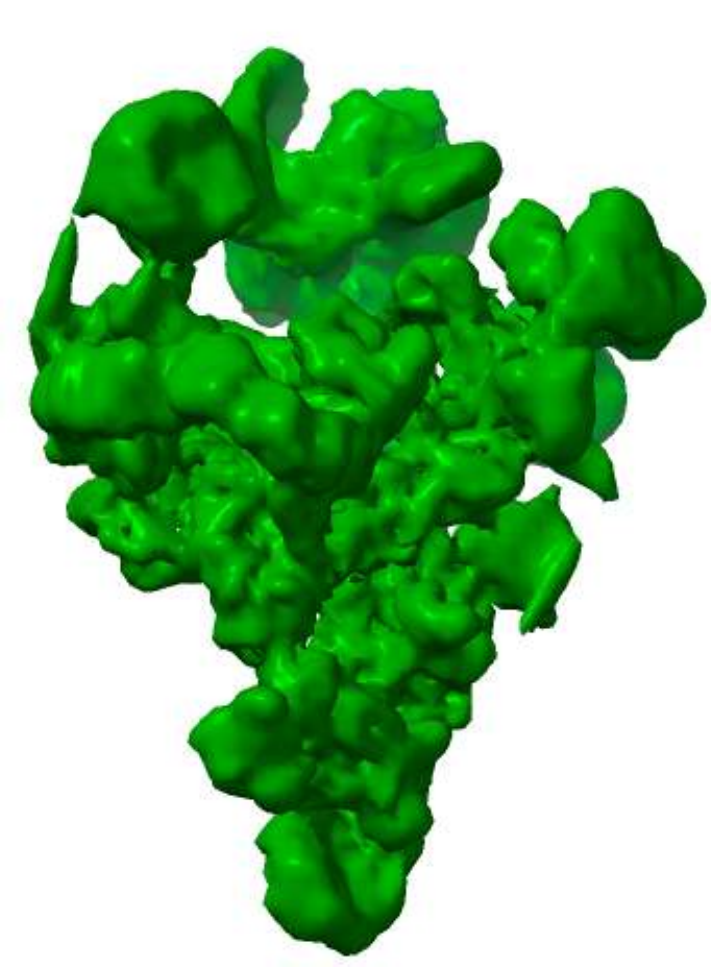}
    \caption{Empiar 10180, DRGN-AI is trained on the latent space of cryoSPHERE. Three Volumes taken evenly along the principal component.}
    \label{fig:empiar10180_drgnai}
\end{figure}

We subsequently train DRGN-AI on the latent space provided by cryoSPHERE, similar to cryoStar Phase II of training. We show volumes taken along the first principal component in Figure \ref{fig:empiar10180_drgnai}. We recover the correct bending of the protein toward its "foot". In addition, this second step detects the compositional heterogeneity and the density is zero in this region. This is a detail that the structural method of cryoSPHERE could not detect. We successfully identified a bias in the base structure.

We provide a movie of the motion recovered by DRGN-AI trained on the latent variable of cryoSPHERE \href{https://drive.google.com/file/d/1YIVhyBJQk4FaTpPRXUiUm9u3PWUVxdVc/view?usp=share_link}{here}.

\subsection{EMPIAR-12093}\label{append:EMPIAR-12093}
This experiment demonstrates that cryoSPHERE is applicable to real data with high noise levels. We applied cryoDRGN, cryoSTAR and cryoSPHERE to a bacterial phytochrome \citep{bodizs_cryo-em_2024} dataset (medium-sized protein, ~120 kDa). This dataset comprises two distinct subsets of 200 000 images of size 400 x 400 each, representing the protein in its light-activated state (red-absorbing state, called Pr) and its resting state (far-red-absorbing state, called Pfr).
The pre-processing steps are detailed in \cite{bodizs_cryo-em_2024}. We downgrade the images to a size of 256 x 256.
We give a computational budget of 24 hours on the same single GPU to cryoDRGN, cryoStar structural method and cryoSphere and compare their results. We subsequently run cryoStar volume method on the latent space obtained by cryoStar and cryoStar volume method on the latent space obtained by cryoSPHERE.

We provide PC 1 traversal movies for both cryoSPHERE and cryoStar structures and volumes for both Pr and Pfr \href{https://drive.google.com/drive/folders/15VvcAn3vh1JuJBkvlBUd4mGmZpBww_TF?usp=share_link}{here}.

\subsubsection{Pfr state}
\begin{figure}[h]
    \centering
    \includegraphics[width=0.3\textwidth]{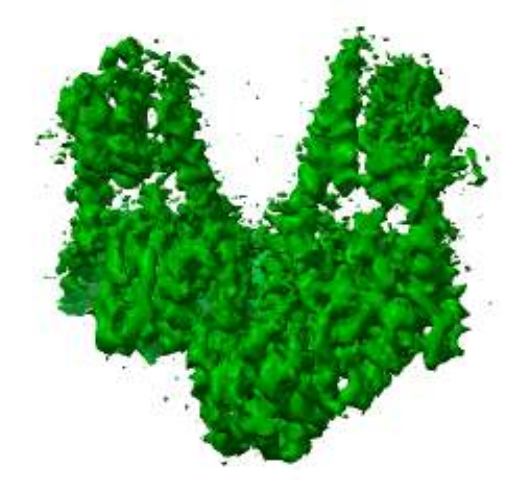}
    \includegraphics[width=0.3\textwidth]{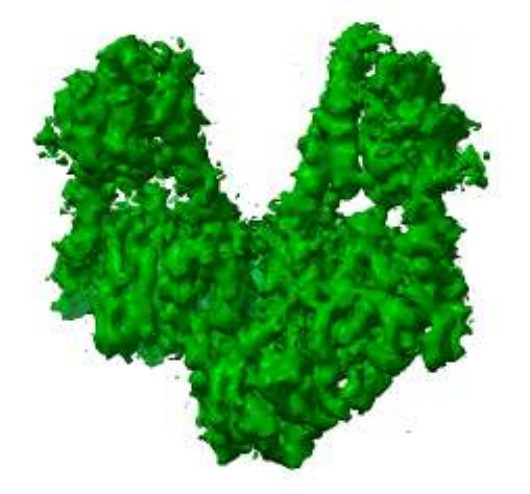}
    \includegraphics[width=0.3\textwidth]{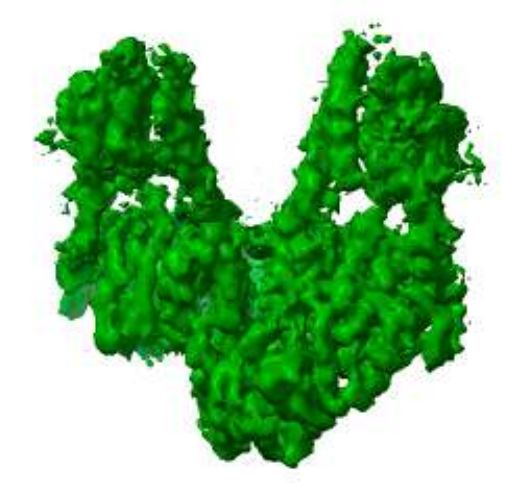}
    \caption{Pfr: 3 volumes taken evenly along the first principal component of cryoDRGN.}\label{fig:Pfr_cryodrgn_pc1}
\end{figure}

\begin{figure}[h]
    \centering
    \includegraphics[width=0.3\textwidth]{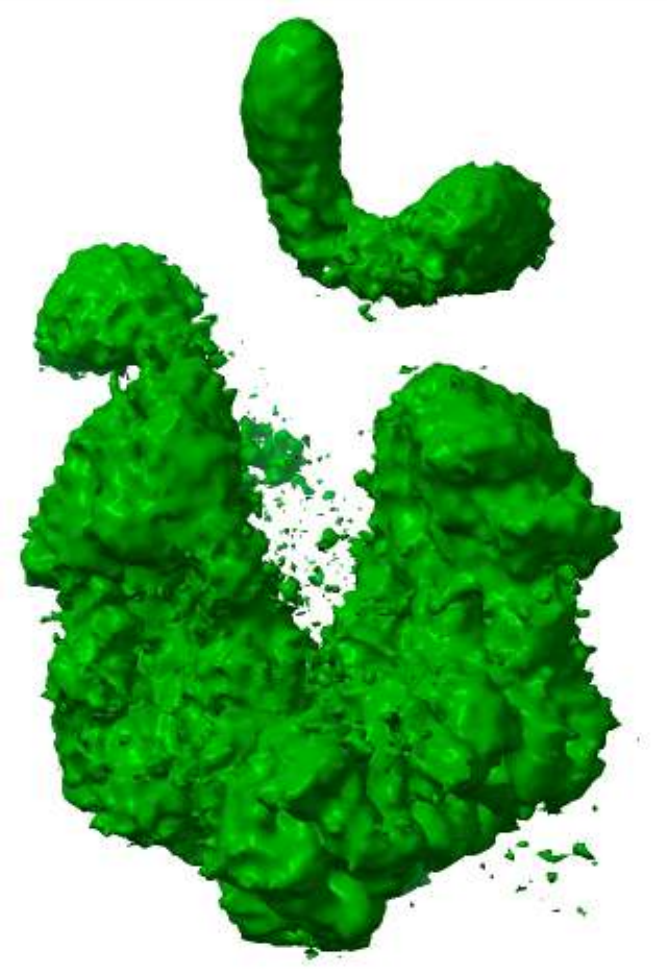}
    \includegraphics[width=0.3\textwidth]{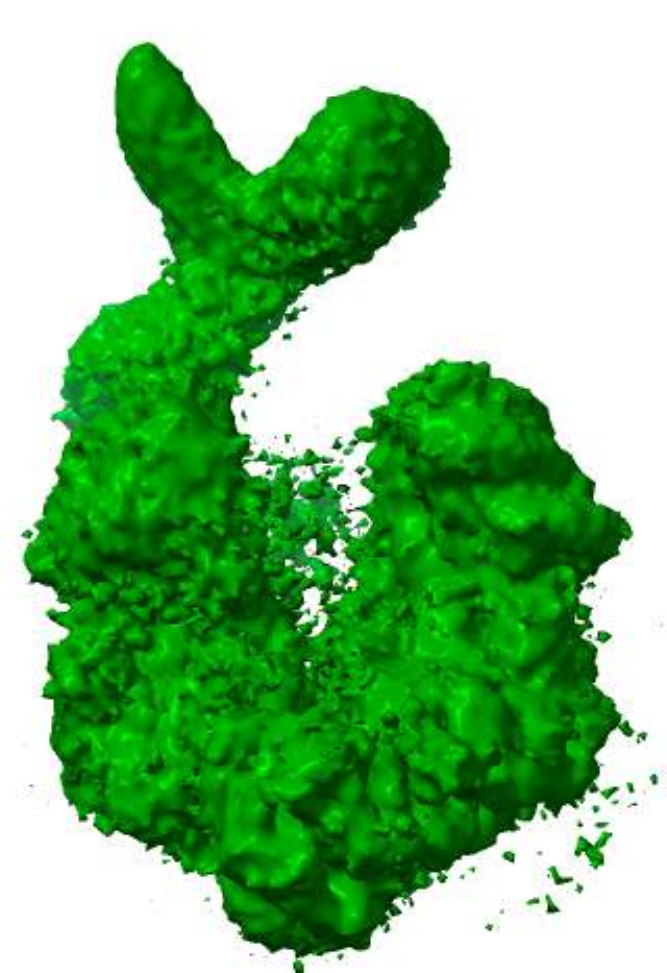}
    \includegraphics[width=0.3\textwidth]{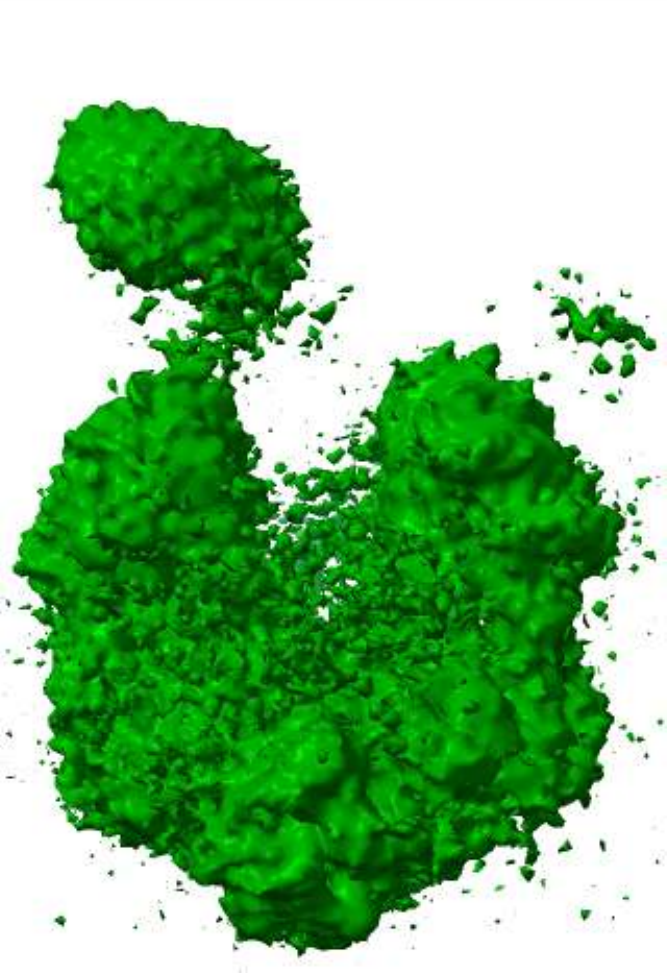}
    \caption{Pfr: 3 volumes taken evenly along the first principal component of cryoStar volume method.}\label{fig:Pfr_cryostar_pc1}
\end{figure}

We provide the first PC traversal for cryoDRGN in Figure \ref{fig:Pfr_cryodrgn_pc1}. CryoDRGN does not recover the upper part of the protein at all. In addition, there is no motion through the principal component. This might indicate that the recovered motion is in fact noise in the top part.
We additionally plot 3 structures taken evenly along the first principal component of the cryoStar volume method in Figure \ref{fig:Pfr_cryostar_pc1}. The method is also unable to reconstruct the very mobile top part, in spite of using the latent space of the structural method. For Pfr, the debiasing technique of cryoStar through a volume method is ineffective for the top part.

\begin{figure}[h]
    \centering
    \includegraphics[width=0.3\textwidth]{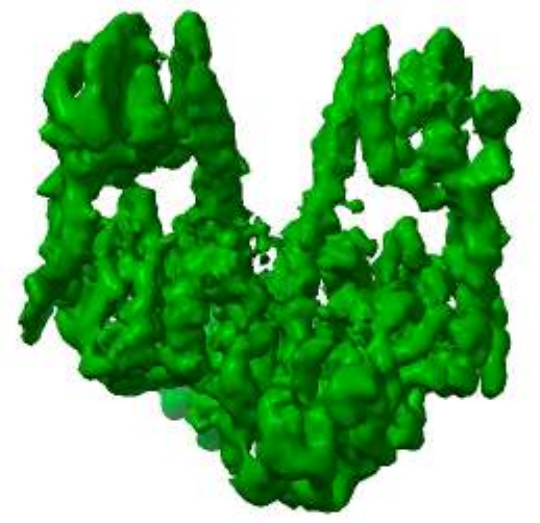}
    \includegraphics[width=0.3\textwidth]{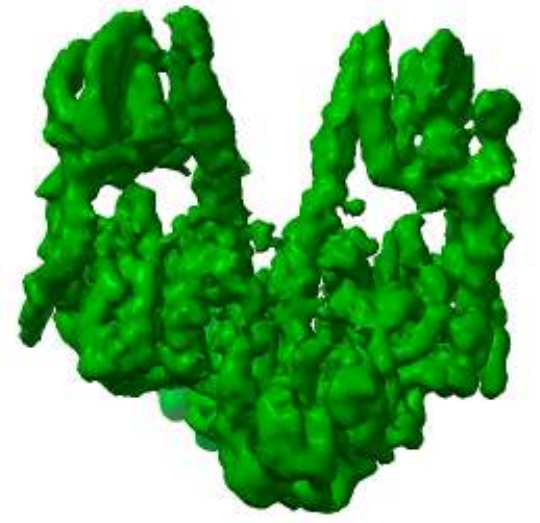}
    \includegraphics[width=0.3\textwidth]{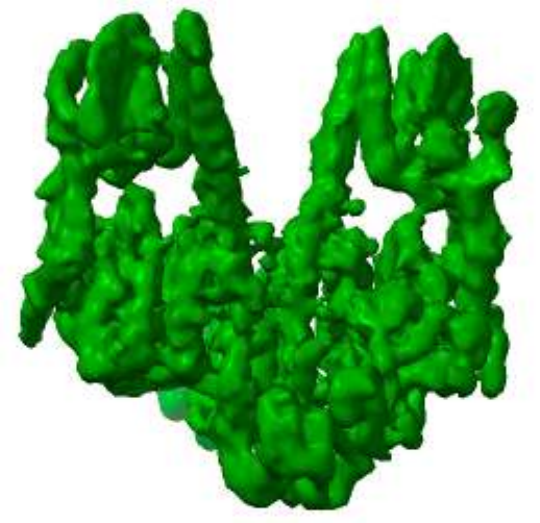}
    \caption{Pfr state. CryoStar volume method is trained on the latent space of cryoSPHERE. From left to right: three volumes taken evenly along the first principal component.}\label{fig:Pfr_cryosphere_phase_II}
\end{figure}

Finally, we run cryoStar volume method on the latent space recovered by cryoSPHERE and show the first principal component traversal in Figure \ref{fig:Pfr_cryosphere_phase_II}. Similar to cryoDRGN and cryoStar Phase II, this procedure is unable to reconstruct the top part of the protein. Hence this allows debiasing on the bottom part of the protein only.

We provide a movie of the first PC traversal of cryoStar volume method with cryoSPHERE latent variables \href{https://drive.google.com/file/d/17bbZHWJfugZtxFE636HioLK_LrLz5Z6R/view?usp=share_link}{here} and the same movie with cryoStar latent variables \href{https://drive.google.com/file/d/1KNSvyleBSNw6LY-YprQsnsWc9JyZuRxs/view?usp=share_link}{here}.

\subsubsection{Pr state}
\begin{figure}[h]
    \centering
    \includegraphics[width=0.3\textwidth]{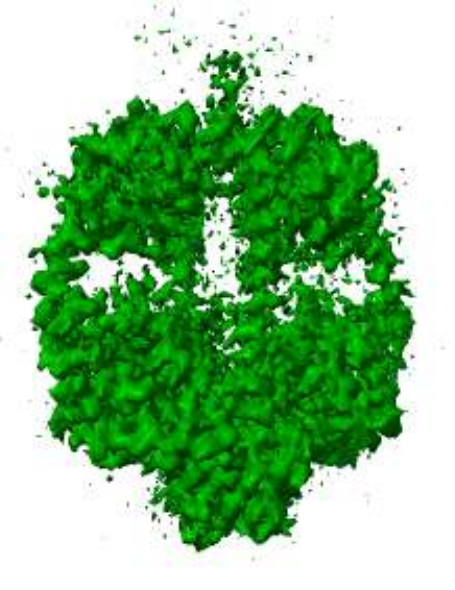}
    \includegraphics[width=0.3\textwidth]{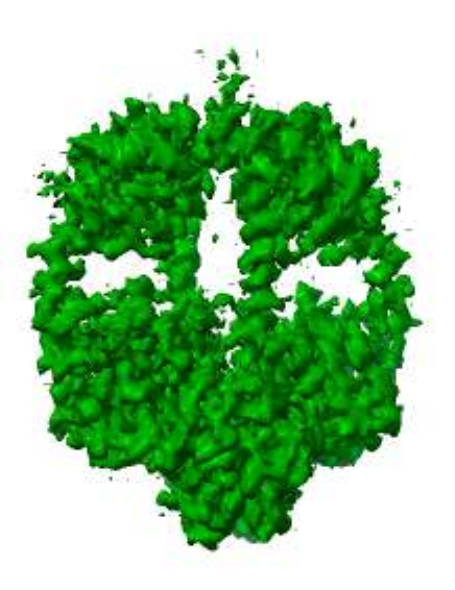}
    \includegraphics[width=0.3\textwidth]{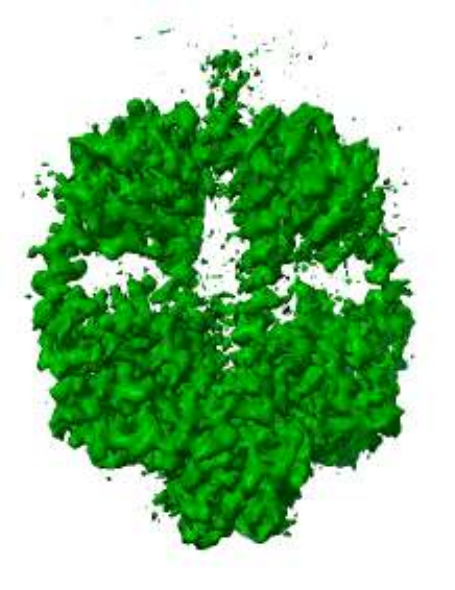}\\
    \includegraphics[width=0.3\textwidth]{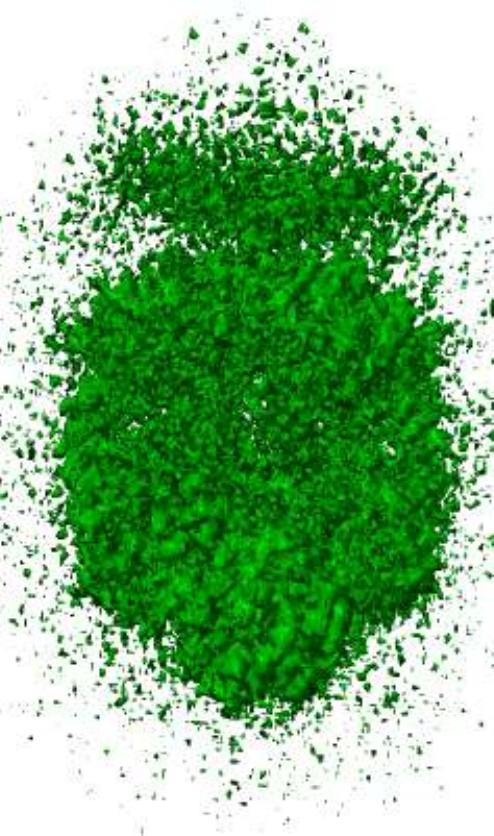}
    \includegraphics[width=0.3\textwidth]{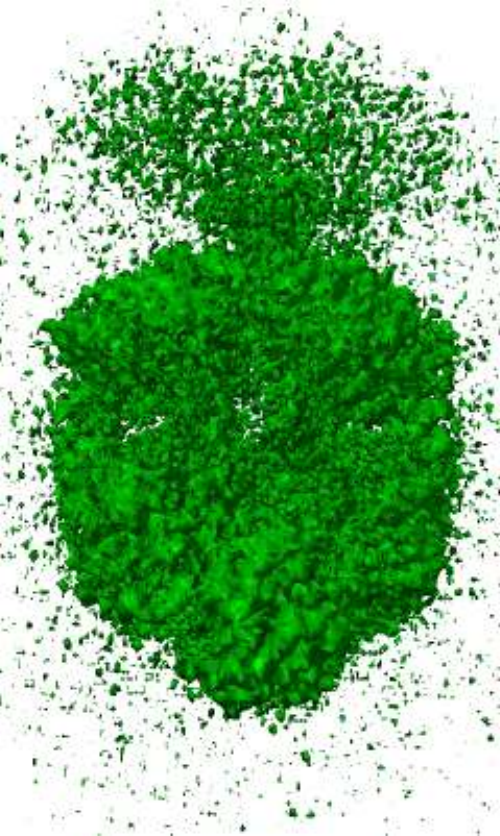}
    \includegraphics[width=0.3\textwidth]{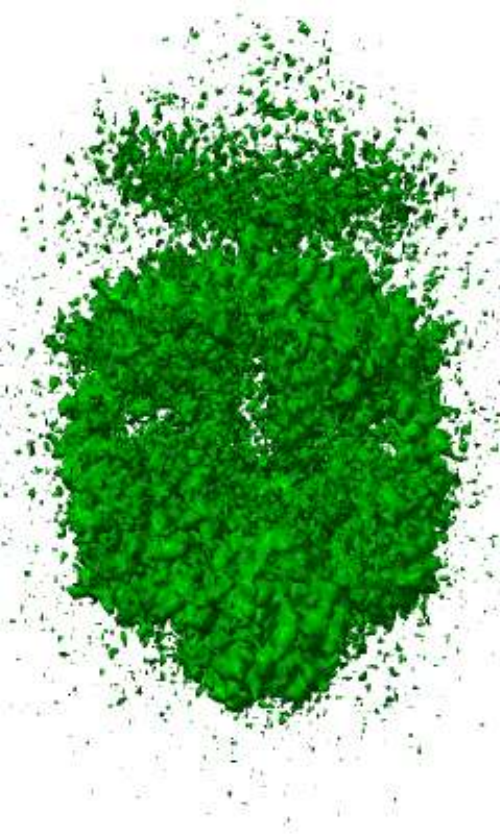}
    \caption{Pr: 3 volumes taken evenly along the first principal component of cryoDRGN volume method. The top and bottom volumes are the same with a different density threshold. }\label{fig:Pr_cryodrgn}
\end{figure}

We provide three structures taken evenly along the first principal component of cryoDRGN in Figure \ref{fig:Pr_cryodrgn}. There is no motion in the bottom part as expected. However, the method is unable to recover the top part.

\begin{figure}[h]
    \centering
    \includegraphics[width=0.3\textwidth]{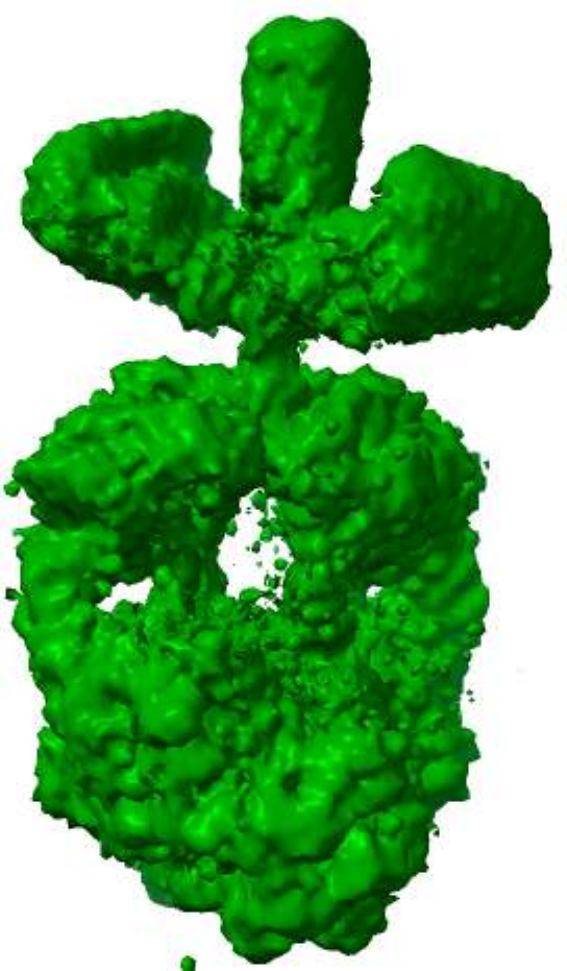}
    \includegraphics[width=0.3\textwidth]{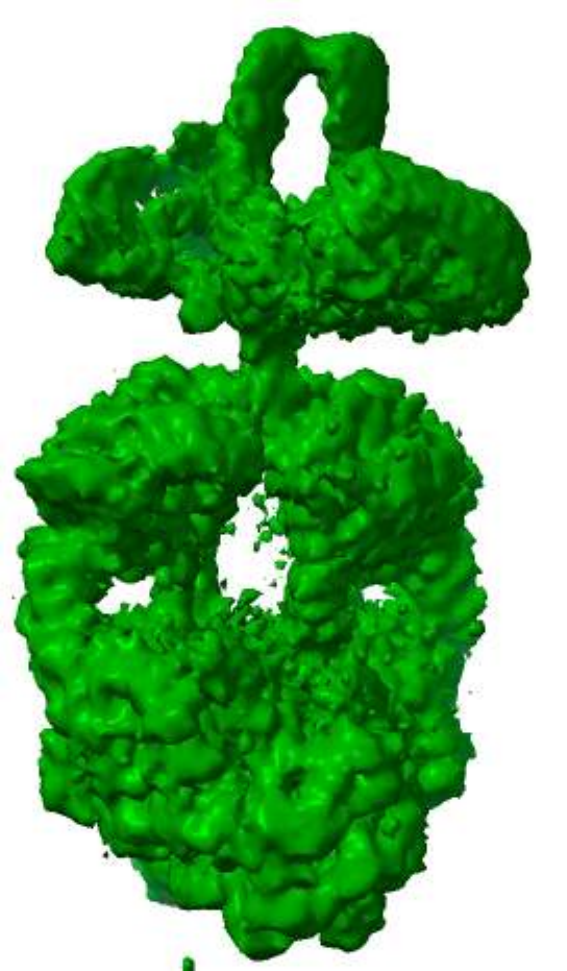}
    \includegraphics[width=0.3\textwidth]{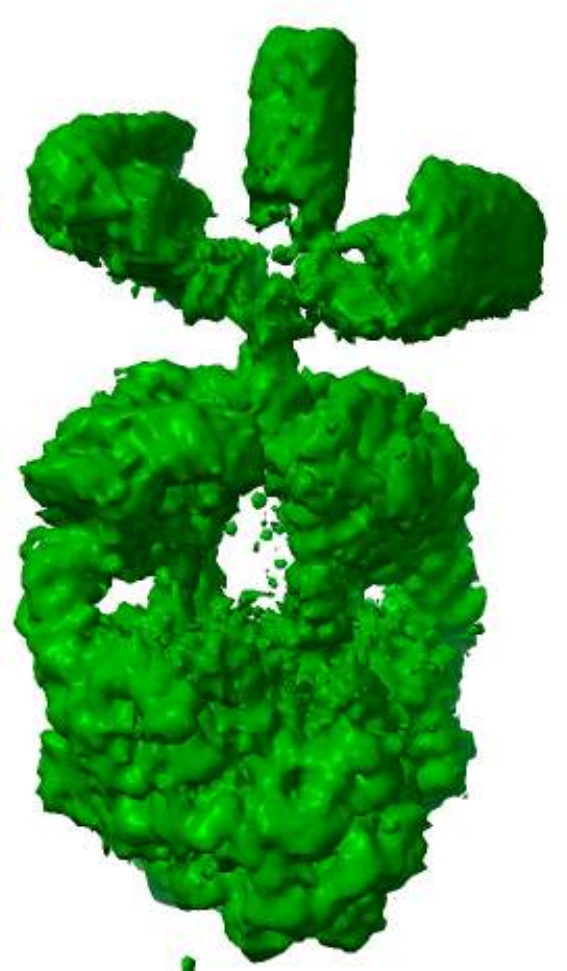}
    \caption{Pr: 3 volumes taken evenly along the first principal component of cryoStar volume method. }\label{fig:Pr_cryostar_volume}
\end{figure}

\begin{figure}[h]
    \centering
    \includegraphics[width=0.3\textwidth]{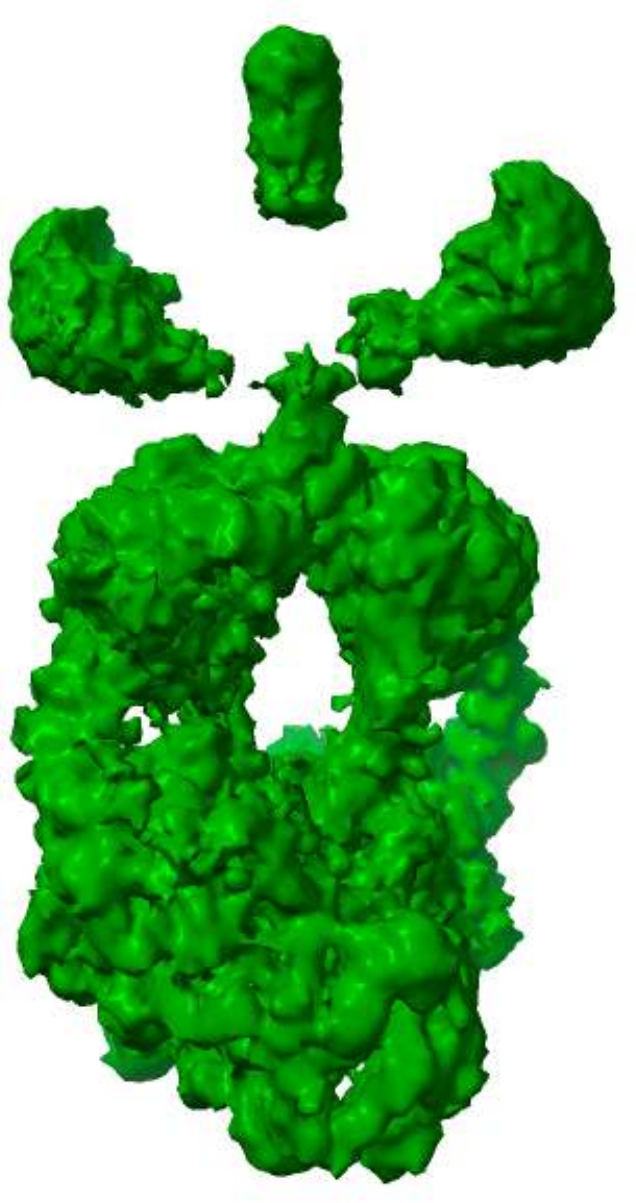}
    \includegraphics[width=0.3\textwidth]{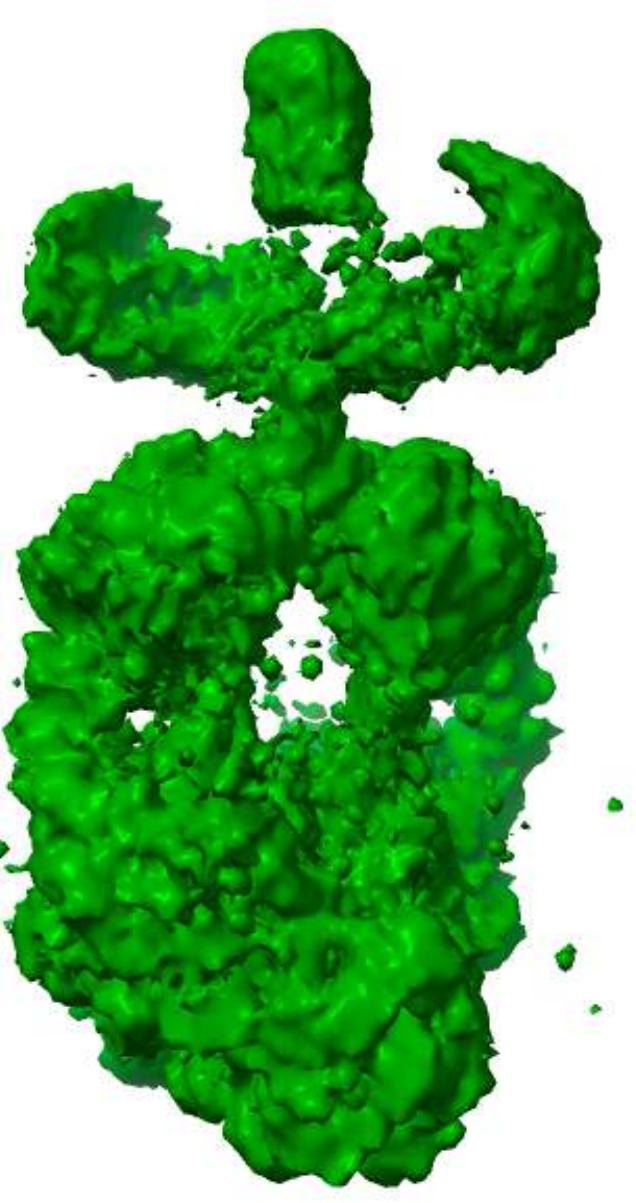}
    \includegraphics[width=0.3\textwidth]{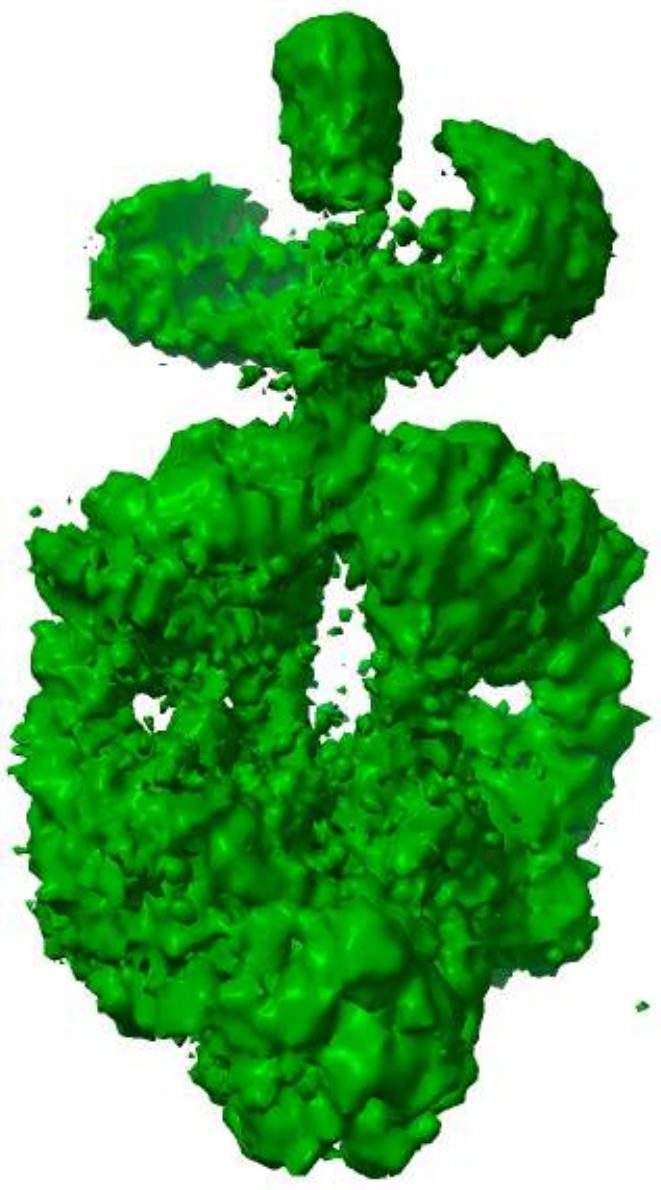}
    \caption{Pr state. CryoStar volume method is trained using the latent space of cryoSPHERE. From left to right: volumes taken evenly along the first principal component.}\label{fig:Pr_cryosphere_phase_II}
\end{figure}

We also run cryoStar volume method on the latent space recovered by cryoSPHERE and display the first principal component traversal in Figure \ref{fig:Pr_cryosphere_phase_II}. We encounter the same difficulties as cryoStar volume method and cryoDRGN, see Figures \ref{fig:Pr_cryodrgn}, \ref{fig:Pr_cryostar_volume}. CryoDRGN is not able to recover the top part of the protein. While cryoStar volume method does recover some motion for this top part, the resolution is too low to debias the base structure. Therefore, this debiasing procedure is only useful for the lower part of the protein.

We provide a movie of the first PC traversal of cryoStar volume method with cryoSPHERE latent variables \href{https://drive.google.com/file/d/1ty-PRHh1vmVK98KuyXmw5LgqAeqQfedn/view?usp=share_link}{here} and the same movie with cryoStar latent variables \href{https://drive.google.com/file/d/1qjyQmmTTWDqq1KMFl0jfpjb0K3LwvUhY/view?usp=share_link}{here}.

\subsection{Computational costs}\label{append:cost}
CryoStar and cryoSPHERE share the same way of turning a structure into a volume. This is a computationally expensive procedure, reduced by the sperability of Gaussian kernels, see e.g \citep{chen_integrating_2023}.For an of size $\Npix\times \Npix $, it involves computing two times the distance of each residues to $\Npix$ pixels and taking the product of these vectors to obtain a matrix representing the images for each residue, then summing over the residues. This is one of the computational bottlenecks of the structural methods.
In spite of sharing the same bottleneck, cryoSPHERE tends to be slightly more computationally demanding. This is because cryoSPHERE needs to compose $\Ndomains$ rotations for each residue, while cryoStar only translates each residue. 
For example, for the experiments of Section \ref{sec:empiar12093}, in 24 hours, cryoStar performs roughly 190 epochs and cryoSPHERE with $\Ndomains=20$ performs roughly 130 epochs.
Similarly, for the experiment of Section \ref{empiar10180}, cryoStar performs 184 epochs while cryoSPHERE with $\Ndomains$ performs 95 epochs.
However, owing to the reduced number of freedom of cryoSPHERE compared to cryoStar, we observe that cryoSPHERE performs as well or better as cryoStar for the same computational budget.

%%%%%%%%%%%%%%%%%%%%%%%%%%%%%%%%%%%%%%%%%%%%%%%%%%%%%%%%%%%%%%%%%%%%%%%%%%%%%%%
%%%%%%%%%%%%%%%%%%%%%%%%%%%%%%%%%%%%%%%%%%%%%%%%%%%%%%%%%%%%%%%%%%%%%%%%%%%%%%%

\end{document}